\documentclass[
    a4paper,
    12pt,
    BCOR=0mm,
    listof=totoc,
    bibliography=totoc,
    numbers=noenddot
]{scrbook}

\ifcsname dm\endcsname\else
\usepackage[utf8]{inputenc} 
\usepackage[T1]{fontenc}    
\usepackage{textcomp}       
\usepackage{lmodern}        
\usepackage{slantsc}        
\usepackage{microtype}      
\usepackage{scrwfile}       
\usepackage{siunitx}        
\usepackage{amssymb}        
\usepackage{amsthm}         
\usepackage{mathtools}      
\usepackage{xcolor}         
\usepackage{xspace}         
\usepackage{caption}        
\usepackage{subcaption}     
\usepackage{float}          
\usepackage{booktabs}       
\usepackage{longtable}      
\usepackage{multirow}       
\usepackage{colortbl}       
\usepackage{enumitem}       
\usepackage{multicol}       
\usepackage{setspace}       
\usepackage{tikz}           
\usepackage{pgfplots}       
\usepackage{drawmatrix}     
\usepackage{tcolorbox}      
\usepackage{listings}       
\usepackage{lstautodedent}  
\usepackage{environ}        
\usepackage{xstring}        
\usepackage{etoolbox}       
\usepackage[
    urldate=comp,
    maxnames=99,
    abbreviate=false,
    defernumbers=true,
    sorting=nty,
]{biblatex}                 
\usepackage{mparhack}       
\usepackage{apptools}       
\usepackage{scrhack}        
\usepackage{hyperref}       
\usepackage[
    capitalize,
    nameinlink,
    noabbrev
]{cleveref}                 
\usepackage[
    pruefung
]{RWTH-Fak1}                

\selectcolormodel{cmyk}

\definecolor{rwthblau}{cmyk}{1,.5,0,0}
\colorlet{rwthblue}{rwthblau}
\definecolor{rwthblau75}{cmyk}{.75,.38,0,0}
\colorlet{rwthblue75}{rwthblau75}
\definecolor{rwthblau50}{cmyk}{.45,.14,0,0}
\colorlet{rwthblue50}{rwthblau50}
\definecolor{rwthblau25}{cmyk}{.23,.07,0,0}
\colorlet{rwthblue25}{rwthblau25}
\definecolor{rwthblau10}{cmyk}{.09,.03,0,0}
\colorlet{rwthblue10}{rwthblau10}
\colorlet{rwthlightblue}{rwthblue50}

\definecolor{rwthschwarz}{cmyk}{0,0,0,1}
\colorlet{rwthblack}{rwthschwarz}

\definecolor{rwthmagenta}{cmyk}{0,1,.25,0}

\definecolor{rwthgelb}{cmyk}{0,0,1,0}
\colorlet{rwthyellow}{rwthgelb}

\definecolor{rwthpetrol}{cmyk}{1,.3,.5,.3}

\definecolor{rwthtuerkis}{cmyk}{1,0,.4,0}
\colorlet{rwthcyan}{rwthtuerkis}

\definecolor{rwthgruen}{cmyk}{.7,0,1,0}
\colorlet{rwthgreen}{rwthgruen}

\definecolor{rwthmaigruen}{cmyk}{.35,0,1,0}
\colorlet{rwthyellowgreen}{rwthmaigruen}

\definecolor{rwthorange}{cmyk}{0,.4,1,0}

\definecolor{rwthrot}{cmyk}{.15,1,1,0}
\colorlet{rwthred}{rwthrot}

\definecolor{rwthbordeaux}{cmyk}{.25,1,.7,.2}
\colorlet{rwthdarkred}{rwthbordeaux}

\definecolor{rwthviolett}{cmyk}{.7,1,.35,.15}
\colorlet{rwthviolet}{rwthviolett}

\definecolor{rwthlila}{cmyk}{.6,.6,0,0}

\definecolor{dfglogoblue}{HTML}{2a3c9a}
\definecolor{telekommagenta}{HTML}{e2007e}

\colorlet{blue}{rwthblue}
\colorlet{lightblue}{rwthlightblue}
\colorlet{black}{rwthblack}
\colorlet{gray}{rwthblack!50}
\colorlet{lightgray}{rwthblack!25}
\colorlet{magenta}{rwthmagenta}
\colorlet{yellow}{rwthyellow}
\colorlet{petrol}{rwthpetrol}
\colorlet{cyan}{rwthcyan}
\colorlet{green}{rwthgreen}
\colorlet{yellowgreen}{rwthyellowgreen}
\colorlet{orange}{rwthorange}
\colorlet{red}{rwthred}
\colorlet{darkred}{rwthdarkred}
\colorlet{violet}{rwthviolet}
\colorlet{lila}{rwthlila}

\colorlet{matrixborder}{black!25}
\colorlet{graybgop}{black!15}
\colorlet{matrixbg}{graybgop}
\colorlet{graybg}{black!5}
\newcommand\graybgop{.3333}

\KOMAoptions{DIV=calc}

\deffootnote[1.3em]{1.3em}{1em}{\textsuperscript{\thefootnotemark} }
\allowdisplaybreaks
\let\paragraphold\paragraph
\renewcommand\paragraph[1]{\paragraphold{#1}\mbox{}\nopagebreak\par\noindent\ignorespaces}

\renewcommand\tt\ttfamily
\renewcommand\sf\sffamily
\renewcommand\it\itshape
\renewcommand\sc\scshape
\renewcommand\bf\bfseries
\renewcommand\rm\mathrm
\AtBeginDocument{%
    \DeclareFontShape{T1}{lmr}{m}{scit}{<->ssub*lmr/m/scsl}{}%
    \DeclareFontShape{T1}{lmr}{bx}{sc}{<->ssub*cmr/bx/sc}{}%
}
\setkomafont{disposition}{\color{blue}\bfseries\boldmath}
\setlist[itemize]{font=\color{blue}}
\setlist[description]{font=\usekomafont{disposition}}

3
3

\makeatletter\renewcommand\@pnumwidth{2em}\makeatother

\makeatletter
\renewcommand\fps@figure t
\renewcommand\fps@table t
\makeatother

\newcommand\subfigwidthratio{.4875}
\newlength\subfigwidth
\newcommand\figdimratio{1.91}
\newcommand\figwidthsub{1.6666cm}
\newcommand\plotwidthsub{0pt}
\newcommand\plotheightsub{0pt}
\newcommand\plotwidth{(\textwidth - \figwidthsub - \plotwidthsub)}
\newcommand\plotheight{(\plotwidth / \figdimratio - \plotheightsub)}
\newcommand\subplotheight{(
    (\textwidth / \subfigwidthratio - \figwidthsub) / \figdimratio -
    \plotheightsub
)}
\newcommand\subfigref[1]{\hyperref[#1]{(\subref*{#1})}}

\arrayrulecolor{blue}

\tcbuselibrary{skins, breakable, listings}
\tcbset{shield externalize}

\crefname{subappendix}%
    {\IfAppendix{Section}{Appendix}}%
    {\IfAppendix{Sections}{Appendices}}

\DeclareSortingScheme{ymdnt}{
    \sort{\field{presort}}
    \sort[final]{\field{sortkey}}
    \sort[direction=descending]{
        \field{sortyear}
        \field{year}
        \literal{9999}
    }
    \sort[direction=descending]{
        \field[padside=left,padwidth=2,padchar=0]{month}
        \literal{99}
    }
    \sort[direction=descending]{
        \field[padside=left,padwidth=2,padchar=0]{day}
        \literal{99}
    }
    \sort{
        \name{sortname}
        \name{author}
        \name{editor}
        \name{translator}
        \field{sorttitle}
        \field{title}
        \field{sorttitle}
    }
    \sort{
        \field[padside=left,padwidth=4,padchar=0]{volume}
        \literal{0000}
    }
}
\DeclareSourcemap{\maps[datatype=bibtex]{
    \map{
        \pernottype{online}
        \step[fieldset=editor, null]
        \step[fieldset=address, null]
        \step[fieldset=location, null]
    }
    \map{
        \pernottype{online}
        \step[fieldsource=doi, final]
        \step[fieldset=url, null]
        \step[fieldset=isbn, null]
        \step[fieldset=issn, null]
    }
    \map{
        \pernottype{online}
        \step[fieldsource=eprint, final]
        \step[fieldset=url, null]
        \step[fieldset=isbn, null]
        \step[fieldset=issn, null]
    }
    \map{
        \pertype{online}
        \step[fieldset=presort, fieldvalue=z]
        \step[fieldset=urldate, null]
    }
    \map[overwrite]{
        \perdatasource{publications.bib}
        \step[fieldset=keywords, fieldvalue={, own}, append]
        \step[fieldset=presort, fieldvalue=a]
    }
}}
\DeclareFieldFormat{eprint:github}{%
    GitHub\addcolon\space \href{http://github.com/#1}{\nolinkurl{#1}}%
}
\nocite{elaps, relem, nsdla, relapack, tensorpred, qrcaching, gwas2, caching,
pmrrr, gwas1, pred, msthesis}
\biburlsetup
\renewcommand\bibfont\small
\renewcommand\bibsetup{\singlespacing}

\SendSettingsToPgf
\DeclareSIUnit\Byte{byte}
\DeclareSIUnit\bytes{bytes}
\DeclareSIUnit\double{double}
\DeclareSIUnit\doubles{doubles}
\DeclareSIUnit\flop{FLOP}
\DeclareSIUnit\flops{FLOPs}
\DeclareSIUnit\cycle{cycle}
\DeclareSIUnit\cycles{cycles}
\DeclareSIUnit\core{core}
\DeclareSIUnit\cores{cores}

\lstdefinelanguage{pseudocode}{
    keywords=[1]{for,traverse},
    keywordstyle=[1]{\bf},
}
\hypersetup{
    breaklinks=true,
    colorlinks=true,
    linkcolor=blue,
    citecolor=green,
    urlcolor=green,
    pdfborder=0 0 0,
    pdfdisplaydoctitle=true,
}
\makeatletter\providecommand*\toclevel@tcolorbox{0}\makeatother

\usetikzlibrary{external, scopes, positioning, fadings, calc, arrows.meta}
\tikzset{
    line width=.5pt,
    external/export=true,
    external/mode=list and make,
    external/prefix=externalized/,
    external/optimize command away=\dm1,
    external/optimize command away=\cite1,
    external/optimize command away=\citeauthor1,
    external/optimize command away=\fullcite1,
    external/verbose IO=false,
    external/only named=true,
    brace/.style={
        decorate,
        decoration={brace, raise=#1},
        auto,
        every node/.append style={midway, outer sep=#1}
    },
    brace/.default=.5\smallskipamount,
    brace2/.style={brace=1.5\smallskipamount},
    bracei/.style={brace, decoration=mirror, swap},
    bracei2/.style={bracei, brace2},
    mat/.style={
        line join=round,
        draw=matrixborder,
        fill=matrixbg,
        fill opacity=\graybgop,
        text opacity=1,
        every node/.append style=midway
    },
    routinecommand/.style={
        baseline=(n.base),
        inner sep=0, outer sep=0,
        every node/.append style={
            inner sep=0,
            anchor=base,
            opacity=1
        }
    },
    hreful/.style={solid, line cap=round, opacity=.5},
    hwul/.style={hreful, red},
    blasul/.style={hreful, green},
    lapackul/.style={hreful, blue},
    mat00/.style={mat, fill=blue},
    mat01/.style={mat, fill=lila},
    mat02/.style={mat, fill=petrol},
    mat10/.style={mat, fill=green},
    mat11/.style={mat, fill=red},
    mat12/.style={mat, fill=yellow},
    mat20/.style={mat, fill=orange},
    mat21/.style={mat, fill=yellowgreen},
    mat22/.style={mat, fill=lightblue},
    qrmat11/.style={mat, fill=red},
    qrmat12/.style={mat, fill=green},
    qrmat21/.style={mat, fill=blue},
    qrmat22/.style={mat, fill=orange},
    qrmatW1/.style={mat, fill=yellowgreen},
    qrmatW2/.style={mat, fill=cyan},
    qrmattau/.style={mat, magenta},
    plot1/.style=blue,
    plot2/.style=red,
    plot3/.style=green,
    plot4/.style=orange,
    plot5/.style=lightblue,
    plot6/.style=yellowgreen,
    plot7/.style=cyan,
    plot8/.style=magenta,
    plot9/.style=yellow,
    plotibacc/.style=violet,
    plotsbopen/.style=lightblue,
    plothwopen/.style=blue,
    plotsbblis/.style=yellowgreen,
    plothwblis/.style=green,
    plotsbmkl/.style=orange,
    plothwmkl/.style=red,
    plotmin/.style=green,
    plotmed/.style=blue,
    plotmax/.style=red,
    plotavg/.style=lightblue,
    plotstd/.style=orange,
    plotstdf/.style={orange, fill=orange, fill opacity=.5},
}

\usepgfplotslibrary{units, fillbetween}
\pgfplotsset{
    compat=1.13,
    cell picture=if necessary,
    scale only axis,
    width=\plotwidth,
    height=\plotheight,
    xmin=0,
    ymin=0,
    xlabel near ticks,
    ylabel near ticks,
    enlarge x limits={value=.05, auto},
    enlarge y limits={value=.05, auto},
    try min ticks=4,
    every x tick label/.append style={inner xsep=0pt},
    every y tick label/.append style={inner ysep=0pt},
    axis background/.style={fill=graybgop, fill opacity=\graybgop},
    ymajorgrids=true,
    grid style=lightgray,
    legend cell align=left,
    legend columns=-1,
    legend style={/tikz/every even column/.append style={column sep=1ex}},
    every axis plot/.append style={thick, line join=round, line cap=round},
    after end axis/.append code={\node at (current axis.north) {\vphantom1};},
    /pgfplots/bar cycle list/.style={
        every axis plot/.append style={fill, fill opacity=.5, mark=none}
    },
    /pgfplots/ybar legend/.style={
        /pgfplots/legend image code/.code={%
            \draw[##1, /tikz/.cd, bar width=3pt, yshift=-.2666em, bar shift=0pt]
            plot coordinates {(0cm, 0.8em) (2*\pgfplotbarwidth, 0.6em)};%
        }
    },
    colormap={greenred}{color=(green!75!graybg) color=(red!75!graybg)},
    colormap={black}{color=(black) color=(black)},
    cycle list={plot1, plot2, plot3, plot4, plot5, plot6, plot7, plot8},
    cycle list/.define={blas}{
        plotsbopen, plotsbblis, plotsbmkl, plothwopen, plothwblis, plothwmkl
    },
    cycle list/.define={stat}{plotstd, plotavg, plotmax, plotmin, plotmed},
    cycle list/.define={statarea}{
        plotstd, plotstd, plotstdf,
        plotavg, plotmax, plotmin, plotmed
    },
    cycle list/.define={sbblas}{plotsbopen, plotsbblis, plotsbmkl},
    cycle list/.define={hwblas}{plothwopen, plothwblis, plothwmkl},
    perfplot/.style={
        ylabel=performance,
        y unit=\si{\giga\flops\per\second},
        after end axis/.append code={
            \draw[very thick, darkred]
                (current axis.north west) -- (current axis.north east);
        },
        ymax/.expanded=#1,
    },
    perfplot/.default=\pgfkeysvalueof{/pgfplots/ymax},
    0line/.style={
        execute at begin axis={
            \draw (\pgfkeysvalueof{/pgfplots/xmin}, 0) --
                  (\pgfkeysvalueof{/pgfplots/xmax}, 0);
        }
    },
    twocolplot/.style={height=\subplotheight},
    twocolhalfplot/.style={
        height=.5 * \subplotheight - 2.5pt,
        ylabel=\mstrut,
    },
    twocolhalfplot1/.style={
        twocolhalfplot,
        at={(0, 0)},
        anchor=south,
        xlabel={},
        xticklabels={},
    },
    twocolhalfplot2/.style={twocolhalfplot, at={(0, -5pt)}, anchor=north},
    threerowplot/.style={height=.6666 * \subplotheight},
    figwidthsub/.code={\renewcommand\figwidthsub{#1}},
    plotheightsub/.code={\renewcommand\plotheightsub{#1}},
    /pgfplots/warning/colorbar CMYK unsupported/.code={},
}
\drawmatrixset{
    exp/.style={exponent=#1},
    inv/.style={exp=-1},
    '/.style={exp=T},
    inv'/.style={exp=-T},
    scale=(\baselineskip - 2\pgflinewidth) / 1cm,
    mat
}

\title{\rmfamily\color{blue}Performance Modeling and Prediction\\for Dense Linear Algebra}
\author{\color{blue}Elmar Peise}
\hypersetup{
    pdftitle={Performance Modeling and Prediction for Dense Linear Algebra},
    pdfauthor={Elmar Peise},
}
\actualpregrade{}
\actualpostgrade{Master of Science}
\birthplace{Aachen, Deutschland.}
\dateoforalexamination{date of oral examination}
\expertI{Universitätsprofessor Paolo Bientinesi, Ph.D.}
\expertII{Universitätsprofessor Zweitgutachter}
\grade{eines Doktors der Naturwissenschaft}

\newcommand\namestyle{\sc}
\newcommand\hwstyle{\sc}
\newcommand\swstyle{\sc}
\newcommand\langstyle{\sc}
\newcommand\codestyle{\sf}
\newcommand\code[1]{\ifmmode\mathsf{#1}\else{\codestyle#1}\fi\xspace}
\newcommand\software[1]{{\swstyle#1}\xspace}

\newcommand\mstrut{\vphantom{\ensuremath(}}

\newcommand\currentchapterlabel{}
\newcommand\chapterlabel[1]{
    \label{ch:#1}\tikzsetexternalprefix{externalized/#1-}
    \renewcommand\currentchapterlabel{#1}
}

\newif\iftableoflist
\tableoflistfalse
\newcommand\newhardware[3]{
    \newcommand#1[1][]{%
        {\hwstyle#2 \tikz[routinecommand]\draw node (n)
            {\hypersetup{hidelinks}\hyperref[hardware:#3]{#3}}
            (0, -1.5pt -| n.west) edge[hwul] (0, -1.5pt -| n.east);%
        }\xspace%
    }
}
\newcommand\newhardwareshort[2]{\newcommand#1{{\hwstyle#2}\xspace}}

\newcommand\newsoftware[2]{\newcommand#1{{\swstyle#2}\xspace}}

\newcommand\newblasroutine[1]{
    \expandafter\providecommand\csname#1\endcsname{}
    \expandafter\renewcommand\csname#1\endcsname[1][]{%
        \tikz[routinecommand]\draw node (n) {%
            \codestyle\hypersetup{hidelinks}%
            \hyperref[routine:#1]{#1\textsubscript{##1}}%
        } (0, -1.5pt -| n.west) edge[blasul] (0, -1.5pt -| n.east);\xspace%
    }
}
\newcommand\newlapackroutine[1]{
    \expandafter\providecommand\csname#1\endcsname{}
    \expandafter\renewcommand\csname#1\endcsname[1][]{%
        \tikz[routinecommand]\draw node (n) {%
            \codestyle\hypersetup{hidelinks}%
            \hyperref[routine:#1]{#1\textsubscript{##1}}%
        } (0, -1.5pt -| n.west) edge[lapackul] (0, -1.5pt -| n.east);\xspace%
    }
}
\newcommand\newlapackroutinenum[1]{
    \expandafter\providecommand\csname#1\endcsname{}
    \expandafter\renewcommand\csname#1\endcsname[2][]{%
        \tikz[routinecommand]\draw node (n) {%
            \codestyle\hypersetup{hidelinks}%
            \hyperref[routine:#1##2]{#1##2\textsubscript{##1}}%
        } (0, -1.5pt -| n.west) edge[lapackul](0, -1.5pt -| n.east);\xspace%
    }
}

\newhardware\harpertown{Harpertown}{E5450}
\newhardware\sandybridge{Sandy Bridge-EP}{E5-2670}
\newhardware\ivybridge{Ivy Bridge-EP}{E5-2680 v2}
\newhardware\haswell{Haswell-EP}{E5-2680 v3}
\newhardware\broadwell{Broadwell}{i7-5557U}
\newhardwareshort\harpertownshort{Harpertown}
\newhardwareshort\sandybridgeshort{Sandy Bridge}
\newhardwareshort\ivybridgeshort{Ivy Bridge}
\newhardwareshort\haswellshort{Haswell}
\newhardwareshort\broadwellshort{Broadwell}

\newsoftware\openmp{OpenMP}
\newsoftware\blas{BLAS}
\newsoftware\blis{BLIS}
\newsoftware\lapack{LAPACK}
\newsoftware\scalapack{ScaLAPACK}
\newsoftware\mkl{MKL}
\newsoftware\openblas{OpenBLAS}
\newsoftware\accelerate{Accelerate}
\newsoftware\atlas{ATLAS}
\newsoftware\essl{ESSL}
\newsoftware\libflame{libFLAME}
\newsoftware\papi{PAPI}
\newsoftware\recsy{RECSY}
\newsoftware\elaps{ELAPS}
\newsoftware\relapack{ReLAPACK}

\newsoftware\sampler{Sampler}
\newsoftware\playmat{PlayMat}
\newsoftware\viewer{Viewer}

\newcommand\blasl[1]{\blas Level~#1\xspace}

\newcommand\rwth{{\namestyle RWTH}\xspace}
\newcommand\github{{\namestyle GitHub}\xspace}
\newcommand\intel{{\namestyle Intel}\xspace}
\newcommand\apple{{\namestyle Apple}\xspace}
\newcommand\turboboost{{\namestyle Turbo Boost}\xspace}

\newcommand\fortran{{\langstyle FORTRAN}\xspace}
\newcommand\clang{{\langstyle C}\xspace}
\newcommand\cpplang{{\langstyle C++}\xspace}
\newcommand\python{{\langstyle Python}\xspace}
\newcommand\matlab{{\langstyle MATLAB}\xspace}

\newblasroutine{dcopy}
\newblasroutine{dswap}
\newblasroutine{ddot}
\newblasroutine{daxpy}

\newblasroutine{dgemv}
\newblasroutine{dger}
\newblasroutine{dtrsv}

\newblasroutine{dgemm}
\newblasroutine{dsymm}
\newblasroutine{dtrmm}
\newblasroutine{strsm}
\newblasroutine{dtrsm}
\newblasroutine{ctrsm}
\newblasroutine{ztrsm}
\newblasroutine{ssyrk}
\newblasroutine{dsyrk}
\newblasroutine{cherk}
\newblasroutine{zherk}
\newcommand\dsyrtk[1][]{%
    \tikz[routinecommand]\draw node (n) {%
        \codestyle\hypersetup{hidelinks}%
        \hyperref[routine:dsyr2k]{dsyr2k\textsubscript{#1}}%
    } [lapackul] (0, -1.5pt -| n.west) -- (0, -1.5pt -| n.east);\xspace%
}

\newlapackroutine{ilaenv}
\newlapackroutine{dlaswp}
\newlapackroutine{dlarft}
\newlapackroutine{dlarfb}
\newlapackroutine{dlauum}
\newlapackroutine{dsygst}
\newlapackroutine{dtrtri}
\newlapackroutine{dpotrf}
\newlapackroutine{dgeqrf}
\newlapackroutine{dtrsyl}
\newlapackroutine{dgetrf}
\newlapackroutine{dgesv}
\newlapackroutine{dsyev}
\newlapackroutine{dsyevd}
\newlapackroutine{dsyevr}
\newlapackroutine{dsyevx}
\newlapackroutine{dposv}
\newlapackroutine{dpotrs}
\newlapackroutine{dtrsyl}

\newlapackroutinenum{dlauu}
\newlapackroutinenum{dsygs}
\newlapackroutinenum{dtrti}
\newlapackroutinenum{spotf}
\newlapackroutinenum{dpotf}
\newlapackroutinenum{cpotf}
\newlapackroutinenum{zpotf}
\newlapackroutinenum{dgetf}
\newlapackroutinenum{dgeqr}

\newcommand\figurestyle{\centering\small}
\newcommand\tablestyle{\centering\small}

\newcommand\mycaption[1]{
    \tableoflisttrue
    \caption[{\ignorespaces#1}]{\tableoflistfalse{\ignorespaces#1}}
    \tableoflistfalse
}
\newcommand\captiondetails[1]{%
    \unskip\iftableoflist\else\mstrut\\\scriptsize(\ignorespaces#1\unskip)\fi%
}

\newcommand\pluseqq{\mathrel{{+}{=}}}

\newcommand\defeqq{\stackrel{\mathrm{def}}{\coloneqq}}
\newcommand\R{\ensuremath{\mathbb R}}
\DeclareMathOperator*{\argmin}{arg\,min}

\newcommand\footnotetextbefore[1]{%
    \stepcounter{footnote}%
    \footnotetext{#1}%
    \addtocounter{footnote}{-1}%
}
\newtcolorbox[
    auto counter,
    list inside=loe,
    number within=chapter,
    crefname={Example}{Examples},
    Crefname={Example}{Examples},
]{example}[3][]{
    blanker,
    breakable,
    left=1.5em,
    borderline west={2pt}{0pt}{lightblue},
    parbox=false,
    before skip=.5\baselineskip,
    fonttitle=\usekomafont{disposition},
    adjusted title={Example~\thetcbcounter: #2},
    attach title to upper={%
        \setlength\parskip{0pt}%
        \setlength\parindent{11.74983pt}%
        \par\nobreak\noindent%
    },
    list text={#2},
    label=ex:#3,
    #1
}
\makeatletter
\pretocmd{\@chapter}{\addtocontents{loe}{\addvspace{10pt}}}{}{}
\makeatother

\newtcblisting{codelisting}{
    listing only,
    listing options={basicstyle=\sf\small},
    shield externalize,
    enhanced,
    breakable,
    left=0ex, right=0pt, top=0pt, bottom=0pt,
    colback=graybgop,
    opacityfill=0, opacityback=\graybgop,
    arc=2pt,
    borderline={.5pt}{1pt}{black},
    width=.9\textwidth,
    left skip=.1\textwidth,
    before skip=.25\baselineskip,
    before upper=\linespread1,
}

\newtcblisting{alglisting}[1][]{
    listing only,
    listing options={basicstyle=\sf\small},
    box align=center,
    shield externalize,
    enhanced,
    left=0ex, right=0pt, top=0pt, bottom=0pt,
    colback=graybgop,
    opacityfill=0, opacityback=\graybgop,
    arc=2pt,
    borderline={.5pt}{1pt}{black},
    nobeforeafter,
    before upper=\linespread1,
    #1
}

\pgfplotsset{every subfigaxis/.style={}}
\NewEnviron{subfigaxis}[1][]{%
    \pgfplotsset{
        fig caption code/.code={},
        fig label code/.code={},
        fig external code/.code={},
        fig caption/.style={fig caption code/.code=\caption{##1}},
        fig label/.style={
            fig external code/.code={
                \StrSubstitute{##1}:-[\subfigfilename]%
                \tikzsetnextfilename\subfigfilename%
            },
            fig label code/.code={\label{fig:\currentchapterlabel:##1}}
        },
        fig legend code/.code={},
        fig legend/.style={fig legend code/.code={%
            \centering\ref*{leg:##1}\par%
        }},
        fig width/.initial=\subfigwidth,
        fig align/.initial=\raggedleft,
        fig vertical align/.initial=c,
        every subfigaxis,%
        #1,
        fig width/.get=\subfigfigwidth,
        fig align/.get=\subfigalign,
        fig vertical align/.get=\subfigvertalign,
    }%
    \begin{subfigure}[\subfigvertalign]\subfigfigwidth%
        \pgfplotsset{fig legend code}%
        \subfigalign%
        \pgfplotsset{fig external code}%
        \def\begintikzpictureaxis{
            \begin{tikzpicture}
                \begin{axis}
        }%
        \expandafter\begintikzpictureaxis\BODY
            \end{axis}
        \end{tikzpicture}
        \pgfplotsset{fig caption code, fig label code}
    \end{subfigure}%
}

\newcommand\definition[2][]{%
    \def\marginlabel{#1}\ifx\marginlabel\empty\def\marginlabel{#2}\fi%
    \marginpar{\raggedright\small\em\textcolor{black}{\marginlabel}}%
    {\em#2}%
    \xspace%
}
\newcommand\definitionp[2][]{%
    \def\marginlabel{#1}\ifx\marginlabel\empty\def\marginlabel{#2}\fi%
    {\em#2}%
    \marginpar{\raggedright\small\em\textcolor{black}{\marginlabel}}%
    \xspace%
}
\renewcommand\definition[2][]{{\em#2}\xspace}
\let\definitionp\definition

\newenvironment{pubitemize}{%
    \bgroup%
        \par\singlespacing\small
        \setlist[itemize]{font=\color{black}}
        \newcommand\pubitem[1]{\item[\cite{##1}] \fullcite{##1}.}
        \begin{itemize}
}{%
        \end{itemize}%
        \onehalfspacing%
    \egroup%
}

\newcommand\savedm[2][]{%
    \def\dmid{#1.\string#2.\the\baselineskip}%
    \ifcsname\dmid\endcsname\else%
        \global\expandafter\newsavebox\csname\dmid\endcsname%
        \global\expandafter\sbox\csname\dmid\endcsname{%
            \ensuremath{\drawmatrix[#1]{#2}}%
        }%
    \fi%
    \expandafter\usebox\csname\dmid\endcsname%
}
\newcommand\dm[2][]{%
    \ensuremath{\mathchoice%
        {\savedm[#1, scale=1.25]{#2}}%
        {\savedm[#1]{#2}}%
        {{\color{red} script style drawmatrix}}
        {{\color{red} scriptscript style drawmatrix}}
    }%
}
\newcommand\dmx[2][]{%
    \ensuremath{\mathchoice%
        {\drawmatrix[#1, scale=1.25]{#2}}%
        {\drawmatrix[#1]{#2}}%
        {{\color{red} script style drawmatrix}}
        {{\color{red} scriptscript style drawmatrix}}
    }%
}
\newcommand\dv[2][]{\dm[width=0, #1]{#2}}
\newcommand\dmstack[2]{%
    \begin{tikzpicture}[
            baseline=(x.base),
            every node/.append style={inner sep=0}
        ]
        \node (b) {#2};
        \node[above=-1pt] (t) at (b.north) {#1};
        \path (b.south) -- (t.north) node[midway] (x) {\mstrut};
    \end{tikzpicture}%
}
\newcommand\matmatsep\;
\newcommand\matvecsep\,
\newcommand\lowerpostsep{\!\xspace}

\newcommand\SIvar[3][]{\SI[parse-numbers=false, number-math-rm=\ensuremath, #1]{#2}{#3}\xspace}

\newlength\capheight
\newcommand\tarrow[2]{%
    \settoheight\capheight{A}%
    \tikz[baseline=(x.base), rotate=#1, scale=#2]
        \draw[-stealth, line cap=round] (0, 0) -- (0, \capheight)
            node[midway, inner sep=0] (x) {\phantom{$\downarrow$}};%
    \xspace%
}
\newcommand\tnarrow{\tarrow0{1.25}}
\newcommand\tnwarrow{\tarrow{45}{1.75}}

\newcommand\tsarrow{\tarrow{180}{1.25}}
\newcommand\tsearrow{\tarrow{225}{1.75}}
\newcommand\tearrow{\tarrow{270}{1.25}}
\newcommand\tnearrow{\tarrow{315}{1.75}}

\newcommand\legendline[1][]{%
    \tikz[/pgfplots/every crossref picture, /pgfplots/legend image
    code={/pgfplots/every axis plot, #1}];%
}

\newcommand\overarg[2]{%
    \tikz[baseline=(x.base)]{
        \node[inner sep=0] (x) {#2\mstrut};
        \path (x.north) node[inner sep=0, above=2pt, anchor=south]
        {\scriptsize\color{lightblue}#1\mstrut};
    }%
}

\newcommand\displaycall[2]{
    \renewcommand\arg[2]{%
        \tikz[baseline=(x.base)]{
            \node[inner sep=0] (x) {##2\mstrut};
            \path (x.north) node[inner sep=0, above=2pt, anchor=south]
            {\scriptsize\color{lightblue}##1\mstrut};
        }%
    }
    \newcommand\varg[2]{\arg{##1}{\it\color{blue}##2}}
    \[\text{#1{}\code{(\ignorespaces#2\unskip)}}\enspace,\]
}

\newcommand\tensoralgname[2]{\mbox{$#1$-\csname d#2\endcsname}}

\newenvironment{drawcube}[1][(0, 0, 0)]{
    \begin{scope}[
        shift={#1},
        inner/.style={
            red,
            thick,
            line join=round,
            line cap=round,
            fill=red,
            fill opacity=.2
        },
        outer/.style={blue, line join=round, fill=blue, fill opacity=.1},
        scale=.5
    ]
        \newcommand\slicenone[1][inner]{\draw[##1]
            +(-1, -1, 1) rectangle +(1, 1, 1) -- +(1, 1, -1)
            +(-1, 1, 1) -- +(-1, 1, -1) -- +(1, 1, -1)
            -- +(1, -1, -1) -- (1, -1, 1);
        }
        \newcommand\sliceA[1][inner]{\draw[##1]
            +(-1,  0, -1) -- +(1, 0, -1) -- +(1, 0, 1) -- +(-1,  0, 1) -- cycle;
        }
        \newcommand\sliceB[1][inner]{\draw[##1]
            +( 0, -1, -1) -- +(0, 1, -1) -- +(0, 1, 1) -- +( 0, -1, 1) -- cycle;
        }
        \newcommand\sliceC[1][inner]{\draw[##1]
            +(-1, -1, 0) rectangle +(1, 1, 0);
        }
        \newcommand\sliceAB[1][inner]{\draw[##1]
            +( 0,  0, -1) -- +(0, 0, 1);
        }
        \newcommand\sliceAC[1][inner]{\draw[##1]
            +(-1,  0,  0) -- +(1, 0, 0);
        }
        \newcommand\sliceBC[1][inner]{\draw[##1]
            +( 0, -1,  0) -- +(0, 1, 0);
        }
        \newcommand\sliceABC[1][inner]{\draw[##1, ultra thick]
            +(0, 0, 0) -- +(0, 0, 0);
        }
        \filldraw[mat]
            +(-1, -1,  1) -- +( 1, -1,  1) -- +( 1, -1, -1) --
            +( 1,  1, -1) -- +(-1,  1, -1) -- +(-1,  1,  1) -- cycle;
        \draw[matrixborder]
            +(-1, -1, -1) -- +( 1, -1, -1)
            +(-1, -1, -1) -- +(-1,  1, -1)
            +(-1, -1, -1) -- +(-1, -1,  1);
}{
        \draw[matrixborder]
            +(1, 1, 1) -- +(-1,  1,  1)
            +(1, 1, 1) -- +( 1, -1,  1)
            +(1, 1, 1) -- +( 1,  1, -1);
        \slicenone[draw=matrixborder]
    \end{scope}
}

\newenvironment{drawsquare}[1][(0, 0)]{
    \begin{scope}[
        shift={#1},
        inner/.style={
            red,
            thick,
            line join=round,
            line cap=round,
            fill=red,
            fill opacity=.2
        },
        outer/.style={blue, line join=round},
        scale=.5
    ]
        \newcommand\slicenone[1][inner]{
            \draw[##1] +(-1, -1) rectangle +(1, 1);
        }
        \newcommand\sliceA[1][inner]{
            \draw[##1] +(-1,  0) -- +(1, 0);
        }
        \newcommand\sliceB[1][inner]{
            \draw[##1] +( 0, -1) -- +(0, 1);
        }
        \newcommand\sliceAB[1][inner]{
            \draw[##1, ultra thick] +(0, 0) -- +(0, 0);
        }
        \filldraw[mat]
            +(-1, -1) rectangle +(1, 1);
}{
    \end{scope}
}

\fi
\tikzifexternalizing{\renewenvironment{alglisting}[1][]{}{}}{}
\tikzifexternalizing{\renewenvironment{example}[2]{}{}}{}

\begin{document}
    \frontmatter

    \maketitle

\chapter*{Abstract}

This dissertation introduces measurement-based performance modeling and
prediction techniques for dense linear algebra algorithms.  As a core
principle, these techniques avoid executions of such algorithms entirely, and
instead predict their performance through runtime estimates for the underlying
compute kernels.  For a variety of operations, these predictions allow to
quickly select the fastest algorithm configurations from available alternatives.
We consider two scenarios that cover a wide range of computations:

To predict the performance of blocked algorithms, we design
algorithm-independent performance models for kernel operations that are
generated automatically once per platform.  For various matrix operations,
instantaneous predictions based on such models both accurately identify the
fastest algorithm, and select a near-optimal block size.

For performance predictions of BLAS-based tensor contractions, we propose
cache-aware micro-benchmarks that take advantage of the highly regular structure
inherent to contraction algorithms.  At merely a fraction of a contraction's
runtime, predictions based on such micro-benchmarks identify the fastest
combination of tensor traversal and compute kernel.

\chapter*{Acknowledgments}

First and foremost, I would like to express my sincere gratitude to my advisor
Paolo Bientinesi.  While guiding me through my studies, he always embraced my
own ideas and helped me shape and develop them in countless discussions.  While
he granted me freedom in many aspects of my work, he always had time for
anything between a quick exchange of thoughts and extensive brainstorming
sessions.  Beyond our professional relationship, we enjoyed twisty puzzles and
board games in breaks from work, long game nights, and annual trips to SPIEL.  I
consider my self lucky to have spend my time as a doctoral student with him and
his research group.

The HPAC group proved to be much more than a collection of researchers working
on remotely associated projects; my colleagues were not only a source of
incredibly valuable discussions and feedback regarding my work, we also indulged
in various unrelated arguments and exchanges over lunch and at many other
occasions.  My thanks go to Edoardo Di~Napoli, Diego Fabregat-Traver, Paul
Springer, Jan Winkelmann, Henrik Barthels, Markus Höhnerbach, Sebastian
Achilles, William McDoniel, and Caterina Fenu, as well as our former group
members Matthias Petschow, Roman Iakymchuk, Daniel Tameling, and Lucas Beyer.

I am grateful for financial support from the {\namestyle Deutsche
Forschungs\-gemeinschaft} (DFG) through grant GSC~111 (the graduate school
AICES) and the {\namestyle Deutsche Telekomstiftung}.  Their programs not only
funded my work, but opened further opportunities in the form of seminars and
workshops, and connected me with like-minded students from various disciplines.

The {\namestyle\rwth IT Center} provided and maintained an extremely reliable
infrastructure central to my work: the {\namestyle \rwth Compute Cluster}.  I
thank its staff not only for ensuring smooth operations but also for their
competent and detailed responses to my many inquiries and requests regarding our
institute's cluster partition.

The AICES service team did their best to shield me from the bureaucracy of
contracts, stipends, and reimbursements.  I am grateful they allowed me to focus
solely on my research.

Even more important than a gratifying work environment is forgetting about it
every once in a while.  My friends played a bigger role in this effort than
probably most of them know, whether we were simply spending time hanging out or
playing games, went swimming, climbing or playing badminton, or taught swimming
and worked as lifeguards.  You are too many to enumerate, but you know who you
are.

Finally, but most importantly, none of this would have been possible without the
endless and uncompromising support of may parents.  You are the reason I grew
into the person I am today.  Danke!

\tableofcontents

    \mainmatter

    \chapter{Introduction}
\chapterlabel{intro}
{
    \tikzsetexternalprefix{externalized/intro-}

    Software developers in scientific computing are often faced with
performance-critical decisions such as the choice of algorithms, configuration
parameters, hardware platforms, and software libraries.  This dissertation
presents novel techniques and tools to guide such decisions for dense linear
algebra computations with accurate yet fast performance predictions.  These
predictions avoid the otherwise common exhaustive execution and timing of all
potential alternatives, and thereby shorten the decision-making process both in
compute time and developer effort.

The task of accurately predicting the performance of dense linear algebra
algorithms is particularly challenging due to the complexity of the
performance-related factors:  The runtime of compute-kernels is not only
non-linear in the problem size due to multi-threading and kernel-internal
caching effects, but is also influenced by data locality and caching in
sequences of such kernels.  As a result, analytical performance predictions are
either extremely rough and complex, or hardware-dependent; in contrast, this
work investigates measurement-based techniques that are tailored to represent
the kernel-specific performance effects.

The goal of measurement-based predictions is to estimate the performance of an
algorithm both accurately and notably faster than the algorithm execution
itself.  These requirements lead to two practical alternatives as the basis for
performance predictions: an algorithm-independent database of performance
models for the building blocks that are automatically generated once per
platform, or micro-benchmarks that execute a fraction of the algorithm's
building blocks and extrapolate their runtime.  Neither of these alternatives is
applicable in all situations, and which one is more suitable depends on the type
algorithm.  By addressing two different types of operations that are at the core
of many dense computations, this work investigates both alternatives:  Blocked
algorithms are predicted through algorithm-independent performance models, and
tensor contraction algorithms are predicted through cache-aware
micro-benchmarks.

\paragraph{Contributions}
The main contributions of this work are the following:
\begin{itemize}
    \item \elaps, a lightweight yet portable and universal performance
        measurement framework for dense linear algebra routines and algorithms,
    \item Methods and tools for the automated generation of highly accurate
        performance models for compute kernels,
    \item Model-based performance predictions of blocked algorithms for optimal
        algorithm selection and configuration,
    \item A study on the influence of caching on kernel invocations within
        blocked algorithms, and
    \item Cache-aware micro-benchmarks to predict \blas-based tensor
        contractions for optimal algorithm selection.
\end{itemize}

\paragraph{Outline}
The remainder of this dissertation is structured as follows:
\begin{itemize}
    \item \Cref{ch:intro} proceeds to introduce blocked algorithms and tensor
        contractions, and motivates our performance prediction goals in
        \Cref{sec:intro:blocked,sec:intro:tensor}.  It concludes with an
        overview of related work in \Cref{sec:intro:relwork}.

    \item \Cref{ch:meas} addresses common performance characteristics of compute
        kernels, and introduces \elaps, a novel framework for performance
        measurements that serves as the basis for the following Chapters.

    \item \Cref{ch:model} presents the design and automatic generation of
        performance models, and analyzes their accuracy.

    \item \Cref{ch:pred} predicts the runtime and performance of blocked
        algorithms based on such models, and uses the predictions to select
        platform-specific optimal algorithm configurations.

    \item \cref{ch:cache} studies the influence of caching on the runtime of
        compute kernels within blocked algorithms and the feasibility of
        integrating caching effects into predictions.

    \item \Cref{ch:tensor} is devoted to the prediction of \blas-based tensor
        contractions.  It describes the creation of cache-aware
        micro-benchmarks that, for a given contraction, allow to identify the
        fastest algorithm(s).

    \item \Cref{ch:conclusion} concludes this dissertation, summarizes the
        presented techniques and results, and gives an overview of potential
        extensions of this work.
\end{itemize}

The main chapters are supplemented by three appendices:
\begin{itemize}
    \item \Cref{app:term} introduces readers new to high-performance computing
        to performance-related terminology and concepts.

    \item \Cref{app:libs} gives an overview of the \blas and \lapack interfaces,
        their kernels used in this work, and relevant implementations.

    \item \Cref{app:hardware} details the hardware used throughout this work.
\end{itemize}

    \section[Performance Modeling for Blocked Algorithms]
            {Performance Modeling\newline for Blocked Algorithms}
    \label{sec:intro:blocked}
    We aim to predict the performance of blocked algorithms with the goals of
1)~selecting the fastest algorithm from a set of mathematically equivalent
alternatives, and 2)~tuning their algorithmic block size.  In the following,
\cref{sec:intro:blocked:algs} introduces the concept of blocked algorithms, and
exposes their inherent optimization challenges, and
\cref{sec:intro:blocked:pred} gives a brief overview of our approach to address
these challenges using on performance models.

Readers familiar with blocked algorithms and the influence of block sizes may
skip the introduction to these concepts in \cref{sec:intro:blocked:algs}, and
focus on our prediction approach in \cref{sec:intro:blocked:pred} on
\cpageref{sec:intro:blocked:pred}.

\subsection{Motivation: Blocked Algorithms}
\label{sec:intro:blocked:algs}

\definitionp[blocked algorithm]{Blocked algorithms} are commonly used to exploit
the performance of optimized \blasl3 kernels\footnote{%
    The {\namestyle Basic Linear Algebra Subprograms} (BLAS) form the basis for
    high-performance in dense linear algebra.  See \cref{app:term,app:libs}.
} in other matrix operations, such as decompositions, inversions, and
reductions.  Every blocked algorithm traverses its input matrix (or matrices) in
steps of a fixed \definition{block size}; in each step of this traversal, it
exposes a set of \definition[sub-matrices\\updates]{sub-matrices} to which it
applies a series of {\em updates}.  Through these updates, it progresses with
the computation and obtains a portion of the operation's result; once the matrix
traversal completes, the entire result is computed.

\begin{figure}[p]\figurestyle
    \begin{subfigure}\subfigwidth\centering
        \begin{tikzpicture}[y={(0, -1)}, scale=.78]
            \def\s{5.5}
            { [shift={(0, -0)}]
                \def\p{0} \def\q{1}
                \filldraw[mat  ] (0,  0 ) rectangle (\s,     \s    );
                \filldraw[mat11] (\p, \p) --        (\q-.05, \q-.05) node {$A_{11}$} -| cycle;
                \filldraw[mat21] (\p, \q) rectangle (\q-.05, \s    ) node {$A_{21}$};
                \filldraw[mat22] (\q, \q) --        (\s,     \s    ) node {$A_{22}$} -| cycle;
                \draw[brace ] (\p,     \p) -- (\q-.05, \p    ) node {$b$}; 
                \draw[brace ] (\q-.05, \p) -- (\q-.05, \q-.05) node {$b$}; 
                \draw[bracei] (0,      \s) -- (\s,     \s    ) node {$n$}; 
                \draw[bracei] (0,      0 ) -- (0,      \s    ) node {$n$}; 
            }
            \draw[-stealth, very thick] (2.75, 6.25) -- ++(0, 1);
            { [shift={(0, 7.5)}]
                \def\p{1} \def\q{2}
                \filldraw[mat  ] (0,  0 ) rectangle (\s,     \s    );
                \filldraw[mat00] (0,  0 ) --        (\p-.05, \p-.05) node {$A_{00}$} -| cycle;
                \filldraw[mat10] (0,  \p) rectangle (\p-.05, \q-.05) node {$A_{10}$};
                \filldraw[mat11] (\p, \p) --        (\q-.05, \q-.05) node {$A_{11}$} -| cycle;
                \filldraw[mat20] (0,  \q) rectangle (\p-.05, \s    ) node {$A_{20}$};
                \filldraw[mat21] (\p, \q) rectangle (\q-.05, \s    ) node {$A_{21}$};
                \filldraw[mat22] (\q, \q) --        (\s,     \s    ) node {$A_{22}$} -| cycle;
                \draw[brace ] (0,      0 ) -- (\p-.05,  0    ) node {$b$}; 
                \draw[brace ] (\p-.05, 0 ) -- (\p-.05, \p-.05) node {$b$}; 
                \draw[brace ] (\p,     \p) -- (\q-.05, \p    ) node {$b$}; 
                \draw[brace ] (\q-.05, \p) -- (\q-.05, \q-.05) node {$b$}; 
                \draw[bracei] (0,      \s) -- (\s,     \s    ) node {$n$}; 
                \draw[bracei] (0,      0 ) -- (0,      \s    ) node {$n$}; 
            }
            \draw[-stealth, very thick] (2.75, 13.75) -- ++(0, 1);
            { [shift={(0, 15.25)}]
                \def\p{2} \def\q{3}
                \filldraw[mat  ] (0,  0 ) rectangle (\s,     \s    );
                \filldraw[mat00] (0,  0 ) --        (\p-.05, \p-.05) node {$A_{00}$} -| cycle;
                \filldraw[mat10] (0,  \p) rectangle (\p-.05, \q-.05) node {$A_{10}$};
                \filldraw[mat11] (\p, \p) --        (\q-.05, \q-.05) node {$A_{11}$} -| cycle;
                \filldraw[mat20] (0,  \q) rectangle (\p-.05, \s    ) node {$A_{20}$};
                \filldraw[mat21] (\p, \q) rectangle (\q-.05, \s    ) node {$A_{21}$};
                \filldraw[mat22] (\q, \q) --        (\s,     \s    ) node {$A_{22}$} -| cycle;
                \draw[brace ] (0,      0 ) -- (\p-.05,  0    ) node {$2b$}; 
                \draw[brace ] (\p-.05, 0 ) -- (\p-.05, \p-.05) node {$2b$}; 
                \draw[brace ] (\p,     \p) -- (\q-.05, \p    ) node {$b$}; 
                \draw[brace ] (\q-.05, \p) -- (\q-.05, \q-.05) node {$b$}; 
                \draw[bracei] (0,      \s) -- (\s,     \s    ) node {$n$}; 
                \draw[bracei] (0,      0 ) -- (0,      \s    ) node {$n$}; 
            }
            \path (2.75, 21.5) -- +(0, 0) node {$\large\mathbf\vdots$};
        \end{tikzpicture}

        \caption{Blocked matrix traversal}
        \label{algs:chol:traversal}
    \end{subfigure}\hfill
    \begin{minipage}\subfigwidth
        \makeatletter\let\tikz@ensure@dollar@catcode=\relax\makeatother
        \newcommand\AzzTi{\dm[mat00, upper, inv']{A_{00}}}
        \newcommand\Aoz{\dm[mat10, height=.5]{A_{10}}}
        \newcommand\AozT{\dm[mat10, width=.5, ']{A_{10}}}
        \newcommand\Aoo{\dm[mat11, size=.5, lower]{A_{11}}}
        \newcommand\AooT{\dm[mat11, size=.5, upper, ']{A_{11}}}
        \newcommand\AooTi{\dm[mat11, size=.5, upper, inv']{A_{11}}}
        \newcommand\Aoofull{\dm[mat11, size=.5]{A_{11}}}
        \newcommand\Atz{\dm[mat20, height=1.25]{A_{20}}}
        \newcommand\Ato{\dm[mat21, width=.5, height=1.25]{A_{21}}}
        \newcommand\AtoT{\dm[mat21, width=1.25, height=.5, ']{A_{21}}}
        \newcommand\Att{\dm[mat22, size=1.25, lower]{A_{22}}}

        \newcommand\traversal{%
            \begin{tikzpicture}[
                    baseline=(x.base),
                    y={(0, -1)}, scale=.1
                ]
                \def\s{5.5} \def\p{2} \def\q{3}
                \filldraw[mat  ] (0,  0 ) rectangle (\s, \s);
                \filldraw[mat00] (0,  0 ) --        (\p, \p) -| cycle;
                \filldraw[mat10] (0,  \p) rectangle (\p, \q);
                \filldraw[mat11] (\p, \p) --        (\q, \q) -| cycle;
                \filldraw[mat20] (0,  \q) rectangle (\p, \s);
                \filldraw[mat21] (\p, \q) rectangle (\q, \s);
                \filldraw[mat22] (\q, \q) --        (\s, \s) -| cycle;
                \path (0, 0) -- (\s, \s) node[midway, black] (x) {$A$};
            \end{tikzpicture}%
        }

        \begin{subfigure}\textwidth
            \begin{alglisting}[]
                traverse $\traversal$ along $\tsearrow$:
                  !\dtrsm[RLTN]:! $\Aoz \coloneqq \Aoz \matmatsep \AzzTi$
                  !\dsyrk[LN]:! $\Aoo \coloneqq \Aoo - \Aoz \matmatsep \AozT$
                  !\dpotf[LN]2:! $\Aoo \AooT \coloneqq \Aoofull$
            \end{alglisting}
            \caption{Algorithm 1}
            \label{alg:chol1}
        \end{subfigure}

        \bigskip

        \begin{subfigure}\textwidth
            \begin{alglisting}[]
                traverse $\traversal$ along $\tsearrow$:
                  !\dsyrk[LN]:! $\Aoo \coloneqq \Aoo - \Aoz \matmatsep \AozT$
                  !\dpotf[LN]2:! $\Aoo \AooT \coloneqq \Aoofull$
                  !\dgemm[NT]:! $\Ato \coloneqq \Ato - \Atz \matmatsep \AozT$
                  !\dtrsm[RLTN]:! $\Ato \coloneqq \Ato \matmatsep \AooTi$
            \end{alglisting}
            \caption{Algorithm 2}
            \label{alg:chol2}
        \end{subfigure}

        \bigskip

        \begin{subfigure}\textwidth
            \begin{alglisting}[]
                traverse $\traversal$ along $\tsearrow$:
                  !\dpotf[LN]2:! $\Aoo \AooT \coloneqq \Aoofull$
                  !\dtrsm[RLTN]:! $\Ato \coloneqq \Ato \matmatsep \AooTi$
                  !\dsyrk[LN]:! $\Att \coloneqq \Att - \Ato \matmatsep \AtoT$
            \end{alglisting}
            \caption{Algorithm 3}
            \label{alg:chol3}
        \end{subfigure}
    \end{minipage}

    \caption{%
        Blocked algorithms for the lower-triangular Cholesky decomposition.
    }
    \label{algs:chol}
\end{figure}

\footnotetextbefore{%
    \Cref{app:libs} gives an overview of the \blas and \lapack routines used
    throughout this work.  When specified, the subscripts indicate the values of
    the flag arguments, which identify the variant of the operation; e.g., in
    \dpotrf[L] the \code L corresponds to the argument \code{uplo} indicating
    a lower-triangular decomposition.
}
\begin{example}{Blocked algorithms for the Cholesky decomposition}{intro:chol}
    \newcommand\Azz{\dm[mat00, lower]{A_{00}}\xspace}%
    \newcommand\Aoz{\dm[mat10, height=.5]{A_{10}}\xspace}%
    \newcommand\Aoo{\dm[mat11, size=.5, lower]{A_{11}}\xspace}%
    \newcommand\Atz{\dm[mat20, height=1.25]{A_{20}}\xspace}%
    \Cref{algs:chol} illustrates blocked algorithms for a simple yet
    representative operation: the lower-triangular Cholesky decomposition
    \[
        \dm[lower]L \dm[upper, ']L \coloneqq \dm A
    \]
    of a symmetric positive definite (SPD) matrix $\dm A \in \R^{n \times n}$ in
    lower-triangular storage (\lapack: \dpotrf[L]\footnotemark).  For this
    operation there exist three different blocked algorithms.  Each algorithm
    traverses \dm A diagonally from the top-left to the bottom-right \tsearrow
    and computes the Cholesky factor~\dm[lower]L in place.  At each step of the
    traversal, the algorithm exposes the sub-matrices shown in
    \cref{algs:chol:traversal} and makes progress by applying the
    algorithm-dependent updates in \cref{alg:chol1,alg:chol2,alg:chol3}.  Before
    these updates, the sub-matrix~\Azz, which in the first step is of size $0
    \times 0$, already contains a portion of the Cholesky factor~\dm[lower]L;
    after the updates, the sub-matrices~\Aoz and~\Aoo also contain their
    portions of~\dm[lower]L, and in the next step become part of~\Azz.  Once the
    traversal reaches the bottom-right corner (i.e., \Azz is now of size $n
    \times n$), the entire matrix is factorized.
\end{example}

Blocked algorithms pose two \definition[optimization challenges:\\alternative
algorithms]{optimization challenges}:
\begin{itemize}
    \item For each operation there typically exist several {\em alternative
        algorithms}, which are mathematically equivalent in exact arithmetic;
        however, even if such algorithms perform the same number of floating
        point operations, they may differ significantly in performance.

    \item For each algorithm, the \definition{block size} influences the number
        of traversal steps and the sizes and shapes of the exposed sub-matrices,
        and thus the performance of the kernels applied to them.
\end{itemize}
What makes matters more complicated is that the optimal choice depends on
various factors, such as the hardware , the number of threads, the kernel
implementations, and the problem size.

\begin{figure}\figurestyle
    \ref*{leg:chols}

    \pgfplotsset{
        twocolplot, perfplot=52.8,
        xlabel={problem size $n$},
        table/search path={pred/figures/data/varchol},
    }

    \begin{subfigaxis}[
            fig caption=1~thread,
            fig label=chol:vars:1,
            legend to name=leg:chols,
        ]
        \addlegendimage{empty legend}\addlegendentry{Algorithm:}
        \foreach \var in {1, 2, 3} {
            \addplot table[y=meas] {chol\var.Haswell.1.OpenBLAS/perf.dat};
            \addlegendentryexpanded{\var}\label{plt:chol\var}
        }
    \end{subfigaxis}\hfill
    \begin{subfigaxis}[
            fig caption=12~threads,
            fig label=chol:vars:12,
            ymax=480
        ]
        \foreach \var in {1, 2, 3}
            \addplot table[y=meas] {chol\var.Haswell.12.OpenBLAS/perf.dat};
    \end{subfigaxis}

    \mycaption{%
        Performance of the three blocked Cholesky decomposition algorithms.
        \captiondetails{$b = 128$, \haswell, \openblas, median of 10~repetitions}
    }
    \label{fig:intro:chol:vars}
\end{figure}

\footnotetextbefore{%
    \Cref{app:hardware} provides an overview of the processors used throughout
    this work.
}
\begin{example}{Performance of alternative algorithms}{intro:chol:var}
    \Cref{fig:intro:chol:vars} shows the performance of the three blocked
    Cholesky decompositions from \cref{algs:chol} with block size~$b = 128$ and
    increasing problem size~$n$ on a 12-core \haswell\footnotemark{} with
    single- and multi-threaded \openblas.

    In both the single- and multi-threaded scenarios,
    algorithm~3~(\ref*{plt:chol3}) is the fastest among the three alternatives
    for all problem sizes.  On a single core and for problem size $n = 4152$, it
    is \SIlist{27.40;12.89}{\percent} faster than, respectively,
    algorithms~1~(\ref*{plt:chol1}) and~2~(\ref*{plt:chol2}), and it reaches up
    to \SI{91.01}{\percent} of the processor's theoretical peak performance (red
    line \legendline[very thick, darkred] at the top of the plot).  On all 12~of
    the processor's cores, algorithm~3~(\ref*{plt:chol3}) still reaches an
    efficiency of~\SI{69.70}\percent, and outperforms
    algorithms~1~(\ref*{plt:chol1}) and~2~(\ref*{plt:chol2}) by, respectively,
    $5.21\times$ and~$1.92\times$.

    Although algorithm~3~(\ref*{plt:chol3}) is clearly the fastest in this and
    many other scenarios, \lapack's \dpotrf[L] implements
    algorithm~2~(\ref*{plt:chol2}).

    For other operations, the choice becomes more complicated, since no single
    algorithm is the fastest for all problem sizes and scenarios.  For instance,
    for the single-threaded inversion of a lower-triangular matrix $\dm[lower]A
    \coloneqq \dm[lower, inv]A$, two different algorithms are the fastest for
    small and large matrices; with the performance differing by up
    to~\SI{13}{\percent} in either direction (\cref{sec:pred:var:trinv}).
\end{example}

\begin{figure}\figurestyle
    \ref*{leg:b}

    \pgfplotsset{
        every axis/.append style={
            twocolplot,
            xlabel={block size $b$},
            restrict x to domain=0:350,
        }
    }

    \begin{subfigaxis}[
            fig caption=1~thread,
            fig label=chol:b:1,
            perfplot=52.8,
            table/search path={pred/figures/data/bchol/chol3.Haswell.1.OpenBLAS/},
            legend to name=leg:b,
        ]
        \foreach \n/\y/\x in {
            1/34.002049227/96,
            2/42.1764395425/144,
            3/45.9288817014/160,
            4/48.1008517845/184
        } {\edef\temp{
            \noexpand\draw[plot\n, dashed] (\x, 0) -- (\x, \y);
            \noexpand\filldraw[plot\n] (\x, \y) circle (2pt);
        }\temp}
        \foreach \n in {1000, 2000, 3000, 4000} {
            \addplot table[y=meas] {perf\n.dat};
            \addlegendentryexpanded{$n = \n$}\label{plt:n\n}
        }
    \end{subfigaxis}\hfill
    \begin{subfigaxis}[
            fig caption=12~threads,
            fig label=chol:b:8,
            perfplot=480,
            table/search path={pred/figures/data/bchol/chol3.Haswell.12.OpenBLAS/},
        ]
        \foreach \n/\y/\x in {
            1/166.859461585/56,
            2/255.299601144/88,
            3/318.889592564/112,
            4/337.446195653/112
        } {\edef\temp{
            \noexpand\draw[plot\n, dashed] (\x, 0) -- (\x, \y);
            \noexpand\filldraw[plot\n] (\x, \y) circle (2pt);
        }\temp}
        \foreach \n in {1000, 2000, 3000, 4000}
            \addplot table[y=meas] {perf\n.dat};
    \end{subfigaxis}

    \mycaption{%
        Performance of the blocked Cholesky decompositions algorithm~3 for
        varying block sizes.
        \captiondetails{\haswell, \openblas, median of 10~repetitions}
    }
    \label{fig:intro:chol:b}
\end{figure}

\begin{example}{Influence of the block size on performance}{intro:chol:b}
    Let us consider the blocked Cholesky decomposition
    algorithm~3~(\ref*{plt:chol3} in \cref{fig:intro:chol:vars}) with fixed
    problem sizes~$n = 1000$, 2000, 3000, and~4000 and varying block size~$b$.
    \cref{fig:intro:chol:b} presents the performance of these algorithm
    executions for 1 and 12~threads on the \haswell using \openblas:
    Single-threaded, the optimal block size increases from~$b = 96$ for~$n =
    1000$ to~$b = 184$ for~$n = 4000$.  On 12~cores, on the other hand, the
    performance is less smooth and the optimal choices for~$b$ are between~56
    and~112.

    \Cref{fig:intro:chol:b} demonstrates the importance of selecting the block
    size dynamically:  If we use~$b = 184$, which is optimal for~$n = 4000$ on
    one core, for~$n = 1000$ on 12~cores we only reach \SI{77.62}{\percent} of
    the algorithm's optimal performance.  On the other hand, \lapack's default
    block size~$b = 64$ (which is close to the optimal~$b = 56$ for~$n = 1000$
    on 12~cores) would reach \SI{95.95}{\percent} of the optimal single-threaded
    performance for~$n = 4000$.
\end{example}

\subsection{Prediction through Performance Models}
\label{sec:intro:blocked:pred}

Naturally, both the best algorithm and its optimal block size for a given
scenario (operation, problem size, hardware, kernel library, multi-threading)
can be determined through exhaustive performance measurements; however, this is
extremely time consuming and thus often impractical.  Instead we aim to
determine the optimal configuration {\em without executing} any of the
alternative algorithms.  For this purpose, we use the hierarchical structure of
blocked algorithms:  Their entire computation is performed in a series of calls
to a few kernel routines; hence, by accurately estimating the runtime of these
kernels, we can predict an entire algorithm's runtime and performance.

In order to estimate the kernel runtimes, let us study how these kernels are
used:  In each algorithm execution, the same set of kernels is invoked
repeatedly---once for each step of the blocked matrix traversal.  Each
invocation, however, works on operands of different size depending on the
progress of the algorithms' traversal, the input problem size, and the block
size.  In short, we need to estimate the performance of only a few kernels, yet
with potentially wide ranges of operand sizes.

Our solution is \definition{performance modeling}, as detailed in
\cref{ch:model}:  Based on a detailed study of how a kernel's arguments (i.e.,
flags, operand sizes, etc.) affect its performance, we design performance models
in the form of piecewise multi-variate polynomials.  These models are generated
automatically once for each hardware and software setup and subsequently provide
accurate performance estimates at a tiny fraction of the kernel's runtime.

Using such estimates, we \definition[performance prediction]{predict} the {\em
performance} of blocked algorithms, as presented in \cref{ch:pred}.  These fast
predictions prove to be highly accurate, and allow us to both rank the blocked
algorithms for a given operation according to their performance, and find
near-optimal values for the algorithmic block sizes.

While our models yield accurate performance estimates for individual kernel
executions, they do not capture the performance influence of
\definition{caching} between kernels.  Prior to the invocation of each compute
kernel in an algorithm, typically only a portion of its operands are in cache,
and loading operands from main memory increases the kernel runtime.
\cref{ch:cache} investigates how caching effects can be accounted for in blocked
algorithms, and attempts to combine pure in- and out-of-cache estimates into
more accurate prediction.  However, while the results look promising on a rather
old \harpertown, the analysis reveals that on modern processors the effect
caching on kernel performance is so complex that accounting for it in
algorithm-independent performance models to further improve our prediction
accuracy is infeasible.

    \section{Micro-Benchmarks for Tensor Contractions}
    \label{sec:intro:tensor}
    Tensor contractions play an increasingly important role in various scientific
computations, such as machine learning~\cite{tensorml}, general
relativity~\cite{generalrelativity, generalrelativity2}, and quantum
chemistry~\cite{ccd2, ccd1}.  Following a brief introduction to \blas-based
tensor contraction algorithms and their performance in
\cref{sec:intro:tensor:algs}, \cref{sec:intro:tensor:pred} gives an overview of
how predictions based on micro-benchmarks are used to rank alternative
algorithms for a given contraction.

\subsection{Motivation: Tensor Contraction Algorithms}
\label{sec:intro:tensor:algs}

Computationally, tensor contractions are generalizations of matrix-vector and
matrix-matrix products to operands of higher dimensionality.  While
\blas covers contractions of up to two-dimensional operands (i.e., matrices),
there are no equivalently established and standardized high-performance
libraries for general tensor contractions.  Fortunately, just as a matrix-matrix
products can be decomposed into sequences of matrix-vector products, higher
dimensional tensor contractions can be cast in terms of matrix-matrix or
matrix-vector kernels.  (A broader overview of alternative approaches is given
in \cref{sec:relwork:tensor}.)

\begin{figure}\figurestyle
    \begin{subfigure}\textwidth\centering
        \begin{alglisting}[width=.5\textwidth]
            for c = 1:$c$
              for a = 1:$a$
                for b = 1:$b$
                  !\ddot:! $C$[a,b,c] += $A$[a,:] $B$[:,b,c]
        \end{alglisting}
        \qquad
        \begin{tikzpicture}[baseline]
            \begin{drawcube}
                \sliceA[outer]
                \sliceB[outer]
                \sliceC[outer]
                \sliceAB[outer]
                \sliceAC[outer]
                \sliceBC[outer]
                \sliceABC
            \end{drawcube}
            \node at (1, 0, 0) {$\pluseqq$};
            \begin{drawsquare}[(2, 0, 0)]
                \sliceA
            \end{drawsquare}
            \begin{drawcube}[(3.5, 0, 0)]
                \sliceB[outer]
                \sliceC[outer]
                \sliceBC
            \end{drawcube}
        \end{tikzpicture}
        \caption{Algorithm \tensoralgname{cab}{dot}}
    \end{subfigure}

    \medskip

    \begin{subfigure}\textwidth\centering
        \begin{alglisting}[width=.5\textwidth]
            for b = 1:$b$
              for c = 1:$c$
                !\dgemv[N]:! $C$[:,b,c] += $A$[:,:] $B$[:,b,c]
        \end{alglisting}
        \qquad
        \begin{tikzpicture}[baseline=(x.center)]
            \begin{drawcube}
                \sliceB[outer]
                \sliceC[outer]
                \sliceBC
            \end{drawcube}
            \node at (1, 0, 0) (x) {$\pluseqq$};
            \begin{drawsquare}[(2, 0, 0)]
                \slicenone
            \end{drawsquare}
            \begin{drawcube}[(3.5, 0, 0)]
                \sliceB[outer]
                \sliceC[outer]
                \sliceBC
            \end{drawcube}
        \end{tikzpicture}
        \caption{Algorithm \tensoralgname{bc}{gemv}}
    \end{subfigure}

    \medskip

    \begin{subfigure}\textwidth\centering
        \begin{alglisting}[width=.5\textwidth]
            for b = 1:$b$
              !\dgemm[NN]:! $C$[:,b,:] += $A$[:,:] $B$[:,b,:]
        \end{alglisting}
        \qquad
        \begin{tikzpicture}[baseline]
            \begin{drawcube}
                \sliceB
            \end{drawcube}
            \node at (1, 0, 0) {$\pluseqq$};
            \begin{drawsquare}[(2, 0, 0)]
                \slicenone
            \end{drawsquare}
            \begin{drawcube}[(3.5, 0, 0)]
                \sliceB
            \end{drawcube}
        \end{tikzpicture}
        \caption{Algorithm \tensoralgname b{gemm}}
    \end{subfigure}

    \mycaption{%
        Sample of algorithms for the tensor contraction $C_{abc} \coloneqq
        A_{ai} B_{ibc}$.
        \iftableoflist\else%
            All slicings are visualized in {\color{blue} blue}; the kernel
            operands (the intersections) are in {\color{red} red}. The name of
            each algorithm stems from the dimensions its \code{\bf for}-loops
            index and its \blas kernel.
        \fi
    }
    \label{fig:intro:tensor:algs}
\end{figure}

\begin{example}{Tensor contraction algorithms}{intro:tensor:algs}
    Let us consider the contraction $C_{abc} \coloneqq A_{ai} B_{ibc}$ (in
    Einstein notation), which is visualized as follows:
    \[
        \begin{tikzpicture}[baseline=(c.base)]
            \begin{drawcube}
                \node[anchor=east] at (-1, 0, 1) {$\scriptstyle a$};
                \node[anchor=north] at (0, -1, 1) {$\scriptstyle b$};
                \node[anchor=north west] at (1, -1, 0) {$\scriptstyle c$};
                \node (c) {$C$};
            \end{drawcube}
        \end{tikzpicture}
        \coloneqq
        \begin{tikzpicture}[baseline=(c.base)]
            \begin{drawsquare}
                \node[anchor=east] at (-1, 0, 0) {$\scriptstyle a$};
                \node[anchor=north] at (0, -1, 0) {$\scriptstyle i$};
                \node {$A$};
            \end{drawsquare}
        \end{tikzpicture}
        \matmatsep
        \begin{tikzpicture}[baseline=(c.base)]
            \begin{drawcube}
                \node[anchor=east] at (-1, 0, 1) {$\scriptstyle i$};
                \node[anchor=north] at (0, -1, 1) {$\scriptstyle b$};
                \node[anchor=north west] at (1, -1, 0) {$\scriptstyle c$};
                \node {$B$};
            \end{drawcube}
        \end{tikzpicture}
        \enspace.
    \]
    The entries~$C$\code{[a,b,c]} of the resulting three-dimensional tensor $C
    \in \R^{a \times b \times c}$ are computed as
    \[
        \forall \code a \forall \code b \forall \code c :\
        C\text{\code{[a,b,c]}} \coloneqq \sum_\code i A\text{\code{[a,i]}}
        B\text{\code{[i,b,c]}}
        \enspace.
    \]
    As further described in \cref{sec:tensor:alggen}, this contraction can be
    performed by a total of 36~alternative algorithms, each consisting of one or
    more \code{\bf for}-loops with a single \blas kernel at its core.  Three
    examples of such algorithms using \blasl1, 2, and~3 kernels are shown in
    \cref{fig:intro:tensor:algs}.  These algorithms use \matlab's ``\code:''
    slicing notation\footnotemark{} to access matrices and vectors within the
    tensors~$A$, $B$, and~$C$; the resulting operand shapes within the tensors
    passed to the \blas kernel are shown alongside the algorithms.
\end{example}
\footnotetext{%
    The index ``\code:'' in a tensor refers to all elements along that
    dimension, e.g., $A$\code{[a,:]} is the \code a-th row of~$A$.
}

Each tensor contraction can be computed via \blas kernels through many---even
hundreds---of algorithms, each with its own performance behavior.  The
\definition[optimization challenge:\\alternative algorithms\\skewed
diensions]{optimization challenge} of identifying the fastest among such a set
of {\em alternative algorithms} is especially difficult due to the in practice
commonly encountered {\em skewed dimensions} (i.e., one or more dimensions are
extremely small) for which most \blas implementations are typically not
optimized.

\begin{figure}\figurestyle
    \ref*{leg:intro:tensor}

    \pgfplotsset{
        twocolplot,
        ylabel=performance,
        y unit=\si{\giga\flops\per\second},
        xlabel={tensor size $a = b = c$},
        every subfigaxis/.style={
            fig vertical align=t,
        },
    }

    \begin{subfigaxis}[
            fig caption={
                Contraction $C_{abc} \coloneqq A_{ai} B_{ibc}$
                \captiondetails{$i = 8$, \sandybridge, 1~thread}
            },
            fig label=tensor:perf1,
            ymax=6,
            table/search path={tensor/figures/data/pred/meas/},
            legend to name=leg:intro:tensor,
        ]
        \addlegendimage{empty legend}\addlegendentry{kernel:}
        \foreach \s/\kernel in {
            green/dgemm,
            lightblue/dger,
            blue/dgemv,
            orange/daxpy,
            red/ddot%
        } {
            \expandafter\addlegendimage\expandafter{\s}
            \addlegendentryexpanded{%
                \expandafter\noexpand\csname\kernel\endcsname%
            }
            \label{plt:intro:tensor:\kernel}
        }
        \foreach \s/\vars in {
            red/{1, ..., 6},
            orange/{7, ..., 24},
            blue/{25, ..., 30},
            lightblue/{31, ..., 34},
            green/{35, ..., 36}%
        } \foreach \var in \vars
            \expandafter\addplot\expandafter[\s] table {var\var.min};
    \end{subfigaxis}\hfill
    \begin{subfigaxis}[
            fig caption={
                Contraction $C_{abc} \coloneqq A_{ija} B_{jbic}$
                \captiondetails{$i = j = 32$, \ivybridge, 10~threads}
            },
            fig label=tensor:perf2,
            table/search path={tensor/figures/data/ijb_jcid10/meas/},
        ]
        \foreach \s/\vars in {
            red/{1, ..., 48},
            orange/{49, ..., 120},
            blue/{121, ..., 156},
            lightblue/{157, ..., 168},
            green/{169, ..., 176}%
        } \foreach \var in \vars
            \expandafter\addplot\expandafter[\s] table {var\var.min};
    \end{subfigaxis}

    \mycaption{%
        Performance of tensor contraction algorithms.
        \captiondetails{\openblas, median of 10~repetitions}
    }
    \label{fig:intro:tensor:perf}
\end{figure}

\begin{example}{Performance of contraction algorithms}{intro:tensor:perf}
    Let us consider the tensor contraction $C_{abc} \coloneqq A_{ai} B_{ibc}$
    from \cref{ex:intro:tensor:algs} with tensors $A \in \R^{n \times 8}$, $B
    \in \R^{8 \times n \times n}$, and thus $C \in \R^{n \times n \times n}$;
    for~$n = 100$, this can be visualized as follows:
    \[
        \begin{tikzpicture}[baseline=(c.base)]
            \begin{drawcube}
                \node[anchor=east] at (-1, 0, 1) {$\scriptstyle a$};
                \node[anchor=north] at (0, -1, 1) {$\scriptstyle b$};
                \node[anchor=north west] at (1, -1, 0) {$\scriptstyle c$};
                \node (c) {$C$};
            \end{drawcube}
        \end{tikzpicture}
        \coloneqq
        \begin{tikzpicture}[baseline=(a.base), x={(.08, 0)}]
            \begin{drawsquare}
                \node[anchor=east] at (-1, 0, 0) {$\scriptstyle a$};
                \node[anchor=north] at (0, -1, 0) {$\scriptstyle i$};
                \node (a) {$A$};
            \end{drawsquare}
        \end{tikzpicture}
        \begin{tikzpicture}[baseline=(a.base), y={(0, .08)}]
            \begin{drawcube}
                \node[anchor=east] at (-1, 0, 1) {$\scriptstyle i$};
                \node[anchor=north] at (0, -1, 1) {$\scriptstyle b$};
                \node[anchor=north west] at (1, -1, 0) {$\scriptstyle c$};
                \node (b) {$B$};
            \end{drawcube}
        \end{tikzpicture}
        \enspace.
    \]

    \Cref{fig:intro:tensor:perf1} presents the performance of all 36~algorithms
    for this contraction on a \harpertown with single-threaded \openblas.  While
    the two \dgemm-based algorithms~(\ref*{plt:intro:tensor:dgemm}) are clearly
    faster than the others, they differ in performance by up to
    \SI{23.32}\percent; with other kernels the difference are even more extreme,
    exceeding a factor of~60 for the \daxpy-based
    algorithms~(\ref*{plt:intro:tensor:daxpy}).

    \Cref{fig:intro:tensor:perf2} showcases the performance of algorithms for
    the more complex contraction $C_{abc} \coloneqq A_{ija} B_{jbic}$ on all
    10~cores of an \ivybridge using multi-threaded \openblas.  In this scenario,
    the performance of the \dgemm-based algorithms alone differs by up
    to~$3\times$.
\end{example}

One could argue that only \dgemm-based algorithms are viable candidates to
achieve the best performance; while for the most part this observation is true,
due to skewed dimensions, even the performance of only these algorithms can
differ dramatically.  Furthermore, some contractions (e.g., $C_a \coloneqq
A_{iaj} B_{ji}$) cannot be implemented via \dgemm in the first place.
Therefore, we aim at the accurate prediction of any \blas-based contraction,
irrespective of which kernel is used.

\subsection{Prediction through Micro-Benchmarks}
\label{sec:intro:tensor:pred}

At first sight the situation seems similar to the selection of blocked
algorithms:  We want to avoid exhaustive performance measurements and select the
best algorithm {\em without executing} any of the alternatives; our strategy is
once again to predict each algorithm's performance by estimating its invoked
kernel's runtime.  However, while performance models accurately estimates the
performance of such kernels for many operand sizes, they perform rather poorly
for operations with skewed dimensions:  For extremely thin or small operands,
\blas kernels exhibit strong size-dependent performance fluctuations, which are
impractical to capture and represent in performance models.

While we cannot rely on performance models, analyzing the structure of tensor
contraction algorithms suggests a different approach:  In contrast to blocked
algorithms, a contraction algorithm performs its entire computation in a series
of calls to a \definition[single kernel\\fixed size\\micro-benchmarks]{single
\blas kernel} of with operands of {\em fixed size}.  Based on this observation,
we estimate the performance of such calls by constructing a small set of {\em
micro-benchmarks} that executes the kernel only a few times, and thus performs
only a fraction of the algorithm's computation.  Since memory locality plays an
especially important role in contractions with skewed dimensions, we carefully
recreate the stat of the processor's caches within the micro-benchmarks to time
the kernel in conditions analogous to those in the actual algorithm.

Based on such micro-benchmarks, we can predict the total runtime of contraction
algorithms for tensors of various shapes and sizes.  These predictions reliably
single out the fastest algorithm from a set of alternatives several orders of
magnitude faster than a single algorithm execution.

    \section{Related Work}
    \label{sec:intro:relwork}
    This overview of related research is structured as follows:
\Cref{sec:relwork:libsalgs} summarizes the history and state-of-the-art of dense
linear algebra (DLA) libraries and algorithms, \cref{sec:relwork:meas} addresses
performance measurements and profiling tools, \cref{sec:relwork:model} presents
performance modeling and prediction efforts, and \cref{sec:relwork:tensor}
discusses developments in high-performance tensor contractions.

\subsection{Dense Linear Algebra Libraries and Algorithms}
\label{sec:relwork:libsalgs}

We begin with a brief history of the fundamental DLA libraries \blas and \lapack
and prominent implementations in \cref{sec:relwork:libs}.  We then focus on
blocked algorithms and their tuning opportunities in \cref{sec:relwork:blocked},
and finally give an overview of alternative algorithms and libraries for
distributed-memory and accelerator hardware in, respectively,
\cref{sec:relwork:altalgs,sec:relwork:dist}.

\subsubsection{\blas and \lapack}
\label{sec:relwork:libs}

The development of standardized DLA libraries began in~1979 with the inception
of the {\namestyle Basic Linear Algebra Subprograms}
(\definition{\blas})~\cite{blasl1}, a \fortran interface specification for,
initially, various ``Level~1'' scalar and vector operations.  It was
subsequently extended to kernels for ``Level~2'' matrix-vector~\cite{blasl2} and
``Level~3'' matrix-matrix~\cite{blasl3} operations in, respectively, 1988
and~1990.  The aim of the \blas specification is to enable performance portable
applications:  DLA codes reach high performance on different hardware by using
architecture-specific \blas implementations.  Although computer architectures
have evolved dramatically in the last~40 years, this principle of performance
portability is still at the core of all current DLA libraries.

The \blas specification is accompanied by a reference
implementation~\cite{blasweb} that, while fully functional and well documented,
is deliberately simple and thus slow; to reach high performance, users instead
link with optimized \definition[open-source implementations]{\blas
implementations}.  The oldest {\em open-source} implementation still in
use is the {\namestyle Automatically Tuned Linear Algebra Software}
(\atlas)~\cite{atlas1, atlas3, atlas2, atlasweb}, first released in 1997; this
auto-tuning based library's main proficiency is to yield decent performance on a
wide range of hardware platforms with little developer and user effort.  The
first major open-source implementation hand-tuned for modern processors with
cache hierarchies was {\swstyle GotoBLAS}~\cite{gotoblas1, gotoblas2,
gotoblasweb}.  It reaches up to around \SI{90}{\percent} of a processor's peak
floating-point performance for both sequential and multi-threaded Level~3
kernels and good bandwidth-bound performance for Level~1 and~2 operations.
After {\swstyle GotoBLAS}'s discontinuation in~2010, its code-base and approach
were picked up and extended to more recent processors in the \openblas
library~\cite{openblasweb}, which is currently the fastest open-source
implementation for many architectures.  Also inspired by {\swstyle GotoBLAS}'s
approach is the fairly recent {\namestyle \blas-like Library Instantiation
Software} (\blis)~\cite{blis3, blis1, blis2, blisweb}, an open-source framework
that provides optimized kernels for basic DLA operations, such as the \blas,
based on one hand-tuned micro-kernel per architecture.

In addition to open-source implementations, many hardware \definition[vendor
implementations]{vendors} maintain and distribute their own high-performance
{\em\blas}, e.g., \intel's {\namestyle Math Kernel Library}
(\mkl)~\cite{mklweb}, \apple's framework \accelerate~\cite{accelerateweb}, and
{\namestyle IBM}'s {\namestyle Engineering and Scientific Subroutine Library}
(\essl)~\cite{esslweb}.

\blas forms the basis for DLA libraries covering more advanced operations.  The
earliest library built on top of first \blasl1 and later Level~2 was {\swstyle
LINPACK}~\cite{linpack, linpackweb}, a package of solvers for linear equations
and least-squares problems from the~1970s and~1980s.  {\swstyle LINPACK}
together with {\swstyle EISPACK}~\cite{eispack, eispackweb}, a collection of
eigenvalue solvers, was superseded by the {\namestyle Linear Algebra PACKage}
(\definition{\lapack})~\cite{lapack, lapackweb} in~1992.  \lapack has since been
extended with new features and algorithms, and is still under active
development.  Just like \blas, \lapack functions as a de-facto standard
interface specification for many advanced DLA operations; libraries such as
\openblas and \mkl adopt its interface and provide tuned implementations of
various routines.

For more details on \blas and \lapack, and their kernels and implementations
used throughout this work, see \cref{app:libs}.

\subsubsection{Blocked Algorithms}
\label{sec:relwork:blocked}

\lapack uses \definition{blocked algorithms} for most of its dense operations.
The core idea behind these algorithms is to leverage a processor's cache
hierarchy by increasing the spacial and temporal locality of operands, as well
as casting most of an operation's computation in terms of \blasl3 kernels.  As a
result, complex operations can reach performance levels close to the hardware's
theoretical peak.

However, for each operation, there typically exist multiple
\definition{alternative blocked algorithms}, of which \lapack offers only one,
but not always the fastest.  The alternative algorithms for a given operation
can be derived from its mathematical formulation
systematically~\cite{derivingbalgs} and automatically~\cite{loopgen, pmegen}.
Based on these principles, \libflame~\cite{libflameref, libflame, libflameweb}
offers many alternative algorithms for each operation, and for several
operations provides more efficient default algorithms than \lapack.  In this
work we consider \libflame's blocked algorithms for various operations, and aim
to predict which of them is most efficient for given scenarios.

Another caveat of blocked algorithms is their \definition[block size
tuning]{block sizes}, which need to be carefully {\em tuned} to maximize
performance.  Since this is a well-known aspect of blocked
algorithms~\cite{rooflinedla, blocksizetuning}, \lapack encapsulates and exposes
all its tuning parameters in \ilaenv, a central routine that is used to
configure the library at compile time; for many operations the block
sizes used by \lapack's reference implementation of \ilaenv (64~for most
algorithms) have been too small on recent hardware for quite some time.
Although the necessity of optimizing block sizes is well understood and taken
care of by implementations such as \mkl, it remains non-trivial, and in fact few
end-users and application-developers are aware of it.  The automated model-based
optimization of the block size for blocked algorithms is the second major goal
of this work.

\subsubsection{Alternatives to Blocked Algorithms}
\label{sec:relwork:altalgs}

An alternative to blocked algorithms is \definition{recursive algorithms},
which avoid both the algorithm selection and block-size optimization.  They are
also known as ``cache oblivious'' algorithms~\cite{cacheoblivious2,
cacheoblivious1} since they minimize the data-movement between cache
levels~\cite{dlarec}.  Recursion has been suggested for many DLA operations,
such as the LU~decomposition~\cite{lurec, lurec2}, the Cholesky
decomposition~\cite{cholrec}, triangular matrix inversion~\cite{trinvrec},
two-sided linear systems~\cite{sygstrec}, tall-and-skinny
QR~factorization~\cite{qrrec}, and Sylvester-type equation solvers~\cite{recsy,
recsyweb}.

However, since no readily-available recursion-based library comparable to
\lapack existed, we developed the {\namestyle Recursive \lapack collection}
(\definition{\relapack})~\cite{relapack, relapackweb}.  \relapack provides
recursive implementations for 48~\lapack routines, and outperforms not only the
reference implementation but in many cases also optimized libraries such as
\openblas and \mkl.

A second alternative to blocked algorithms tailored to shared-memory systems are
task-based \definition{algorithms-by-blocks}, also known as ``block algorithms''
or ``tiled algorithms''.  However, these algorithms not only introduce a
specialized storage scheme of matrices ``by block'', but also require custom
task scheduling mechanisms.  Implementations of such schedulers include
{\namestyle QUARK}~\cite{quark} as part of {\namestyle PLASMA}~\cite{plasma},
{\namestyle DAGuE}~\cite{dague}, {\namestyle SMPSs}~\cite{smpssdla}, and
{\namestyle SuperMatrix}~\cite{supermatrix}.

\subsubsection{Distributed-Memory and Accelerators}
\label{sec:relwork:dist}

\definition[distributed memory]{Distributed-memory} systems and super-computers
are indispensable for large-scale DLA computations.  The first noteworthy
extension of the \blas and the \lapack to this domain was the {\namestyle
Scalable Linear Algebra PACKage} (\scalapack)~\cite{scalapack, scalapackweb},
written in \fortran and based on \blas, \lapack, and the {\namestyle Message
Passing Interface} (MPI).  However, {\namestyle ScaLAPACK} is only sparingly
updated (last in~2012), and, instead, the state of the art for
distributed-memory DLA is {\namestyle Elemental}~\cite{elemental, elementalweb},
an actively developed \cpplang~library, based on \libflame's methodology in and
object-oriented and templated programming techniques.

Since \definition{accelerators} such as {\namestyle Xeon-Phi} coprocessors and
graphics processors lend themselves well to compute-intensive operations, they
are a natural target for DLA codes.  While some classic \blas implementations
such as \atlas, \blis, and \mkl, can be used on the x68-based {\namestyle Xeon
Phi}s, separate libraries are required for graphics processors: {\namestyle
NVIDIA}'s {\namestyle cuBLAS}~\cite{cublasweb} provides high-performance \blas
kernels for {\langstyle CUDA}-enabled graphics cards, and {\namestyle
clBLAS}~\cite{clblasweb} targets {\langstyle OpenCL}-capable devices.
Furthermore, {\namestyle Matrix Algebra on GPU and Multicore Architectures}
({\namestyle MAGMA})~\cite{magma, magmaweb} targets \blas and \lapack operations
on heterogeneous systems (e.g., CPU + GPU).

\subsection{Performance Measurements and Profiling}
\label{sec:relwork:meas}

Runtime measurements of both application codes and algorithms are crucial in the
investigation of performance behaviors, bottlenecks, as well as optimization
and tuning in general; hence, numerous tools facilitate such
measurements.  Simple timers are accessible in virtually any language and
environment: e.g., \code{time} in Unix, \code{rdtsc} in x86~assembly,
\code{gettimeofday()} in~\clang, \code{omp\_get\_wtime()} in {\namestyle
OpenMP}, \code{tic} and \code{toc} in \matlab, and \code{timeit} in \python.
Several more advanced tools \definition[profiling]{profile} executions of
functions and communications in applications by tracing or sampling: e.g.,
{\namestyle gprof}~\cite{gprof, gprofweb}, {\namestyle VAMPIR}~\cite{vampirweb},
{\namestyle TAU}~\cite{tau, tauweb}, {\namestyle Scalasca}~\cite{scalasca,
scalascaweb}, and \intel's {\namestyle VTune}~\cite{vtuneweb}.  While such tools
are invaluable in the performance analysis of application codes, their
generality makes them somewhat unwieldy for our purposes of investigating DLA
kernel performance.  Therefore, we designed {\namestyle Experimental Linear
Algebra Performance Studies} (\definition{\elaps})~\cite{elaps, elapsweb}, a
framework for performance measurements and analysis of DLA routines and
algorithms, further detailed in \cref{sec:meas:elaps}.

\subsection{Performance Modeling and Predictions}
\label{sec:relwork:model}

Predicting and modeling application performance is an important aspect of
high-performance computing, and the term ``performance modeling'' is used to
describe many different techniques and approaches.  This section gives a brief
overview of such approaches with focus on methods for DLA algorithms.

The well-established \definition{Roofline model}~\cite{roofline1} does not
predict performance, but relates an algorithm's attained performance to the
hardware's potential:  As detailed in \ref{sec:term:roofline}, it allows to
evaluate an execution's resource efficiency by relating its algorithm's
arithmetic intensity and int performance relative to the hardware's peak
main-memory bandwidth and floating-point performance.  It has been applied,
implemented, and extended in numerous publications, such as~\cite{rooflinecache,
rooflinetoolkit, roofline2}.  Notably, \citeauthor*{rooflinedla} use the
roofline model (the arithmetic intensity in particular) to optimize the block
size for a blocked matrix inversion algorithm~\cite{rooflinedla}.

Model-based performance tuning of \blas implementations was suggested for both
\atlas~\cite{atlasmodel} and \blis~\cite{blismodel}, showing that near-optimal
\blas performance can be reached without measurement-based autotuning:  Instead
they, e.g., select blocking sizes according to the \blas implementation and the
target processor's cache sizes.  Note that these approaches are used to tune
\blas kernels, and do not actually predict their performance; hence they cannot
serve as a basis for our predictions.

Previous work in our research group by \citeauthor*{roman1} constructed accurate
\definition[analytical models]{analytical performance models} for small DLA
kernels~\cite{romandis, roman1}.  These models target problems that fit within a
\harpertown's last-level cache (L2), and are based on the number of
memory-stalls and arithmetic operations as well as their overlap incurred by
specific kernel implementations.  As such, they require not only a deep
understanding of the processor architecture, but also a detailed analysis of the
kernel implementation.  While the resulting models yield accurate predictions
within a few percent of reference measurements, they are not easily extended to
larger problems and other operations.  Therefore, this work instead
considers automatically generated, measurement-based models.

\Citeauthor*{blis3model} construct \definition[piecewise models]{piecewise}
runtime and energy {\em models}---somewhat similar to those presented in this
work---for the \blis implementations of \dgemm and \dtrsm~\cite{blis3model} on a
{\hwstyle Sandy Bridge-EP E5-2620}.  However, their approach is based on
extensive knowledge of \blis~\cite{blismodel}, and their models only represent
one degree of freedom (by considering only square matrices or operations on
panel matrices with fixed width/height).  Their average runtime model accuracy
for \dgemm and \dtrsm is, respectively, \SIlist{1.5;4.5}\percent, with local
errors of up to, respectively, \SIlist{4.5;7}\percent.
\citeauthor*{blischolmodel} extend this work to multi-threaded \dgemm, \dtrsm,
and \dsyrk in order to predict the performance of a blocked Cholesky
decomposition algorithm with fixed block size~\cite{blischolmodel}; their
average runtime prediction errors are \SIlist{3.7;2.4}\percent, depending on the
parallelization within \blis.  In contrast to these publications, the modeling
framework presented in this work, which was developed around the same time, is
fully automated, applicable to any \blas- or \lapack-like routine, not limited
to one implementation and hardware, and offers models with multiple degrees of
freedom.

In a separate effort \citeauthor*{tridiagmodel} constructs measurement-based,
yet hardware- and \definition{implementation-independent models} in the form of
a series of univariate polynomials (one kernel argument is represented by the
polynomial, the other varied in the series) for several \blasl3
kernels~\cite{tridiagmodel, qrmodel}.  These models are used to predict the
performance of both a blocked reduction to tridiagonal form~\cite{tridiagmodel}
and a blocked multishift QR~algorithm~\cite{qrmodel}.  The resulting prediction
error on an unspecified {\namestyle AMD Opteron} is reported to be
below~\SI{10}{\percent} for the single-threaded tridiagonalization, and is on
average around~\SI{10}{\percent} for the QR~algorithm using multi-threaded
\blas.  In contrast, the more general piecewise models proposed in this work
yield considerable smaller prediction errors for various blocked algorithms.

Several research projects model the performance of \definition[distributed
memory]{distributed-memory} applications.  A general purpose approach by
\citeauthor*{alex1} builds basic performance models for kernels in application
codes based on performance profiling~\cite{alex2, alex1}, allowing to
investigate the complexity and scalability of application components.  In the
field of distributed-memory DLA, most modeling efforts target \scalapack using
domain-specific knowledge through, e.g., polynomial
fitting~\cite{scalapackpolfit} or hierarchical modeling of
kernels~\cite{scalapckhierarchmodel}.

\subsection{Tensor Contractions}
\label{sec:relwork:tensor}

Tensor contractions are at the core of scientific computations, such as machine
learning~\cite{tensorml}, general relativity~\cite{generalrelativity,
generalrelativity2}, and quantum chemistry~\cite{ccd2, ccd1}.  Since generally
speaking such contractions are high-dimensional matrix-matrix multiplications,
they are closely related to \blasl3 operations, and in fact most contractions
can be cast in terms of one or more calls to \dgemm, either by adding loops or
transpositions; this is implemented in many frameworks, such as the {\namestyle
Tensor Contraction Engine} (TCE)~\cite{tce, tceweb}, the {\namestyle Cyclops
Tensor Framework} (CTF)~\cite{cyclops, cyclopsweb}, the \matlab{} {\namestyle
Tensor Toolbox}~\cite{matlabtt, matlabttweb}, and {\namestyle
libtensor}~\cite{libtensor, libtensorweb}.

In contrast to these implementations, which rely on a single algorithm for each
contraction (potentially selected through heuristics), previous work in our
group by \citeauthor*{tensorgen} investigated the automated generation of all
alternative \blas-based algorithms~\cite{tensorgen}.  \Cref{ch:tensor} picks up
this work and presents a performance prediction framework for such algorithms
that allow to automatically identify the fastest algorithm~\cite{tensorpred}.

More recent and ongoing work in our group by \citeauthor*{gett} attempts to go
break the barrier between contraction algorithms and \dgemm implementations.
Following the structured design of \blis~\cite{blis1}, they propose code
generators that provide high-performance algorithms tailored to specific
contraction problems that reach close to optimal performance~\cite{gett}.  Their
tools construct numerous alternative implementations, and identify the fastest
through a combination of heuristics and micro-benchmarks.

}

    \chapter[Performance Effects and Measurements]
        {Performance Effects\newline and Measurements}
\label{ch:meas}
{
    \tikzsetexternalprefix{externalized/meas-}

    This work is concerned with predicting the performance of dense linear algebra
routines and algorithms through measurement-based performance models and
micro-benchmarks.  To fully focus on modeling and prediction in the following
chapters, we here establish how accurate runtime measurements are obtained, and
address common influences on such measurements and their effects.  Furthermore,
we presents a performance measurement tool and framework tailored to dense
linear algebra routines that we developed to serve as the foundation for the
experiments, models, and benchmarks throughout this work.

In detail, this chapter covers the following material:
\begin{itemize}
    \item \Cref{sec:meas:effects} presents common effects observed when
        measuring the runtime of dense linear algebra routines.  In particular,
        it addresses {\em library initialization overhead}, {\em fluctuations}
        (e.g., due to {\em system noise} and {\em varying processor frequency}),
        {\em thread pinning}, and {\em caching}.

    \item \Cref{sec:meas:elaps} introduces the {\em\elaps Framework} that
        evolved from the performance measurement tools developed for this work.
        \elaps provides the {\em\sampler}, a low-level tool for measurements of
        \blas- and \lapack-like dense linear algebra routines, as well as a
        {\em\python framework} with a graphical user interface and various
        utility functions to set up experiments and process their results.
\end{itemize}

Additionally, for readers new to performance studies, \cref{app:term} provides
an introduction into the terminology and concepts of topics such as
computational workload, timings, performance, hardware limitations, and
efficiency.

    \section{Performance Effects for Dense Linear Algebra Kernels}
    \label{sec:meas:effects}
    At the core of any study on performance are accurate runtime measurements.
However, while in principle, timing a computation is as simple as ``start
timer--compute--stop timer'', obtaining reliable and stable timings is not
trivial.  In this section, we present the most relevant effects and influences
on measurements of dense linear algebra routines; in particular, we address
initialization overhead (\cref{sec:meas:effects:init}), different types of
fluctuations (\cref{sec:meas:effects:fluct}), thread pinning
(\cref{sec:meas:effects:pin}), and caching (\cref{sec:meas:effects:caching}).

        \subsection{Library Initialization Overhead}
        \label{sec:meas:effects:init}
        Many high-performance dense linear algebra libraries, such as optimized
implementations of \blas and \lapack, perform of initializations (e.g., hardware
detection, buffer allocation, etc.) the first time one of their kernels is
invoked.  These \definition[initialization overhead]{initializations} imply an
{\em overhead} that can significantly increase the first library invocation's
runtime.

\begin{table}\tablestyle
    \begin{tabular}{rcccc}
        \toprule
                        &\openblas      &\blis          &\mkl           &reference \\
        \midrule
        1st \dgemm[]    &\SI{1.10}\ms   &\SI{1.32}\ms   &\SI{8.14}\ms   &\SI{37.96}\ms \\
        2nd \dgemm[]    &\SI{0.90}\ms   &\SI{0.95}\ms   &\SI{0.86}\ms   &\SI{37.93}\ms \\
        overhead        &\SI{0.20}\ms   &\SI{0.38}\ms   &\SI{7.28}\ms   &\hphantom0\SI{0.04}\ms\\
        \bottomrule
    \end{tabular}

    \mycaption{%
        \blas library initialization overhead for two identical \dgemm[NN]{}s.
        \captiondetails{$m = n = k = 200$, \sandybridge, 1~thread}
    }
    \label{tbl:meas:effects:init}
\end{table}

\begin{example}{Library initialization overhead}{meas:init}
    \Cref{tbl:meas:effects:init} presents the runtime of two consecutive
    matrix-matrix multiplications $\dm{C_1} \coloneqq \dm{A_1} \matmatsep
    \dm{B_1} + \dm{C_1}$ and $\dm{C_2} \coloneqq \dm{A_2} \matmatsep \dm{B_2} +
    \dm{C_2}$ (\dgemm[NN]) with disjoint $\dm{A_1},\allowbreak
    \dm{A_2},\allowbreak \dm{B_1},\allowbreak \dm{B_2},\allowbreak
    \dm{C_1},\allowbreak \dm{C_2} \in \R^{200 \times 200}$ on a \sandybridge
    with single-threaded \openblas, \blis, and \mkl; the two calls to \dgemm are
    the first and only invocations of \blas in program.

    The timings show that the libraries have substantially different overheads:
    \begin{itemize}
        \item The reference \blas implementation has a negligible overhead but
            is around $40\times$ slower than the optimized libraries.

        \item \openblas and \blis are optimized for the \sandybridgeshort, and
            when first invoked, these libraries perform some initializations,
            such allocating auxiliary buffers, that introduce an overhead of,
            respectively, \SI{.20}{\ms} and \SI{.38}\ms.

        \item In addition to the allocation of auxiliary buffers, \mkl
            dynamically detect the processor architecture to accordingly select
            optimized kernels.  Hence it has by far the largest overhead of
            \SI{7.28}\ms, which dominates its first invocation's runtime.
    \end{itemize}
\end{example}

Since we mostly use optimized libraries such as \openblas, \blis, and \mkl, we
counter the initialization overhead by simply preceding any set of measurements
with an unrelated kernel invocation.

        \subsection{Fluctuations}
        \label{sec:meas:effects:fluct}
        Once the initialization overhead is overcome, repeated timings of the same
kernel on the same data may still exhibit significant \definition{performance
fluctuations}.  Such fluctuations can be caused by a variety of effects, such as
background applications and system noise (\cref{sec:meas:fluct:noise}),
\intel{} \turboboost (\cref{sec:meas:fluct:turbo}), or other changes in
processor frequency (\cref{sec:meas:fluct:longterm}).

\subsubsection{Background and System Noise}
\label{sec:meas:fluct:noise}

The potentially most disturbing, yet also quite easily avoidable source of
fluctuations are other \definition{background processes} competing for the
processor's resources.

\begin{figure}\figurestyle
    \ref*{leg:meas:fluct}

    \raggedleft
    \tikzsetnextfilename{fluct}
    \begin{tikzpicture}
        \begin{axis}[
                ymax=.4,
                xlabel=repetition,
                ylabel=runtime, y unit=\ms,
                every axis plot/.append style={
                    mark=*, only marks
                },
                tick label style={/pgf/number format/fixed},
                table/search path={meas/figures/data/fluct},
                legend to name=leg:meas:fluct,
                legend columns=1,
            ]
            \addlegendimage{plotibacc}
            \addlegendentry{\broadwell, \accelerate, heavy noise}
            \label{plt:ibacc:circ}
            \addlegendimage{plotsbmkl}
            \addlegendentry{\sandybridge, \mkl, minimal noise}
            \addplot[plotibacc, mark size=1pt] table {macbook.dat};
            \addplot[plotsbmkl, mark size=1pt] table {cluster.dat};
        \end{axis}
    \end{tikzpicture}

    \mycaption{%
        Runtime fluctuations \dgemm[NN] caused by background processes and
        system noise.
        \captiondetails{$m = n = k = 100$, 1~thread}
    }
    \label{fig:meas:fluct}
\end{figure}

\begin{example}{Influence of background noise}{meas:fluct}
    \Cref{fig:meas:fluct} presents the runtime of 1000~repetitions of the
    matrix-matrix multiplication $\dm C \coloneqq \dm A \matmatsep \dm B + \dm
    C$ (\dgemm[NN]) with $\dm A, \dm B, \dm C \in \R^{100 \times 100}$ on a
    \broadwell (as part of {\namestyle MacBook Pro} with \apple's framework
    \accelerate and a \sandybridge (as part of \rwth's computing cluster) with
    \mkl.

    On the \broadwell~(\ref*{plt:ibacc:circ}) with various other applications
    running in the background (e.g., browser and music player), the fluctuations
    are enormous:  The measurement standard deviation is over $4\times$~the mean
    runtime.  On the \sandybridge~(\ref*{plt:sbmkl:circ}) with no other user
    applications running during measurements, the fluctuations are already much
    smaller at \SI{2.36}{\percent}~of the average time.  For larger problem
    sizes, the fluctuations are considerably smaller, and quickly fall below
    \SI{.1}\percent.
\end{example}

While these type of fluctuations can be avoided to some extend by ensuring that
no other applications run during measurements, they cannot be avoided altogether
even with exclusive access to dedicated high-performance hardware---the
remaining fluctuations are known as \definition{system noise}.  Hence, for our
experiments, models, and micro-benchmarks all our measurements are repeated at
least five times and \definition{summary statistics} of the runtime (or
performance) are presented, such as the minimum or median.

\subsubsection{\intel{} \turboboost}
\label{sec:meas:fluct:turbo}

Compute-bound dense linear algebra computations, such as \blasl3 and
\lapack-level routines, benefit directly from increased processing frequencies.
Therefore, they usually trigger \intel{} \turboboost and constantly run at the
maximum turbo frequency if possible.  Since this frequency cannot be sustained
indefinitely on most machines, the processor frequency is eventually lowered and
henceforth fluctuates to keep the hardware within its power and thermal limits.

\begin{figure}\figurestyle
    \ref*{leg:meas:turbo}
    \renewcommand\plotheightsub{.25\textwidth}
    \pgfplotsset{
        every axis/.append style={
            anchor=north,
            xmax=790,
            ymin={},
            name=tmp
        },
        table/search path={meas/figures/data/turbo}
    }

    \tikzsetnextfilename{turbo}
    \begin{tikzpicture}
        \begin{axis}[
                ylabel=temperature, y unit=\celsius,
                xticklabels={},
                legend to name=leg:meas:turbo,
            ]
            \addlegendimage{plot2}
            \addlegendentry{temperature}
            \label{plt:meas:turbo:temp}
            \addlegendimage{plot3}
            \addlegendentry{frequency}
            \label{plt:meas:turbo:freq}
            \addlegendimage{plot1, mark=*, only marks, mark size=2.5pt}
            \addlegendentry{runtime}
            \label{plt:meas:turbo:time}
            \addplot[red] table[x expr=750 * \coordindex / 330, y=temp] {power.dat};
        \end{axis}
        \path (tmp.south) ++(0, -2ex) coordinate (tmp);
        \begin{axis}[
                at=(tmp),
                ylabel=frequency, y unit=\GHz,
                xticklabels={},
                ymin=2, ymax=3.5,
            ]
            \addplot[green] table[x expr=750 * \coordindex / 330, y expr=\thisrow{freq} / 1000] {power.dat};
        \end{axis}
        \path (tmp.south) ++(0, -2ex) coordinate (tmp);
        \begin{axis}[
                at=(tmp),
                xlabel=repetition,
                ylabel=runtime, y unit=\ms,
                ymax=110,
            ]
            \addplot[blue, mark=*, only marks, mark size=1pt] table {time.dat};
        \end{axis}
    \end{tikzpicture}

    \mycaption{%
        Effect of \turboboost on the runtime of \dgemm[NN].
        \iftableoflist\else\\(Note: $y$-axes are not 0-based.)\fi
        \captiondetails{$m = n = k = 1300$, \broadwell, 2~threads, \accelerate}
    }
    \label{fig:meas:turbo}
\end{figure}

\begin{example}{\turboboost}{meas:turbo}
    \Cref{fig:meas:turbo} presents the runtime of repeated matrix-matrix
    multiplications $\dm C \coloneqq \dm A \matmatsep \dm B + \dm C$
    (\dgemm[NN]) with $\dm A, \dm B, \dm C \in \R^{1300 \times 1300}$ alongside
    the processor's temperature and frequency\footnotemark{} on both cores of a
    \broadwell with multi-threaded \accelerate; in this experiment, no other
    resource intensive programs run in the background.

    In the beginning, the processor is at a cool
    \SI{53}{\celsius}~(\ref*{plt:meas:turbo:temp}) and each \dgemm[NN] takes
    about \SI{60}{\ms}~(\ref*{plt:meas:turbo:time}) at the maximum turbo
    frequency of \SI{3.4}{\GHz}~(\ref*{plt:meas:turbo:freq}).  The processor
    temperature increases steadily up to \SI{105}{\celsius} around
    repetition~200 (\SI{12}{\second} into the experiment); at this point the
    frequency is reduced and continuously adjusted between \SIlist{3;3.2}{\GHz}
    such that this temperature threshold is not exceeded.  This change in
    frequency, as well as its fluctuations towards the end have a direct effect
    on the \dgemm[NN]'s runtime:  It increases by about~\SI{10}{\percent} to
    roughly~\SI{67}\ms.
\end{example}
\footnotetext{%
    Obtained through the \intel {\namestyle Power Gadget}.
}

The behavior of \turboboost depends enormously on the computation environment:
While on a work-station or laptop system the processor temperature increases
rapidly and the maximum turbo frequency is not sustained for long, on dedicated
high-performance compute clusters, efficient cooling allows for the processor to
operate at the maximum turbo frequency for much longer, if not indefinitely.
However, even in our main computing facilities at the {\namestyle\rwth IT
Center}, we observed notable fluctuations of the frequency below its maximum
with negative impacts on our measurement quality and stability.

Throughout this work, we consider processors with and without enabled
\turboboost.  While the performance of these two cases is not directly
comparable, we consider our methodologies for both scenarios.  In particular,
\turboboost is disabled on our \sandybridge (unless otherwise stated) and
enabled on our \haswell---an overview of all hardware configurations is given in
\cref{app:hardware}.

\subsubsection{Distinct Long-Term Performance Levels}
\label{sec:meas:fluct:longterm}

Even with \turboboost disabled, a processor's speed is not always fixed to its
base frequency and we instead observed jumps between two or more
\definition{performance levels}.

\begin{figure}\figurestyle
    \ref*{leg:meas:longterm}

    \tikzsetnextfilename{longterm}
    \begin{tikzpicture}
        \begin{axis}[
                xlabel=repetitions,
                ylabel=runtime, y unit=\ms,
                ymax=400,
                legend to name=leg:meas:longterm,
                every axis plot/.append style={mark=*, only marks},
                table/search path={meas/figures/data/longterm}
            ]
            \addlegendimage{plotsbopen}
            \addlegendentry{\sandybridge}\label{plt:meas:longterm:sandy}
            \addlegendimage{plothwopen}
            \addlegendentry{\haswell}\label{plt:meas:longterm:haswell}
            \addplot[plotsbopen, mark size=1pt] table {sandy.dat};
            \addplot[plothwopen, mark size=1pt] table {haswell.dat};
        \end{axis}
    \end{tikzpicture}

    \mycaption{%
        Varying runtime for a skewed \dgemm[NN] over a period of time.
        \captiondetails{$m = k = 4000$, $n = 200$, 1~thread, \openblas}
    }
    \label{fig:meas:longterm}
\end{figure}

\begin{example}{Performance levels}{meas:longterm}
    \Cref{fig:meas:longterm} presents the runtime of 1000~repetitions of the
    matrix-matrix multiplication $\dm[width=.05]C \coloneqq \dm A \matvecsep
    \dm[width=.05]B + \dm[width=.05]C$ (\dgemm[NN]) with $\dm A \in \R^{4000
    \times 4000}$ and $\dm[width=.05]B, \dm[width=.05]C \in \R^{4000 \times
    200}$ on a \sandybridge and a \haswell (both with \turboboost disabled) with
    single-threaded \openblas.

    On both systems, we can clearly make out two distinct runtime levels: on the
    \sandybridgeshort, the measurements jump between \SIlist{354;359}\ms, which
    are \SI{1.4}{\percent}~apart, and on the \haswellshort with twice the
    floating-point performance per cycle, the two levels
    at~\SIlist{205;213}{\ms} differ by~\SI{3.9}\percent.  There is no
    discernible pattern to the jumps between these levels and the processors
    commonly stay at the same level for~\SI{10}{\second} or longer
    (50~repetitions at \SI{200}{\ms} each).
\end{example}

Since we found no means to eradicate this type of fluctuations, we adopt our
measurement setups to account for them:  Whenever we have more than one
measurement point (e.g., varying the routines or problem sizes), we not only
repeat each measurement several times in isolation, but also shuffle the
repetitions.  As a result, the repetitions for each data point are spread across
the entire experiment duration and summary statistics such as the minimum and
median yield a stable runtime estimate for only one performance level.

In summary, we can avoid or account for various types of fluctuations within our
measurements.

        \subsection{Thread Pinning}
        \label{sec:meas:effects:pin}
        Which processor cores a program runs on is generally controlled by the
operating system, and in fact most system schedulers every now and then move
threads between cores at runtime.  However, since dense linear algebra kernels
immensely rely on temporal data locality within the cache hierarchy and caches
shared across multiple cores, moving or physically separating threads may
significantly decrease a computation's efficiency.  Counteracting these effects
by restricting threads to physical cores is called \definition{thread pinning}.

\begin{figure}\figurestyle
    \ref*{leg:meas:pin}

    \raggedleft
    \tikzsetnextfilename{pin}
    \begin{tikzpicture}
        \begin{axis}[
                ybar, every axis plot/.append style=ybar legend,
                xlabel=\#threads,
                ylabel=efficiency, y unit=\percent,
                xmin=.5, xmax=8.5,
                xtick={1, ..., 8},
                ymax=100,
                legend to name=leg:meas:pin,
                table/search path={meas/figures/data/pin},
                nodes near coords style={text opacity=1},
            ]
            \addplot+[red] table[y=nopin] {results.dat};
            \addlegendentry{without pinning}\label{plt:nopin}
            \addplot+[green] table[y=pin] {results.dat};
            \addlegendentry{with pinning}\label{plt:pin}
            \addplot[
                sharp plot, only marks,
                nodes near coords={%
                    \SI[round-mode=places, round-precision=2]
                       \pgfplotspointmeta\percent%
                },
                point meta=explicit symbolic
            ] table[y=pin, meta=speedup] {results.dat};
        \end{axis}
    \end{tikzpicture}

    \mycaption{%
        Effects of thread pinning on the compute-bound efficiency of
        a multi-threaded \dgemm[TN].
        \iftableoflist\else Annotations: speedup of \ref*{plt:pin} over
        \ref*{plt:nopin}.\fi
        \captiondetails{$m = 64$, $n = k = 2000$, \sandybridge, \openblas,
        median of 100~repetitions}
    }
    \label{fig:meas:pin}
\end{figure}

\begin{example}{Thread pinning}{meas:pin}
    \Cref{fig:meas:pin} presents the compute-bound efficiency (see
    \cref{sec:term:eff}) of the matrix-matrix multiplication $\dm[height=.032]C
    \coloneqq \dm[height=.032, ']A \matmatsep \dm B + \dm[height=.032]C$ with
    $\dm[height=.032]A, \dm[height=.032]C \in \R^{64 \times 2000}$ and $\dm B
    \in \R^{2000 \times 2000}$ (an example taken from within \lapack's blocked
    \dlauum) using \openblas with an increasing number of threads on a
    two-socket \sandybridge system with and without thread pinning.

    While the single-threaded \dgemm is not affected by pinning, with two
    threads, the execution pinned to two cores of one socket~(\ref*{plt:pin}) is
    \SI{7.47}{\percent}~faster than the unpinned version~(\ref*{plt:nopin});
    this difference increases with the number of threads up
    to~\SI{28.08}{\percent} on 8~cores.
\end{example}

To ensure that \blas implementations reach their full potential, throughout this
work all measurements are performed with threads pinned to the cores of a single
processor.

        \subsection{Caching}
        \label{sec:meas:effects:caching}
        The location of operands in a computer's memory hierarchy---also referred to as
the \definition{cache precondition}---can have significant influence on a
routine's performance; an operation whose operands already reside in the
processor's cache (called an \definition{in-cache} scenario or operating on
``warm'' data) is faster than the same operation that has to load its operands
from the slow main memory (\definition{out-of-cache}, ``cold'' data).  This
effect is strongest for memory bound operations that cannot amortize memory
stalls with computations.

\begin{table}\tablestyle
    \begin{tabular}{rcccc}
        \toprule
                        &\openblas      &\blis          &\mkl           &reference \\
        \midrule
        out-of-cache    &\SI{0.60}\ms   &\SI{1.27}\ms   &\SI{0.68}\ms   &\SI{6.81}\ms \\
        in-cache        &\SI{0.33}\ms   &\SI{1.02}\ms   &\SI{0.41}\ms   &\SI{6.63}\ms \\
        overhead        &\SI{0.27}\ms   &\SI{0.25}\ms   &\SI{0.27}\ms   &\SI{0.18}\ms \\
        \bottomrule
    \end{tabular}

    \mycaption{%
        Influence of caching on the execution time of \dgemv.
        \captiondetails{$m = n = 1000$, \sandybridge, 1~thread, median of 100~repetitions}
    }
    \label{tbl:meas:effects:cache}
\end{table}

\begin{example}{Caching}{meas:cache}
    \Cref{tbl:meas:effects:cache} presents the runtime of the matrix-vector
    multiplication $\dv y \coloneqq \dm A \matvecsep \dv x + \dv y$ (\dgemv)
    with $\dm A \in \R^{1000 \times 1000}$ either in- or
    out-of-cache\footnotemark{} and the same $\dv x, \dv y \in \R^{1000}$ on one
    core of a \sandybridge with different \blas implementations.

    Even though the implementations differ by more than $10\times$ in runtime,
    the overhead of loading \dm A from main memory is comparable between
    \SIlist{.18;.27}\ms; for \openblas, this corresponds to a runtime increase
    of over~\SI{80}\percent.  Furthermore, the overhead is identical for the two
    fastest implementations \mkl and \openblas, a little lower for the less
    optimal \blis, and lowest for the totally unoptimized reference
    implementation.
\end{example}
\footnotetext{
    To place $A$ out of cache, each repetition uses a different memory location
    for it.
}

The cache precondition of an operation, i.e., which of its operands are where in
the memory hierarchy, largely depends on the operation's context within an
algorithm or application.  \Cref{ch:cache,ch:tensor} address caching in more
detail.

        \subsection{Summary}
        \label{sec:meas:effects:sum}
        This section studied various effects on the performance of dense linear algebra
computations.  While some can be avoided altogether, others can be accounted for
by specific measurement setups.  In the remainder of this work, all measurements
are accordingly configured to yield stable results.

    \section{Measurements and Experiments: \elaps}
    \label{sec:meas:elaps}
    This section introduces {\em Experimental Linear Algebra Performance Studies}
(\definition{\normalfont\elaps}), the performance measurement framework that
serves as the basis for all experiments, modeling procedures, and benchmarks
throughout this work.  \elaps was initially developed specifically for our
modeling and benchmarking applications, but has since evolved into a versatile
general purpose tool-set for various dense linear algebra performance
experiments.  It is available as an open-source project on
\github~\cite{elapsweb}.

\elaps consists of two layers:  The bottom layer offers the \sampler, a
low-level tool for runtime and performance counter measurements
(\cref{sec:meas:sampler}); the top layer is a \python framework that, among
other features, offers user-friendly access to performance experiments and a
graphical user interface (\cref{sec:meas:elapslib}).

\paragraph{Publication}
The work presented in section is in parts based on research published in:
\begin{pubitemize}
    \pubitem{elaps}
\end{pubitemize}

        \subsection{The \sampler}
        \label{sec:meas:sampler}
        The \definition{\sampler} is a command-line performance measurement tool written
in \clang/\cpplang; it essentially times arbitrary executions of dense linear
algebra routines.  Each \sampler instance typically provides access to all \blas
and \lapack routines from one---potentially machine-specific---implementation
(e.g., \openblas, \blis, or \mkl), but it is easily extended to other routines
with similar interfaces at compile time.

At runtime, the input to the \sampler determines which routine invocations are
executed and timed.  The interface provides the following work-flow:
\begin{enumerate}
    \item Read from standard input a list of \definition[call]{calls}, i.e.,
        routine names with corresponding lists of arguments.

    \item Execute the specified calls, and measuring their runtime  in terms of
        processor cycles; optionally track further performance counters through
        the {\namestyle Performance Application Programming Interface}
        (\papi)~\cite{papi, papiweb}.

    \item Print the measured performance numbers to standard output.
\end{enumerate}

The \sampler provides configuration options and commands that enable a wide
range of performance studies:
\begin{itemize}
    \item Routine operands can be individually allocated, subdivided, and
        initialized; this allows to create specific preconditions for calls,
        such as symmetric positive definite matrices and the placement of
        operands in the cache hierarchy.

    \item Any routine that follow the interface conventions of \blas and \lapack
        (see \cref{app:libs}) can be sampled.

    \item Parallel regions allow to execute several routines in parallel through
        \openmp.  Within such regions, sequential blocks allow run parallel
        sequences of calls instead.

    \item Hardware counters (e.g., for cache misses or stalls) can be analyzed
        through \papi.
\end{itemize}

We conclude this section with an example of simple performance experiments in
the \sampler.  A more detailed presentation of the sampler is given
in~\cite{elaps}, and a complete specification of its interface can be found in
its documentation~\cite{elapsweb}.

\begin{example}{The \sampler}{meas:elaps:sampler}
    We interactively start a \sampler linked with \openblas on a \haswell.  To
    measure the runtime of the matrix-matrix multiplication $\dm C \coloneqq \dm
    A \matmatsep \dm B + \dm C$ (\dgemm[NN]) with $\dm A, \dm B, \dm C \in
    \R^{1000 \times 1000}$, we first allocate three double-precision operands of
    size $1000 \times 1000 = \SI{1000000}\doubles$ as follows:

    \begin{codelisting}
        dmalloc A 1000000
        dmalloc B 1000000
        dmalloc C 1000000
    \end{codelisting}

    To also study the number of Level~3 cache misses, we enable the \papi
    counter \code{PAPI\_L3\_TCM}:

    \begin{codelisting}
        set_counters PAPI_L3_TCM
    \end{codelisting}

    Next, we pass five repeated \dgemm-calls to the \sampler and start the
    measurements with the command \code{go}:

    \begin{codelisting}
        dgemm N N 1000 1000 1000 1 A 1000 B 1000 1 C 1000
        dgemm N N 1000 1000 1000 1 A 1000 B 1000 1 C 1000
        dgemm N N 1000 1000 1000 1 A 1000 B 1000 1 C 1000
        dgemm N N 1000 1000 1000 1 A 1000 B 1000 1 C 1000
        dgemm N N 1000 1000 1000 1 A 1000 B 1000 1 C 1000
        go
    \end{codelisting}

    After roughly \SI{340}\ms, we receive the following output:

    \begin{codelisting}
        146867632   47155
        143853672   10981
        143771180   7144
        143439224   6764
        143589228   6542
    \end{codelisting}

    Here, each line corresponds to one of the five \dgemm invocations, while the
    first and second entry, respectively, report the number of cycles and
    Level~3 cache misses.  The first \dgemm[NN] causes considerable more cache
    misses than the following and has a slightly higher runtime.

    Next, we measure $\dv y \coloneqq 1.5 \dv x + \dv y$ (\daxpy) with $\dv x,
    \dv y \in \R^{\num{100000}}$ using ad-hoc memory locations for the vectors:

    \begin{codelisting}
        daxpy 100000 1.5 [100000] 1 [100000] 1
        daxpy 100000 1.5 [100000] 1 [100000] 1
        daxpy 100000 1.5 [100000] 1 [100000] 1
        daxpy 100000 1.5 [100000] 1 [100000] 1
        daxpy 100000 1.5 [100000] 1 [100000] 1
    \end{codelisting}

    We end the input stream (\code{ctrl+D}) and the \sampler produces the
    following output before terminating:

    \begin{codelisting}
        209740	760
        157047	0
        156753	0
        157022	0
        157088	0
    \end{codelisting}

    Of the five \daxpy{}s only the first caused 760~cache misses because it
    needs to load the kernel itself (the operands were randomized prior to the
    measurements and thus are still in cache); as a result, the first execution
    of the inherently memory-bound \blasl1 kernel took about~\SI{27}{\percent}
    longer than the following.
\end{example}

While the \sampler can be used interactively, its interface mainly intended
for scripting, which allows its use in various components throughout this work.
For interactive use, the \elaps{} \python Framework offers a user-friendly
interface and tools.

        \subsection{The \elaps{} \python Framework}
        \label{sec:meas:elapslib}
        The \definition{\elaps{} \python Framework} provides a comprehensive set of
tools to facilitate easy and fast, yet powerful performance experimentation in
dense linear algebra.  It covers various aspects of performance studies:
\begin{itemize}
    \item Users can easily design \definition{experiments} either
        through \python scripts or a specialized graphical user interface (GUI):
        the \definition{\playmat}. Such experiments allow to investigate how
        performance and efficiency vary depending on factors such as caching,
        algorithmic parameters, problem size, and parallelism.  The experiment
        design is assisted by features such as built-in knowledge of \blas and
        \lapack signatures and the automatic propagation of problem sizes to
        various operands within and across routine calls.

    \item With a simple click (or a method call), an experiment's measurements
        are \definition[local or remote execution]{executed} using a compiled
        \sampler.  Here, a wide range of execution setups are possible, ranging
        from {\em local} executions on laptops, workstations, or interactive
        nodes to {\em remote executions} on accelerators or clusters and
        super-computers through batch-job schedulers.

    \item The measurements result in experiment \definition{reports} that can be
        evaluated through further tools and a separate GUI: the
        \definition{\viewer}.  These cover the core aspects of performances
        analyses, such as applying different metrics (e.g., runtime~[\ms{}],
        performance~[\si{\giga\flops\per\second}], efficiency~[\percent{}]),
        combining measurement repetitions into summary statistics (e.g.,
        minimum, median, mean), generating publication-quality
        \definition[plots\\data exports]{plots}, and {\em exporting raw data}.
\end{itemize}

Since we are concerned with performance modeling and prediction, covering
\elaps's whole spectrum of features for performance experimentation would exceed
this work's focus and scope---interested readers are referred to~\cite{elaps}
and encouraged to clone the project from \github~\cite{elapsweb}.  At this
point, we limit the presentation of \elaps to two examples: one that
demonstrates the installation process, and another that shows a typical workflow
of designing and evaluating a performance experiment through the GUIs.

\begin{example}{\elaps installation}{meas:elaps:install}
    In this example, we work on a dedicated \sandybridge remotely through
    \code{ssh}; \openblas, \python~2.7, {\swstyle PyQt4}, and {\swstyle
    matplotlib} are already available.  We begin by cloning \elaps:

    \begin{codelisting}
        !\color{red}{\$}! git clone https://github.com/elmar-peise/ELAPS.git
        !\it [...]!
        !\color{red}{\$}! cd ELAPS
    \end{codelisting}

    Next, we create a \sampler configuration \code{Sampler/cfg/OpenBLAS.cfg}
    (from the provided template) to compile a \sampler with \openblas:

    \begin{codelisting}
        !\color{red}{\$}! cd Sampler
        !\color{red}{\$}! cat cfg/OpenBLAS.cfg
        . ./gathercfg.sh
        DFLOPS_PER_CYCLE=8
        LINK_FLAGS="-L!\it /path/to/openblas!/lib/ \
            -lopenblas -lgfortran"
        BACKEND_PREFIX="OPENBLAS_NUM_THREADS={nt}"
        !\color{red}{\$}! ./make.sh cfgs/OpenBLAS.cfg
        !\it [...]!
        !\color{red}{\$}! cd ..
    \end{codelisting}

    As part of the configuration file, \code{gathercfg.sh} automatically detects
    various hardware properties, such as the processor model and frequency, and
    number of available sockets, cores, and (hyper-)threads.

    Now \elaps is ready for experimentation.
\end{example}

\begin{figure}\figurestyle

    \includegraphics[width=\textwidth]{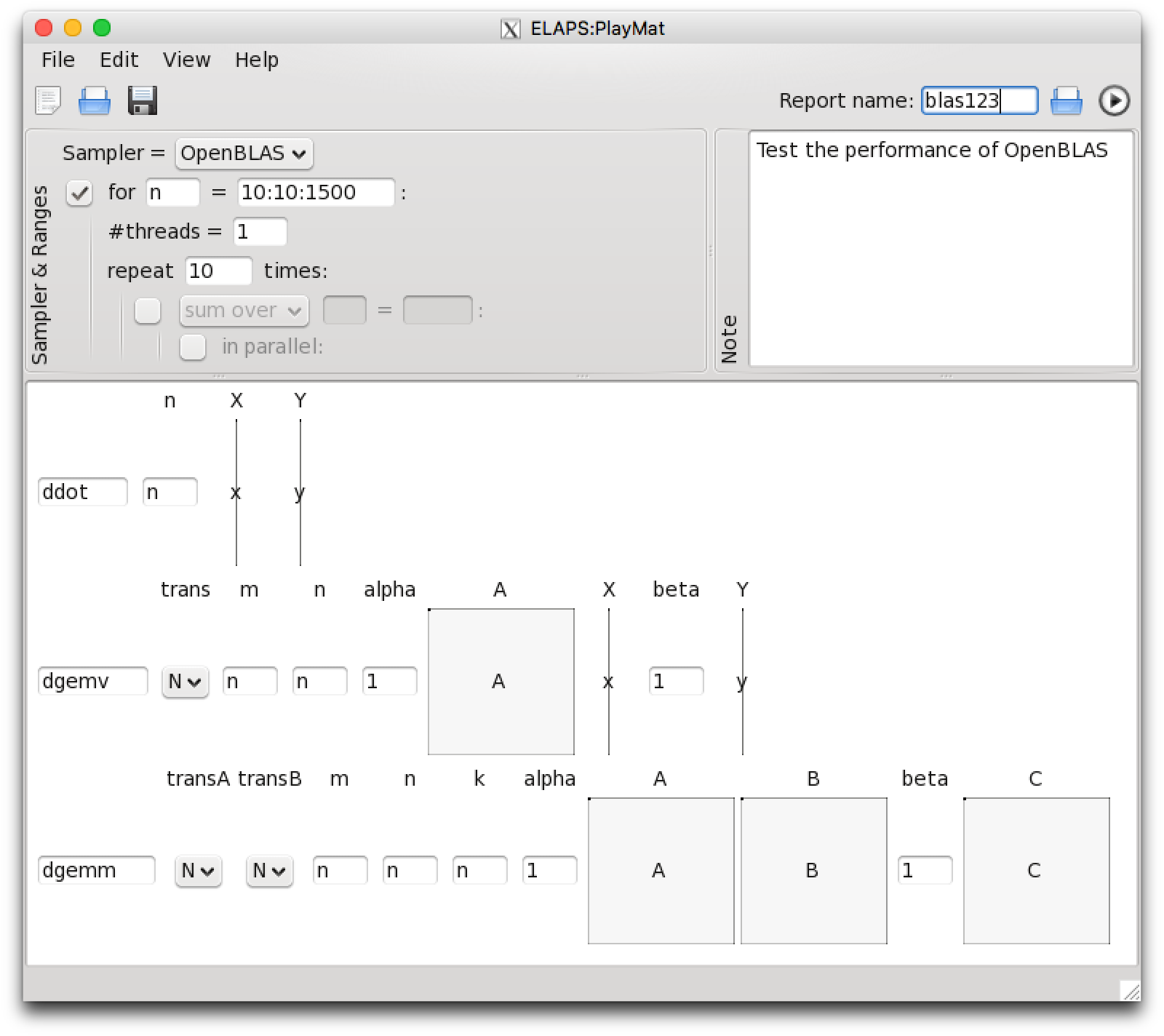}

    \caption{%
        Setting up an \elaps experiment in the \playmat via \software{X11}.
    }
    \label{fig:meas:elaps:playmat}
\end{figure}

\begin{figure}\figurestyle

    \includegraphics[width=\textwidth]{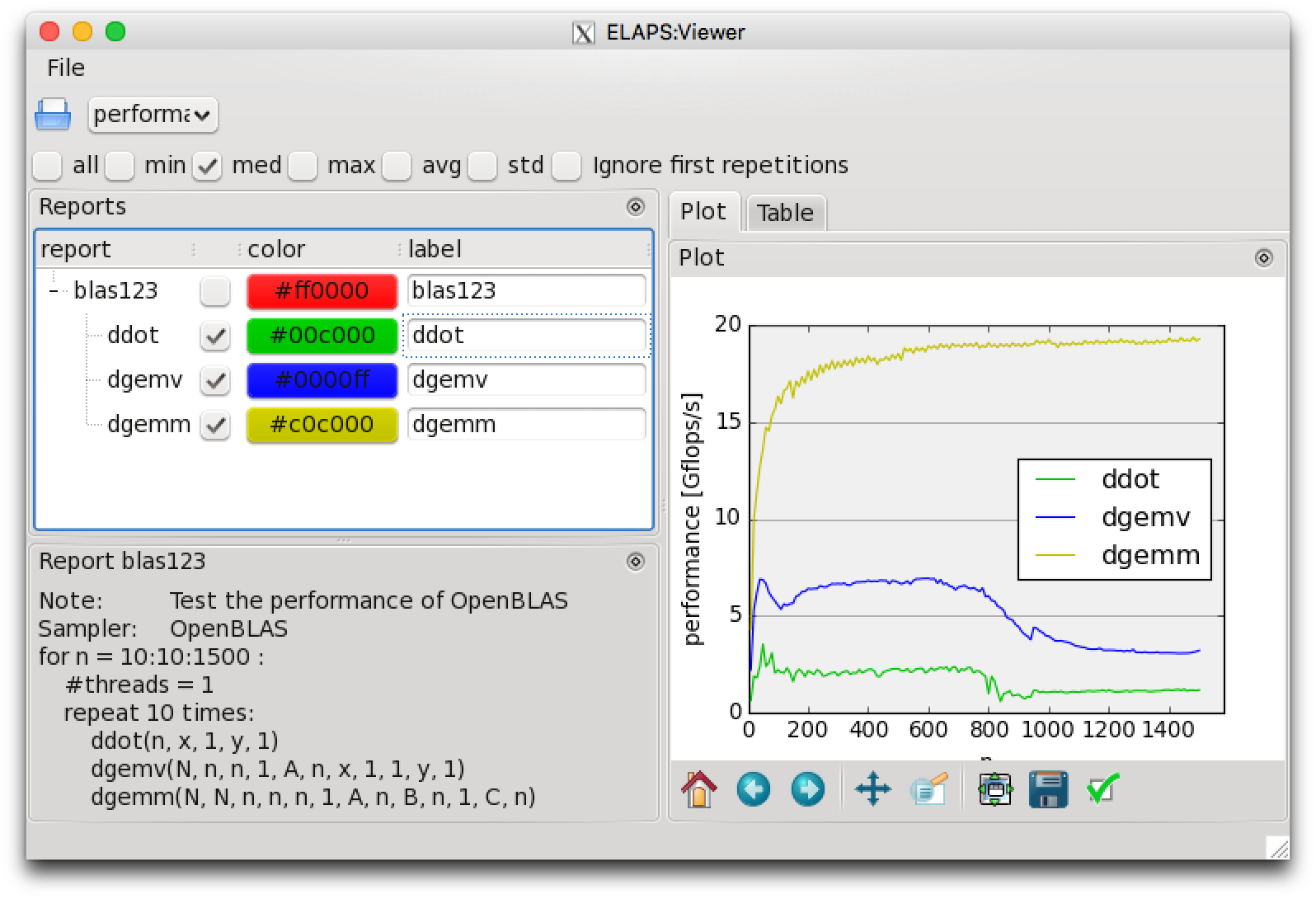}

    \mycaption{The \elaps{} \viewer showing a performance plot.}
    \label{fig:meas:elaps:viewer}
\end{figure}

\begin{example}{\elaps workflow}{meas:elaps:exp}
    To evaluate the \openblas library on our \sandybridge, we measure the
    performance of the representative \blasl1, 2, and~3 kernels \ddot, \dgemv,
    and \dgemm.  We start the \playmat (\code{bin/PlayMat}) and through a few
    clicks construct the experiment shown in \cref{fig:meas:elaps:playmat}. It
    consists of the three operations $\alpha \coloneqq \dm[height=0, ']x \dv y$
    (\ddot), $\dv y \coloneqq \dm A \matvecsep \dv x + \dv y$ (\dgemv[N]), and
    $\dm C \coloneqq \dm A \matmatsep \dm B + \dm C$ (\dgemm[NN]) with $\dm A,
    \dm B, \dm C \in \R^{n \times n}$ and $\dv x, \dv y \in \R^n$, and
    increasing problem size~$n = 10, 20, \ldots, 1500$; for each problem size
    the three operations are repeated 10 times.

    A further click starts the experiment execution on the \sampler compiled in
    \cref{ex:meas:elaps:install}.  We open the resulting report in the
    \viewer and quickly obtain a plot of the three routines' median performance
    as seen in \cref{fig:meas:elaps:viewer}.

    The results show that the performance of the compute-bound \dgemm[NN]
    quickly increases with the problem size and plateaus around
    \SI{19.3}{\giga\flops\per\second}; considering the \sandybridgeshort's
    single-threaded peak floating-point performance of
    \SI{20.8}{\giga\flops\per\second} (\turboboost disabled), this corresponds
    to an efficiency of~\SI{92.79}\percent.  The performance of the memory-bound
    \dgemv[N] and \ddot on the other hand is considerably lower and only
    reaches, respectively, \SIlist{6.7;2.3}{\giga\flops\per\second}.  However,
    from problem size~$n = 800$ to~1000, the performance of these kernels drops
    by roughly a factor of~2, because their operands (\SIvar{3 n^2 + 2
    n}\doubles) are larger than the last-level cache (L3) of \SI{20}{\mebi\byte}
    beyond~$n = 935$.
\end{example}

    \section{Summary}
    \label{sec:meas:conclusion}
    This chapter covered the basic phenomena and tools encountered throughout this
work:  It gave an overview of important effects on the performance of dense
linear algebra kernels, including overheads, fluctuations, thread pinning, and
caching.  It then introduced the runtime and performance measurement and
analysis framework \elaps, which serves as the basis for all experiments,
modeling procedures, and benchmarks throughout this work.

}

    \chapter{Performance Modeling}
\chapterlabel{model}
{
    \newcommand\dmA[1][]{\dm[lower, #1]A\xspace}
\newcommand\dmAi[1][]{\dm[lower, inv, #1]A\xspace}
\newcommand\dmB[1][]{\dm[#1]B\xspace}

\newcommand\x{\ensuremath{\boldsymbol x}}

\newcommand\Q[3]{\ensuremath{#1^\mathrm{#2}_\mathrm{#3}}}
\newcommand\Qi[3]{\ensuremath{#1^{#2}_\mathrm{#3}}}

    Many dense linear algebra operations, such as matrix decompositions, reductions,
and inversions are commonly implemented as blocked algorithms.  Since such
algorithms generally cast their entire computation as a sequence of calls to
\blasl3 and unblocked \lapack kernels, we predict their runtime by estimating and
summing the runtime of these calls.  To motivate how we obtain such estimates
for the underlying kernels, recall (from~\cref{sec:intro:blocked:algs}) that
every blocked algorithm traverses the input matrix (or matrices) with a fixed
block size, and in each traversal step it performs the same kernel operations on
the exposed sub-matrices.  The sizes of these sub-matrices depend on three
factors: the input problem size, the block size, and the traversal progress.
Therefore, in order to predict blocked algorithms, we seek a procedure to
estimate the runtime of a few compute kernels with potentially widely varying
operand sizes.

Our solution to obtain such estimates is measurement-based performance models:
For each hardware and software setup and each compute kernel, we construct a
separate performance model that represents the kernel's runtime as a function of
its arguments.  To efficiently obtain highly accurate models, we tailor them
specifically to dense linear algebra computations.

The remainder of this chapter is concerned with the design and automated
generation of such models:
\begin{itemize}
    \item To guide the development of our models, \cref{sec:model:args} studies
        how the runtime of dense linear algebra kernels depends on their
        arguments.  The study reveals the effects of different argument types:
        While some have little to no effect and can thus be safely ignored in
        our models to reduce their complexity (i.e., their dimensionality),
        others require careful treatment.

    \item Based on these insights, \cref{sec:model:generation} introduces the
        structure of our performance models and their automated
        adaptive-refinement-based generation.

    \item \Cref{sec:model:config} presents the configuration options of the
        modeling process and analyzes the resulting models.  It studies the
        trade-off between low model generation cost versus high accuracy, and
        determines a suitable configuration to generate all models for our
        predictions.
\end{itemize}
Following the design and generation of our models, \cref{ch:pred} employs them
to predict the performance of blocked algorithms and evaluate the predictions'
accuracy and practical value.

\paragraph{Publication}
The work presented in this chapter is in parts based on research previously
published in:
\begin{pubitemize}
    \pubitem{caching}
    \pubitem{pred}
    \pubitem{msthesis}
\end{pubitemize}

    \section{Kernel Argument Analysis}
    \label{sec:model:args}
    Although maximizing our models' accuracy is our primary focus, we aim to avoid
unnecessary complexity and generation cost.  For this purpose, we base our model
design on domain-specific knowledge regarding the performance influence of
various kernel arguments, which is built up and illustrated in this section.

While dense linear algebra kernels typically have between 5 and 15~arguments,
these arguments' semantics divide them among a small set of \definition{argument
types}.  These argument types play distinct roles in the kernel operation,
and heave significantly different effects on the attained performance.  In the
following we study each argument type, and then use the obtained knowledge to
design performance models to best represent the observed features.

We consider the following argument types, which cover all \blas and most \lapack
routines:
\begin{itemize}
    \item \definitionp[flag]{Flag} arguments identify the form of the operation,
        such as the order of operands and transpositions
        (\cref{sec:model:args:flag}).

    \item \definitionp{size} arguments specify the operand sizes
        (\cref{sec:model:args:size}).

    \item \definitionp[scalar]{Scalar} arguments contain real or complex scalars
        that typically multiply (parts of) an operation
        (\cref{sec:model:args:scalar}).

    \item \definitionp[data]{Data} arguments are (pointers to) vector and matrix
        operands (\cref{sec:model:args:data}).

    \item \definitionp[leading dimension]{Leading dimension} arguments accompany
        matrix arguments and specify the distance in memory between two
        consecutive entries in each matrix row (\cref{sec:model:args:ld}); they
        allow algorithms to operate not only on contiguously stored matrices but
        also on sub-matrices.

    \item \definitionp[increment]{Increment} arguments similarly accompany
        vectors and specify the distance between consecutive entries
        (\cref{sec:model:args:inc}); they allow to operate not only on
        contiguous (column) vectors but, e.g., on rows of matrices.
\end{itemize}

\begin{example}{Argument types}{model:args:types}
    \renewcommand\dmB[1][]{\dm[width=.7, #1]B\xspace}
    Let us consider \dtrsm, the double-precision triangular linear system solver
    with  multiple right-hand-sides (e.g., $\dmB \coloneqq \dmAi \dmB$).  This
    representative \blasl3 kernel contains most of the above argument types, and
    is a key component for many \lapack-level algorithms; hence it is an ideal
    candidate to illustrate both the semantics of the argument types in this
    example and their performance effects in the following sections.

    \dtrsm is invoked with 11~arguments:
    \[\text{%
        \newcommand\argnode[1]{%
            \hspace{-.3333em}%
            \tikz[remember picture, baseline=(#1.base)]
                \node (#1) {#1\mstrut};%
            \hspace{-.3333em}%
        }%
        \dtrsm\code{(%
            \argnode{side}, \argnode{uplo}, \argnode{transA}, \argnode{diag},
            \argnode m, \argnode n, \argnode{alpha},
            \argnode A, \argnode{ldA}, \argnode B, \argnode{ldB}%
        )}%
        \hspace{-.3333em}%
        \tikz[baseline=(x.base)]\path node (x) {\mstrut}
            (x.base) node[above=\baselineskip, anchor=base] {\mstrut}
            (x.base) node[below=1.5\baselineskip, anchor=base] {\mstrut};%
        \hspace{-.3333em}%
        \begin{tikzpicture}[
                overlay, remember picture,
                every edge/.append style={thick, blue, -Stealth},
                every node/.append style={
                    midway,
                    below=1.5\baselineskip,
                    anchor=base
                }
            ]
            \path (side.base) -- (diag.base) node {flags}
                edge (side) edge (uplo) edge (transA) edge (diag);
            \path (m.base) -- (n.base)
                node[above=1.5\baselineskip, anchor=base] {sizes}
                edge (m) edge (n);
            \path (alpha.base) -- (alpha.base) node {scalar} edge (alpha);
            \path (A.base) -- (B.base) node {data} edge (A) edge (B);
            \path (ldA.base) -- (ldB.base)
                node[above=1.5\baselineskip, anchor=base] {leading dimensions}
                edge (ldA) edge (ldB);
        \end{tikzpicture}%
    } \enspace.\]
    The semantics of these arguments are as follows:
    \begin{itemize}
        \item \code{side}, \code{uplo}, \code{transA}, and \code{diag} are flag
            arguments.
            \begin{itemize}
                \item $\code{side} \in \{\code L, \code R\}$ determines from
                    which side \dmB is multiplied with \dmAi, i.e., the {\em
                    left} ($\dmB \coloneqq \dmAi \dmB$) or {\em right} ($\dmB
                    \coloneqq \dmB \matmatsep \dmAi[size=.7]$),
                \item $\code{uplo} \in \{\code L, \code U\}$ indicates a {\em
                    lower}- or {\em upper}-triangular system matrix
                    (\dmA\lowerpostsep or~\lowerpostsep\dm[upper]A),
                \item $\code{transA} \in \{\code N, \code T\}$ specifies whether
                    \dmA\lowerpostsep appears {\em non-transposed} or {\em
                    transposed}, and
                \item $\code{diag} \in \{\code N, \code U\}$ determines whether
                    the diagonal entries of~\dmA\lowerpostsep are stored {\em
                    normally} or all implicitly equal to~1, making
                    \dmA\lowerpostsep{} ,,{\em unit triangular}''.
            \end{itemize}

            All $2^4 = 16$~combinations of these four flag arguments are
            possible.  For instance, $(\code{side}, \code{uplo}, \code{transA},
            \code{diag}) = (\code L, \code U, \code N, \code N)$ identifies the
            operation $\dmB \coloneqq \dm[upper, inv]A \dmB$, and $(\code R,
            \code L, \code T, \code N)$ yields $\dmB \coloneqq \dmB \matmatsep
            \dm[lower, size=.7, inv']A$.

        \item \code m and~\code n are size arguments; they determine the size of
            $\dmB \in \R^{\code m \times \code n}$ and accordingly
            $\dmA\lowerpostsep \in \R^{\code m \times \code m}$ if $\code{side}
            = \code L$ and $\dmA[size=.7] \in \R^{\code n \times \code n}$ if
            $\code{side} = \code R$.

        \item \code{alpha} is a scalar argument; it multiplies the whole linear
            system, i.e., $\dmB \coloneqq \alpha \dmAi \dmB$.

        \item \code A and~\code B are data arguments; they represent the
            operands \dmA\lowerpostsep and~\dmB (as pointers to their first
            entries).

        \item \code{ldA} and~\code{ldB} are leading dimension arguments for,
            respectively, \code A and~\code B.
    \end{itemize}

    A brief overview of not only \dtrsm but all \blas and \lapack routines used
    throughout this work and their arguments is given in \cref{app:libs}.
\end{example}

In the following, we consider the influence of each argument type on the
performance of kernels, and determine how they shall be handled in our models.

        \subsection{Flag Arguments}
        \label{sec:model:args:flag}
        Flag arguments accept only a few discrete values---in most cases two.  However,
since they specify which form of the operation is performed, they may trigger
entirely different execution branches in kernel implementations, and thus result
in independent runtimes.

\begin{figure}\figurestyle
    \ref*{leg:blas:circ}

    \raggedleft
    \tikzsetnextfilename{flags}
    \begin{tikzpicture}
        \begin{axis}[
            ylabel=runtime, y unit=\ms,
            xmin=-.5, xmax=15.5,
            ymax=1.6,
            xtick={0, ..., 15},
            xticklabels={
                {\code L \code L \code N \code N},
                {\code L \code L \code N \code U},
                {\code L \code L \code T \code N},
                {\code L \code L \code T \code U},
                {\code L \code U \code N \code N},
                {\code L \code U \code N \code U},
                {\code L \code U \code T \code N},
                {\code L \code U \code T \code U},
                {\code R \code L \code N \code N},
                {\code R \code L \code N \code U},
                {\code R \code L \code T \code N},
                {\code R \code L \code T \code U},
                {\code R \code U \code N \code N},
                {\code R \code U \code N \code U},
                {\code R \code U \code T \code N},
                {\code R \code U \code T \code U}
            },
            every x tick label/.append style={text width=.3cm, align=center},
            every axis plot/.append style={only marks, ultra thick},
            legend columns=4,
            legend to name=leg:blas:circ,
            table/search path={model/figures/data/flags},
        ]
            \addlegendimage{empty legend}
            \addlegendentry{\sandybridge:}
            \addplot[plotsbopen] table {sandy_open.dat};
            \addlegendentry{\openblas}\label{plt:sbopen:circ}
            \addplot[plotsbblis] table {sandy_blis.dat};
            \addlegendentry{\blis}\label{plt:sbblis:circ}
            \addplot[plotsbmkl] table {sandy_mkl.dat};
            \addlegendentry{\mkl}\label{plt:sbmkl:circ}
            \addlegendimage{empty legend}
            \addlegendentry{\haswell:}
            \addplot[plothwopen] table {haswell_open.dat};
            \addlegendentry{\openblas}\label{plt:hwopen:circ}
            \addplot[plothwblis] table {haswell_blis.dat};
            \addlegendentry{\blis}\label{plt:hwblis:circ}
            \addplot[plothwmkl] table {haswell_mkl.dat};
            \addlegendentry{\mkl}\label{plt:hwmkl:circ}
        \end{axis}
        \draw (0, 0) node[anchor=north east, align=left]
            {\code{side}\\\code{uplo}\\\code{transA}\\\code{diag}};
    \end{tikzpicture}

    \mycaption{%
        Runtime of \dtrsm as a function of its flag arguments.
        \captiondetails{$m = n = 256$, 1~thread, median of 100~repetitions}
    }
    \label{fig:model:flags}
\end{figure}

\begin{example}{Flag arguments}{model:args:flags}
    \Cref{fig:model:flags} shows the runtime of
    \displaycall\dtrsm{
        \varg{side}{side}, \varg{uplo}{uplo},
        \varg{transA}{transA}, \varg{diag}{diag},
        \arg m{256}, \arg n{256}, \arg{alpha}{1.0},
        \arg AA, \arg{ldA}{256}, \arg BB, \arg{ldB}{256}
    }
    i.e., an operation like $\dmB \coloneqq \dmAi \dmB$ with $\dmA\lowerpostsep,
    \dmB \in \R^{256 \times 256}$, for all 16~combinations of the flag arguments
    \code{side}, \code{uplo}, \code{transA}, and \code{diag} on a \sandybridge
    and a \haswell with single-threaded \openblas, \blis, and \mkl.

    Across all systems and libraries, we encounter a large spectrum of
    performance dependencies, which cannot be summarized in a single pattern.
    In particular, each argument influences the runtime of the
    implementations differently:
    \begin{itemize}
        \item For non-square $\dmB[width=.7] \in \R^{m \times n}$, \code{side}
            affects the \dtrsm's minimal \flop-count:  While for $\code{side} =
            \code L$ its cost is \SIvar{m^2 n}\flops, for $\code{side} = \code
            R$ it is \SIvar{m n^2}\flops.  Hence changing the value of
            \code{side} will generally lead to an entirely different runtime.

            Since this example uses $m = n = 256$, the \dtrsm requires at least
            $\SIvar{256^3}\flops$ for both values of \code{side}.  However, in
            our measurements, \code{side} still has the largest impact on
            performance, which is most evident for \openblas:  While on the
            \sandybridgeshort~(\ref*{plt:sbopen:circ}) the \dtrsm takes on
            average \SI{104.52}{\micro\second} (\SI{8.35}\percent) longer for
            $\code{side} = \code L$ than with $\code{side} = \code R$, on the
            \haswell~(\ref*{plt:hwopen:circ}) $\code{side} = \code L$ is
            \SI{82.845}{\micro\second} (\SI{9.06}\percent) slower than
            $\code{side} = \code R$.

        \item The effects of \code{uplo} and \code{transA} are closely related,
            which is most evident in \blis~(\ref*{plt:sbblis:circ},
            \ref*{plt:hwblis:circ}).  Possibly due to the similarity of the
            operations, $(\code{uplo}, \code{transA}) = (\code L, \code N)$ and
            $(\code U, \code T)$ commonly share a runtime that is different from
            $(\code L, \code T)$ and $(\code U, \code N)$.

        \item \code{diag} has almost no influence on the runtime of most
            implementations.  Only \mkl~(\ref*{plt:sbmkl:circ},
            \ref*{plt:hwmkl:circ})---the fastest implementation across all
            setups---takes advantage of $\code{diag} = \code U$, and avoids the
            division instructions.
    \end{itemize}

    Note that both the magnitude of the flag arguments' influence as well as the
    type of the resulting runtime characteristics vary both from one
    architecture to another and between implementations.
\end{example}

Since flag arguments can have a decisive impact on a kernel's runtime with no
general discernible patters across architectures and implementations, we will
generate a separate performance (sub-)model for each different combination of
flags.  However, note that in our target range of algorithms, we encounter only
a limited set of such combinations, and will therefore not generate models for
all possibilities.

        \subsection{Scalar Arguments}
        \label{sec:model:args:scalar}
        At first sight, scalar arguments should not have any effect on a kernel's
runtime---after all, they only scale a kernel operand independent of the
argument's value.  However, at closer inspection, we find that for certain
values---namely $-1$, 0, and~1---this multiplication can be avoided altogether.
Since in applications and algorithms, scalar arguments to kernels are almost
exclusively $-1$, 0, and~1, most kernel implementations feature optimized
execution branches for these values.  Just as for flag arguments, such branches
can noticeably impact a kernel's runtime and performance.

\begin{figure}\figurestyle
    \ref*{leg:model:scalar}

    \raggedleft
    \tikzsetnextfilename{scalar}
    \begin{tikzpicture}
        \begin{axis}[
                ybar, every axis plot/.append style=ybar legend,
                ylabel=runtime, y unit=\ms,
                xmin=-.75, xmax=6.25,
                xtick={0, 1, 2, 3.5, 4.5, 5.5},
                xticklabels={\openblas, \blis, \mkl, \openblas, \blis, \mkl},
                legend to name=leg:model:scalar,
                table/search path={model/figures/data/scalar},
            ]
            \addplot table {0.6.dat};
            \addlegendentry{$\alpha = 0.6$}\label{plt:model:size:0.6}
            \addplot+[plot4] table {-1.dat};
            \addlegendentry{$\alpha = -1$}\label{plt:model:size:-1}
            \addplot+[plot3] table {1.dat};
            \addlegendentry{$\alpha = 1$}\label{plt:model:size:1}
            \addplot+[plot2] table {0.dat};
            \addlegendentry{$\alpha = 0$}\label{plt:model:size:0}
            \coordinate (sandy) at (axis cs: 1, 0);
            \coordinate (haswell) at (4.5, 0);
        \end{axis}
        \path (sandy) ++(0, -1) node {\sandybridge};
        \path (haswell) ++(0, -1) node {\haswell};
    \end{tikzpicture}

    \mycaption{%
        Runtime of \dtrsm[LLNN] with different values for~$\alpha$.
        \captiondetails{$m = 100$, $n = 800$, 1~thread, median of
        100~repeititons}
    }
    \label{fig:model:scalar}
\end{figure}

\begin{example}{Scalar arguments}{model:args:scalar}
    \renewcommand\dmAi{\dm[size=.125, lower, inv]A\xspace}%
    \renewcommand\dmA{\dm[size=.125, lower]A\xspace}%
    \renewcommand\dmB{\dm[height=.125]B\xspace}%
    \Cref{fig:model:scalar} shows the runtime of
    \displaycall\dtrsm{
        \arg{side}L, \arg{uplo}L, \arg{transA}N, \arg{diag}N,
        \arg m{100}, \arg n{800}, \varg{alpha}{alpha},
        \arg AA, \arg{ldA}{100}, \arg BB, \arg{ldB}{100}
    }
    i.e., $\dmB \coloneqq \alpha \dmAi \dmB$ with $\dmA \in
    \R^{100 \times 100}$ and $\dmB \in \R^{100 \times 800}$, for $\alpha \in
    \{0.6, 0, -1, 1\}$ on a \sandybridge and a \haswell with single-threaded
    \openblas, \blis, and \mkl.  While $\alpha = 0.6$ represents the ``general
    case'', $\alpha = 0$, $-1$, and~1 are special values for which
    implementations can avoid multiplications---in algorithms and applications
    $\alpha = 1$ and~$-1$ are the most common values.

    All implementations take advantage of $\alpha =
    0$~(\ref*{plt:model:size:0}).  In this case, the \dtrsm[LLNN] only sets
    $\dmB \coloneqq \dm[height=.125]0$ and no computations are performed.
    Furthermore, all implementations treat $\alpha =
    -1$~(\ref*{plt:model:size:-1}) just like the general
    case~(\ref*{plt:model:size:0.6}) resulting in the same runtime.

    $\alpha = 1$~(\ref*{plt:model:size:1}) is handled differently by the three
    implementations:  While \blis attains the same performance as for $\alpha =
    0.6$ and~$-1$, \openblas and \mkl are on average \SI{9.66}{\percent} faster
    compared to these cases, indicating optimizations that avoid multiplications
    with~1.  While we can appreciate that \openblas and \mkl are faster for
    $\alpha = 1$, put into perspective the increase in runtime for other values
    of~$\alpha$ is surprisingly high:  In our example, scaling \dmB accounts for
    only~\SI1{\percent} of the \dtrsm[LLNN]'s minimal \flop-count, yet makes the
    operation almost~\SI{10}{\percent} slower.
\end{example}

To represent the influence of scalar arguments on kernel performance in our
models, we will treat them like flag arguments with the four possible values
$-1$, 0, 1, and ``any other value''.  Since blocked algorithms almost
exclusively use the values $-1$ and~1, we will not observe a four-fold increase
in the complexity of our models.

        \subsection{Leading Dimension Arguments}
        \label{sec:model:args:ld}
        Leading dimension arguments determine the memory access strides of kernels that
load multiple columns of a matrix simultaneously.  They only have a small
influence on kernel performance, but we need to be aware of certain patterns to
avoid undesirable effects when generating our performance models.

\subsubsection{Alignment to Cache-Lines}

Data is moved through the memory hierarchy in blocks of \SI{64}{\bytes} ($=
\SI8\doubles$) called \definition{cache-lines}.\footnote{%
    The cache-line size is generally not fixed but for most processors it is
    \SI{64}{\Byte}.
}  Hence using multiples of the cache-lines size as memory access strides
typically shows a more regular and often better performance compared to other
strides.

\begin{figure}\figurestyle
    \ref*{leg:blas}

    \raggedleft
    \tikzsetnextfilename{ld8}
    \begin{tikzpicture}
        \begin{axis}[
            xlabel=leading dimension $ld$,
            ylabel=time, y unit=\ms,
            xmin={},
            ymax=1.5,
            xtick={256, 264, ..., 320},
            cycle list name=blas,
            legend columns=4,
            legend to name=leg:blas,
            table/search path={model/figures/data/ld8},
        ]
            \draw[dotted] foreach \x in {256, 264, ..., 320}
                {(\x, 0) -- (\x, 1.5)};
            \addlegendimage{empty legend}
            \addlegendentry{\sandybridge:}
            \addplot table {sandy_open.dat};
            \addlegendentry{\openblas}\label{plt:sbopen}
            \addplot table {sandy_blis.dat};
            \addlegendentry{\blis}\label{plt:sbblis}
            \addplot table {sandy_mkl.dat};
            \addlegendentry{\mkl}\label{plt:sbmkl}
            \addlegendimage{empty legend}
            \addlegendentry{\haswell:}
            \addplot table {haswell_open.dat};
            \addlegendentry{\openblas}\label{plt:hwopen}
            \addplot table {haswell_blis.dat};
            \addlegendentry{\blis}\label{plt:hwblis}
            \addplot table {haswell_mkl.dat};
            \addlegendentry{\mkl}\label{plt:hwmkl}
        \end{axis}
    \end{tikzpicture}

    \mycaption{%
        Runtime of \dtrsm as a function of its leading dimension arguments on a
        small scale.
        \iftableoflist\else Dotted lines: multiples of~8.\fi
        \captiondetails{$m = n = 256$, 1~thread, median of 100~repetitions}
    }
    \label{fig:model:ld8}
\end{figure}

\footnotetextbefore{%
    Since $A$ and~$B$ have 256~rows, the leading dimensions are at least~256.
}
\begin{example}{Aligning leading dimensions to cache-lines}{model:args:ld:8}
    \Cref{fig:model:ld8} shows the runtime of
    \displaycall\dtrsm{
        \arg{side}L, \arg{uplo}L,
        \arg{transA}N, \arg{diag}N,
        \arg m{256}, \arg n{256},
        \arg{alpha}{1.0},
        \arg AA, \arg{ldA}{\it\color{blue}ld},
        \arg BB, \arg{ldB}{\it\color{blue}ld}
    }
    i.e., $\dmB \coloneqq \dmAi \dmB$ with $\dmA\lowerpostsep, \dmB \in \R^{256
    \times 256}$, for leading dimensions\footnotemark{} $ld = 256, \ldots, 320$
    in steps of~1 on a \sandybridge and a \haswell with single-threaded
    \openblas, \blis, and \mkl.

    For all setups, the \dtrsm[LLNN]'s runtime exhibits some regular pattern in
    terms of the leading dimension arguments---with an average amplitude
    of~\SI{2.19}\percent.  However the patterns are quite different:  While
    \openblas's runtime on the \sandybridgeshort~(\ref*{plt:sbopen}) drops
    equally at every even leading dimension, \mkl on the
    \haswellshort~(\ref*{plt:hwmkl}) dips only at multiples of~4, and on the
    \sandybridgeshort~(\ref*{plt:sbmkl}) it has stronger dips at multiples of~8.
    \blis on the other hand shows the exact opposite behavior:  On both
    platforms~(\ref*{plt:sbblis}, \ref*{plt:hwblis}) its runtime spikes slightly
    at multiples of~8.

    Independent of the specific behavior of each setup, a smooth runtime curve
    is obtained when only multiples of~8 are considered as leading dimensions.
\end{example}

To avoid small performance irregularities, we will generate our models using
\definition[use multiples of the cache-line size]{multiples of the cache-line
size} for leading dimensions---in double-precision: multiples of~8.

\subsubsection{Set-Associative Cache Conflicts}
\label{sec:model:args:ld512}

The Level~1 and~2 caches in our processors are \definition{8-way
set-associative}:  They are divided into sets of 8~cache-lines, and when a
cache-line is loaded, its address's least significant bits determine which of the
sets it is assigned to; within the set, an architecture-dependent cache
replacement policy determines in which of the 8~slots it is stored.  When the
address space is accessed contiguously, consecutive cache-lines are loaded into
consecutive sets, and the cache is filled evenly.  In the worst case, however,
the address space is accessed with a stride equal to the number of sets, and
all loaded cache-lines are associated to the same set:  Only 8~cache-lines are
cached, and each additional line results in a \definition{cache conflict miss}
causing a recently loaded line to be evicted.  This effect should be avoided
whenever possible.

On recent \intel{} {\namestyle Xeon} processors, the Level~1 data cache~(L1d)
fits \SI{32}{\kibi\byte} organized as 64~sets of 8~cache-lines.  A memory
location with address~$a$ is a part of cache-line~$\lfloor a / 64 \rfloor$ (due
to the size of \SI{64}{\Byte} per line) and assigned to set $\lfloor a / 64
\rfloor \bmod 64$ (due to the capacity of 64~sets).  The Level~2 cache (L2) in
turn fits \SI{256}{\kibi\byte} in 1024~sets; here address~$a$ is assigned to set
$\lfloor a / 64 \rfloor \bmod 1024$.

In a double-precision matrix stored with leading dimension~$ld$, consecutive
elements in each row are $8 ld$~\bytes apart ($\SI1\double = \SI8\bytes$).
Hence, for $ld = 512$, the consecutive row elements starting at address~$a_0$
are stored at~$a_i = a_0 + 8 ld \cdot i  = a_0 + 4096 i$, and associated to the
same set in the L1d~cache:
\begin{align*}
    \left\lfloor \frac{a_i}{64} \right\rfloor \bmod 64
    &= \left\lfloor \frac{a_0 + 4096 i}{64} \right\rfloor \bmod 64 \\
    &= \left(\left\lfloor \frac{a_0}{64} \right\rfloor + 64 i \right) \bmod 64 \\
    &= \left\lfloor \frac{a_0}{64} \right\rfloor \bmod 64.
\end{align*}
The same problem occurs for leading dimensions that are multiples of~512, and
even below~512 powers of~2 have a similar effect:  E.g., with $ld = 256$ the
elements of a row are associated to only two of the cache's 64~sets.  Similarly,
for the L2~cache with 1024~sets, consecutive row-elements are associated to the
same cache set for leading dimensions that are multiples of~8192, and multiples
of~4096 utilize only two sets.

\begin{figure}\figurestyle
    \ref*{leg:blas}

    \raggedleft
    \tikzsetnextfilename{ld512}
    \begin{tikzpicture}
        \begin{axis}[
            xlabel=leading dimension $ld$,
            ylabel=runtime, y unit=\ms,
            ymax=1.5,
            xtick={0, 1024, ..., 8192},
            cycle list name=blas,
            table/search path={model/figures/data/ld512},
        ]
            \draw[dotted] foreach \x in {512, 1024, ..., 8192}
                {(\x, 0) -- (\x, 1.5)};
            \foreach \system in {sandy, haswell}
                \foreach \blas in {open, blis, mkl}
                    \addplot table {\system_\blas.dat};
        \end{axis}
    \end{tikzpicture}

    \mycaption{%
        Runtime of \dtrsm[LLNN] as a function of its leading dimension arguments
        on a large scale.
        \iftableoflist\else Dotted lines: multiples of~512.\fi
        \captiondetails{$m = n = 256$, 1~thread, median of 100~repetitions}
    }
    \label{fig:model:ld512}
\end{figure}

\begin{example}{Cache conflict misses caused by leading
    dimensions}{model:args:ld:512}
    \Cref{fig:model:ld512} shows the runtime of
    \displaycall\dtrsm{
        \arg{side}L, \arg{uplo}L, \arg{transA}N, \arg{diag}N,
        \arg m{256}, \arg n{256}, \arg{alpha}{1.0},
        \arg AA, \varg{ldA}{ld}, \arg BB, \varg{ldB}{ld}
    }
    i.e., $\dmB \coloneqq \dmAi \dmB$ with $\dmA\lowerpostsep, \dmB \in \R^{256
    \times 256}$, for leading dimensions $ld = 256, \ldots, 8320$ in steps
    of~128 on a \sandybridge and a \haswell with single-threaded \openblas,
    \blis, and \mkl.

    For most setups the runtime spikes above the baseline at multiples of~512.
    However, the average magnitude of these spikes ranges
    from~\SI{.14}{\percent} for \blis on the
    \sandybridgeshort~(\ref*{plt:sbblis}) to~\SI{8.37}{\percent} for \openblas
    on the \haswellshort~(\ref*{plt:hwopen}).  Especially for
    \openblas~(\ref*{plt:sbopen}, \ref*{plt:hwopen}), there are additional, yet
    lower spikes of \SI{1.40}{\percent} at multiples of~256.  Furthermore, on
    the \haswellshort for both \openblas~(\ref*{plt:hwopen}) and
    \blis~(\ref*{plt:hwblis}) the spikes are especially high at $ld = 4096$
    and~8192, exceeding the baseline by, respectively,
    \SIlist{6.55;11.24}\percent.
\end{example}

To prevent distortions from unfortunate leading dimensions in our model
generation altogether, we will \definition{avoid multiples of~256} for these
arguments.

Note that by using leading dimensions that are multiples of~8, yet not of~256 in
our measurements, our models will not yield accurate predictions for kernel
invocations that do not follow this pattern.  However, predicting the
performance of such unfortunate invocations, which can be systematically
avoided, is not part of our models' purpose and would exceed the scope of this
work.

        \subsection{Increment Arguments}
        \label{sec:model:args:inc}
        With our focus on predicting algorithms that primarily use \blasl3
(matrix-matrix operations) and unblocked \lapack kernels, the performance of
vector operations is not our primary focus.  However, to make our performance
modeling technique applicable to all types of operations, this section briefly
studies the influence of increment arguments on kernel performance.

Increment arguments directly determine the memory access strides of vector
operands.  In algorithms and applications, they are typically either~1 to access
contiguous vectors (e.g., columns of matrices) or the leading dimension of a
matrix, i.e., $\gg 1$, to access matrix rows.  While in the first case, a vector
of length~$n$ occupies $\lfloor n / 8 \rfloor$~cache-lines,\footnote{%
    Assuming the first entry is aligned to the beginning of a cache-line.
} in the second case it is spread across $n$~cache-lines.  As a result,
increments of~1 cause less data movement and are thus favorable in terms of
performance.

Beyond the ideal increment of~1, the influences of increment arguments on
performance exhibit periodic patterns similar to those for leading dimensions.
However, in comparison the resulting effects are commonly far more severe
because cache misses directly increase the runtime for bandwidth-bound
(matrix-)vector operations.

\begin{figure}\figurestyle
    \ref*{leg:blas}

    \pgfplotsset{
        twocolplot,
        xlabel={vector increment $inc$},
        ylabel={time}, y unit=\si{\micro\second},
        cycle list name=blas,
        xtick={0, 16, ..., 96},
    }

    \begin{subfigaxis}[
            fig caption={\daxpy (\blasl1) \captiondetails{$n = 1024$}},
            fig label=inc:blas1,
            table/search path={model/figures/data/inc1},
            ymax=6
        ]
        \draw[dotted] foreach \x in {16, 32, ..., 96}
            {(\x, 0) -- (\x, 6)};
        \foreach \system in {sandy, haswell}
            \foreach \blas in {open, blis, mkl}
                \addplot table {\system_\blas.dat};
    \end{subfigaxis}\hfill
    \begin{subfigaxis}[
            fig caption={\dtrsv (\blasl2) \captiondetails{$n = 512$}},
            fig label=inc:blas2,
            table/search path={model/figures/data/inc2},
            ymax=300
        ]
            \draw[dotted] foreach \x in {16, 32, ..., 96}
                {(\x, 0) -- (\x, 300)};
            \foreach \system in {sandy, haswell}
                \foreach \blas in {open, blis, mkl}
                    \addplot table {\system_\blas.dat};
    \end{subfigaxis}

    \mycaption{%
        Runtime of \daxpy and \dtrsv[LNN] as a function of their increment
        arguments.
        \iftableoflist\else Dotted lines: multiples of~16.\fi
        \captiondetails{1~thread, median of 100~repetitions}
    }
    \label{fig:mnodel:inc}
\end{figure}

\begin{example}{Increment arguments in \blasl1}{model:args:blas1}
    \Cref{fig:model:inc:blas1} shows the runtime of the \blasl1 calls
    \displaycall\daxpy{
        \arg n{1024}, \arg{alpha}{2.0},
        \arg XX, \varg{incX}{inc}, \arg YY, \varg{incY}{inc}
    }
    i.e., $\dv y \coloneqq 2 \dv x + \dv y$ with $\dv x, \dv y \in \R^{1024}$,
    for increments $inc = 1, \ldots, 100$ in steps of~1 on a \sandybridge and a
    \haswell with single-threaded \openblas, \blis, and \mkl.

    The results for all three implementations are similar on both systems: The
    \daxpy's runtime is shortest for $inc = 1$, and increases steadily until $inc
    = 8$; the difference in performance between these two cases lies between
    $3.53\times$ for \blis on the \sandybridgeshort~(\ref*{plt:sbblis}) (whose
    \blasl1 is not optimized for our architectures) and $18.29\times$ for \mkl
    on the \haswellshort~(\ref*{plt:hwmkl}).

    Beyond $inc = 8$, the runtime spikes above a steady baseline of
    \SI{3.20}{\micro\second} on the \sandybridgeshort~(\ref*{plt:sbopen},
    \ref*{plt:sbblis}, \ref*{plt:sbmkl}) and \SI{2.73}{\micro\second} on the
    \haswellshort~(\ref*{plt:hwopen}, \ref*{plt:hwblis}, \ref*{plt:hwmkl}) by up
    to~\SI{95.88}{\percent} at each multiple of~$32$ and slightly less
    by~\SI{16.80}{\percent} for other multiples of~$16$.
\end{example}

\begin{example}{Increment arguments in \blasl2}{model:args:inc:blas2}%
    \Cref{fig:model:inc:blas2} shows the runtime of the \blasl2 calls
    \displaycall\dtrsv{
        \arg{uplo}L, \arg{trans}N, \arg{diag}N, \arg n{512},
        \arg AA, \arg{ldA}{1000}, \arg XX, \varg{incX}{inc}
    }
    i.e., $\dv x \coloneqq \dmAi \dv x$ with $\dmA\lowerpostsep \in \R^{512
    \times 512}$ and $\dv x \in \R^{512}$, for increments $inc = 1, \ldots, 100$
    in steps of~1 on a \sandybridge and a \haswell with single-threaded
    \openblas, \blis, and \mkl.

    We immediately notice that \blis on the \haswellshort~(\ref*{plt:hwblis})
    has runtime spikes similar to those for \daxpy, which hints at an
    implementation of \dtrsv in terms of  \blasl1 kernels.  For all other
    setups, the runtime is considerably smoother with the exception of
    \mkl~(\ref*{plt:sbmkl}, \ref*{plt:hwmkl}), which shows small spikes
    of~\SI{6.03}{\percent} at multiples of~16.
\end{example}

Since in practice increments are either~1 or equal to the leading dimension of a
matrix, we will treat them in our models like flag arguments that take the
values~1 and ``any large value'', for which we \definition{avoid multiples
of~16} to avoid outlier measurements.

        \subsection{Size Arguments}
        \label{sec:model:args:size}
        A kernel's size arguments determine its minimal \flop-count and thus directly
influence on its runtime.  In the following, we study this influence first for
small changes in the operand sizes (\cref{sec:model:args:size:small}) and then
on a larger scale (\cref{sec:model:args:size:large}).

\subsubsection{Smalls Scale Behavior}
\label{sec:model:args:size:small}

Optimizations of compute kernels commonly involve vectorization and loop
unrolling of length~4 or~8.  These optimizations typically have a direct
influence on a kernel's runtime for small variations of the size arguments.

\begin{figure}\figurestyle
    \ref*{leg:blas}

    \raggedleft
    \tikzsetnextfilename{size8}
    \begin{tikzpicture}
        \begin{axis}[
                xlabel={matrix size $m = n$},
                ylabel=runtime, y unit=\ms,
                xmin={},
                xtick={256, 264, ..., 320},
                cycle list name=blas,
                table/search path={model/figures/data/size8},
            ]
            \draw[dotted] foreach \x in {256, 264, ..., 320}
                {(\x, 0) -- (\x, 3)};
            \foreach \system in {sandy, haswell}
                \foreach \blas in {open, blis, mkl}
                    \addplot table {\system_\blas.dat};
        \end{axis}
    \end{tikzpicture}

    \mycaption{%
        Runtime of \dtrsm[LLNN] as a function of its size arguments on a small
        scale.
        \iftableoflist\else Dotted lines: multiples of~8.\fi
        \captiondetails{1~thread, median of~100 repetitions}
    }
    \label{fig:model:size8}
\end{figure}

\begin{example}{Small variations of size arguments}{model:args:size:8}
    \Cref{fig:model:size8} shows the runtime of
    \displaycall\dtrsm{
        \arg{side}L, \arg{uplo}L, \arg{transA}N, \arg{diag}N,
        \varg mn, \varg nn, \arg{alpha}{1.0},
        \arg AA, \arg{ldA}{400}, \arg BB, \arg{ldB}{400}
    }
    i.e., $\dmB \coloneqq \dmAi \dmB$ with $\dmA\lowerpostsep, \dmB \in \R^{n
    \times n}$, for $n = 256, \ldots, 320$ in steps of~1 on a \sandybridge and a
    \haswell with single-threaded \openblas, \blis, and \mkl.

    All setups show periodic patterns in their runtimes.  While these patterns
    differ between the implementations, most have local runtime minima at
    multiples of~4, and all of them have minima at multiples of~8.
\end{example}

To avoid runtime artefacts introduced by vectorization and loop unrolling, we
will build our models on measurements that \definition{use multiples of~8} for
all size arguments.

\subsubsection{Piecewise Polynomial Behavior}
\label{sec:model:args:size:large}

Since an operation's minimal \flop-count is generally a (multivariate)
polynomial function of the size arguments, one might expect that (for
compute-bound kernels) it translates directly into an equally polynomial
runtime.  However, since a kernel's performance is generally not constant for
varying operand sizes, a single polynomial is often insufficient to accurately
represent a kernel's runtime for large ranges of problem sizes.

\begin{figure}[p]\figurestyle
    \ref*{leg:blas}

    \pgfplotsset{
        xlabel={problem size $n$},
        ymin=-15, ymax=15,
        table/search path={model/figures/data/size},
        plotheightsub=1.11pt
    }

    \begin{subfigaxis}[
            fig caption=Runtime,
            fig label=size,
            twocolplot,
            ylabel=runtime, y unit=\ms,
            ymin=0, ymax={},
            cycle list name=blas
        ]
        \foreach \system in {sandy, haswell}
            \foreach \blas in {open, blis, mkl}
                \addplot table {\system_\blas.dat};
    \end{subfigaxis}\hfill
    \begin{subfigure}\subfigwidth\raggedleft
        \tikzsetnextfilename{size_err1}
        \begin{tikzpicture}
            \begin{axis}[twocolhalfplot1, cycle list name=sbblas]
                \foreach \blas in {open, blis, mkl}
                    \addplot table[y=err1] {sandy_\blas.dat};
                \draw[dashed] foreach \x in {24, 536} {(\x, -15) -- (\x, 15)};
                \coordinate (label1) at (yticklabel cs:.5);
            \end{axis}
            \begin{axis}[twocolhalfplot2, cycle list name=hwblas]
                \draw[dashed] foreach \x in {24, 536}
                    {(\x, -15) -- (\x, 15)};
                \foreach \blas in {open, blis, mkl}
                    \addplot table[y=err1] {haswell_\blas.dat};
                \coordinate (label2) at (yticklabel cs:.5);
            \end{axis}
            \path (label1) -- (label2)
                node[midway, rotate=90, midway, anchor=south]
                    {relative error [\percent{}]};
        \end{tikzpicture}
        \caption{Error for one polynomial}
        \label{fig:model:size:err1}
    \end{subfigure}

    \begin{subfigure}\subfigwidth\raggedleft
        \tikzsetnextfilename{size_err2}
        \begin{tikzpicture}
            \begin{axis}[twocolhalfplot1, cycle list name=sbblas]
                \draw[dashed] foreach \x in {24, 280, 536}
                    {(\x, -15) -- (\x, 15)};
                \foreach \blas in {open, blis, mkl}
                    \addplot table[y=err2] {sandy_\blas.dat};
                \coordinate (label1) at (yticklabel cs:.5);
            \end{axis};
            \begin{axis}[twocolhalfplot2, cycle list name=hwblas]
                \draw[dashed] foreach \x in {24, 280, 536}
                    {(\x, -15) -- (\x, 15)};
                \foreach \blas in {open, blis, mkl}
                    \addplot table[y=err2] {haswell_\blas.dat};
                \coordinate (label2) at (yticklabel cs:.5);
            \end{axis};
            \path (label1) -- (label2)
                node[midway, rotate=90, midway, anchor=south]
                    {relative error [\percent{}]};
        \end{tikzpicture}
        \caption{Error for two polynomials}
        \label{fig:model:size:err2}
    \end{subfigure}\hfill
    \begin{subfigure}\subfigwidth\raggedleft
        \tikzsetnextfilename{size_err3}
        \begin{tikzpicture}
            \begin{axis}[twocolhalfplot1, cycle list name=sbblas]
                \draw[dashed] foreach \x in {24, 152, 280, 536}
                    {(\x, -15) -- (\x, 15)};
                \foreach \blas in {open, blis, mkl}
                    \addplot table[y=err3] {sandy_\blas.dat};
                \coordinate (label1) at (yticklabel cs:.5);
            \end{axis};
            \begin{axis}[twocolhalfplot2, cycle list name=hwblas]
                \draw[dashed] foreach \x in {24, 152, 280, 536}
                    {(\x, -15) -- (\x, 15)};
                \foreach \blas in {open, blis, mkl}
                    \addplot table[y=err3] {haswell_\blas.dat};
                \coordinate (label2) at (yticklabel cs:.5);
            \end{axis};
            \path (label1) -- (label2)
                node[midway, rotate=90, midway, anchor=south]
                    {relative error [\percent{}]};
        \end{tikzpicture}
        \caption{Error for three polynomials}
        \label{fig:model:size:err3}
    \end{subfigure}

    \mycaption{%
        Runtime and error of piecewise cubic polynomial fits \dtrsm[LLNN].
        \iftableoflist\else\\Dashed lines: polynomial boundaries.\fi
        \captiondetails{1~thread, median of 100~repetitions}
    }
\end{figure}

\begin{example}{Polynomial fitting for size arguments}{model:args:size}
    \Cref{fig:model:size} shows the runtime of
    \displaycall\dtrsm{
        \arg{side}L, \arg{uplo}L, \arg{transA}N, \arg{diag}N,
        \varg m{n}, \varg n{n}, \arg{alpha}{1.0},
        \arg AA, \arg{ldA}{1000}, \arg BB, \arg{ldB}{1000}
    }
    i.e., $\dmB \coloneqq \dmAi \dmB$ with $\dmA\lowerpostsep, \dmB \in \R^{n
    \times n}$, with $n = 24, \ldots, 536$ in steps of~16 on a \sandybridge and
    a \haswell with single-threaded \openblas, \blis, and \mkl.

    At first sight, the runtime for all setups follows a smooth cubic
    behavior---perfectly in line with the operation's minimal cost of
    \SIvar{n^3}\flops.  However, if for each setup we fit the measurements with
    a single cubic polynomial that minimizes the least-squares relative error
    (details in~\cref{sec:model:fit}), we are left with the approximation error
    shown in~\cref{fig:model:size:err1}.  The absolute relative approximation
    error\footnotemark{} lies between \SI{.86}{\percent} for \blis on the
    \sandybridgeshort~(\ref*{plt:sbblis}) and \SI{11.22}{\percent} for \openblas
    on the \haswellshort~(\ref*{plt:hwopen}); on average it
    is~\SI{5.30}\percent.

    If we look closer at the approximation errors in
    \cref{fig:model:size:err1}---especially for \openblas on the
    \haswellshort~(\ref*{plt:hwopen})---we observe a piecewise smooth(er)
    behavior.  Motivated by this observation, we now fit not one polynomial to
    each data-set but two: one for the first half ($n \leq 280$) and one for the
    second half ($n \geq 280$).  For this two-split polynomial fit the
    approximation error is shown in~\cref{fig:model:size:err2}: The largest
    error is now reduced to~\SI{5.25}{\percent} for \mkl on the
    \haswellshort~(\ref*{plt:hwmkl}), and the average error
    is~\SI{2.55}{\percent}---less than half of the original approximation error.
    (Based on a more detailed analysis, a better splitting point than
    $\frac{24+536}2 = 280$ could have been chosen, but as
    \cref{fig:model:size:err1} shows such choices would be notably different for
    each setup.)  Within the new approximation, the error for the second
    polynomial ($n \geq 280$) is already quite low---on
    average~\SI{.38}\percent.  Hence, in a second step, we further subdivide
    only the first half of the domain ($n \leq 280$) at~$n = 152$, and generate
    a new approximation consisting of three polynomials.  As
    \cref{fig:model:size:err3} shows, the error of this approximation is
    below~\SI{1.28}{\percent}~(\ref*{plt:hwmkl}) in all cases and on
    average~\SI{.71}\percent.
\end{example}
\footnotetext{%
    For a polynomial~$p(x)$ fit to measurements~$y_1, \ldots, y_N$ in
    points~$x_1, \ldots, x_N$ we consider the error $1 / N \sum_{i=1}^N \lvert
    y_i - p(x_i) \rvert / y_i$.  Note that the least-squares fitting minimizes
    not this sum of absolute relative errors but the sum of squared relative
    errors.
}

To account for the not purely polynomial influence of a kernel's size arguments
on its runtime, we will represent it in our models through \definition{piecewise
polynomials}.  Details on the such piecewise polynomial representations and
their automated generation are given in
\cref{sec:model:fit,sec:model:adaptive,sec:model:config}.

        \subsection{Data Arguments}
        \label{sec:model:args:data}
        With few exceptions (such as eigensolvers), the executed instructions and thus
the runtime of kernels do not depend on their operands' numerical values.
However, the runtime may depend on where these operands are located within the
memory hierarchy:  Kernels whose operands reside in cache prior to their
invocation run faster.

\begin{figure}\figurestyle
    \ref*{leg:model:datacache}

    \raggedleft
    \tikzsetnextfilename{datacache}
    \begin{tikzpicture}
        \begin{axis}[
                ybar, every axis plot/.append style=ybar legend,
                ylabel=runtime, y unit=\ms,
                xmin=-.75, xmax=6.25,
                xtick={0, 1, 2, 3.5, 4.5, 5.5},
                xticklabels={\openblas, \blis, \mkl, \openblas, \blis, \mkl},
                legend to name=leg:model:datacache,
                table/search path={model/figures/data/datacache},
            ]
            \addplot[green] table {ic.dat};
            \label{plt:model:data:ic}
            \addlegendentry{in-cache}
            \addplot[blue] table {Aic.dat};
            \label{plt:model:data:Aic}
            \addlegendentry{$A$ in-cache}
            \addplot[orange] table {Bic.dat};
            \label{plt:model:data:Bic}
            \addlegendentry{$B$ in-cache}
            \addplot[red] table {oc.dat};
            \label{plt:model:data:oc}
            \addlegendentry{out-of-cache}
            \coordinate (sandy) at (axis cs: 1, 0);
            \coordinate (haswell) at (4.5, 0);
        \end{axis}
        \path (sandy) ++(0, -1) node {\sandybridge};
        \path (haswell) ++(0, -1) node {\haswell};
    \end{tikzpicture}

    \mycaption{%
        Runtime of \dtrsm[LLNN] with in-cache and out-of-cache operands.
        \captiondetails{$m = n = 256$, 1~thread, median of 100~repetitions}
    }
    \label{fig:model:datacache}
\end{figure}

\begin{example}{Data arguments}{model:args:data}
    \Cref{fig:model:datacache} shows the runtime of
    \displaycall\dtrsm{
        \arg{side}L, \arg{uplo}L, \arg{transA}N, \arg{diag}N,
        \arg m{256}, \arg n{256}, \arg{alpha}{1.0},
        \varg A{A}, \arg{ldA}{256}, \varg B{B}, \arg{ldB}{256}
    }
    \noindent i.e., $\dmB \coloneqq \dmAi \dmB$ with $\dmA\lowerpostsep, \dmB
    \in \R^{256 \times 256}$, for \dmA\lowerpostsep and \dmB a-priori either in-
    or out-of-cache on a \sandybridge and a \haswell with single-threaded
    \openblas, \blis, and \mkl.

    Across all setups, the pure in-cache scenario~(\ref*{plt:model:data:ic}) is
    consistently faster than out-of-cache~(\ref*{plt:model:data:oc}) by between
    \SI{7.75}{\percent}~(\openblas on the \sandybridgeshort) and
    \SI{45.08}{\percent}~(\mkl on the \haswellshort).  While the scenarios where
    either only \dmA\lowerpostsep or only \dmB is
    in-cache~(\ref*{plt:model:data:Aic}, \ref*{plt:model:data:Bic}) are always
    between these extremes, which of the two is faster depends on both the
    architectures and the \blas implementation.
\end{example}

The exact effects of caching on kernel runtime and performance are hard to
predict.  However, since blocked algorithms operate on matrices with high
locality, we will generate our models with in-cache operands where possible:  By
repeating each measurement twice, the most-recently-used portions of a kernel's
operands (the entire operands for small operations) from the first repetition
are in-cache prior to the second repetition.  Only these second repetitions'
measurements are used to construct our models.

We will revisit caching in more detail in \cref{ch:cache,ch:tensor}.

        \subsection{Summary}
        \label{sec:model:args:sum}
        This section studied the effects of various argument types on kernel runtime.
In summary, these effects and our decisions on how to represent them in our
models are as follows:
\begin{itemize}
    \item Flag arguments (\cref{sec:model:args:flag}) can invoke separate
        execution branches within kernel implementations.  Hence we will
        generate a separate sub-model for each relevant combination of flag
        arguments.

    \item Scalar arguments (\cref{sec:model:args:scalar}) affect the performance
        of kernels only for the special values that allow to avoid certain
        arithmetic operations.  Hence we will scalars them just like flags with
        the possible values~$-1$, 0, 1, and ``any other value''.

    \item Size arguments (\cref{sec:model:args:size}) greatly influence a
        kernel's runtime by determining its minimal \flop-count.  While this
        \flop-count is usually polynomial in the operand sizes, a kernel's
        runtime can typically not be represented accurately by a single
        polynomial.  Hence, we will model the effect of size arguments on
        runtime as piecewise polynomials.  Furthermore, to avoid small-scale
        runtime artefacts, we will ensure that in all measurements all size
        arguments are multiples of~8.

    \item Data arguments (\cref{sec:model:args:data}) do not affect the runtime
        of targeted kernels through their numeric values.  However, the
        operand's location in the processor's memory hierarchy prior to a kernel
        invocation may lead to different performance.  While we could account
        for this effect by generating separate models for specific memory
        preconditions (such as in- and out-of-cache), we will focus on models
        based on repeated measurements that correspond to in-cache data for
        operands smaller than the cache.

    \item Leading dimension arguments (\cref{sec:model:args:ld}) generally have
        only a minor effect on kernel runtime, but they should be choose as
        multiples of~8, yet not of~512.  To generate our models, we will hence
        set all leading dimensions to a constant value, such as~5000.

    \item Increment arguments (\cref{sec:model:args:inc}) are typically~1 or
        equal to a matrix's leading dimension.  We will hence treat them as flag
        arguments with the two values~1 and ``any large value''.  Since
        multiples of~16 as leading dimensions can incur runtime spikes,
        especially in \blasl1 kernels, we will choose a fixed large value for
        the second case that is not a multiple of~16, such as~5000.
\end{itemize}

Based on these decisions on how to represent the influence of various argument
types on kernel runtime in our measurement-based performance models, the
following section describes our models' structure and their automated
generation.

    \section{Model Generation}
    \label{sec:model:generation}
    After analyzing the performance effects of various argument types on dense
linear algebra kernels in the previous section, we now turn to the design and
generation of our performance models.

\Cref{sec:model:structure} introduces the model structure and how their coverage
is configured.  The following sections detail how each model (and sub-model) is
generated based on measurements:  \Cref{sec:model:grids} describes the selection
of measurement points in the kernel's argument space; \cref{sec:model:stat}
discusses how repeated measurements at these points are used to compute summary
statistics of the expected kernel runtime; \cref{sec:model:fit} specifies how
set of measurements is least-squares fitted with a single polynomial; and
finally \cref{sec:model:adaptive} introduces the adaptive refinement approach
that covers the range of problem sizes with piecewise polynomials.

        \subsection{Model Structure}
        \label{sec:model:structure}
        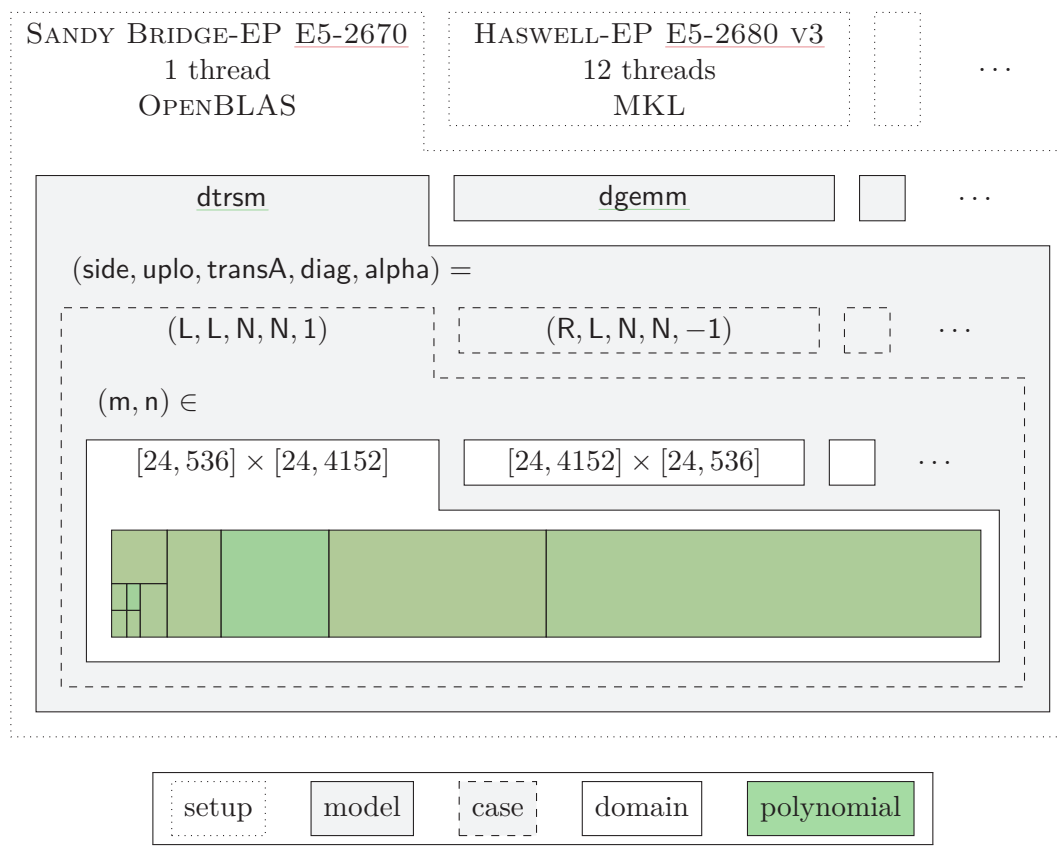
\begin{figure}\figurestyle
    \newcommand\sep{2ex}
    \newcommand\halfsep{1ex}
    \newcommand\doublesep{4ex}
    \newcommand\tabheight{.6cm}
    \begin{tikzpicture}
        \coordinate (tl) at (0, 0);
        \coordinate (bl) at (0, -209.96326pt - 8ex - 1.5 * \tabheight);
        \path (tl) ++(\textwidth, 0) coordinate (tr);
        \path (bl) ++(\textwidth, 0) coordinate (br);
        \path (tl) -- (tr) coordinate[pos=.4] (tm);
        \path (tl) -- (tr) coordinate[pos=.8] (tm2);
        \path (bl) -- (br) coordinate[midway] (bm);

        \draw[dotted] (tm) ++(\halfsep, -2.5 * \tabheight) coordinate(x) (tm2) ++(-\halfsep, 0) rectangle (x) node[midway, align=center] {\haswell\\12~threads\\\mkl};
        \draw[dotted] (tm2) ++(\halfsep, -2.5 * \tabheight) rectangle ++(\tabheight, 2.5 * \tabheight);
        \path (tm2) ++(\tabheight+\halfsep, -2.5 * \tabheight) -- (tr) node[midway] {\ldots};
        \draw[dotted] (tm) ++(-\halfsep, -2.5 * \tabheight-\sep) coordinate (corner) -- (corner |- tl) -- (tl) -- (bl) -- (br) -- (tr |- corner) -- cycle;
        \path (corner) ++(0, \sep) -- (tl) node[midway, align=center] {\sandybridge\\1~thread\\\openblas};

        \path (tl) ++(\sep, -2.5 * \tabheight-\doublesep) coordinate (tl);
        \path (tr) ++(-\sep, -2.5 * \tabheight-\doublesep) coordinate (tr);
        \path (bl) ++(\sep, \sep) coordinate (bl);
        \path (br) ++(-\sep, \sep) coordinate (br);

        \path (tl) -- (tr) coordinate[pos=.4] (tm);
        \path (tl) -- (tr) coordinate[pos=.8] (tm2);

        \filldraw[fill=graybg] (tm) ++(\halfsep, -\tabheight) coordinate(x) (tm2) ++(-\halfsep, 0) rectangle (x) node[midway] {\dgemm};
        \filldraw[fill=graybg] (tm2) ++(\halfsep, -\tabheight) rectangle ++(\tabheight, \tabheight);
        \path (tm2) ++(\tabheight+\halfsep, -\tabheight) -- (tr) node[midway] {\ldots};
        \filldraw[fill=graybg] (tm) ++(-\halfsep, -\tabheight-\sep) coordinate (corner) -- (corner |- tl) -- (tl) -- (bl) -- (br) -- (tr |- corner) -- cycle;
        \path (corner) ++(0, \sep) -- (tl) node[midway] {\dtrsm};

        \path (tl) ++(\sep, -\tabheight-\sep) coordinate (tl);
        \path (bl) ++(\sep, \sep) coordinate (bl);
        \path (br) ++(-\sep, \sep) coordinate (br);

        \node[anchor=north west] (label) at (tl) {$(\code{side}, \code{uplo}, \code{transA}, \code{diag}, \code{alpha}) = $};

        \path (label.south west) ++ (0, -\halfsep) coordinate (tl);
        \coordinate (tr) at (tl -| br);
        \path (tl) -- (tr) coordinate[pos=.4] (tm);
        \path (tl) -- (tr) coordinate[pos=.8] (tm2);

        \draw[dashed] (tm) ++(\halfsep, -\tabheight) coordinate(x) (tm2)
        ++(-\halfsep, 0) rectangle (x) node[midway] {$(\code R, \code L, \code N, \code N, \code{-1})$};
        \draw[dashed] (tm2) ++(\halfsep, -\tabheight) rectangle ++(\tabheight, \tabheight);
        \path (tm2) ++(\tabheight+\halfsep, -\tabheight) -- (tr) node[midway] {\ldots};
        \draw[dashed] (tm) ++(-\halfsep, -\tabheight-\sep) coordinate (corner) -- (corner |- tl) -- (tl) -- (bl) -- (br) -- (tr |- corner) -- cycle;
        \path (corner) ++(0, \sep) -- (tl) node[midway] {$(\code L, \code L, \code N, \code N, \code1)$};

        \path (tl) ++(\sep, -\tabheight-\sep) coordinate (tl);
        \path (bl) ++(\sep, \sep) coordinate (bl);
        \path (br) ++(-\sep, \sep) coordinate (br);

        \node[anchor=north west] (label) at (tl) {$(\code m, \code n) \in $};

        \path (label.south west) ++ (0, -\halfsep) coordinate (tl);
        \coordinate (tr) at (tl -| br);
        \path (tl) -- (tr) coordinate[pos=.4] (tm);
        \path (tl) -- (tr) coordinate[pos=.8] (tm2);

        \filldraw[fill=white] (tm) ++(\halfsep, -\tabheight) coordinate(x) (tm2) ++(-\halfsep, 0) rectangle (x) node[midway] {$[24, 4152] \times [24, 536]$};
        \filldraw[fill=white] (tm2) ++(\halfsep, -\tabheight) rectangle ++(\tabheight, \tabheight);
        \path (tm2) ++(\tabheight+\halfsep, -\tabheight) -- (tr) node[midway] {\ldots};
        \filldraw[fill=white] (tm) ++(-\halfsep, -\tabheight-\sep) coordinate (corner) -- (corner |- tl) -- (tl) -- (bl) -- (br) -- (tr |- corner) -- cycle;
        \path (corner) ++(0, \sep) -- (tl) node[midway] {$[24, 536] \times [24, 4152]$};

        \path (tl) ++(\sep, -\tabheight-\doublesep) coordinate (tl);
        \path (tr) ++(-\sep, -\tabheight-\doublesep) coordinate (tr);
        \path (bl) ++(\sep, \sep) coordinate (bl);
        \path (br) ++(-\sep, \sep) coordinate (br);

        \pgfpointdiff{\pgfpointanchor{tl}{center}}{\pgfpointanchor{br}{center}}
        \makeatletter
        \pgfmathparse{\the\pgf@x/4.096cm}\let\scale\pgfmathresult
        \makeatother
        {
            \path (tl) ++([scale=\scale] -.024, -.512) coordinate (bl);
            { [shift={(bl)}, scale=\scale]
                \fill[graybgop, fill opacity=.5] (.024, .024) rectangle (4.152, .536);
                \foreach \lx/\ly/\ux/\uy/\c in {
                    0.024/0.024/0.096/0.152/12.4064791435,
                    0.096/0.024/0.16/0.152/12.8336964497,
                    0.024/0.152/0.096/0.28/11.6060049092,
                    0.096/0.152/0.16/0.28/4.01573270566,
                    0.16/0.024/0.288/0.28/13.9479394764,
                    0.024/0.28/0.288/0.536/17.7385455275,
                    0.288/0.024/0.544/0.536/15.4712857129,
                    0.544/0.024/1.056/0.536/3.33591905182,
                    1.056/0.024/2.088/0.536/16.6796736462,
                    2.088/0.024/4.152/0.536/11.7575433615
                }
                    \filldraw[fill=red!\c!green, fill opacity=.5]
                    (\lx, \ly) rectangle (\ux, \uy);
            }
        }

        \path (bm) ++(0, -\sep) node[anchor=north] {
            \fbox{
                \tikzset{every node/.append style={
                    inner sep=\halfsep, outer sep=\halfsep, draw
                }}%
                \tikz\node[dotted] {\mstrut setup};
                \hspace{\sep}
                \tikz\node[fill=graybg] {\mstrut model};
                \hspace{\sep}
                \tikz\node[fill=graybg, dashed] {\mstrut case};
                \hspace{\sep}
                \tikz\node {\mstrut domain};
                \hspace{\sep}
                \tikz\node[fill=green, fill opacity=.5, text opacity=1] {\mstrut polynomial};
            }
        };
    \end{tikzpicture}

    \caption{Structure of the performance models.}
    \label{fig:model:structure}
\end{figure}

Based on the analyses of how a kernel's different argument types affect its
performance in \cref{sec:model:args}, we arrive at the structure for our
\definition{performance models} depicted in \cref{fig:model:structure}.

For each setup~\dm[fill=white, draw=black, dotted]{~} consisting of the hardware
platform, number of threads, and the \blas implementation, a separate set of
models is generated.  Independent kernels can be modeled for each setup.

Each model \dm[draw=black]{~} represents the runtime of one kernel (e.g.,
\dtrsm or \dgemm): It is essentially a function of the kernel's arguments that
returns runtime estimates.\footnote{%
    Optionally further performance counters provided by the \sampler or derived
    metrics can be modeled.  However, throughout this work we solely focus on
    runtime models.
} To account for variations in kernel runtime for fixed arguments, each estimate
is not a single number, but a set of basic summary statistics, such as minimum,
median, average, and standard deviation.

Each model takes two sets of kernel arguments into account:
\begin{itemize}
    \item Flag and scalar arguments (and increment arguments for vector
        operations) are limited to a few discrete values: the distinct options
        for flags and the values $-1$, 0, 1, and ``any other value'' for scalars
        (and either~1 or a ``any large value'' for increments).  For a given
        kernel invocation, the combination of these argument values identifies
        one of several discrete \definition{cases} \dm[draw=black, dashed]{~}.
        To best match the application scenario, each model can be configured to
        represent only a subset of these cases.

    \item Size arguments take values from potentially large ranges of problem
        sizes.  In our models, these represented ranges are specified as
        (collections of) rectangular (generally: hyper-cuboidal)
        \definition{domains} \dm[draw=black, fill=white]{~}.  For each model and
        case, these domains can be separately selected.
\end{itemize}
All other arguments, such as data arguments and leading dimensions, are not
represented in our models.

For each case and domain, we generate a separate \definition{sub-model} that
represents the kernel runtime as a \definition{piecewise polynomial}.  Each
polynomial piece \dm[fill=green, draw=black]{~} actually consists of a small
list of polynomials corresponding to the modeled runtime summary statistics.

Since implementing the composition of models from sub-models and the
corresponding separation and treatment of argument types is fairly straight
forward, the following sections focuses on the generation of a single sub-model.

        \subsection{Sample Distribution}
        \label{sec:model:grids}
        For a fixed setup, discrete case, and rectangular domain, we model a kernel's
runtime by taking a series of measurements---referred to as samples---and
fitting a polynomial to the measured runtime.  The first step is to select a
\definition{sampling point distribution}, i.e., a set of points in the domain at
which the kernel runtime is measured.

An intuitive option would be to (pseudo-)randomly distribute the sampling points
within the domain.  However, this approach does not guarantee that, e.g., points
close to the domain's boundary are well represented in the sampling set, which
in these areas greatly reduces the accuracy of polynomials fitted to such data.
Hence we do not use random sampling point distributions, and instead consider
two regular grid patterns:
\begin{itemize}
    \item The simplest structured pattern is a regular \definition{Cartesian
        grid} that covers the whole domain evenly with points.  In one
        dimension, a Cartesian grid of $n$~points $x_0, \ldots, x_{n-1}$
        between~0 and~1 is defined as
        \[
            x_i = \frac i{n - 1}\enspace.
        \]

        With regards to the adaptive refinement approach (see
        \cref{sec:model:adaptive}), the Cartesian grid's advantage is its high
        \definition{sample reuse}:  When the domain is divided in two along one
        dimension, all points of the original grid are also points in the two
        new grids.  Hence, the number of points in which new measurement are
        required is reduced significantly.

    \item However, fitting a polynomial behavior with an even distribution of
        samples is not ideal.  A better alternative is to use
        \definition{Chebyshev nodes}~\cite[Section~8.3]{numa}, which minimize
        the approximation error by essentially moving the sampling points closer
        to the region's boundaries.  In one dimension, the $n$~Chebyshev nodes
        $x_0, \ldots, x_{n-1}$ between~$-1$ and~1 are given by
        \[
            x_i = \cos\left(\frac{2i+1}{2n} \pi\right) \enspace.
        \]

        In contrast to the Cartesian grid with perfect sample reuse,
        the Chebyshev nodes offer no reuse at all.  Furthermore, they do not
        include points on the domain's boundary.  We hence use a slightly
        modified configuration that moves the Chebyshev nodes to include the
        boundary:
        \[
            x_i = \cos\left(\frac i{n-1} \pi\right) \enspace.
        \]
        We refer do these points as a \definition{Chebyshev grid}.
\end{itemize}

\begin{figure}\figurestyle
    \tikzset{
        x={(3, 0)}, y={(0, 3)}, z={(0, 0, 3)},
        baseline={(0, .5)}
    }

    \begin{tabular}{lcccc}
        &1D &\shortstack{re-\\use} &2D &3D\\ \\
        \parbox{\widthof{ (b) Cartesian}}{%
            \subcaption{Caresian\label{fig:model:grids:cart}}%
        }
        &\begin{tikzpicture}
            \draw (0, 0) -- (0, 1);
            \foreach \y in {0, .333, .666, 1}
                \filldraw[blue] (0, \y) circle (2.5pt);
        \end{tikzpicture}
        &\begin{tikzpicture}
            \draw (0, 0) -- (0, 1);
            \draw[gray] foreach \y in {0, .5, 1} {(-5pt, \y) -- (5pt, \y)};
            \foreach \y in {0, .166, .333, .5, .666, .833, 1}
                \filldraw[blue] (0, \y) circle (2.5pt);
            \foreach \y in {0, .333, .666, 1}
                \filldraw[blue, fill=green, thick] (0, \y) circle (2.5pt);
        \end{tikzpicture}
        &\begin{tikzpicture}
            \filldraw[fill=graybg, fill opacity=.5] (0, 0) rectangle (1, 1);
            \foreach \y in {0, .333, .666, 1}
                \foreach \x in {0, .25, ..., 1}
                    \filldraw[blue] (\x, \y) circle (2.5pt);
        \end{tikzpicture}
        &\begin{tikzpicture}[baseline={(0, .5, .5)}]
            \fill[fill=graybg, fill opacity=.5]
                (0, 0, 1) -- (1, 0, 1) -- (1, 0, 0) --
                (1, 1, 0) -- (0, 1, 0) -- (0, 1, 1) -- cycle;
            \draw
                (0, 0, 0) -- (0, 0, 1)
                (0, 1, 0) -- (0, 1, 1)
                (1, 0, 0) -- (1, 0, 1)
                (1, 1, 0) -- (1, 1, 1)
                (0, 0, 0) rectangle (1, 1, 0)
                (0, 0, 1) rectangle (1, 1, 1);
            \foreach \y in {0, .333, .666, 1}
                \foreach \x in {0, .25, ..., 1}
                    \foreach \z in {0, .5, 1} {
                        \pgfmathparse{100*\z}
                        \filldraw[lightblue!\pgfmathresult!blue]
                            (\x, \y, \z) circle (2.5pt);
                    }
        \end{tikzpicture}\\ \\
        \parbox{\widthof{ (a) Chebyshev }}{%
            \subcaption{Chebyshev\label{fig:model:grids:gauss}}%
        }
        &\begin{tikzpicture}
            \draw (0, 0) -- (0, 1);
            \foreach \ya in {0, .333, .666, 1} {
                \pgfmathsetmacro\y{.5 - .5 * cos(pi * \ya r)}
                \filldraw[blue] (0, \y) circle (2.5pt);
            }
        \end{tikzpicture}
        &\begin{tikzpicture}
            \draw (0, 0) -- (0, 1);
            \draw[gray] foreach \y in {0, .5, 1} {(-5pt, \y) -- (5pt, \y)};
            \foreach \ya in {0, .333, .666, 1} {
                \pgfmathsetmacro\y{.5 - .5 * cos(pi * \ya r)}
                \draw[red] (0, \y) circle (2.5pt);
                \pgfmathsetmacro\y{.25 - .25 * cos(pi * \ya r)}
                \filldraw[blue] (0, \y) circle (2.5pt);
                \pgfmathsetmacro\y{.75 - .25 * cos(pi * \ya r)}
                \filldraw[blue] (0, \y) circle (2.5pt);
            }
            \foreach \y in {0, 1}
                \filldraw[blue, fill=green, thick] (0, \y) circle (2.5pt);
        \end{tikzpicture}
        &\begin{tikzpicture}
            \filldraw[fill=graybg, fill opacity=.5] (0, 0) rectangle (1, 1);
            \foreach \ya in {0, .333, .666, 1} {
                \pgfmathsetmacro\y{.5 - .5 * cos(pi * \ya r)}
                \foreach \xa in {0, .25, ..., 1} {
                    \pgfmathsetmacro\x{.5 - .5 * cos(pi * \xa r)}
                    \filldraw[blue] (\x, \y) circle (2.5pt);
                }
            }
        \end{tikzpicture}
        &\begin{tikzpicture}[baseline={(0, .5, .5)}]
            \fill[fill=graybg, fill opacity=.5]
                (0, 0, 1) -- (1, 0, 1) -- (1, 0, 0) --
                (1, 1, 0) -- (0, 1, 0) -- (0, 1, 1) -- cycle;
            \draw
                (0, 0, 0) -- (0, 0, 1)
                (0, 1, 0) -- (0, 1, 1)
                (1, 0, 0) -- (1, 0, 1)
                (1, 1, 0) -- (1, 1, 1)
                (0, 0, 0) rectangle (1, 1, 0)
                (0, 0, 1) rectangle (1, 1, 1);
            \foreach \ya in {0, .333, .666, 1} {
                \pgfmathsetmacro\y{.5 - .5 * cos(pi * \ya r)}
                \foreach \xa in {0, .25, ..., 1} {
                    \pgfmathsetmacro\x{.5 - .5 * cos(pi * \xa r)}
                    \foreach \za in {0, .5, 1} {
                        \pgfmathsetmacro\z{.5 - .5 * cos(pi * \za r)}
                        \pgfmathsetmacro\grade{100 * \z}
                        \filldraw[lightblue!\grade!blue]
                            (\x, \y, \z) circle (2.5pt);
                    }
                }
            }
        \end{tikzpicture}
    \end{tabular}

    \mycaption{Sampling point distributions and reuse.}
    \label{fig:model:grids}
\end{figure}

\begin{example}{Sampling point distributions}{model:grid}
    \newcommand\pointb{%
        \tikz[baseline=-4pt] \filldraw[blue] circle (2.5pt);\xspace%
    }%
    \newcommand\pointg{%
        \tikz[baseline=-4pt]
            \filldraw[blue, fill=green, thick] circle (2.5pt);%
        \xspace%
    }%
    \newcommand\pointr{\tikz[baseline=-4pt]\draw[red] circle (2.5pt);\xspace}%
    \Cref{fig:model:grids} visualizes the two alternative sampling point
    distributions for 1D, 2D, and 3D~domains.  We select 4~points along
    the first dimension, 5~along the second, and 3~along the third.

    The point reuse is shown for the 1D~case:  When the domain is split in half,
    all points from the original Cartesian grid~\pointg are reused in the
    refined grid, and only three new points~\pointb are generated; for the
    Chebyshev grid, however, only the two outermost points~\pointg are reused,
    while the other two~\pointr are not matched by points in the refined grid,
    and five new points~\pointb are generated.
\end{example}

Once the sampling points are chosen, we avoid implementation-dependent
performance artefacts of size argument increments in steps of~1 (see
\cref{sec:model:args:size:small}) by \definition[round to\\multiples
of~8]{rounding} all generated grid points {\em to multiples of~8} along each
dimension.

        \subsection{Repeated Measurements and Summary Statistics}
        \label{sec:model:stat}
        Based on the kernel and the modeled cases, each sampling point is turned into a
\definition[measurement calls]{measurement call}:  While the flag, size,
and scalar arguments are determined by the case and the point, the leading
dimensions are set to a fixed large value (such as 5000), and the operand sizes
are deduced automatically.

To both avoid outliers and represent measurement fluctuations in our models, each
such constructed measurement call is then executed by the \sampler (see
\cref{sec:meas:sampler}) not only once, but
\definition[repetitions]{repeatedly}---typically between 5 and 20~times.  To
avoid the effects of frequency fluctuations (see \cref{sec:meas:fluct:turbo}
and~\ref{sec:meas:fluct:longterm}), the repetitions for each measurement call
are not executed in a single batch but shuffled among all calls' repetitions to
obtain measurements across the whole \sampler execution for each call.
Furthermore, each repetition, executes the measurement call twice in a row, to
ensure consistent cache preconditions, which offer high temporal locality
(``warm'' data) for small operations.

Once obtained, the collected measurement results for each call are turned into
\definition{summary statistics}: minimum, median, maximum, average, and standard
deviation.  In the next step, each of these statistics is fitted with a separate
polynomial.

        \subsection{Relative Least-Squares Polynomial Fitting}
        \label{sec:model:fit}
        {

The starting point for the polynomial fitting procedure is a set of sampling
\definition[points and values]{points} $\x_1, \ldots, \x_N \in \R^d$ (from the
$d$-dimensional range of size arguments) and corresponding measurement {\em
values} $y_i \in \R$ (i.e., per summary statistic).\footnote{%
    Technically, we have $\x_i \in \mathbb N_0^d$ and $y \in \mathbb N_0$;
    however, for the fitting procedure these points are treated as
    floating-point tuples.
}  As the set of polynomial basis functions, we use \definition[monomial
degree]{monomials} $m_1, \ldots, m_M \colon \R^d \to \R$ whose maximum {\em
degree} is determined by the kernel's asymptotic complexity (given by its
minimal \flop-count), yet may be further increased.  The polynomial~$p$ is
constructed as a linear combination of these monomials with \definition{weights}
$\beta_1, \ldots, \beta_M \in \R$:
\[
    p(\x) = \sum_{j=1}^N \beta_j m_j(\x) \enspace.
\]

\begin{example}{Polynomial basis functions}{model:fit}
    If we model the runtime of \dtrsm[LLNN] by letting its cost of \SIvar{m^2
    n}{\flops} determine the maximum monomial degree, we use a bivariate
    polynomial in $\x = (x_1, x_2)$ of the form
    \[
        p(\x)
        = \beta_1
        + \beta_2 x_1
        + \beta_3 x_2
        + \beta_4 x_1^2
        + \beta_5 x_1 x_2
        + \beta_6 x_1^2 x_2 \\
        = \sum_{j=1}^6 \beta_j m_j(\x) \enspace,
    \]
    i.e., with the monomial basis
    \begin{align*}
         m_1(\x) &= 1
        &m_2(\x) &= x_1
        &m_3(\x) &= x_2 \\
         m_4(\x) &= x_1^2
        &m_5(\x) &= x_1 x_2
        &m_6(\x) &= x_1^2 x_2 \enspace.
    \end{align*}
    Had we chosen to increase the monomial degree in each dimension by one, we
    would use a polynomial with the 12~basis monomials:
    \begin{align*}
         m_1(\x) &= 1
        &m_2(\x) &= x_1
        &m_3(\x) &= x_2 \\
         m_4(\x) &= x_1^2
        &m_5(\x) &= x_1 x_2
        &m_6(\x) &= x_2^2 \\
         m_7(\x) &= x_1^3
        &m_8(\x) &= x_1^2 x_2
        &m_9(\x) &= x_1 x_2^2 \\
         m_{10}(\x) &= x_1^3 x_2
        &m_{11}(\x) &= x_1^2 x_2^2
        &m_{12}(\x) &= x_1^3 x_2^2 \enspace.
    \end{align*}
\end{example}

\newcommand\dvb{\dv[height=.5]\beta\xspace}
\newcommand\dvbT{\dv[height=0, width=.5, ']\beta\xspace}
\newcommand\dmX{\dm[width=.5]X\xspace}
\newcommand\dmXT{\dm[height=.5, ']X\xspace}
The weights~$\beta_j$ are chosen by \definition[minimize the squared relative
error]{minimizing the squared relative error}
\[
    S(\beta_1, \ldots, \beta_M)
    \defeqq \sum_{i=1}^N \left(\frac{y_i - p(\x_i)}{y_i}\right)^2
    = \sum_{i=1}^N \left(1 - \sum_{j=1}^M \frac{\beta_j m_j(\x_i)}{y_i}\right)^2
    \enspace.
\]
With
\[
    \dvb \defeqq \begin{pmatrix}
        \beta_1\\\beta_2\\\vdots\\\beta_M
    \end{pmatrix} \text{ and }
    \dmX \defeqq \begin{pmatrix}
         \dfrac{m_1(\x_1)}{y_1}
        &\dfrac{m_2(\x_1)}{y_1} &\cdots
        &\dfrac{m_M(\x_1)}{y_1}
        \\
         \dfrac{m_1(\x_2)}{y_2}
        &\dfrac{m_2(\x_2)}{y_2} &\cdots
        &\dfrac{m_M(\x_2)}{y_2}
        \\
        \vdots &\vdots &\ddots &\vdots
        \\
         \dfrac{m_1(\x_N)}{y_N}
        &\dfrac{m_2(\x_N)}{y_N} &\cdots
        &\dfrac{m_M(\x_N)}{y_N}
    \end{pmatrix} \enspace,
\]
this error can be expressed as
\begin{align*}
    S(\dvb)
    &= \bigl\lVert\dv1 - \dmX \matvecsep \dvb\bigr\rVert^2 \\
    &= \bigl(\dv1 - \dmX \matvecsep \dvb\bigr)^T
       \bigl(\dv1 - \dmX \matvecsep \dvb\bigr) \\
    &= 1 - 2 \dvbT \matvecsep \dmXT \matvecsep \dv1
         + \dvbT \matvecsep \dmXT \matmatsep \dmX \matvecsep \dvb \enspace .
\end{align*}
Since~$S(\dvb)$ is convex, we can find its minimum by setting its derivative to
zero:
\[
    \frac{\partial S\bigl(\dvb\bigr)}{\partial \dvb}
    = -2 \dmXT \matvecsep \dv1 + 2 \dmXT \matmatsep \dmX \dvb = 0 \enspace.
\]
Rewritten as
\[
    \left(\dmXT \matmatsep \dmX\right) \dvb = \dmXT \matvecsep \dv1 \enspace,
\]
this is known as the \definition{normal equations}, which have a unique solution
because, since the~$m_j$ are linearly independent, \dmX has full rank.  To
obtain a numerically stable solution of the normal equations, we use {\swstyle
numpy}'s \code{linalg.lstsq}, which is based on the singular value decomposition
of~\dmX.
}

        \subsection{Adaptive Refinement}
        \label{sec:model:adaptive}
        So far, we have determined how sampling points are chosen in a given rectangular
(generally: hyper-cuboidal) domain, how summary statistics are computed from
repeated measurements in these points, and how a multivariate polynomial is
fitted to one of these statistics.  We now describe how a domain is adaptively
subdivided and fitted with a piecewise function consisting of such polynomials.

The basis for this adaptive subdivision is an \definition{error measure} for the
approximation accuracy.  To compute this measure, we consider the polynomial fit
of a selected \definition{reference statistic}; typical choices are the minimum
or median since they are insensitive to fluctuations.  For the selected
reference statistic, we compute the point-wise absolute relative error~$e_i$ for
the polynomial approximation~$p$ in each measurement point~$\x_i$ with respect
to the measurement statistic~$y_i$:
\[
    e_i \defeqq \left\lvert\frac{y_i - p(\x_i)}{y_i}\right\rvert
    \enspace.
\]
Next, the error measure is computed from the set of errors $\{e_1, \ldots,
e_N\}$ as its average, maximum, or ninetieth percentile.

Based on this error measure, the \definition{adaptive refinement} process
subdivides the initial domain as follows:  It starts by sampling the entire
domain and fits one polynomial to all measurements (for each statistic).  If
either the error measure for this approximation is below a specified
\definition{error bound} (i.e., a threshold value) or the size of the domain
along each dimension is below a configurable \definition{minimum width}, the
process terminates.  Otherwise, the domain is split in half along its
\definition{relatively largest dimension}:  If along each dimension~$i$ the
domain spans the interval~$[l_i, u_i]$, we choose the dimension for which $u_i /
l_i$ is the largest.  Along this dimension~$s$, the new interval is split in
half\footnote{%
    We choose the interval's center, since it guarantees the most regular
    subdivision.  A more guided choice would require either advanced knowledge
    of the kernel implementation or a significantly higher sampling resolution.
} (rounded to the nearest multiple of~8) at
\[
    m_s \defeqq \operatorname{round}\left(\frac{l_s + u_s}2, 8\right) = 8
    \left\lfloor \frac{l_s + u_s + 8}{16} \right\rfloor \enspace,
\]
and the new domains are defined by the intervals~$[l_s, m_s]$ and~$[m_s, u_s]$.
The process is applied recursively to both new domains until either the error
bound or the minimum width is reached.

Note that the resulting performance models are not smooth because the polynomial
pieces are not required to match at the boundaries.  Since our applications do
not require any continuity in our models, this does not pose a problem.  Hence,
we do not apply, e.g., splines to generate smooth models at increased cost.

\begin{figure}[p]\figurestyle
    \pgfplotsset{
        ymajorgrids=false,
        xlabel=$n$,
        ylabel=$m$,
        xmin=24, xmax=4152,
        ymin=24, ymax=536,
        ytick={24, 536},
        scale mode=scale uniformly,
        clip=false,
        cell picture=true,
        every subfigaxis/.style={
            fig width=\textwidth,
            fig align=\centering,
        },
    }
    \tikzset{
        acc/.style={
            midway, fill=white, fill opacity=.5, text opacity=1
        }
    }

    \noindent error measure:
    \raisebox{-1.23\baselineskip}{\ref*{leg:model:adaptive}}

    \pgfmathsetmacro\grade{100 * 4.21411059742 / 5}
    \begin{subfigaxis}[
            fig caption=Initial sampling points and fit,
            fig label=adaptive:1,
            xtick={24, 4152},
            colorbar horizontal, colormap name=greenred,
            point meta min=0, point meta max=5,
            colorbar style={
                width=.5\textwidth,
                xtick={0, ..., 5},
                xticklabel={\SIvar{\pgfmathprintnumber{\tick}}\percent},
                scale mode=auto,
            },
            colorbar to name=leg:model:adaptive,
        ]
        \filldraw[red!\grade!green, opacity=.75] (24, 24) rectangle (4152, 536);
        \fill[blue]
            foreach \m in {24, 56, 152, 280, 408, 504, 536} {
                foreach \n in {24, 416, 1448, 2728, 3760, 4152}
                    {(\n, \m) circle (2.5pt)}};
        \path (24, 24) -- (4152, 536) node[acc] {error: \SI{4.21}\percent};
    \end{subfigaxis}

    \pgfmathsetmacro\gradea{100 * 3.36274563217 / 5}
    \pgfmathsetmacro\gradeb{100 * .587877168075 / 5}
    \begin{subfigaxis}[
            fig caption=First refinement step,
            fig label=adaptive:2,
            xtick={24, 2088, 4152}
        ]
        \filldraw[red!\gradea!green, fill opacity=.75] (24, 24) rectangle (2088, 536);
        \filldraw[red!\gradeb!green, fill opacity=.75] (2088, 24) rectangle (4152, 536);
        \filldraw[blue]
            foreach \m in {24, 56, 152, 280, 408, 504, 536} {
                foreach \n in {224, 736, 1376, 1888, 2088, 2288, 2800, 3440, 3952}
                    {(\n, \m) circle (2.5pt)}};
        \filldraw[thick, blue, fill=green]
            foreach \m in {24, 56, 152, 280, 408, 504, 536} {
                foreach \n in {24, 4152}
                    {(\n, \m) circle (2.5pt)}};
        \path (24, 24) -- (2088, 536) node[acc] {\SI{3.36}\percent};
        \path (2088, 24) -- (4152, 536) node[acc] {\SI{.59}\percent};
    \end{subfigaxis}

    \pgfmathsetmacro\gradea{100 * 1.13453566361 / 5}
    \pgfmathsetmacro\gradeb{100 * .83398368231 / 5}
    \pgfmathsetmacro\gradec{100 * .587877168075 / 5}
    \begin{subfigaxis}[
            fig caption=Second refinement step,
            fig label=adaptive:3,
            xtick={24, 1056, 2088, 4152}
        ]
        \filldraw[red!\gradea!green, fill opacity=.75] (2088, 24) rectangle (4152, 536);
        \filldraw[red!\gradeb!green, fill opacity=.75] (24, 24) rectangle (1056, 536);
        \filldraw[red!\gradec!green, fill opacity=.75] (1056, 24) rectangle (2088, 536);
        \filldraw[blue]
            foreach \m in {24, 56, 152, 280, 408, 504, 536} {
                foreach \n in {120, 384, 696, 960, 1056, 1152, 1416, 1728, 1992}
                    {(\n, \m) circle (2.5pt)}};
        \filldraw[thick, blue, fill=green]
            foreach \m in {24, 56, 152, 280, 408, 504, 536} {
                foreach \n in {24, 2088}
                    {(\n, \m) circle (2.5pt)}};
        \path (24, 24) -- (1056, 536) node[acc] {\SI{1.13}\percent};
        \path (1056, 24) -- (2088, 536) node[acc] {\SI{.83}\percent};
        \path (2088, 24) -- (4152, 536) node[acc] {\SI{.59}\percent};
    \end{subfigaxis}

    \medskip

    \begin{subfigure}\textwidth\centering
        \begin{tikzpicture}[scale=1.6]
            \path (0, .636);
            { [shift={(0, 0)}]
                \filldraw[fill=graybgop, fill opacity=\graybgop]
                    (.024, .024) rectangle (4.152, .536);
                \foreach \lx/\ly/\ux/\uy/\c in {
                    0.024/0.024/0.544/0.536/22.7760673256,
                    0.544/0.024/1.056/0.536/3.33591905182,
                    1.056/0.024/2.088/0.536/16.6796736462,
                    2.088/0.024/4.152/0.536/11.7575433615
                }
                    \filldraw[fill=red!\c!green, fill opacity=.75]
                        (\lx, \ly) rectangle (\ux, \uy);
            }
            { [shift={(4.252, 0)}]
                \filldraw[fill=graybgop, fill opacity=\graybgop]
                    (.024, .024) rectangle (4.152, .536);
                \foreach \lx/\ly/\ux/\uy/\c in {
                    0.024/0.024/0.288/0.536/28.4587908413,
                    0.288/0.024/0.544/0.536/15.4712857129,
                    0.544/0.024/1.056/0.536/3.33591905182,
                    1.056/0.024/2.088/0.536/16.6796736462,
                    2.088/0.024/4.152/0.536/11.7575433615
                }
                    \filldraw[fill=red!\c!green, fill opacity=.75]
                        (\lx, \ly) rectangle (\ux, \uy);
            }
            { [shift={(0, -.636)}]
                \filldraw[fill=graybgop, fill opacity=\graybgop]
                    (.024, .024) rectangle (4.152, .536);
                \foreach \lx/\ly/\ux/\uy/\c in {
                    0.024/0.024/0.288/0.28/22.9480019755,
                    0.024/0.28/0.288/0.536/17.7385455275,
                    0.288/0.024/0.544/0.536/15.4712857129,
                    0.544/0.024/1.056/0.536/3.33591905182,
                    1.056/0.024/2.088/0.536/16.6796736462,
                    2.088/0.024/4.152/0.536/11.7575433615
                }
                    \filldraw[fill=red!\c!green, fill opacity=.75]
                        (\lx, \ly) rectangle (\ux, \uy);
            }
            { [shift={(4.252, -.636)}]
                \filldraw[fill=graybgop, fill opacity=\graybgop]
                    (.024, .024) rectangle (4.152, .536);
                \foreach \lx/\ly/\ux/\uy/\c in {
                    0.024/0.024/0.16/0.28/22.5978096893,
                    0.16/0.024/0.288/0.28/13.9479394764,
                    0.024/0.28/0.288/0.536/17.7385455275,
                    0.288/0.024/0.544/0.536/15.4712857129,
                    0.544/0.024/1.056/0.536/3.33591905182,
                    1.056/0.024/2.088/0.536/16.6796736462,
                    2.088/0.024/4.152/0.536/11.7575433615
                }
                    \filldraw[fill=red!\c!green, fill opacity=.75]
                        (\lx, \ly) rectangle (\ux, \uy);
            }
            { [shift={(0, -1.272)}]
                \filldraw[fill=graybgop, fill opacity=\graybgop]
                    (.024, .024) rectangle (4.152, .536);
                \foreach \lx/\ly/\ux/\uy/\c in {
                    0.024/0.024/0.16/0.152/20.5845333257,
                    0.024/0.152/0.16/0.28/23.6786212,
                    0.16/0.024/0.288/0.28/13.9479394764,
                    0.024/0.28/0.288/0.536/17.7385455275,
                    0.288/0.024/0.544/0.536/15.4712857129,
                    0.544/0.024/1.056/0.536/3.33591905182,
                    1.056/0.024/2.088/0.536/16.6796736462,
                    2.088/0.024/4.152/0.536/11.7575433615
                }
                    \filldraw[fill=red!\c!green, fill opacity=.75]
                        (\lx, \ly) rectangle (\ux, \uy);
            }
            { [shift={(4.252, -1.272)}]
                \filldraw[fill=graybgop, fill opacity=\graybgop]
                    (.024, .024) rectangle (4.152, .536);
                \foreach \lx/\ly/\ux/\uy/\c in {
                    0.024/0.024/0.096/0.152/12.4064791435,
                    0.096/0.024/0.16/0.152/12.8336964497,
                    0.024/0.152/0.096/0.28/11.6060049092,
                    0.096/0.152/0.16/0.28/4.01573270566,
                    0.16/0.024/0.288/0.28/13.9479394764,
                    0.024/0.28/0.288/0.536/17.7385455275,
                    0.288/0.024/0.544/0.536/15.4712857129,
                    0.544/0.024/1.056/0.536/3.33591905182,
                    1.056/0.024/2.088/0.536/16.6796736462,
                    2.088/0.024/4.152/0.536/11.7575433615
                }
                    \filldraw[fill=red!\c!green, fill opacity=.75]
                        (\lx, \ly) rectangle (\ux, \uy);
            }
        \end{tikzpicture}
        \caption{Further refinement steps}
        \label{fig:model:adaptive:x}
    \end{subfigure}

    \mycaption{%
        Modeling through adaptive refinement for \dtrsm[LLNN].
        \captiondetails{\sandybridge, 1~thread, \openblas}
    }
    \label{fig:model:adaptive}
\end{figure}

\begin{example}{Adaptive refinement}{model:adaptive}
    \Cref{fig:model:adaptive} illustrates the adaptive refinement process for
    \displaycall\dtrsm{
        \arg{side}L, \arg{uplo}L, \arg{transA}N, \arg{diag}N,
        \varg mm, \varg nn, \arg{alpha}{1.0},
        \arg AA, \arg{ldA}{5000}, \arg BB, \arg{ldB}{5000}
    }
    i.e., $\dm[height=.7]B \coloneqq \dmAi[size=.7] \dm[height=.7]B$ with
    $\dmA[size=.7]A \in \R^{m \times m}$ and $\dm[height=.7]B \in \R^{m \times
    n}$, with $m \in [24, 536]$ and $n \in [24, 4152]$ on a \sandybridge with
    single-threaded \openblas.  We use a Chebyshev sampling point distributions
    with 6 and 5~values along, respectively, dimensions~$m$ and~$n$, and apply
    adaptive refinement to fit a piecewise polynomial to the minimum of
    15~measurement repetitions until either the maximum error across the
    sampling points falls below~\SI1{\percent} or all domain dimensions fall
    below~64.

    The initial distribution of sampling points \legendline[blue, only marks,
    mark=*, mark size=2.5pt] is shown in \cref{fig:model:adaptive:1}.  The
    polynomial fit to samples in these points has an error measure
    of~\SI{4.21}\percent.  Since this exceeds the error bound of~\SI1\percent,
    the domain is split in half along the (relatively) larger dimension~$n$
    at~$n = \frac{4152 + 24}2 = 2088$.

    The sampling points for the two new domains are displayed in
    \cref{fig:model:adaptive:2}; the error measure for their newly fitted
    polynomials is \SI{3.36}{\percent} ($n \leq 2088$) and \SI{.59}{\percent}
    ($n \geq 2088$).  While the latter is already below the error bound
    of~\SI1\percent, the approximation for $n \leq 2088$ is further refined.

    After the next refinement step (\cref{fig:model:adaptive:3}) the error is
    reduced to~\SI{1.13}{\percent} ($n \leq 1056$) and \SI{.83}{\percent} ($1056
    \leq n \leq 2088$).  As illustrated in \cref{fig:model:adaptive:x}, further
    steps are applied until the error measure is globally below~\SI1{\percent}
    after a total 8~refinements---the process was solely terminated by
    globally reaching the target error bound and not the minimum width of~64.

    While the sampler configuration in this example was chosen to demonstrate
    the adaptive refinement process, the increased number of polynomial pieces
    for smaller problem sizes is typical and in practice commonly triggers the
    minimum-width termination criterion.  However, for kernels with a cubic
    asymptotic complexity (such as \blasl3), generating models for such small
    problem sizes is quite cheep compared to larger sizes.
\end{example}

With the adaptive refinement procedure, we can now generate models for a wide
range of dense linear algebra kernels, and proceed to take a closer look at the
generated models.

    \section{Model Generator Configuration}
    \label{sec:model:config}
    We now discuss the configuration options of the adaptive refinement process, and
examine how they affect the model accuracy and generation cost.  We then select
a default configuration to generate the models used for our performance
predictions in \cref{ch:pred}.

\subsection{Configuration Parameters}

The adaptive refinement is controlled by a total of eight
\definition{configuration parameters}.  They allow to control the model
accuracy, but also affect the time spent for the required measurements.  The
eight parameters regulate the model generation as follows:
\begin{itemize}
    \item To represent the runtime of a kernel, the monomial basis for the
        fitted polynomials needs to at least cover the kernel's asymptotic
        complexity (i.e., its minimal \flop-count).  To better represent
        performance variations, however, the maximum degree of the monomials can
        be increased in each each dimension (i.e., size argument).  We refer to
        this increase as \definition[overfitting:\\between 0
        and~2]{overfitting}; practical values are {\em between 0 and~2}.

    \item To fit a polynomial to a routine's runtime, the number of sampling
        points along each dimension needs to be at least one more than the
        corresponding polynomial degree.  However, since this minimal number of
        points yields a polynomial that fits the measurements perfectly, we
        cannot use it to compute an approximation error.  We hence increase the
        number of sampling points per dimension by at least one, and to further
        improve the approximation accuracy, further points can be added; we
        refer to the total number of points added as
        \definition[oversampling:\\between 1 and~10]{oversampling}; practical
        values are values {\em between 1 and~10}.

    \item We introduced two alternatives to \definition[distribution
        grid:\\Cartesian or Chebyshev]{distribute} sampling points on {\em
        grids} that cover the domains of problem sizes: a {\em Cartesian} grid
        and a {\em Chebyshev} grid.

    \item For each sampling point, we perform several \definition[measurement
        repetitions:\\between 5 and~20]{measurement repetitions}; practical
        values are {\em between 5 and~20}.

    \item From the repetitions, we compute several runtime summary statistics:
        minimum, median, maximum, average, and standard deviation.  One of these
        is selected as the \definition[reference statistic:\\minimum or
        median]{reference statistic}; practical choices are the {\em minimum and
        median}.

    \item From the absolute relative errors in the reference statistic for all
        sampling points, we compute the \definition[error measure:\\average,
        maximum, or 90th~percentile]{error measure} which is these relative
        errors' {\em average, maximum, or 90th~percentile}.

    \item The first termination criterion for the adaptive refinement process is
        the approximation accuracy:  The refinement stops when the computed
        error measure is below a \definition[target error bound:\\between
        {\SIlist[detect-all=true]{1;5}\percent}]{target error bound}; practical
        values for this bound are {\em between
        \SIlist[detect-all=true]{1;5}\percent}.

    \item The second termination criterion is the size of the domains:  The
        refinement stops when a new domain is smaller than a \definition[minimum
        width:\\32 or~64]{minimum width} along all dimensions; typical values
        are {\em 32 and~64}.
\end{itemize}

\subsection{Trade-Off and Configuration Selection}

In the following, we analyze the accuracy of our models and their generation
cost, and select a configuration to generate the models for the performance
predictions in the \cref{ch:pred}.

We consider the model generation for
\displaycall\dtrsm{
    \arg{side}L, \arg{uplo}L, \arg{transA}N, \arg{diag}N,
    \arg m{\it\color{blue}m}, \arg n{\it\color{blue}n}, \arg{alpha}{1.0},
    \arg AA, \arg{ldA}{5000}, \arg BB, \arg{ldB}{5000}
}
i.e., $\dmB[height=.5] \coloneqq \dmAi[size=.5] \dmB[height=.5]$ with
$\dmA[size=.5] \in \R^{m \times m}$ and $\dmB[height=.5] \in \R^{m \times n}$,
for sizes $m \in [24, 536]$ and $n \in [24, 4152]$ on a \sandybridge and a
\haswell using single-threaded \openblas, \blis, and \mkl.

For each setup, our first step is to exhaustively measure the \dtrsm[LLNN]'s
runtime 15~times in all points $(m, n)$ in the domain $[24, 536] \times [24,
4152]$ at which both~$m$ and~$n$ are multiples of~8---a total of \num{504075}
measurements.   These measurements are used both as the basis for our model
generation and to evaluate the model accuracy across the entire domain (contrary
to the model generation, which can only evaluate the error in its sampling
points).

\begin{table}\tablestyle
    \begin{tabular}{ll}
        \toprule
        parameter               &values \\
        \midrule
        overfitting             &0, 1, 2\\
        oversampling            &1, 2, 3, 4, 5, 6, 7, 8, 9, 10\\
        distribution grid       &Cartesian, Chebyshev\\
        measurement repetitions &5, 10, 15\\
        reference statistic     &minimum, median\\
        error measure           &90th~percentile, maximum\\
        target error bound      &\SI1\percent, \SI2\percent\\
        minimum width           &32, 64\\
        \bottomrule
    \end{tabular}

    \caption{
        Configuration parameters for the model generation and their studied
        values.
    }
    \label{tbl:model:config}
\end{table}

We generate models for all 2880~configurations obtained from combining the
parameter values shown in \cref{tbl:model:config}.  These configurations result
in a wide range of models with significantly different accuracies and generation
costs.  To evaluate them, we quantify the \definition{model error} as the
averaged relative error of the predicted minimum runtime~$p(\x_i)$ relative to
the measured minimum~$y_i$ across all $N = \num{33605}$ points~$\x_i$ of the
domain:
\[
    \text{model error} \defeqq
    \frac1N \sum_{i=1}^N \frac{\lvert p(\x_i) - y_i \rvert}{y_i} \enspace;
\]
furthermore, we define the \definition{model cost} as the total runtime of the
required measurements used as samples.

\begin{figure}[p]\figurestyle
    \pgfplotsset{
        ymajorgrids=false,
        xlabel=$n$,
        ylabel=$m$,
        xmin=20, xmax=4156,
        ymin=20, ymax=540,
        xtick={24, 1056, 2088, 3120, 4152},
        ytick={24, 536},
        scale mode=scale uniformly,
        cell picture=true,
        every subfigaxis/.style={
            fig width=\textwidth,
            fig align=\centering,
        }
    }

    \ref*{leg:model:modelplots}

    \begin{subfigaxis}[
            fig caption={
                Least accurate
                (model error: \SI{.92}\percent; cost: \SI{1.68}\second)
            },
            fig label=modelplots:maxerr,
        ]
        \addplot[opacity=.75] graphics[xmin=20, xmax=4156, ymin=20, ymax=540]
            {model/figures/data/modelplots/maxerr.png};
        \draw[blue] foreach \ly/\lx/\uy/\ux in {24/24/536/4152}
            {(\lx, \ly) rectangle (\ux, \uy)};
    \end{subfigaxis}

    \begin{subfigaxis}[
            fig caption={
                Most accurate
                (model error: \SI{.12}\percent; cost: \SI{5.48}\minute)
            },
            fig label=modelplots:minerr,
        ]
        \addplot[opacity=.75] graphics[xmin=20, xmax=4156, ymin=20, ymax=540]
            {model/figures/data/modelplots/minerr.png};
        \draw[blue] foreach \ly/\lx/\uy/\ux in {
            280/2088/536/4152, 280/1056/536/2088, 152/2088/280/4152,
            280/544/536/1056, 152/1056/280/2088, 88/2088/152/4152,
            280/24/536/288, 280/288/536/544, 152/544/280/1056, 88/1056/152/2088,
            152/288/280/544, 88/544/152/1056, 24/1056/88/1576, 152/24/280/160,
            152/160/280/288, 88/288/152/544, 24/24/152/96, 88/160/152/288,
            24/96/88/160, 88/96/152/160, 24/160/88/224, 24/224/88/288,
            24/288/88/352, 24/352/88/416, 24/416/88/480, 24/480/88/544,
            24/544/88/608, 24/608/88/672, 24/672/88/736, 24/736/88/800,
            24/800/88/864, 24/864/88/928, 24/928/88/992, 24/992/88/1056,
            24/1576/88/1640, 24/1640/88/1704, 24/1704/88/1768, 24/1768/88/1832,
            24/1832/88/1896, 24/1896/88/1960, 24/1960/88/2024, 24/2024/88/2088,
            24/2088/88/2160, 24/2160/88/2224, 24/2224/88/2288, 24/2288/88/2352,
            24/2352/88/2416, 24/2416/88/2480, 24/2480/88/2544, 24/2544/88/2608,
            24/2608/88/2672, 24/2672/88/2736, 24/2736/88/2800, 24/2800/88/2864,
            24/2864/88/2928, 24/2928/88/2992, 24/2992/88/3056, 24/3056/88/3120,
            24/3192/88/3256, 24/3256/88/3320, 24/3320/88/3384, 24/3384/88/3448,
            24/3448/88/3512, 24/3512/88/3576, 24/3576/88/3640, 24/3640/88/3704,
            24/3704/88/3768, 24/3768/88/3832, 24/3832/88/3896, 24/3896/88/3960,
            24/3960/88/4024, 24/4024/88/4088, 24/4088/88/4152, 24/3120/88/3160,
            24/3160/88/3192
        } {(\lx, \ly) rectangle (\ux, \uy)};
    \end{subfigaxis}

    \begin{subfigaxis}[
            fig caption={
                Least expensive
                (model error: \SI{.73}\percent; cost: \SI{.96}\second)
            },
            fig label=modelplots:mincost,
            colormap name=greenred, colorbar horizontal,
            point meta min=0, point meta max=5,
            colorbar style={
                width=.5\textwidth,
                xtick={0, ..., 5},
                xlabel={absolute relative error}, x unit=\percent,
                scale mode=auto,
            },
            colorbar to name=leg:model:modelplots,
        ]
        \addplot[opacity=.75] graphics[xmin=20, xmax=4156, ymin=20, ymax=540]
            {model/figures/data/modelplots/mincost.png};
        \draw[blue] foreach \ly/\lx/\uy/\ux in {24/24/536/4152}
            {(\lx, \ly) rectangle (\ux, \uy)};
    \end{subfigaxis}

    \begin{subfigaxis}[
            fig caption={
                Most expensive
                (model error: \SI{.22}\percent; cost: \SI{15.49}\minute)
            },
            fig label=modelplots:maxcost,
        ]
        \addplot[opacity=.75] graphics[xmin=20, xmax=4156, ymin=20, ymax=540]
            {model/figures/data/modelplots/maxcost.png};
        \draw[blue] foreach \ly/\lx/\uy/\ux in {
            152/2088/280/4152, 280/544/536/1056, 152/1056/280/2088,
            88/2088/152/4152, 280/3120/408/4152, 152/544/280/1056,
            88/1056/152/2088, 408/3120/536/3640, 280/160/536/288,
            152/288/280/544, 88/544/152/1056, 408/1320/536/1576,
            280/2608/344/3120, 280/24/536/96, 280/416/408/544, 408/288/536/416,
            280/1056/344/1320, 344/1320/408/1576, 472/1056/536/1320,
            280/1576/344/1832, 472/1576/536/1832, 408/1832/472/2088,
            280/2088/344/2352, 280/2352/344/2608, 344/2088/408/2352,
            344/2352/408/2608, 344/2864/408/3120, 408/2088/472/2352,
            472/2352/536/2608, 408/2608/472/2864, 472/2608/536/2864,
            408/3896/472/4152, 472/3640/536/3896, 152/160/216/288,
            88/416/152/544, 280/288/344/416, 344/288/408/416, 408/416/472/544,
            344/1192/408/1320, 280/1448/344/1576, 408/1192/472/1320,
            280/1960/344/2088, 344/1704/408/1832, 344/1832/408/1960,
            408/1704/472/1832, 472/1832/536/1960, 344/2736/408/2864,
            408/2352/472/2480, 472/2224/536/2352, 408/2992/472/3120,
            472/2992/536/3120, 408/3640/472/3768, 472/4024/536/4152,
            24/96/88/160, 88/96/152/160, 152/64/280/96, 152/96/216/160,
            216/96/280/160, 24/160/88/224, 24/224/88/288, 88/160/152/224,
            88/224/152/288, 216/160/280/224, 216/224/280/288, 280/96/344/160,
            344/96/408/160, 408/96/472/160, 472/96/536/160, 24/288/88/352,
            24/352/88/416, 24/416/88/480, 24/480/88/544, 88/288/152/352,
            88/352/152/416, 472/416/536/480, 472/480/536/544, 24/544/88/608,
            24/608/88/672, 24/672/88/736, 24/736/88/800, 24/800/88/864,
            24/864/88/928, 24/928/88/992, 24/992/88/1056, 24/1128/88/1192,
            24/1192/88/1256, 24/1256/88/1320, 24/1320/88/1384, 24/1384/88/1448,
            24/1448/88/1512, 24/1512/88/1576, 24/1576/88/1640, 24/1640/88/1704,
            24/1704/88/1768, 24/1768/88/1832, 24/1832/88/1896, 24/1896/88/1960,
            24/1960/88/2024, 24/2024/88/2088, 344/1056/408/1128,
            344/1128/408/1192, 280/1320/344/1384, 280/1384/344/1448,
            408/1056/472/1128, 408/1128/472/1192, 280/1832/344/1896,
            280/1896/344/1960, 344/1576/408/1640, 344/1640/408/1704,
            344/1960/408/2024, 344/2024/408/2088, 408/1576/472/1640,
            408/1640/472/1704, 472/1960/536/2024, 472/2024/536/2088,
            24/2160/88/2224, 24/2224/88/2288, 24/2288/88/2352, 24/2352/88/2416,
            24/2416/88/2480, 24/2480/88/2544, 24/2544/88/2608, 24/2608/88/2672,
            24/2672/88/2736, 24/2736/88/2800, 24/2800/88/2864, 24/2864/88/2928,
            24/2928/88/2992, 24/2992/88/3056, 24/3056/88/3120, 24/3192/88/3256,
            24/3256/88/3320, 24/3320/88/3384, 24/3384/88/3448, 24/3448/88/3512,
            24/3512/88/3576, 24/3576/88/3640, 24/3640/88/3704, 24/3704/88/3768,
            24/3768/88/3832, 24/3832/88/3896, 24/3896/88/3960, 24/3960/88/4024,
            24/4024/88/4088, 24/4088/88/4152, 344/2608/408/2672,
            344/2672/408/2736, 408/2480/472/2544, 408/2544/472/2608,
            472/2160/536/2224, 408/2864/472/2928, 408/2928/472/2992,
            472/2864/536/2928, 472/2928/536/2992, 408/3768/472/3832,
            408/3832/472/3896, 472/3896/536/3960, 472/3960/536/4024,
            24/24/88/64, 24/64/88/96, 88/24/152/64, 88/64/152/96, 152/24/216/64,
            216/24/280/64, 24/1056/88/1096, 24/1096/88/1128, 24/2088/88/2128,
            24/2128/88/2160, 24/3120/88/3160, 24/3160/88/3192,
            472/2088/536/2128, 472/2128/536/2160
        } {(\lx, \ly) rectangle (\ux, \uy)};
    \end{subfigaxis}

    \mycaption{%
        Accuracy and structure of models for \dtrsm[LLNN].
        \captiondetails{\sandybridge, 1~thread, \openblas}
    }
    \label{fig:model:modelplots}
\end{figure}

\begin{table}\tablestyle
    \begin{tabular}{lllll}
        \toprule
                                &\multicolumn2c{model error}            &\multicolumn2c{model cost} \\
                                &minimum            &maximum            &minimum            &maximum \\
        \midrule
        overfitting             &1                  &0                  &0                  &1 \\
        oversampling            &10                 &2                  &1                  &9 \\
        distribution grid       &Cartesian          &Chebyshev          &Cartesian          &Cartesian \\
        measurement repetitions &15                 &5                  &5                  &5 \\
        reference statistic     &median             &minimum            &minimum            &median \\
        error measure           &maximum            &90th perc.         &maximum            &maximum \\
        target error bound      &\SI1\percent       &\SI2\percent       &\SI2\percent       &\SI1\percent \\
        minimum width           &32                 &32                 &64                 &32 \\
        \midrule
        model error             &\SI{.12}\percent   &\SI{.92}\percent   &\SI{.73}\percent   &\SI{.22}\percent \\
        model cost              &\SI{5.48}\minute   &\SI{1.68}\second   &\SI{.96}\second    &\SI{15.49}\minute \\
        \bottomrule
    \end{tabular}

    \mycaption{%
        Model configuration parameters for minimum and maximum error and cost.
        \captiondetails{\sandybridge, 1~thread, \openblas}
    }
    \label{tbl:model:modelplots}
\end{table}

\begin{example}{Model accuracy}{model:acc}
    \Cref{fig:model:modelplots} shows the structure and point-wise accuracy of
    the four models with minimum and maximum accuracy and cost for
    single-threaded \openblas on a \sandybridge; \cref{tbl:model:modelplots}
    lists the corresponding configurations.  Both the cheapest and least
    accurate model use only a single polynomial for the entire domain but also
    offer only poor accuracy.  The expensive and accurate models on the other
    hand subdivide the domain repetitively, and thus find a better fitting
    piecewise polynomial.
\end{example}

\begin{figure}[p]\figurestyle
    \ref*{leg:blas:circ}

    \pgfplotsset{
        every axis/.append style={
            xlabel=model error, x unit=\percent,
            ylabel=model cost, y unit=\minute,
            clip=false,
        },
        table/search path={model/figures/data/tradeoff},
        plotheightsub=2.5pt
    }
    \pdfsuppresswarningpagegroup=1

    \def\xmax{3.663033722941479}
    \def\ymin{-.6028058423922666}\def\ymax{63.294613451188}
    \begin{subfigaxis}[
            fig width=\textwidth,
            fig caption=All 2880 configurations,
            fig label=tradeoff:full,
            xmax=\xmax, ymax=\ymax
        ]
        \addplot graphics[points={(0, \ymin) (\xmax, \ymax)}]
            {model/figures/data/tradeoff/full.png};
    \end{subfigaxis}

    \def\xmax{1.6138456091630848}
    \def\ymin{-.6028058423922666}\def\ymax{63.294613451188}
    \begin{subfigaxis}[
            fig caption=Within $2 \times$ of most accurate,
            fig label=tradeoff:within2err,
            twocolplot, xmax=\xmax, ymax=\ymax
        ]
        \addplot graphics[points={(0, \ymin) (\xmax, \ymax)}]
            {model/figures/data/tradeoff/within2err.png};
    \end{subfigaxis}\hfill
    \def\xmax{1.516014196735335}%
    \def\ymin{-.39395012207313335}\def\ymax{41.364762817679}%
    \begin{subfigaxis}[
            fig caption=Below 10th percentile in cost,
            fig label=tradeoff:belowmedcost,
            twocolplot, xmax=\xmax, ymax=\ymax
        ]
        \addplot graphics[points={(0, \ymin) (\xmax, \ymax)}]
            {model/figures/data/tradeoff/belowmedcost.png};
        \node[align=center, text width=3cm] (def) at (.5, 25)
            {default\\configuration\par};
        \foreach \x/\y/\c in {
            0.197359936492/1.55262269872/plotsbopen,
            0.214630012174/1.81293255854/plotsbblis,
            0.311360191981/1.24348791526/plotsbmkl,
            0.312787995073/0.99623371806/plothwopen,
            1.12916774706/6.15191690081/plothwblis,
            0.792048206807/3.19212656229/plothwmkl%
        } {\edef\temp{
            \noexpand\filldraw[thick, \c, draw=white]
                (\x, \y) circle (2pt) coordinate (x);
            \noexpand\draw[\c, -stealth, shorten >=2.5pt] (def) -- (x);
        }\temp}
    \end{subfigaxis}

    \mycaption{%
        Model configuration trade-off in accuracy versus cost and steps towards
        selecting a default configuration.
        \captiondetails{1~thread, error in the minimum measure}
    }
    \label{fig:model:tradeoff}
\end{figure}

The accuracy and cost of all 2880~generated models for each setup are presented
in \cref{fig:model:tradeoff:full}; in this plot, the preferable models with low
error and cost are found close to the origin.  All setups share the same general
trend:  Models with low accuracy are quite cheap, while models with high
accuracy are more expensive.  Hence we are faced with a
\definition[trade-off:\\accuracy vs.~cost]{trade-off between accuracy and cost}.
However, the configuration selection is not straight-forward:  Models with
practically identical accuracy are up to a factor of~16 apart in generation
cost, and a cheap and accurate configuration for one setup may be neither for
other setups.  In the following, we describe how we approach the search-space of
all considered configurations, and identify a desirable default configuration
that we subsequently use to generate the models for all setups and kernels
needed for our performance predictions in \cref{ch:pred}.

Before we begin to reduce our search space, we notice that on the \haswellshort,
the models for both \blis~(\ref*{plt:hwblis:circ}) and
\mkl~(\ref*{plt:hwmkl:circ}) are on average less than half as accurate than for
the other setups.  The cause is a rather jagged performance behavior, which is
difficult to represent accurately.  Hence, to identify a good default
configuration, we consider only the \sandybridgeshort~(\ref*{plt:sbopen:circ},
\ref*{plt:sbblis:circ}, \ref*{plt:sbmkl:circ}) and \openblas on the
\haswellshort~(\ref*{plt:hwopen:circ}).

Our first step is to \definition{prune by accuracy}:  We discard any
configuration that for any of the considered setups yields a model error larger
than $1.5\times$ the minimum error for that setup;  in other words, all
remaining configurations generate models that are at most \SI{50}{\percent}
less accurate than the most accurate model.  This step reduces the number of
potential configurations from 2880 to~163; all remaining configurations use an
oversampling value of~3 or higher, and a target error bound of~\SI1\percent.
\cref{fig:model:tradeoff:within2err} shows the 163~remaining models' accuracy
and cost.

\begin{table}\tablestyle
    \sisetup{detect-weight}
    \tikzset{
        every node/.append style={
            rotate=90, inner sep=0, align=left, text width=2.2cm
        }
    }
    \renewcommand{\arraystretch}{1.1}
    \begin{tabular}{rcccccccc}
        \toprule
        &\tikz\node{overfitting};
        &\tikz\node{oversampling};
        &\tikz\node{distribution grid};
        &\tikz\node{measurement repetitions};
        &\tikz\node{reference statistic};
        &\tikz\node{error\\measure};
        &\tikz\node{target error bound};
        &\tikz\node{minimum width};\\
        \midrule
        (1)     &    0  &\bf 4  &\bf Chebyshev  &\bf 10         &    median     &\bf maximum    &\bf \SI1\percent   &\bf 32\\
        (2)     &    0  &\bf 4  &\bf Chebyshev  &    15         &\bf minimum    &\bf maximum    &\bf \SI1\percent   &\bf 32\\
        (3)     &    1  &    5  &\bf Chebyshev  &\bf 10         &    median     &\bf maximum    &\bf \SI1\percent   &\bf 32\\
        (4)     &    1  &    5  &\bf Chebyshev  &\bf 10         &\bf minimum    &\bf maximum    &\bf \SI1\percent   &\bf 32\\
        (5)     &    1  &    8  &    Cartesian  &\hphantom15    &\bf minimum    &\bf maximum    &\bf \SI1\percent   &\bf 32\\
        (6)     &\bf 2  &\bf 4  &    Cartesian  &\hphantom15    &    median     &\bf maximum    &\bf \SI1\percent   &    64\\
        (7)     &\bf 2  &\bf 4  &    Cartesian  &\bf 10         &    median     &\bf maximum    &\bf \SI1\percent   &\bf 32\\
        (8)     &\bf 2  &\bf 4  &\bf Chebyshev  &\bf 10         &    median     &\bf maximum    &\bf \SI1\percent   &\bf 32\\
        (9)     &\bf 2  &\bf 4  &\bf Chebyshev  &\bf 10         &    median     &\bf maximum    &\bf \SI1\percent   &    64\\
        \color{blue}\bf (10)
                &\color{blue}\bf 2
                        &\color{blue}\bf 4
                                &\color{blue}\bf Chebyshev
                                                &\color{blue}\bf 10
                                                                &\color{blue}\bf minimum
                                                                                &\color{blue}\bf maximum
                                                                                                &\color{blue}\bf \SI1\percent
                                                                                                                    &\color{blue}\bf 32\\
        (11)    &\bf 2  &\bf 4  &\bf Chebyshev  &\bf 10         &\bf minimum    &\bf maximum    &\bf \SI1\percent   &    64\\
        (12)    &\bf 2  &\bf 4  &\bf Chebyshev  &    15         &\bf minimum    &\bf maximum    &\bf \SI1\percent   &\bf 32\\
        (13)    &\bf 2  &\bf 4  &\bf Chebyshev  &    15         &\bf minimum    &\bf maximum    &\bf \SI1\percent   &    64\\
        (14)    &\bf 2  &    7  &    Cartesian  &\bf 10         &\bf minimum    &\bf maximum    &\bf \SI1\percent   &\bf 32\\
        \bottomrule
    \end{tabular}

    \mycaption{%
        Model generator configurations remaining after pruning.
        \iftableoflist\else\\
            {\bf Bold}: majority value.
            {\color{blue}\bf Blue}: default configuration.
        \fi
    }
    \label{tbl:model:tradeoff:final}
\end{table}

Our second step is to similarly \definition{prune by cost}:  We discard any
configuration that for any considered setup takes longer than the first quartile
in generation time for that setup; in other words, the remaining models are all
within the \SI{25}{\percent} that are generated the fastest.  This step further
reduces the number of potential configurations from 163 to~14, as shown in
\cref{fig:model:tradeoff:belowmedcost}.

The parameter values for the 14~remaining configurations are shown in
\cref{tbl:model:tradeoff:final}.  For each parameter, we can find one value that
is common to at least 8~of the 14~configurations (highlighted in {\bf bold}).
We choose our \definition{default configuration} by selecting this most common
value for each parameter.  It corresponds to line~(10) in
\cref{tbl:model:tradeoff:final} (highlighted in {\bf\color{blue}blue}), and is
marked for each setup in \cref{fig:model:tradeoff:belowmedcost}.  Note that it
also serves as a good choice between accuracy and cost for
\blis~(\ref*{plt:hwblis:circ}) and \mkl~(\ref*{plt:hwmkl:circ}) on the
\haswellshort, which were not included in the pruning process.

\subsection{Variations of the Default Configuration}

While the configuration was found to yield good accuracies at reasonable costs
for almost all encountered kernels, it proves to be quite expensive for kernels
with \definition[3D case (\dgemm)]{three degrees of freedom}, which for the
predictions in \cref{ch:pred} only applies to {\em\dgemm} with its three size
arguments~\code m, \code n, and~\code k.  To reduce the modeling cost for this
kernel, we adjust the default configuration by reduce the overfitting from~2
to~0, and increasing the minimum width from~32 to~64.

Furthermore, the performance of \blas kernels becomes less smooth when we bring
\definition{multi-threading} into the picture.  Hence, to avoid excessive
partitioning as seen in \cref{fig:model:modelplots:maxcost}, we increase the
minimum width for all models to~64, and for \dgemm to~256.

    \section{Summary}
    \label{sec:model:sum}
    This chapter first studied the effects of various kernel argument types on
performance, and then introduced the structure of our performance models and
their automated measurement-based generation.  Since this generation process
offers various configuration parameters, we studied the trade-off between the
resulting models' accuracy and generation cost, and concluded with the selection
of default configurations, which are used to generate all models for the
following chapter's performance predictions.

}

    \chapter[Model-Based Predictions for Blocked Algorithms]
        {Model-Based Predictions\newline for Blocked Algorithms}
\chapterlabel{pred}
{
    \newcommand\Q[3]{\ensuremath{#1^\mathrm{#2}_\mathrm{#3}}\xspace}
\newcommand\Qi[3]{\ensuremath{#1^{#2}_\mathrm{#3}}\xspace}
\newcommand\bpred{\ensuremath{b_\mathrm{pred}}\xspace}
\newcommand\bopt{\ensuremath{b_\mathrm{opt}}\xspace}
\newcommand\bdef{\ensuremath{b_\mathrm{def}}\xspace}
\pgfplotsset{
    dashdotted 2 color/.style 2 args={
        #1, postaction={draw, densely dashdotted, #2}
    }
}

    With accurate performance models at hand, we predict the runtime and performance
of blocked algorithms in order to both select the fastest algorithm for a
given operation from available alternatives and tune its block size.  We thereby
arrive at a near-optimal solution entirely without executing any of the
potential algorithms and configurations; compared to tuning through empirical
measurements, accurate model-based performance predictions are orders of
magnitude faster.

For this chapter, we generated performance models to predict all studied
algorithms with problem sizes up to~$n = 4152$ and block sizes between~$b = 24$
and~536.  E.g., our models for \dtrsm each cover 4 cases (combinations of flag
argument values) and domains (ranges of problem sizes) of size $[24, 536] \times
[24, 4152]$.

We begin by introducing runtime, performance, and efficiency predictions for
executions of blocked algorithms in \cref{sec:pred:pred}, followed by accuracy
metrics for such predictions in \cref{sec:pred:acc}.  Next, we present a
detailed study on the prediction accuracy for a blocked Cholesky decomposition
under various conditions in \cref{sec:pred:chol} and a broader accuracy
evaluation for a range of blocked \lapack algorithms in \cref{sec:pred:lapack}.
We then apply our predictions to identify the fastest blocked algorithms for
different operations in \cref{sec:pred:var}, and finally determine near-optimal
block sizes for a range of algorithms in \cref{sec:pred:b}.

\paragraph{Publication}
The work presented in this chapter is in parts based on research previously
published in:
\begin{pubitemize}
    \pubitem{relapack}
    \pubitem{caching}
    \pubitem{pred}
    \pubitem{msthesis}
\end{pubitemize}

    \section{Performance Prediction}
    \label{sec:pred:pred}
    Based on our performance models, we now predict the runtime and performance of
individual blocked algorithm executions.  For each algorithm, the problem size
and the block size uniquely determine the exact \definition{sequence of calls}
(i.e., kernel invocations).  For each call~$\mathcal C$ in this sequence and a
selected hardware and software setup, our performance models provide a
\definition[runtime estimate and prediction]{runtime
estimate}~$t_\mathrm{est}(\mathcal C)$.  Summing these estimates yields our {\em
runtime prediction}
\begin{equation}
    t_\mathrm{pred} \defeqq \sum_{\text{calls } \mathcal C}
    t_\mathrm{est}(\mathcal C) \enspace.
\end{equation}

\begin{table}\tablestyle
    \renewcommand\A[3]{{\color{blue}\it A\textsubscript{#1#2}}}

    \begin{tabular}{clr}
        \toprule
        step
            &\multicolumn1c{call $\mathcal C$}
                &\multicolumn1c{$t_\text{est}(\mathcal C)$}\\
        \midrule
        \arrayrulecolor{lightblue}
        &\dtrmm{}\code{(%
            \overarg{side}R,\overarg{uplo}L,\overarg{transA}N,\overarg{diag}N,
            \overarg m{300}, \overarg n{\color{darkred}0},
            \overarg{alpha}{1.0},
            \overarg A{\A001}, \overarg{ldA}{800},
            \overarg B{\A101}, \overarg{ldB}{800})}
                &\SI{0.00}\ms \\
        1
        &\dtrsm{}\code{(%
            \overarg{side}L,\overarg{uplo}L,\overarg{transA}N,\overarg{diag}N,
            \overarg m{300}, \overarg n{\color{darkred}0},
            \overarg{alpha}{-1.0},
            \overarg A{\A111}, \overarg{ldA}{800},
            \overarg B{\A101}, \overarg{ldB}{800})}
                &\SI{0.00}\ms \\
        &\dtrti2{}\code{(%
            \overarg{uplo}L, \overarg n{300},
            \overarg A{\A111}, \overarg{ldA}{800})}
                &\SI{2.64}\ms \\
        \midrule
        &\dtrmm{}\code{(%
            \overarg{side}R,\overarg{uplo}L,\overarg{transA}N,\overarg{diag}N,
            \overarg m{300}, \overarg n{300},
            \overarg{alpha}{1.0},
            \overarg A{\A002}, \overarg{ldA}{800},
            \overarg B{\A102}, \overarg{ldB}{800})}
                &\SI{1.71}\ms \\
        2
        &\dtrsm{}\code{(%
            \overarg{side}L,\overarg{uplo}L,\overarg{transA}N,\overarg{diag}N,
            \overarg m{300}, \overarg n{300},
            \overarg{alpha}{-1.0},
            \overarg A{\A112}, \overarg{ldA}{800},
            \overarg B{\A102}, \overarg{ldB}{800})}
                &\SI{2.07}\ms \\
        &\dtrti2{}\code{(%
            \overarg{uplo}L, \overarg n{300},
            \overarg A{\A112}, \overarg{ldA}{800})}
                &\SI{2.64}\ms \\
        \midrule
        &\dtrmm{}\code{(%
            \overarg{side}R,\overarg{uplo}L,\overarg{transA}N,\overarg{diag}N,
            \overarg m{200}, \overarg n{600},
            \overarg{alpha}{1.0},
            \overarg A{\A002}, \overarg{ldA}{800},
            \overarg B{\A102}, \overarg{ldB}{800})}
                &\SI{4.15}\ms \\
        3
        &\dtrsm{}\code{(%
            \overarg{side}L,\overarg{uplo}L,\overarg{transA}N,\overarg{diag}N,
            \overarg m{200}, \overarg n{600},
            \overarg{alpha}{-1.0},
            \overarg A{\A112}, \overarg{ldA}{800},
            \overarg B{\A102}, \overarg{ldB}{800})}
                &\SI{2.17}\ms \\
        &\dtrti2{}\code{(%
            \overarg{uplo}L, \overarg n{200},
            \overarg A{\A112}, \overarg{ldA}{800})}
                &\SI{0.85}\ms \\
        \arrayrulecolor{blue}
        \midrule
        &\multicolumn1r{$t_\text{pred}$:}
                &\SI{16.22}\ms \\
        \bottomrule
    \end{tabular}

    \mycaption{%
        Sequence of calls, runtime estimates, and accumulated prediction for the
        inversion of a lower-triangular matrix with blocked algorithm~1.
        \captiondetails{$n = 800$, $b = 300$, \sandybridge, \openblas, 1~thread,
        statistic: median}
    }
    \label{tbl:pred:pred}
\end{table}

\begin{example}{Runtime prediction}{pred:pred}
    \Cref{tbl:pred:pred} lists the sequence of calls to invert a
    lower-triangular matrix of size~$n = 800$ (i.e., $\dm[lower]A\lowerpostsep
    \coloneqq \dm[lower, inv]A$ with $\dm[lower]A\lowerpostsep \in \R^{800
    \times 800}$) using blocked algorithm~1 (\cref{alg:chol1} on
    \cpageref{alg:chol1}) with block size~$b = 300$; for each call the table's
    last column presents median runtime estimates from performance models for a
    \sandybridge with single-threaded \openblas.  The sum of these estimates is
    our runtime prediction for the entire algorithm: $t_\mathrm{pred} =
    \SI{16.22}\ms$.

    Note that with block size $b = 300$, the algorithm traverses the input
    matrix of size~$n = 800$ in three steps, and in each step the
    sub-matrices~$A_{00}$, $A_{10}$, and~$A_{11}$ refer to different portions of
    \dm[lower]A\lowerpostsep, i.e., after every three calls in
    \cref{tbl:pred:pred}.  As a result, the first two calls perform no
    operations since their size arguments~\code n are~0 (i.e., their operand
    operand~$A_{10}$ has a width of~0); hence their estimated runtime
    is~\SI0\ms.
\end{example}

Our performance models estimate the runtime  of kernel invocations not as a
single number but as a range of summary statistics: minimum~\Q t{min}{est},
median~\Q t{med}{est}, maximum~\Q t{max}{est}, mean (average)~\Q t\mu{est}, and
standard deviation~\Q t\sigma{est}.  Each of these \definition[prediction
statistics]{statistics} is also available for our {\em prediction}:
\begin{gather}
    \Qi ts{pred}
    \defeqq \sum_{\text{calls } \mathcal C} \Qi ts{est}(\mathcal C)
        \quad \text{for } s \in
            \{\mathrm{min}, \mathrm{med}, \mathrm{max}, \mu\}
        \enspace,\\
    \Q t\sigma{pred}
    \defeqq \sqrt{\sum_{\text{calls } \mathcal C} \Q t\sigma{est}(\mathcal C)^2}
    \enspace.
\end{gather}
Note that the definition for the standard deviation~\Q t\sigma{pred} assumes
uncorrelated estimates~$\Q t\sigma{est}(\mathcal C)$.

\begin{example}{Prediction summary statistics}{pred:stat}
    For the algorithm execution in \cref{ex:pred:pred}, our predictions yield
    the following summary statistics:
    \begin{gather*}
        \Q t{min}{pred}  = \SI{16.18}\ms \qquad
        \Q t{med}{pred}  = \SI{16.22}\ms \qquad
        \Q t{max}{pred}  = \SI{16.46}\ms \enspace\\
        \Q t\mu{pred}    = \SI{16.25}\ms \qquad
        \Q t\sigma{pred} = \SI{95.88}{\micro\second} \enspace.
    \end{gather*}

    The predictions indicate only minimal runtime fluctuations:  The predicted
    standard deviation~\Q t\sigma{pred} is only \SI{.59}{\percent} of the
    mean~\Q t\mu{pred}.
\end{example}

Predictions for derived metrics, such as performance and (compute-bound)
efficiency, are obtained from the runtime prediction in combination with
properties of the operation and the execution hardware (see \cref{app:term}):
\begin{itemize}
    \item The \definition[performance\\prediction]{performance
        prediction}~$p_\mathrm{pred}$ is computed from the runtime prediction
        and the operation's cost (i.e., minimal \flop-count):
        \newcommand\cost{\text{cost}}
        \begin{gather}
            \Q p{min}{pred}  \defeqq \frac\cost{\Q t{max}{pred}} \qquad
            \Q p{med}{pred}  \defeqq \frac\cost{\Q t{med}{pred}} \qquad
            \Q p{max}{pred}  \defeqq \frac\cost{\Q t{min}{pred}} \enspace\\
            \Q p\mu{pred}    \defeqq \frac\cost{\Q t\mu{pred}}
                \left(1 + \frac{{\Q t\sigma{pred}}^2}{{\Q t\mu{pred}}^2}\right)
                \qquad
            \Q p\sigma{pred}
            \defeqq \cost \times \frac{\Q t\sigma{pred}}{{\Q t\mu{pred}}^2}
                \enspace.
        \end{gather}
        Note that the definitions of the performance prediction's mean~\Q
        p\mu{pred} and standard deviation~\Q p\sigma{pred} are, respectively,
        second- and first-order approximations through Taylor
        expansions~\cite[Section~4.3.2]{avgstdexpansion}.

    \item The \definition{efficiency prediction}~$e_\mathrm{pred}$ is obtained
        from the performance prediction and the processor's peak
        (floating-point) performance:
        \begin{equation}
            \Qi es{pred} \defeqq \frac{\Qi ps{pred}}{\text{peak performance}}
                \quad \text{for } s \in
                    \{\mathrm{min}, \mathrm{med}, \mathrm{max}, \mu, \sigma\}
            \enspace.
        \end{equation}
\end{itemize}

\begin{example}{Performance and efficiency predictions}{pred:perf}
    Following \cref{ex:pred:pred,ex:pred:stat}, we consider that the
    inversion of a triangular matrix of size~$n = 800$ has a minimal cost of
    $\SIvar{\frac 16 n (n + 1) (2 n + 1)}\flops = \SI{170986800}\flops$ and
    obtain the following performance prediction:
    \begin{gather*}
        \Q p{min}{pred}  = \SI{10.39}{\giga\flops\per\second} \qquad
        \Q p{max}{pred}  = \SI{10.57}{\giga\flops\per\second} \enspace\\
        \Q p{med}{pred}  = \SI{10.54}{\giga\flops\per\second} \enspace\\
        \Q p\mu{pred}    = \SI{10.52}{\giga\flops\per\second} \qquad
        \Q p\sigma{pred} = \SI{62.09}{\mega\flops\per\second} \enspace.
    \end{gather*}

    If we compare this prediction to the \sandybridge's theoretical
    single-threaded peak performance of \SI{20.8}{\giga\flops\per\second}, we
    arrive at the following efficiency prediction:
    \begin{gather*}
        \Q e{min}{pred}  = \SI{49.93}\percent \qquad
        \Q e{med}{pred}  = \SI{50.68}\percent \qquad
        \Q e{max}{pred}  = \SI{50.81}\percent \enspace\\
        \Q e\mu{pred}    = \SI{50.59}\percent \qquad
        \Q e\sigma{pred} = \SI{.30}\percent \enspace.
    \end{gather*}
\end{example}

    \section{Accuracy Quantification}
    \label{sec:pred:acc}
    We evaluate the accuracy of our performance models by comparing their
predictions to \definition{measurements}.  For this purpose, we time the
predicted algorithm ten times (with the \sampler), and compute the summary
statistics minimum~\Q t{min}{meas}, median~\Q t{med}{meas}, maximum~\Q
t{max}{meas}, mean~\Q t\mu{meas}, and standard deviation~\Q t\sigma{meas}.  In
contrast to our predictions, measurement statistics for other metrics, such as
performance~$p_\mathrm{meas}$ and efficiency~$e_\mathrm{meas}$, are obtained by
first computing the metric value for each individual data-point, and then
applying the corresponding statistic.

\begin{example}{Algorithm performance measurements}{pred:meas}
    Measuring the runtime of the triangular matrix inversion from
    \cref{ex:pred:pred} ten times yields the following results:
    \begin{gather*}
        \SI{16.25}\ms \quad
        \SI{16.27}\ms \quad
        \SI{16.26}\ms \quad
        \SI{16.27}\ms \quad
        \SI{16.26}\ms \hphantom.\\
        \SI{16.26}\ms \quad
        \SI{16.28}\ms \quad
        \SI{16.27}\ms \quad
        \SI{16.26}\ms \quad
        \SI{16.26}\ms \mathrlap{\enspace.}
    \end{gather*}
    From these repetitions, we obtain the following summary statistics:
    \begin{gather*}
        \Q t{min}{meas}  = \SI{16.25}\ms \qquad
        \Q t{med}{meas}  = \SI{16.26}\ms \qquad
        \Q t{max}{meas}  = \SI{16.25}\ms \\
        \Q t\mu{meas}    = \SI{16.26}\ms \qquad
        \Q t\sigma{meas} = \SI{7.61}{\micro\second} \enspace.
    \end{gather*}

    These measurements exhibit even less fluctuations than our models predicted
    (\cref{ex:pred:stat}):  The runtime standard deviation~\Q t\sigma{meas} is
    only \SI{.05}{\percent}~of the mean~\Q t\mu{meas}.
\end{example}

We compute the \definition{prediction error}~$x_\mathrm{err}$ for any metric~$x$
as the difference between the prediction and the measurement:
\[
    \Qi xs{err} \defeqq \Qi xs{pred} - \Qi xs{meas}
        \quad \text{for } x \in \{t, p, e\}, s \in
        \{\mathrm{min}, \mathrm{med}, \mathrm{max}, \mu, \sigma\} \enspace.
\]

To compare the prediction error for different algorithms and problem sizes, we
relate it to the predicted metric (e.g., the median measured runtime).  For this
purpose, we compute the \definition{relative error (RE)}~$x_\mathrm{RE}$ with
respect to the measurement:
\[
    \Qi xs{RE} \defeqq \frac{\Qi xs{err}}{\Qi xs{meas}}
    \quad \text{for } x \in \{t, p, e\}, s \in
    \{\mathrm{min}, \mathrm{med}, \mathrm{max}, \mu, \sigma\} \enspace.
\]
Furthermore, to average errors across multiple data-points (e.g., problem sizes
or setups), we use the \definition{absolute relative error
(ARE)}~$x_\mathrm{ARE}$:
\[
    \Qi xs{ARE} \defeqq \left\lvert \Qi xs{RE} \right\rvert
    \quad \text{for } x \in \{t, p, e\}, s \in
    \{\mathrm{min}, \mathrm{med}, \mathrm{max}, \mu, \sigma\} \enspace.
\]

\begin{example}{Prediction error}{pred:err}
    The error of our runtime predictions from \cref{ex:pred:stat} with respect
    to the measurements from \cref{ex:pred:meas} is as follows:
    \begin{gather*}
        \Q t{min}{err}  = \SI{-76.99}{\micro\second} \qquad
        \Q t{med}{err}  = \SI{-38.38}{\micro\second} \qquad
        \Q t{max}{err}  = \SI{208.65}{\micro\second} \\
        \Q t\mu{err}    = \SI{-13.15}{\micro\second} \qquad
        \Q t\sigma{err} = \SI{88.27}{\micro\second} \enspace.
    \end{gather*}
    The corresponding relative error is
    \begin{gather*}
        \Q t{min}{RE}  = \SI{-.47}\percent \qquad
        \Q t{med}{RE}  = \SI{-.24}\percent \qquad
        \Q t{max}{RE}  = \SI{1.28}\percent \\
        \Q t\mu{RE}    = \SI{-.08}\percent \qquad
        \Q t\sigma{RE} = \SI{1160}\percent \enspace.
    \end{gather*}

    Note that the median, minimum, and mean runtimes are slightly
    under-predicted, yet well within~\SI1{\percent} of the measurements.
    However, the prediction for the maximum is somewhat less accurate; this is
    to be expected since it is inherently more susceptible to fluctuations.
    Finally, since the standard deviation was predicted as only
    \SI{.59}{\percent} of the mean but measured even lower at only
    \SI{.05}\percent, its relative error is gigantic; while this observation is
    confirmed in the following section, it does not diminish the otherwise high
    accuracy of our predictions.
\end{example}

    \section[Accuracy Case Study: Cholesky Decomposition]
            {Accuracy Case Study:\newline Cholesky Decomposition}
    \label{sec:pred:chol}
    This section presents an in-depth evaluation of the prediction accuracy for
various execution scenarios of a single algorithm on a fixed hardware and
software setup:  We consider the lower-triangular Cholesky decomposition
\[
    \dm[lower]L \dm[upper, ']L \coloneqq \dm A
\]
of a symmetric positive definite (SPD) matrix $\dm A \in \R^{n \times n}$
(\lapack: \dpotrf[L]) using blocked algorithm~3 (also known as ``right looking''
or ``greedy'').  \Cref{fig:pred:cholalg} recapitulates this algorithm, which was
previously detailed alongside algorithms~1 and~2 in \cref{ex:intro:chol} on
\cpageref{ex:intro:chol}.  We focus on algorithm~3 because, as already seen in
\cref{ex:intro:chol:var}, it is the fastest among the three alternatives.

\begin{figure}\figurestyle

    \begin{subfigure}\subfigwidth\centering
        \begin{tikzpicture}[y={(0, -1)}, scale=.7]
            \def\s{5.5} \def\p{2} \def\q{3}
            \filldraw[mat  ] (0,  0 ) rectangle (\s,     \s    );
            \filldraw[mat00] (0,  0 ) --        (\p-.05, \p-.05) node {$A_{00}$} -| cycle;
            \filldraw[mat10] (0,  \p) rectangle (\p-.05, \q-.05) node {$A_{10}$};
            \filldraw[mat11] (\p, \p) --        (\q-.05, \q-.05) node {$A_{11}$} -| cycle;
            \filldraw[mat20] (0,  \q) rectangle (\p-.05, \s    ) node {$A_{20}$};
            \filldraw[mat21] (\p, \q) rectangle (\q-.05, \s    ) node {$A_{21}$};
            \filldraw[mat22] (\q, \q) --        (\s,     \s    ) node {$A_{22}$} -| cycle;
            \draw[brace ] (\p,     \p) -- (\q-.05, \p    ) node {$b$}; 
            \draw[brace ] (\q-.05, \p) -- (\q-.05, \q-.05) node {$b$}; 
            \draw[bracei] (0,      \s) -- (\s,     \s    ) node {$n$}; 
            \draw[bracei] (0,      0 ) -- (0,      \s    ) node {$n$}; 
            \draw[ultra thick, lightgray, -stealth]
                (1.5 + .5, 1.5 - 1) -- (\s - 1.5 + 1, \s - 1.5 - .5);
        \end{tikzpicture}
        \caption{Matrix partitioning}
        \label{fig:pred:choltraversal}
    \end{subfigure}\hfill
    \begin{subfigure}\subfigwidth
        \makeatletter\let\tikz@ensure@dollar@catcode=\relax\makeatother
        \newcommand\Aoo{\dm[mat11, size=.5, lower]{A_{11}}}
        \newcommand\AooT{\dm[mat11, size=.5, upper, ']{A_{11}}}
        \newcommand\Aoofull{\dm[mat11, size=.5]{A_{11}}}
        \newcommand\Ato{\dm[mat21, width=.5, height=1.25]{A_{21}}}
        \newcommand\AtoT{\dm[mat21, width=1.25, height=.5, ']{A_{21}}}
        \newcommand\Att{\dm[mat22, size=1.25, lower]{A_{22}}}
        \newcommand\traversal{%
            \begin{tikzpicture}[
                    baseline=(x.base),
                    y={(0, -1)}, scale=.1
                ]
                \def\s{5.5} \def\p{2} \def\q{3}
                \filldraw[mat  ] (0,  0 ) rectangle (\s, \s);
                \filldraw[mat00] (0,  0 ) --        (\p, \p) -| cycle;
                \filldraw[mat10] (0,  \p) rectangle (\p, \q);
                \filldraw[mat11] (\p, \p) --        (\q, \q) -| cycle;
                \filldraw[mat20] (0,  \q) rectangle (\p, \s);
                \filldraw[mat21] (\p, \q) rectangle (\q, \s);
                \filldraw[mat22] (\q, \q) --        (\s, \s) -| cycle;
                \path (0, 0) -- (\s, \s) node[midway, black] (x) {$A$};
            \end{tikzpicture}%
        }
        \begin{alglisting}[]
            traverse $\traversal$ along $\tsearrow$:
              !\dpotf[L]2:! $\Aoo \AooT \coloneqq \Aoofull$
              !\dtrsm[RLTN]:! $\Ato \coloneqq \Ato \matmatsep \Aoo^{-1}$
              !\dsyrk[LN]:! $\Att \coloneqq \Att - \Ato \matmatsep \AtoT$
        \end{alglisting}
        \caption{Algorithm 3}
        \label{alg:chol3_2}
    \end{subfigure}

    \mycaption{%
        Blocked algorithm~3 for the lower-triangular Cholesky decomposition.
    }
    \label{fig:pred:cholalg}
\end{figure}

We perform our study on a \sandybridge using \openblas, and begin with
single-threaded predictions for double-precision matrices of varying size
(\cref{sec:pred:chol:n}), then consider different block sizes
(\cref{sec:pred:chol:b,sec:pred:chol:nb}), other data-types
(\cref{sec:pred:chol:dt}), and finally multi-threaded \blas kernels
(\cref{sec:pred:chol:mt}).

\subsection{Varying Problem Size}
\label{sec:pred:chol:n}

\begin{figure}[p]\figurestyle
    \ref*{leg:statf}

    \pgfplotsset{
        twocolplot,
        xlabel=problem size $n$,
        y unit=\second,
        table/search path={pred/figures/data/chol},
        ymax=1.5,
        cycle list name=statarea,
    }

    \begin{subfigaxis}[
            fig caption=Runtime predictions,
            fig label=chol:time:pred,
            ylabel=runtime $t_\mathrm{pred}$
        ]
        \addplot+[name path=mps]
            table[y expr=\thisrow{avg} - \thisrow{std}] {pred128t.stat};
        \addplot+[name path=mms]
            table[y expr=\thisrow{avg} + \thisrow{std}] {pred128t.stat};
        \addplot fill between[of=mps and mms];
        \foreach \stat in {avg, max, min, med}
            \addplot table[y=\stat] {pred128t.stat};
    \end{subfigaxis}\hfill
    \begin{subfigaxis}[
            fig caption=Runtime measurements,
            fig label=chol:time:meas,
            ylabel=runtime $t_\mathrm{meas}$,
            legend to name=leg:statf
        ]
        \addlegendimage{plotmin}\addlegendentry{min\mstrut}\label{plt:min}
        \addlegendimage{plotmed}\addlegendentry{med\mstrut}\label{plt:med}
        \addlegendimage{plotmax}\addlegendentry{max\mstrut}\label{plt:max}
        \addlegendimage{plotavg}\addlegendentry{$\mu$\mstrut}\label{plt:avg}
        \addlegendimage{plotstdf, area legend}
        \addlegendentry{$\mu \pm \sigma$\mstrut}\label{plt:stdf}
        \addplot+[name path=mps]
            table[y expr=\thisrow{avg} - \thisrow{std}] {meas128t.stat};
        \addplot+[name path=mms]
            table[y expr=\thisrow{avg} + \thisrow{std}] {meas128t.stat};
        \label{plt:std}
        \addplot fill between[of=mps and mms];
        \foreach \stat in {avg, max, min, med}
            \addplot table[y=\stat] {meas128t.stat};
    \end{subfigaxis}

    \pgfplotsset{perfplot=20.8}

    \begin{subfigaxis}[
            fig caption=Performance prediction,
            fig label=chol:perf:pred,
            ylabel=performace $p_\mathrm{pred}$
        ]
        \addplot+[name path=mps]
            table[y expr=\thisrow{avg} - \thisrow{std}] {pred128p.stat};
        \addplot+[name path=mms]
            table[y expr=\thisrow{avg} + \thisrow{std}] {pred128p.stat};
        \addplot fill between[of=mps and mms];
        \foreach \stat in {avg, max, min, med}
            \addplot table[y=\stat] {pred128p.stat};
    \end{subfigaxis}\hfill
    \begin{subfigaxis}[
            fig caption=Performance measurements,
            fig label=chol:perf:meas,
            ylabel=performace $p_\mathrm{meas}$
        ]
        \addplot+[name path=mps]
            table[y expr=\thisrow{avg} - \thisrow{std}] {meas128p.stat};
        \addplot+[name path=mms]
            table[y expr=\thisrow{avg} + \thisrow{std}] {meas128p.stat};
        \addplot fill between[of=mps and mms];
        \foreach \stat in {avg, max, min, med}
            \addplot table[y=\stat] {meas128p.stat};
    \end{subfigaxis}

    \mycaption{%
        Measurements and predictions for the Cholesky decomposition.
        \captiondetails{blocked algorithm~3, $b = 128$, \sandybridge, 1~thread,
        \openblas}
    }
    \label{fig:pred:chol:time_perf}
\end{figure}

In our first analysis, we use only one of the \sandybridgeshort's 8~cores and
vary the problem size between~$n = 56$ and~4152 in steps of~64 while keeping the
block size fixed at~$b = 128$.  \Cref{fig:pred:chol:time_perf} shows the runtime
and performance of predictions and measurements for this setup side-by-side.
(Since the red line \legendline[very thick, darkred] at the top of the
performance plots indicates the processor's theoretical peak performance, such
plots can also be interpreted as compute-bound efficiencies with
\SI0{\percent}~at the bottom and \SI{100}{\percent}~at the top.)  The
predictions give a good idea of the algorithm behavior:  While the runtime
increases cubically with the problem size~$n$, the performance is low for small
matrices and increases steadily towards \SI{18}{\giga\flops\per\second}.  At
first sight, the predictions match the measurements well.

\begin{figure}[p]\figurestyle
    \ref*{leg:stat}

    \pgfplotsset{
        twocolplot,
        xlabel=problem size $n$,
        ymin={}, 0line,
        cycle list name=stat,
        table/search path={pred/figures/data/chol},
        figwidthsub=1.8cm
    }

    \begin{subfigaxis}[
            fig caption=Runtime prediction error,
            fig label=chol:err:time,
            ylabel=error $t_\mathrm{err}$, y unit=\ms,
            legend to name=leg:stat
        ]
        \addlegendimage{plotmin}\addlegendentry{min\mstrut}
        \addlegendimage{plotmed}\addlegendentry{med\mstrut}
        \addlegendimage{plotmax}\addlegendentry{max\mstrut}
        \addlegendimage{plotavg}\addlegendentry{$\mu$\mstrut}
        \addlegendimage{plotstd}\addlegendentry{$\sigma$\mstrut}
        \foreach \stat in {std, avg, max, min, med}
            \addplot table[y=\stat] {err128t.stat};
    \end{subfigaxis}\hfill
    \begin{subfigaxis}[
            fig caption=Performance prediction error,
            fig label=chol:err:perf,
            ylabel=error $p_\mathrm{err}$,
            y unit=\si{\mega\flops\per\second},
        ]
        \foreach \stat in {std, avg, max, min, med}
            \addplot table[y=\stat] {err128p.stat};
    \end{subfigaxis}

    \pgfplotsset{
        y unit=\percent,
        ymin=-5, ymax=5
    }

    \begin{subfigaxis}[
            fig caption={
                Relative runtime prediction error\\
                Average \Q t\sigma{ARE} (\ref*{plt:std}): \SI{194.70}\percent
            },
            fig label=chol:re:time,
            ylabel=relative error $t_\mathrm{RE}$
        ]
        \foreach \stat in {std, avg, max, min, med}
            \addplot table[y=\stat] {re128t.stat};
    \end{subfigaxis}\hfill
    \begin{subfigaxis}[
            fig caption={
                Relative perf. prediction error\\
                Average \Q p\sigma{ARE} (\ref*{plt:std}): \SI{189.44}\percent
            },
            fig label=chol:re:perf,
            ylabel=relative error $p_\mathrm{RE}$
        ]
        \foreach \stat in {std, avg, max, min, med}
            \addplot table[y=\stat] {re128p.stat};
    \end{subfigaxis}

    \mycaption{%
        Prediction accuracy for the Cholesky decomposition.
        \captiondetails{
            blocked algorithm~3, $b = 128$,
            \sandybridge, 1~thread, \openblas
        }
    }
    \label{fig:pred:chol:err}
\end{figure}

To further study the accuracy of our predictions, the top half of
\cref{fig:pred:chol:err} presents the prediction errors.  As one might expect,
\cref{fig:pred:chol:err:time} indicates that with increasing problem size, the
magnitude of the runtime prediction error increases for all summary
statistics---most notably for the maximum~(\ref*{plt:max}).  Since in contrast
the performance prediction error~(\cref{fig:pred:chol:err:perf}) is not affected
by the decomposition's cubic runtime, we instead observe the largest prediction
errors for the smallest problem size~$n = 56$.  Furthermore, we find that the
minimum performance prediction error~(\ref*{plt:min}) seems to alternate between
two separate levels: one around \SI0{\mega\flops\per\second} and one close to
\SI{200}{\mega\flops\per\second}.  This behavior, which is also already somewhat
visible in \cref{fig:pred:chol:perf:meas,fig:pred:chol:err:time}, is caused by
measurement fluctuations as discussed in \cref{sec:meas:fluct:longterm}.

We gain more insights from the prediction errors when we compare it to the
predicted quantities.  For this purpose, the bottom half of
\cref{fig:pred:chol:err} presents the relative runtime and performance
prediction errors.  These relative errors for these metrics are almost identical
up to a change in the sign---since the runtime is generally slightly
underestimated, the performance is somewhat overestimated.  Focusing on the
runtime in \cref{fig:pred:chol:re:time}, we notice that the average standard
deviation ARE is~\SI{194.70}\percent~(\ref*{plt:std}), which, as in
\cref{ex:pred:err}, exceeds the error of the other prediction statistics by far.
Furthermore, the previously addressed measurement fluctuations are also clearly
visible in the maximum~(\ref*{plt:max}) as variations with a magnitude
of~\SI{1.5}\percent.  The minimum~(\ref*{plt:min}), median~(\ref*{plt:med}), and
mean~(\ref*{plt:avg}) AREs on the other hand quickly fall below~\SI2{\percent}
for matrices larger than~$n = 200$ and further below below~\SI1{\percent}
beyond~$n \approx 1000$;  across all chosen problem sizes, the average AREs for
the minimum, median and mean runtime are, respectively,
\SIlist{.78;.91;.90}\percent.

Among the eight metrics presented in
\cref{fig:pred:chol:time_perf,fig:pred:chol:err}, we gained the most insight
from 1)~the performance prediction (\cref{fig:pred:chol:perf:pred}), which gives
a good idea of both the algorithm's performance and efficiency, and 2)~the
relative runtime prediction error (\cref{fig:pred:chol:re:time}), which provides
not only an accuracy measure independent of the operation, the algorithm, and
the actual performance, but also indicates whether the runtime is under- or
overestimated.  Hence,  we use these two types of plots in our following
analyses.

\subsection{Varying Block Size}
\label{sec:pred:chol:b}

\begin{figure}\figurestyle
    \ref*{leg:statf}

    \pgfplotsset{
        twocolplot,
        xlabel=block size $b$,
        table/search path={pred/figures/data/chol},
    }

    \begin{subfigaxis}[
            fig caption={Performance prediction\\\vphantom{\Q t\sigma A}},
            fig label=chol:b:perf,
            perfplot=20.8,
            ylabel=performace $p_\mathrm{pred}$,
            cycle list name=statarea,
        ]
        \addplot+[name path=mps] table[y expr=\thisrow{avg} - \thisrow{std}]
            {pred3000p.stat};
        \addplot+[name path=mms] table[y expr=\thisrow{avg} + \thisrow{std}]
            {pred3000p.stat};
        \addplot fill between[of=mps and mms];
        \foreach \stat in {avg, max, min, med}
            \addplot table[y=\stat] {pred3000p.stat};
    \end{subfigaxis}\hfill
    \begin{subfigaxis}[
            fig caption={
                Relative runtime prediction error\\Average \Q t\sigma{ARE}
                (\ref*{plt:std}): \SI{582.41}\percent
            },
            fig label=chol:b:re,
            ylabel=relative error $t_\mathrm{RE}$, y unit=\percent,
            ymin=-5, 0line, ymax=5,
            cycle list name=stat
        ]
        \foreach \stat in {std, avg, max, min, med}
            \addplot table[y=\stat] {re3000t.stat};
    \end{subfigaxis}

    \mycaption{%
        Predictions and prediction accuracy for the Cholesky decomposition with
        varying block size.
        \captiondetails{
            blocked algorithm~3, $n = 3000$,
            \sandybridge, 1~thread, \openblas
        }
    }
    \label{fig:pred:chol:b}
\end{figure}

In our next analysis, we fix the problem size to~$n = 3000$ and vary the block
size between~$b = 24$ and~536 in steps of~8.  \Cref{fig:pred:chol:b} presents
the performance prediction and the relative runtime prediction error for this
scenario using single-threaded \openblas on the \sandybridgeshort.

The performance prediction (\cref{fig:pred:chol:b:perf}) exhibits the typical
trade-off for any blocked algorithm:  While for both small and large block sizes
the algorithm attains rather poor performance, in between it reaches up to
\SI{17.91}{\giga\flops\per\second}, which corresponds to an efficiency
of~\SI{85.10}\percent.  The cause for this trade-off and the selection of block
sizes are addressed in detail in \cref{sec:pred:b}.

Compared to our previous performance predictions
(\cref{fig:pred:chol:perf:pred}), \cref{fig:pred:chol:b:perf} exhibits a far
wider spread of the summary statistics for large block sizes.  In particular,
the predicted minimum performance~(\ref*{plt:min}) drops drastically, which
immediately causes the mean performance~(\ref*{plt:avg}) to decrease and an
enormous increase in the predicted standard deviation~(\ref*{plt:stdf}).

The relative runtime prediction error (\cref{fig:pred:chol:b:re}) indicates that
the predicted performance fluctuations are not present in the performance
measurements:  The maximum and mean relative errors (\ref*{plt:max} and
\ref*{plt:avg}) increase drastically for large problem size, suggesting that the
model generation was influenced by large outlier measurements.  (A repetition of
the generation process would likely encounter different outliers and distort
these metrics statistics for other problem sizes.)  The minimum~(\ref*{plt:min})
and median~(\ref*{plt:med}), on the other hand, are with few exceptions
predicted within~\SI1\percent; their average prediction AREs are
\SI{.36}{\percent} (minimum \ref*{plt:min}) and \SI{.42}{\percent} (median
\ref*{plt:med}).

\subsection{Varying Problem Size and Block Size}
\label{sec:pred:chol:nb}

\begin{figure}[p]\figurestyle
    \pgfplotsset{
        xlabel=problem size $n$,
        ylabel=block size $b$,
        xmax=4156,
        ymax=540,
        scale mode=scale uniformly,
        every colorbar to name picture/.append style=baseline,
        every subfigaxis/.append style={
            fig width=\textwidth,
            fig align=\centering,
        },
        cell picture=true, 
    }

    \ref*{leg:pred:chol:heatmap}
    \quad
    \ref*{leg:pred:chol:heatmapblack}

    \vspace{-.5\baselineskip}

    ARE [\percent{}]

    \medskip

    \begin{subfigaxis}[
            fig caption={
                Minimum runtime prediction ARE \Q t{min}{ARE}
                (average: \SI{.45}\percent)
            },
            fig label=chol:heatmap:min,
            colorbar horizontal,
            colormap name=greenred,
            point meta min=0, point meta max=5,
            colorbar style={
                width=.5\textwidth,
                xtick={0, ..., 5},
                scale mode=auto,
            },
            colorbar to name=leg:pred:chol:heatmap,
        ]
        \addplot graphics[xmin=20, xmax=4156, ymin=20, ymax=540]
            {pred/figures/data/chol/min.png};
    \end{subfigaxis}

    \begin{subfigaxis}[
            fig width=\textwidth,
            fig caption={
                Median runtime prediction ARE \Q t{med}{ARE}
                (average: \SI{.45}\percent)
            },
            fig label=chol:heatmap:med,
            colorbar horizontal,
            colormap name=black,
            colorbar style={
                width=.5cm,
                xmin=0, xmax=1,
                xtick=.5, xticklabels=\smash{$> 5$}\vphantom{$5$},
                tickwidth=0
            },
            colorbar to name=leg:pred:chol:heatmapblack,
        ]
        \addplot graphics[xmin=20, xmax=4156, ymin=20, ymax=540]
            {pred/figures/data/chol/med.png};
    \end{subfigaxis}

    \begin{subfigaxis}[
            fig caption={
                Maximum runtime prediction ARE \Q t{max}{ARE}
                (average: \SI{3.02}\percent)
            },
            fig label=chol:heatmap:max
        ]
        \addplot graphics[xmin=20, xmax=4156, ymin=20, ymax=540]
            {pred/figures/data/chol/max.png};
    \end{subfigaxis}

    \begin{subfigaxis}[
            fig caption={
                Mean runtime prediction ARE \Q t\mu{ARE}
                (average: \SI{.83}\percent)
            },
            fig label=chol:heatmap:avg
        ]
        \addplot graphics[xmin=20, xmax=4156, ymin=20, ymax=540]
            {pred/figures/data/chol/avg.png};
    \end{subfigaxis}

    \mycaption{%
        Prediction accuracy for the Cholesky decomposition.
        \iftableoflist\else\\Average \Q t\sigma{ARE}: \SI{346.87}{\percent}\fi
        \captiondetails{blocked algorithm~3, \sandybridge, 1~thread, \openblas}
    }
    \label{fig:pred:chol:heatmap}
\end{figure}

If we vary both the problem size~$n$ and the block size~$b$, we can visualize
the runtime prediction ARE as a set of heat-maps as shown in
\cref{fig:pred:chol:heatmap}.  Note that these plots are based on a total of
\num{39690}~measurements of the algorithm's runtime (65~problem sizes, $\approx
65$~block sizes, 10~repetitions) that took over 4~hours.  The performance models
for the kernels needed for the predictions (\dpotf[L]2, \dtrsm[RLTN], and
\dsyrk[LN]), on the other hand, were generated in just under 10~minutes,
produced our predictions in under \SI{20}\second.

The standard deviation ARE is once again too large to fit the chosen scale and
is hence not shown.  Furthermore, as already seen in \cref{fig:pred:chol:b}, the
maximum prediction becomes rather inaccurate for large~$n$ and~$b$, which also
has a negative impact on the mean prediction.  On the other hand, both the
minimum and median predictions are overall quite accurate with an average ARE of
only~\SI{.45}\percent.

Since in the following we compare multiple alternative algorithms and
hardware/software setups, we limit our focus to a single statistic.
While in the previous analysis the runtime minimum or median were predicted with
equivalent accuracy, in practice the expected performance is better represented
by the median runtime.\footnote{%
    In scenarios other than our considered single-node computations different
    measures might be preferable; e.g., the 90th~percentile runtime.
}  Hence, from now on we use the \definition[accuracy
measure: relative median runtime prediction error]{relative median runtime
prediction error}~\Q t{med}{RE} as our {\em prediction accuracy measure}.

\subsection{Other Data-Types}
\label{sec:pred:chol:dt}

\begin{table}\tablestyle
    \begin{tabular}{llll}
        \toprule
        \multicolumn1c{data-type}  &\multicolumn3c{kernels}\\
        \midrule
        single-precision real       &\spotf[L]2 &\strsm[RLTN] &\ssyrk[LN]\\
        double-precision real       &\dpotf[L]2 &\dtrsm[RLTN] &\dsyrk[LN]\\
        single-precision complex    &\cpotf[L]2 &\ctrsm[RLTN] &\cherk[LN]\\
        double-precision complex    &\zpotf[L]2 &\ztrsm[RLTN] &\zherk[LN]\\
        \bottomrule
    \end{tabular}
    \caption{Kernels in the Cholesky decomposition for different data-types.}
    \label{tab:pred:chol:fp}
\end{table}

\begin{figure}\figurestyle
    \ref*{leg:pred:cholfp}

    \pgfplotsset{
        twocolplot,
        xlabel=problem size $n$,
        table/search path={pred/figures/data/chol},
        cycle list={blue, green, red, orange},
    }

    \begin{subfigaxis}[
            fig caption={
                Performance prediction\\Dashed lines: theoretical peak
            },
            fig label=chol:fp:perf,
            after end axis/.append code={
                \draw[thick, blue, postaction={draw, green, dashed}]
                    (current axis.north west) -- (current axis.north east);
            },
            ylabel=performace \Q p{med}{pred},
            y unit=\si{\giga\flops\per\second},
            ymax=2*20.8,
        ]
        \draw[thick, red, postaction={draw, orange, dashed}]
            (axis description cs: 0, .5) -- (axis description cs:1, .5);
        \addplot table[y=med] {schol/pred128p.stat};
        \addplot table[y=med] {cchol/pred128p.stat};
        \addplot table[y=med] {pred128p.stat};
        \addplot table[y=med] {zchol/pred128p.stat};
    \end{subfigaxis}\hfill
    \begin{subfigaxis}[
            fig caption={Relative runtime prediction error\\\mstrut},
            fig label=chol:fp:are,
            ylabel=relative error \Q t{med}{RE}, y unit=\percent,
            ymin=-5, 0line, ymax=5,
            legend to name=leg:pred:cholfp, legend columns=2
        ]
        \addplot table[y=med] {schol/re128t.stat};
        \addlegendentry{single-precision real}\label{plt:dt:s}
        \addplot table[y=med] {cchol/re128t.stat};
        \addlegendentry{single-precision complex}\label{plt:dt:c}
        \addplot table[y=med] {re128t.stat};
        \addlegendentry{double-precision real}\label{plt:dt:d}
        \addplot table[y=med] {zchol/re128t.stat};
        \addlegendentry{double-precision complex}\label{plt:dt:z}
    \end{subfigaxis}

    \mycaption{%
        Predictions and prediction accuracy for the Cholesky decomposition with
        different data-types.
        \captiondetails{algorithm~3, $b = 128$, \sandybridge, 1~thread,
        \openblas, median}
    }
    \label{fig:pred:chol:fp}
\end{figure}

So far, we have considered the Cholesky decomposition of real double-precision
matrices; however, the same algorithm is also applicable to other data-types.
For the four de-facto standard numerical data-types (real and complex\footnote{%
    For the complex cases, the Cholesky decomposition is of the form $L L^H
    \coloneqq A$, where $A$~must be Hermitian positive definite (HPD).
} floating-point numbers in single- and double-precision)
\cref{tab:pred:chol:fp} summarizes the algorithm's \blas and \lapack kernels,
and \Cref{fig:pred:chol:fp} presents our model's performance predictions and
their accuracy.  (For each data-type, we generated a separate set of performance
models.)

In the performance predictions (\cref{fig:pred:chol:fp:perf}), we observe that
the real double-precision version~(\ref*{plt:dt:d}) is most efficient (with
respect to its theoretical peak performance); this was to be expected because
\openblas is most optimized for this data-type.  In contrast, it is somewhat
surprising that, while single-precision complex~(\ref*{plt:dt:c}) is noticeably
more performant than single-precision real~(\ref*{plt:dt:s}), double-precision
complex~(\ref*{plt:dt:z}) does not exceed an efficiency of~\SI{50}\percent.

Although the algorithm's performance for the four data-types differs
significantly, \cref{fig:pred:chol:fp:perf} reveals that our models predict the
runtime for all of them equally well.  Moreover, for the in comparison
inefficient double-precision complex variant~(\ref*{plt:dt:z}), the prediction
is already notably accurate small problem sizes below~$n = 1000$.

With equally accurate predictions demonstrated for four data-types, we will in
the following focus on real operations in double-precision.

\subsection{Multi-Threaded \blas}
\label{sec:pred:chol:mt}

\begin{figure}\figurestyle
    \ref*{leg:nt8}

    \pgfplotsset{
        twocolplot,
        xlabel=problem size $n$,
        table/search path={pred/figures/data/chol},
        cycle list={red, orange, green, blue},
    }

    \begin{subfigaxis}[
            fig caption={
                Performance prediction\\Dashed lines: theoretical peak
            },
            fig label=cholp:perf,
            ylabel=performace \Q p{med}{pred},
            y unit=\si{\giga\flops\per\second},
            ymax=166.4,
            after end axis/.append code={
                \draw[dashed, very thick, blue]
                    (current axis.north west) -- (current axis.north east);
            }
        ]
        \foreach \p/\c in {.5/green, .25/orange, .125/red} {\edef\temp{
            \noexpand\draw[dashed, very thick, \c]
                (axis description cs:0, \p) --
                (axis description cs:1, \p);
        }\temp}
        \addplot table[y=med] {pred128p.stat};
        \addplot table[y=med] {mt2/pred128p.stat};
        \addplot table[y=med] {mt4/pred128p.stat};
        \addplot table[y=med] {mt8/pred128p.stat};
    \end{subfigaxis}\hfill
    \begin{subfigaxis}[
            fig caption={Relative runtime prediction error\\\mstrut},
            fig label=cholp:are,
            ylabel=relative error \Q t{med}{RE}, y unit=\percent,
            ymin=-5, 0line, ymax=5,
            legend to name=leg:nt8
        ]
        \addplot table[y=med] {re128t.stat};
        \addlegendentry{1 core}\label{plt:nt:1}
        \addplot table[y=med] {mt2/re128t.stat};
        \addlegendentry{2 cores}\label{plt:nt:2}
        \addplot table[y=med] {mt4/re128t.stat};
        \addlegendentry{4 cores}\label{plt:nt:4}
        \addplot table[y=med] {mt8/re128t.stat};
        \addlegendentry{8 cores}\label{plt:nt:8}
    \end{subfigaxis}

    \mycaption{%
        Predictions and prediction accuracy for the Cholesky decomposition with
        multi-threaded \openblas.
        \captiondetails{algorithm~3, $b = 128$, \sandybridge, median}
    }
    \label{fig:pred:cholp}
\end{figure}

Finally, we consider how multi-threading (through \openblas) impacts the
algorithm's performance and our predictions' accuracy.  For this purpose,
\cref{fig:pred:cholp} presents the predicted performance of the Cholesky
decomposition and the prediction accuracy with 1, 2, 4, and 8~threads on the
8-core \sandybridgeshort.  (For each of these four levels of parallelism, a
separate set of performance models was generated.)

The predictions show that, while the performance grows with the number of
threads, the efficiency decreases from~\SI{87.74}{\percent} with one thread to a
maximum of~\SI{70.78}{\percent} with eight threads.  Furthermore, the
performance curves become less smooth with increased parallelism.

Considering our prediction's accuracy, we notice that for small problem sizes
below~$n = 500$, the prediction ARE increases significantly when more threads
are added.  Beyond this point however, the prediction for 1~(\ref*{plt:nt:1})
and 2~threads~(\ref*{plt:nt:2}) are both highly accurate with an average ARE
of~\SI{.46}{\percent}; the predictions for 4~(\ref*{plt:nt:4}) and
8~threads~(\ref*{plt:nt:8}) are slightly less accurate and the AREs fluctuate
around~\SI1\percent.  Note that the large fluctuations within the ARE for the
multi-threaded algorithms are caused by the combination of the block size~$b =
128$ and the chosen problem sizes in steps of~64.  While with
8~threads~(\ref*{plt:nt:8}) these fluctuations are represented by our
predictions to some degree, with 2~(\ref*{plt:nt:2}) and
4~threads~(\ref*{plt:nt:4}), they are most striking for large problem sizes,
where our models do not predict such fluctuations.

\subsection{Summary}
\label{sec:pred:chol:sum}

We studied the blocked Cholesky decomposition algorithm~3 on a \sandybridge
using \openblas with varying problem and block sizes, data-types, and kernel
parallelism.  We analyzed this algorithm's measured and predicted runtime and
performance to evaluate the accuracy of our predictions, and selected the
relative median runtime prediction error~\Q t{med}{RE} as our primary accuracy
measure.

    \section[Accuracy Study: Blocked \lapack Algorithms]
            {Accuracy Study:\newline Blocked \lapack Algorithms}
    \label{sec:pred:lapack}
    \begin{figure}[p]\figurestyle

    \makeatletter\let\tikz@ensure@dollar@catcode=\relax\makeatother

    \newcommand\Azz{\dm[mat00, lower]{A_{00}}}
    \newcommand\Aoz{\dm[mat10, height=.5]{A_{10}}}
    \newcommand\AozT{\dm[mat10, width=.5, ']{A_{10}}}
    \newcommand\Aoo{\dm[mat11, size=.5, lower]{A_{11}}}
    \newcommand\AooT{\dm[mat11, size=.5, upper, ']{A_{11}}}
    \newcommand\Aooi{\dm[mat11, size=.5, upper, inv]{A_{11}}}
    \newcommand\Aoofull{\dm[mat11, size=.5]{A_{11}}}
    \newcommand\AooU{\dm[mat11, size=.5, upper]{A_{11}}}
    \newcommand\Aot{\dm[mat12, width=1.25, height=.5]{A_{12}}}
    \newcommand\Atz{\dm[mat20, height=1.25]{A_{20}}}
    \newcommand\Ato{\dm[mat21, width=.5, height=1.25]{A_{21}}}
    \newcommand\AtoT{\dm[mat21, width=1.25, height=.5, ']{A_{21}}}
    \newcommand\Att{\dm[mat22, size=1.25, lower]{A_{22}}}
    \newcommand\Attfull{\dm[mat22, size=1.25]{A_{22}}}

    \newcommand\Looi{\dm[mat11, size=.5, upper, inv, fill opacity=.2]{L_{11}}}
    \newcommand\LooiT{\dm[mat11, size=.5, upper, inv', fill opacity=.2]{L_{11}}}
    \newcommand\Lto{\dm[mat21, width=.5, height=1.25, fill opacity=.2]{L_{21}}}
    \newcommand\LtoT{\dm[mat21, width=1.25, height=.5, ', fill opacity=.2]{L_{21}}}
    \newcommand\Ltti{\dm[mat22, size=1.25, lower, inv, fill opacity=.2]{L_{22}}}

    \newcommand\traversal[2][]{%
        \begin{tikzpicture}[
                baseline=(x.base),
                y={(0, -1)}, scale=.1
            ]
            \def\s{5.5} \def\p{2} \def\q{3}
            \filldraw[mat,   #1] (0,  0 ) rectangle (\s, \s);
            \filldraw[mat00, #1] (0,  0 ) --        (\p, \p) -| cycle;
            \filldraw[mat10, #1] (0,  \p) rectangle (\p, \q);
            \filldraw[mat11, #1] (\p, \p) --        (\q, \q) -| cycle;
            \filldraw[mat20, #1] (0,  \q) rectangle (\p, \s);
            \filldraw[mat21, #1] (\p, \q) rectangle (\q, \s);
            \filldraw[mat22, #1] (\q, \q) --        (\s, \s) -| cycle;
            \path (0, 0) -- (\s, \s) node[midway, black] (x) {$#2$};
        \end{tikzpicture}%
    }

    \newcommand\traversalfull{%
        \begin{tikzpicture}[
                baseline=(x.base),
                y={(0, -1)}, scale=.1
            ]
            \def\s{5.5} \def\p{2} \def\q{3}
            \filldraw[mat  ] (0,  0 ) rectangle (\s, \s);
            \filldraw[mat00] (0,  0 ) rectangle (\p, \p);
            \filldraw[mat01] (\p, 0 ) rectangle (\q, \p);
            \filldraw[mat02] (\q, 0 ) rectangle (\s, \p);
            \filldraw[mat10] (0,  \p) rectangle (\p, \q);
            \filldraw[mat11] (\p, \p) rectangle (\q, \q);
            \filldraw[mat12] (\q, \p) rectangle (\s, \q);
            \filldraw[mat20] (0,  \q) rectangle (\p, \s);
            \filldraw[mat21] (\p, \q) rectangle (\q, \s);
            \filldraw[mat22] (\q, \q) rectangle (\s, \s);
            \path (0, 0) -- (\s, \s) node[midway, black] (x) {$A$};
        \end{tikzpicture}%
    }

    \begin{subfigure}[b]{.4375\textwidth}
        \begin{alglisting}[]
            traverse $\traversal A$ along $\tsearrow$:
              !\dtrmm[LLTN]:! $\Aoz \coloneqq \AooT \Aoz$
              !\dlauu[L]2:! $\Aoofull \coloneqq \Aoo \AooT$
              !\dgemm[TN]:! $\Aoz \coloneqq \Aoz + \AtoT \Atz$
              !\dsyrk[LT]:! $\Aoo \coloneqq \Aoo + \AtoT \Ato$
        \end{alglisting}
        \caption{\dlauum[L]: $\dm A \coloneqq \dm[upper, ']A \dm[lower]A$}
        \label{alg:dlauum}
    \end{subfigure}\hfill
    \begin{subfigure}[b]{.5375\textwidth}
        \begin{alglisting}[]
            traverse $\traversal A$ and $\traversal[fill opacity=.2]L$ along $\tsearrow$:
              !\dsygs[1L]2:! $\Aoofull \coloneqq \Looi \Aoofull \LooiT$
              !\dtrsm[RLTN]:! $\Ato \coloneqq \Ato \LooiT$
              !\dsymm[RL]:! $\Ato \coloneqq \Ato - \frac12 \Lto \Aoofull$
              !\dsyrtk[LN]:! $\!\Att \coloneqq \Att - \Ato \LtoT - \Lto \AtoT$
              !\dsymm[RL]:! $\Ato \coloneqq \Ato - \frac12 \Lto \Aoofull$
              !\dtrsm[LLNN]:! $\Ato \coloneqq \Ltti \Ato$
        \end{alglisting}
        \caption{%
            \dsygst[1L]: $\dm A \coloneqq \dm[lower, inv]L \dm A \matmatsep
            \dm[upper, inv']L$
        }
        \label{alg:dsygst}
    \end{subfigure}

    \medskip

    \begin{subfigure}[b]{.4375\textwidth}
        \begin{alglisting}[]
            traverse $\traversal A$ along $\tnwarrow$:
              !\dtrmm[LLNN]:! $\Ato \coloneqq \Att \Ato$
              !\dtrsm[RLNN]:! $\Ato \coloneqq -\Ato \Aooi$
              !\dtrti[LN]2:! $\Aoo \coloneqq \Aooi$
        \end{alglisting}
        \caption{\dtrtri[LN]: $\dm[lower]A \coloneqq \dm[lower, inv]A$}
        \label{alg:dtrtri}
    \end{subfigure}\hfill
    \begin{subfigure}[b]{.5375\textwidth}
        \begin{alglisting}[]
            traverse $\traversal A$ along $\tsearrow$:
              !\dsyrk[LN]:! $\Aoo \coloneqq \Aoo - \Aoz \matmatsep \AozT$
              !\dpotf[L]2:! $\Aoo \AooT \coloneqq \Aoofull$
              !\dgemm[NT]:! $\Ato \coloneqq \Ato - \Atz \matmatsep \AozT$
              !\dtrsm[RLTN]:! $\Ato \coloneqq \Ato \matmatsep \Aooi$
        \end{alglisting}
        \caption{\dpotrf[L]: $\dm[lower]A \dm[upper, ']A \coloneqq \dm A$}
        \label{alg:dpotrf}
    \end{subfigure}

    \medskip

    \begin{subfigure}[b]\subfigwidth
        \begin{alglisting}[]
            $\dm[dashed]P \coloneqq \dm[diag]I$
            traverse $\traversalfull$ along $\tsearrow$:
              !\dgetf2:! $\dmstack\Aoo\Ato \AooU \coloneqq \dmstack\Aoofull\Ato$, update $\dm[dashed]P$
              !\dlaswp:! apply $\dm[dashed]P$ to $\dmstack\Aoz\Atz$
              !\dlaswp:! apply $\dm[dashed]P$ to $\dmstack\Aot\Attfull$
              !\dtrsm[LLNU]:! $\Aot \coloneqq \Aoo \Aot$
              !\dgemm[NN]:! $\Attfull \coloneqq \Attfull - \Ato \Aot$
        \end{alglisting}
        \caption{%
            \dgetrf: $\dm[dashed]P \matmatsep \dm[lower]L \dm[upper]U \coloneqq
            \dm A$
        }
        \label{alg:dgetrf}
    \end{subfigure}\hfill
    \begin{subfigure}[b]\subfigwidth\centering
        \begin{tikzpicture}[y={(0, -1)}, scale=.7]
            \def\s{5.5} \def\p{2} \def\q{3}
            \filldraw[mat  ] (0,  0 ) rectangle (\s,     \s    );
            \filldraw[mat00] (0,  0 ) rectangle (\p-.05, \p-.05) node {$A_{00}$};
            \filldraw[mat01] (\p, 0 ) rectangle (\q-.05, \p-.05) node {$A_{01}$};
            \filldraw[mat02] (\q, 0 ) rectangle (\s,     \p-.05) node {$A_{02}$};
            \filldraw[mat10] (0,  \p) rectangle (\p-.05, \q-.05) node {$A_{10}$};
            \filldraw[mat11] (\p, \p) rectangle (\q-.05, \q-.05) node {$A_{11}$};
            \filldraw[mat12] (\q, \p) rectangle (\s,     \q-.05) node {$A_{12}$};
            \filldraw[mat20] (0,  \q) rectangle (\p-.05, \s    ) node {$A_{20}$};
            \filldraw[mat21] (\p, \q) rectangle (\q-.05, \s    ) node {$A_{21}$};
            \filldraw[mat22] (\q, \q) rectangle (\s,     \s    ) node {$A_{22}$};
            \draw[brace ] (\p, 0 ) -- (\q-.05, 0     ) node {$b$}; 
            \draw[brace ] (\s, \p) -- (\s,     \q-.05) node {$b$}; 
            \draw[bracei] (0,  \s) -- (\s,     \s    ) node {$n$}; 
            \draw[bracei] (0,  0 ) -- (0,      \s    ) node {$n$}; 
        \end{tikzpicture}
        \caption{Matrix partitioning}
        \label{fig:pred:acc:matpart}
    \end{subfigure}

    \mycaption{%
        \lapack's blocked algorithms for \dlauum[L], \dsygst[1L], \dtrtri[LN],
        \dpotrf[L], and \dgetrf.
    }
    \label{fig:pred:acc:lapackalgs}
\end{figure}

We now extend our analysis from the previous case study to a larger group of
algorithms and a wider range of hardware and software setups.  We consider six
of \lapack's blocked algorithms:
\begin{description}
    \item[{\dlauum[L]}] Lower-triangular matrix multiplication with its
        transpose:
        \[
            \dm A \coloneqq \dm[upper, ']L \matmatsep \dm[lower]L
        \]
        with $\dm[lower]L\lowerpostsep \in \R^{n \times n}$ and $\dm A \in \R^{n
        \times n}$ symmetric.  The algorithm, outlined in \cref{alg:dlauum},
        overwrites \dm[lower]L\lowerpostsep with~\dm A in lower-triangular
        storage.

        \dlauum arises as part of the inversion of symmetric positive definite
        (SPD) matrices ($\dm A \coloneqq \dm[inv]A$).

    \item[{\dsygst[1L]}] Two-sided symmetric lower-triangular linear system
        solve\footnote{
            \dsygst's first flag argument indicates whether (\code 1) $L^{-1}$
            or (\code 2) $L$ is applied.
        }:
        \[
            \dm A \coloneqq \dm[lower, inv]L \dm A \matmatsep \dm[upper, inv']L
        \]
        with $\dm[lower]L\lowerpostsep \in \R^{n \times n}$ and $\dm A \in \R^{n
        \times n}$ symmetric in lower-triangular storage.  The algorithm is
        outlined in \cref{alg:dsygst}.

        \dsygst is used to reduce generalized SPD eigenvalue problems (e.g.,
        $\dm A \matvecsep \dv x = \lambda \dm B \matvecsep \dv x$) to the
        standard form ($\dm A \matvecsep \dv x = \lambda \dv x$).

    \item[{\dtrtri[LN]}] Inversion of a lower-triangular matrix:
        \[
            \dm[lower]A\lowerpostsep \coloneqq \dm[lower, inv]A
        \]
        with $\dm[lower]A\lowerpostsep \in \R^{n \times n}$.  The algorithm is
        outlined in \cref{alg:dtrtri}.

        \dtrtri is a building block for the inversion of general and SPD
        matrices, which are used when, instead of the solution of a linear
        system, the actual numeric entries in the inverse matrix are required.

    \item[{\dpotrf[L]}] Lower-triangular Cholesky decomposition:
        \[
            \dm[lower]L \dm[upper, ']L \coloneqq \dm A
        \]
        with $\dm A \in \R^{n \times n}$ SPD in lower-triangular storage and
        $\dm[lower]L\lowerpostsep \in \R^{n \times n}$.  The algorithm, outlined
        in \cref{alg:dpotrf}, overwrites \dm A with~\dm[lower]L\lowerpostsep.

        \dpotrf is central to many operations on SPD matrices, for instance:
        inversion, solution of linear systems ($\dv x \coloneqq \dm[inv]A \dv
        x$), and reduction of generalized eigenvalue problems to standard form.

    \item[\dgetrf] LU~decomposition with partial pivoting:
        \[
            \dm[dotted]P \matmatsep \dm[width=.75, lower]L \dm[size=.75, bbox
            height=1, upper]U \coloneqq \dm[width=.75]A
        \]
        with $\dm[width=.75]A \in \R^{m \times n}$, $\dm[width=.75, lower]L \in
        \R^{m \times \min(m, n)}$ unit triangular, and $\dm[size=.75, upper]U
        \in \R^{\min(m, n) \times n}$.  The algorithm, outlined in
        \cref{alg:dgetrf}, overwrites \dm[width=.75]A with~\dm[width=.75,lower]L
        and~\dm[size=.75, bbox height=1, upper]U, and returns \dm[dotted]P as a
        permutation vector.

        \dgetrf is used to solve linear systems and invert general matrices.

    \item[\dgeqrf] QR~decomposition:
        \[
            \dm[width=.75]Q \matmatsep \dm[size=.75, bbox height=1, upper]R
            \coloneqq \dm[width=.75]A
        \]
        with $\dm[width=.75]A \in \R^{m \times n}$, $\dm[width=.75]Q \in \R^{m
        \times \min(m, n)}$, and $\dm[size=.75, upper]R \in \R^{\min(m, n)
        \times n}$.  The algorithm, outlined in \cref{alg:dgeqrf}, overwrites
        \dm[width=.75]A's upper-triangular (or \mbox{-trapezoidal}) part with
        \dm[size=.75, upper]R, and represents \dm[width=.75]Q as a product of
        elementary reflectors stored in \dm[width=.75]A's strictly
        lower-triangular (or -trapezoidal) part and a vector of scalar
        factors~\dv\tau.

        \dgeqrf is used in several eigensolvers ($\dm Q \matvecsep
        \dm[diag]\Lambda \matvecsep \dm[inv]Q \coloneqq \dm A$), the
        singular-value decomposition ($\dm[width=.75]U \matvecsep \dm[size=.75,
        bbox height=1, diag]\Sigma \matvecsep \dm[size=.75, bbox height=1, ']V
        \coloneqq \dm[width=.75]A$), and least-squares solvers ($\dm[width=.75]X
        \coloneqq \argmin \lVert \dm[width=.75, height=.8]B - \dm[height=.8]A
        \matmatsep \dm[width=.75]X \rVert$).
\end{description}
For \dgetrf and \dgeqrf, we consider the square case with $m = n$.

\begin{figure}\figurestyle

    \begin{subfigure}{.53\textwidth}\centering
        \tikzset{
            every picture/.append style={
                y={(0, -1)}, scale=.7,
                baseline={(0, 0)}
            }
        }
        \def\sm{6} \def\sn{5.5} \def\p{2} \def\q{3}
        \begin{tikzpicture}
            \filldraw[mat  ] (0,  0 ) rectangle (\sn,    \sm   );
            \filldraw[mat  ] (0,  0 ) rectangle (\p-.05, \p-.05) node {$A_{00}$};
            \filldraw[mat  ] (\p, 0 ) rectangle (\q-.05, \p-.05) node {$A_{01}$};
            \filldraw[mat  ] (\q, 0 ) rectangle (\sn,    \p-.05) node {$A_{02}$};
            \filldraw[mat  ] (0,  \p) rectangle (\p-.05, \q-.05) node {$A_{10}$};
            \filldraw[mat11] (\p, \p) rectangle (\q-.05, \q-.05) node {$A_{11}$};
            \filldraw[mat12] (\q, \p) rectangle (\sn,    \q-.05) node {$A_{12}$};
            \filldraw[mat  ] (0,  \q) rectangle (\p-.05, \sm   ) node {$A_{20}$};
            \filldraw[mat21] (\p, \q) rectangle (\q-.05, \sm   ) node {$A_{21}$};
            \filldraw[mat22] (\q, \q) rectangle (\sn,    \sm   ) node {$A_{22}$};
            \draw[brace ] (\p,  0  ) -- (\q-.05, 0     ) node {$b$}; 
            \draw[brace ] (\sn, \p ) -- (\sn,    \q-.05) node {$b$}; 
            \draw[bracei] (0,   \sm) -- (\sn,    \sm   ) node {$n$}; 
            \draw[bracei] (0,   0  ) -- (0,      \sm   ) node {$m$}; 
        \end{tikzpicture}
        \begin{tikzpicture}
            \draw[mat   ] (0, 0 ) -- (0, \sn);
            \draw[blue] (0, \p) -- (0, \q ) node[midway] {$\tau_1$};
        \end{tikzpicture}
        \begin{tikzpicture}
            \filldraw[mat  ] (0, 0) rectangle (1, \sn   );
            \filldraw[mat10] (0, 0) --        (1, 1     ) node {$W_1$} |- cycle;
            \filldraw[mat20] (0, 1) rectangle (1, \sn-\p) node {$W_2$};
            \draw[brace ] (1, 0) -- (1, \sn) node {$n$};
            \draw[brace ] (0, 0) -- (1, 0  ) node {$b$};
            \draw[brace2] (1, 0) -- (1, 1  ) node {$b$};
        \end{tikzpicture}

        \caption{Matrix partitioning}
    \end{subfigure}\hfill
    \begin{subfigure}{.46\textwidth}\centering
        \newcommand\traversalA{
            \begin{tikzpicture}[
                    baseline=(x.base),
                    y={(0, -1)}, scale=.1
                ]
                \def\sm{6} \def\sn{5.5} \def\p{2} \def\q{3}
                \filldraw[mat]   (0,  0 ) rectangle (\sn, \sm);
                \filldraw[mat]   (0,  0 ) rectangle (\p,  \p);
                \filldraw[mat]   (\p, 0 ) rectangle (\q,  \p);
                \filldraw[mat]   (\q, 0 ) rectangle (\sn, \p);
                \filldraw[mat]   (0,  \p) rectangle (\p,  \q);
                \filldraw[mat11] (\p, \p) rectangle (\q,  \q);
                \filldraw[mat12] (\q, \p) rectangle (\sn, \q);
                \filldraw[mat]   (0,  \q) rectangle (\p,  \sm);
                \filldraw[mat21] (\p, \q) rectangle (\q,  \sm);
                \filldraw[mat22] (\q, \q) rectangle (\sn, \sm);
                \path (0, 0) -- (\sn, \sm) node[midway] (x) {$A$};
            \end{tikzpicture}
        }
        \newcommand\traversaltau{
            \begin{tikzpicture}[
                    baseline=(x.base),
                    y={(0, -1)}, scale=.1
                ]
                \def\sm{5.5} \def\sn{5} \def\p{2} \def\q{3}
                \draw[mat   ] (0,  0) rectangle (0, 5);
                \draw[blue] (0,  2) rectangle (0, 3);
                \path (0, 0) -- (0, 5) node[midway] (x) {$\tau$};
            \end{tikzpicture}
        }

        \makeatletter\let\tikz@ensure@dollar@catcode=\relax\makeatother
        \newcommand\Aoo{\dm[mat11, size=.5]{A_{11}}}
        \newcommand\AooL{\dm[mat11, size=.5, lower]{A_{11}}}
        \newcommand\AooLT{\dm[mat11, size=.5, upper, ']{A_{11}}}
        \newcommand\Aot{\dm[mat12, width=1.25, height=.5]{A_{12}}}
        \newcommand\AotT{\dm[mat12, width=.5, height=1.25, ']{A_{12}}}
        \newcommand\Ato{\dm[mat21, width=.5, height=1.5]{A_{21}}}
        \newcommand\Att{\dm[mat22, width=1.25, height=1.5]{A_{22}}}
        \newcommand\AttT{\dm[mat22, width=1.5, height=1.25, ']{A_{22}}}
        \newcommand\Wo{\dm[mat10, size=.5, upper]{W_1}}
        \newcommand\Wt{\dm[mat20, width=.5, height=1.25]{W_2}}
        \newcommand\WtT{\dm[mat20, width=1.25, height=.5, ']{W_2}}
        \newcommand\tauo{\dv[blue]\tau}
        \begin{alglisting}[]
            traverse $\traversalA$ along $\tsearrow$, $\traversaltau$ along $\tsarrow$:
              !\dgeqr2:! $\dmstack\Aoo\Ato, \tauo \coloneqq$ QR($\dmstack\Aoo\Ato$)
              !\dlarft[FC]:! $\Wo \coloneqq$ T($\dmstack\Aoo\Ato, \tauo$)
              !\dlarfb[LTFC]:!
                !$b\times$\dcopy:! $\Wt \coloneqq \AotT$
                !\dtrmm[RLNU]:! $\Wt \coloneqq \Wt \AooL$
                !\dgemm[TN]:! $\Wt \coloneqq \Wt + \AttT \Ato$
                !\dtrmm[RUNN]:! $\Wt \coloneqq \Wt \Wo$
                !\dgemm[NT]:! $\Att \coloneqq \Att - \Ato \WtT$
                !\dtrmm[RLTU]:! $\Wt \coloneqq \Wt \AooLT$
                !inline:! $\Aot \coloneqq \Aot - \WtT$
        \end{alglisting}
        \caption{Algorithm}
    \end{subfigure}

    \mycaption{\lapack's blocked algorithm for \dgeqrf.}
    \label{alg:dgeqrf}
\end{figure}

We study a total of six hardware and software setups:  An 8-core \sandybridge
and a 12-core \haswell with \openblas, \blis, and \mkl.  We consider both the
single-threaded case and the scenario where all processor cores are used by the
\blas implementation (with the exception of \blis, which did not offer a
user-friendly threading model at the time of writing).  For all of these
operations, we both predict and measure the runtime for problem sizes between $n
= 56$ and~4152 in steps of~64.

\subsection{Single-Threaded \blas}
\label{sec:pred:acc:st}

We begin with a study of the single-threaded prediction accuracy with \lapack's
default block size ($b = 64$, except for \dgeqrf with~$b = 32$).  While these
are generally sub-optimal configurations and often even sub-optimal algorithms
for the performed operations, this configuration is unfortunately still
encountered frequently in application codes that use the reference \lapack
implementation.  As such, it forms a quite canonical reference for the
evaluation of our predictions.

\begin{figure}[p]\figurestyle
    \ref*{leg:blas}

    \newcommand\accplot[2]{%
        \begin{subfigaxis}[
                fig label=accst:#1,
                fig caption={#2},
                twocolplot, threerowplot, 0line,
                xlabel={problem size $n$},
                ylabel=relative error \Q t{med}{RE}, y unit=\percent,
                ymin=-10, ymax=10,
                table/search path={pred/figures/data/acc},
                cycle list name=blas,
                plotheightsub=25.6pt,
            ]
            \foreach \system in {SandyBridge, Haswell}
                \foreach \blas in {OpenBLAS, BLIS, MKL}
                    \addplot table {#1.\system.1.\blas/are.dat};
        \end{subfigaxis}%
    }

    \accplot{dlauum}{
        \dlauum[L]: $\dm A \coloneqq \dm[upper, ']L \dm[lower]L$
    }\hfill
    \accplot{dsygst}{
        \dsygst[1L]: $\dm A
        \coloneqq \dm[lower, inv]L \dm A \matmatsep \dm[upper, inv']L$
    }

    \accplot{dtrtri}{
        \dtrtri[LN]: $\dm[lower]A \coloneqq \dm[lower, inv]A$
    }\hfill
    \accplot{dpotrf}{
        \dpotrf[L]: $\dm[lower]L \dm[upper, ']L \coloneqq \dm A$
    }

    \accplot{dgetrf}{
        \dgetrf: $\dm[dotted]P \matmatsep \dm[lower]L \dm[upper]U
        \coloneqq \dm A$
    }\hfill
    \accplot{dgeqrf}{
        \dgeqrf: $\dm Q \matmatsep \dm[upper]R \coloneqq \dm A$
    }

    \mycaption{%
        Single-threaded prediction accuracy for \lapack algorithms.
        \captiondetails{$b = 64$, except \dgeqrf: $b = 32$}
    }
    \label{fig:pred:lapack:st}
\end{figure}

\begin{table}\tablestyle
    \begin{tabular}{lccccccc}
        \toprule
        &\multicolumn3c\sandybridge
        &\multicolumn3c\haswell
        &\multirow3*{\begin{tabular}{@{}c@{}}aver-\\age\end{tabular}} \\
        &\openblas              &\blis                  &\mkl
        &\openblas              &\blis                  &\mkl \\
        &(\ref*{plt:sbopen})    &(\ref*{plt:sbblis})    &(\ref*{plt:sbmkl})
        &(\ref*{plt:hwopen})    &(\ref*{plt:hwblis})    &(\ref*{plt:hwmkl}) \\
        \midrule
        \dlauum[L] \kern-12pt &\SI{1.23}\percent &\SI{2.70}\percent &\SI{1.40}\percent &\SI{0.92}\percent &\SI{0.75}\percent &\SI{2.19}\percent &\SI{1.53}\percent \\
\dsygst[1L]\kern-12pt &\SI{1.05}\percent &\SI{2.05}\percent &\SI{3.31}\percent &\SI{3.58}\percent &\SI{2.44}\percent &\SI{3.35}\percent &\SI{2.63}\percent \\
\dtrtri[LN]\kern-12pt &\SI{0.71}\percent &\SI{2.02}\percent &\SI{1.31}\percent &\SI{2.09}\percent &\SI{1.67}\percent &\SI{1.69}\percent &\SI{1.58}\percent \\
\dpotrf[L] \kern-12pt &\SI{1.44}\percent &\SI{1.03}\percent &\SI{1.44}\percent &\SI{2.05}\percent &\SI{1.52}\percent &\SI{2.44}\percent &\SI{1.65}\percent \\
\dgetrf    \kern-12pt &\SI{1.01}\percent &\SI{0.96}\percent &\SI{0.80}\percent &\SI{1.13}\percent &\SI{1.63}\percent &\SI{1.67}\percent &\SI{1.20}\percent \\
\dgeqrf    \kern-12pt &\SI{1.85}\percent &\SI{2.05}\percent &\SI{3.55}\percent &\SI{3.64}\percent &\SI{3.93}\percent &\SI{2.22}\percent &\SI{2.87}\percent \\
\midrule
average    \kern-12pt &\SI{1.22}\percent &\SI{1.80}\percent &\SI{1.97}\percent &\SI{2.24}\percent &\SI{1.99}\percent &\SI{2.26}\percent &\SI{1.91}\percent \\

        \bottomrule
    \end{tabular}
    \mycaption{%
        Single-threaded runtime prediction ARE~\Q t{med}{ARE} for blocked
        \lapack algorithms averaged across problem sizes.
        \captiondetails{$n = 56, \ldots, 4152$ in steps of~64; $b = 64$ except
        \dgeqrf: $b = 32$}
    }
    \label{tbl:pred:acc:st}
\end{table}

\Cref{fig:pred:lapack:st} presents the relative runtime prediction error~\Q
t{med}{RE} for this scenario.  For all algorithms and setups, our
predictions are mostly within \SI5{\percent}~of the measured runtime, and in
many situations considerably closer.  The runtime prediction ARE averaged across
all problem sizes for each routine and setup is summarized in
\cref{tbl:pred:acc:st}: It ranges from~\SIrange{.71}{3.93}\percent, and its
average and median are, respectively, \SIlist{1.91;1.69}\percent.  Overall, the
predictions are slightly more accurate on the \sandybridge (average $\Q
t{med}{ARE} = \SI{1.66}\percent$) with the lowest average $\Q t{med}{ARE} =
\SI{1.22}\percent$ for \openblas~(\ref*{plt:sbopen}); on the \haswell (average
$\Q t{med}{ARE} = \SI{2.16}\percent$), the predictions are least accurate for
\mkl~(\ref*{plt:hwmkl}) with an average of $\Q t{med}{ARE} = \SI{2.26}\percent$.

Most routines are predicted equally well (with an average \Q t{med}{ARE} around
\SI{1.5}\percent) with two exceptions: \dsygst[1L] (average $\Q t{med}{ARE} =
\SI{2.63}\percent$) and \dgeqrf (average $\Q t{med}{ARE} = \SI{2.87}\percent$).
\begin{itemize}
    \item For the two-sided linear system solver \dsygst,
        \cref{fig:pred:accst:dsygst} reveals that for most setups, the
        predictions consistently underestimate the algorithm runtime for large
        problem sizes~$n$.

        A quick calculation shows that this effect is related to the size of the
        last-level cache~(L3):  On the \haswellshort, the problem emerges
        beyond~$n \approx 2000$ at which point the two operands~\dm A (symmetric
        in lower-triangular storage) and~\dm[lower]L\lowerpostsep take up
        $\SIvar{2 \times \frac{2000^2}2}\doubles \approx
        \SI{30.52}{\mebi\byte}$---slightly more than the L3~cache of
        \SI{30}{\mebi\byte}.  On the \sandybridgeshort with \SI{20}{\mebi\byte}
        of L3~cache, the effect is accordingly already visible beyond~$n \approx
        1600$.

        The cause for the underestimation of large problems is as follows:  Our
        models are based on repeated kernel measurements, which operate on
        cached (``warm'') data as long as all of the kernel's arguments fit in
        the cache.  However, each traversal step of \dsygst[1L]
        (\cref{alg:dsygst}) uses two separate kernels (namely \dsyrtk[LN] and
        \dtrsm[LLNN]) that operate on the trailing parts of \dm A and
        \dm[lower]L\lowerpostsep{}---since these do not fit in the cache
        simultaneously, they are mutually evicted by these kernels, and hence
        have to be loaded from main memory repeatedly (``cold'' data).  To
        summarize, our models estimate fast operations on cached data, while in
        the algorithm the operations are slower due to cache misses.

        A more detailed study of caching effects within blocked algorithms and
        attempts to account for them are presented in \cref{ch:cache}.

        Note that only \dsygst is affected by caching effects on this scale
        because all other routines involve only one dense operand.

    \item For the QR~decomposition \dgeqrf, \cref{fig:pred:accst:dgeqrf} reports
        that the runtime for almost all setups is consistently
        underestimated---especially for small problems.

        The cause is the transposed matrix copy and addition (see
        \cref{alg:dgeqrf}), which account for about~\SI4{\percent} of the
        runtime for small problems ($n \approx 250$) and \SI1{\percent} for
        large problems ($n \approx 4000$):  The copy, performed by a sequence of
        $b = 32$~\dcopy{}s, is underestimated by~$2\times$ to~$7\times$ because
        our models do not account for caching effects; the addition, which
        inlined as two nested loops, is not accounted for at all.
\end{itemize}

\subsection{Multi-Threaded \blas}
\label{sec:pred:acc:mt}

We study the multi-threaded prediction accuracy for the same six \lapack
algorithms using all available cores of the processors, i.e., 8~threads on the
\sandybridge and 12~threads on the \haswell.  In contrast to the single-threaded
predictions, we use a block size of~$b = 128$ for all algorithms---while this
configuration is certainly not optimal for all algorithms and problem sizes, it
generally yields better performance than \lapack's default values.

\begin{figure}[p]\figurestyle
    \ref*{leg:pred:accmt}

    \pgfplotsset{
        twocolplot, threerowplot, 0line,
        xlabel={problem size $n$},
        ylabel=relative error \Q t{med}{RE}, y unit=\percent,
        ymin=-20, ymax=20,
        table/search path={pred/figures/data/acc},
        cycle list={plotsbopen, plotsbmkl, plothwopen, plothwmkl},
        plotheightsub=24.8pt,
    }

    \newcommand\accplot[2]{%
        \begin{subfigaxis}[
                fig label=accmt:#1,
                fig caption={#2},
            ]
            \foreach \system in {SandyBridge.8, Haswell.12}
                \foreach \blas in {OpenBLAS, MKL}
                    \addplot table {#1.\system.\blas/are.dat};
        \end{subfigaxis}%
    }

    \begin{subfigaxis}[
            fig label=accmt:dlauum,
            fig caption={
                \dlauum[L]: $\dm A \coloneqq \dm[upper, ']L \dm[lower]L$
            },
            legend columns=3,
            legend to name=leg:pred:accmt
        ]
        \addlegendimage{empty legend}
        \addlegendentry{\sandybridge (8 cores):}
        \addplot table {dlauum.SandyBridge.8.OpenBLAS/are.dat};
        \addlegendentry{\openblas}
        \addplot table {dlauum.SandyBridge.8.MKL/are.dat};
        \addlegendentry{\mkl}
        \addlegendimage{empty legend}
        \addlegendentry{\haswell (12 cores):}
        \addplot table {dlauum.Haswell.12.OpenBLAS/are.dat};
        \addlegendentry{\openblas}
        \addplot table {dlauum.Haswell.12.MKL/are.dat};
        \addlegendentry{\mkl}
    \end{subfigaxis}\hfill
    \accplot{dsygst}{
        \dsygst[1L]: $\dm A
        \coloneqq \dm[lower, inv]L \dm A \matmatsep \dm[upper, inv']L$
    }

    \accplot{dtrtri}{
        \dtrtri[LN]: $\dm[lower]A \coloneqq \dm[lower, inv]A$
    }\hfill
    \accplot{dpotrf}{
        \dpotrf[L]: $\dm[lower]L \dm[upper, ']L \coloneqq \dm A$
    }

    \accplot{dgetrf}{
        \dgetrf: $\dm[dotted]P \matmatsep \dm[lower]L \dm[upper]U
        \coloneqq \dm A$
    }\hfill
    \accplot{dgeqrf}{
        \dgeqrf: $\dm Q \matmatsep \dm[upper]R \coloneqq \dm A$
    }

    \mycaption{%
        Multi-threaded prediction accuracy for \lapack algorithms.
        \captiondetails{$b = 128$}
    }
    \label{fig:pred:lapack:mt}
\end{figure}

\begin{table}\tablestyle
    \begin{tabular}{lccccc}
        \toprule
        &\multicolumn2c\sandybridge
        &\multicolumn2c\haswell
        &\multirow3*{\begin{tabular}{@{}c@{}}aver-\\age\end{tabular}} \\
        &{\openblas}            &{\mkl}
        &{\openblas}            &{\mkl} \\
        &(\ref*{plt:sbopen})    &(\ref*{plt:sbmkl})
        &(\ref*{plt:hwopen})    &(\ref*{plt:hwmkl}) \\
        \midrule
        \dlauum[L]  &\hphantom0\SI{9.42}\percent &\hphantom0\SI{2.29}\percent &\hphantom0\SI{3.73}\percent &\hphantom0\SI{1.93}\percent &\hphantom0\SI{4.34}\percent \\
\dsygst[1L] &\hphantom0\SI{1.83}\percent &\hphantom0\SI{4.55}\percent &\hphantom0\SI{7.17}\percent &\hphantom0\SI{5.03}\percent &\hphantom0\SI{4.65}\percent \\
\dtrtri[LN] &\hphantom0\SI{1.91}\percent &\hphantom0\SI{5.28}\percent &\hphantom0\SI{3.18}\percent &\hphantom0\SI{7.05}\percent &\hphantom0\SI{4.35}\percent \\
\dpotrf[L]  &\hphantom0\SI{6.89}\percent &\hphantom0\SI{7.46}\percent &\hphantom0\SI{3.00}\percent &\hphantom0\SI{4.65}\percent &\hphantom0\SI{5.50}\percent \\
\dgetrf     &\hphantom0\SI{1.07}\percent &\hphantom0\SI{2.81}\percent &\hphantom0\SI{1.87}\percent &\hphantom0\SI{3.28}\percent &\hphantom0\SI{2.26}\percent \\
\dgeqrf     &\hphantom0\SI{6.89}\percent &\hphantom0\SI{6.37}\percent &\SI{10.32}\percent &\hphantom0\SI{8.42}\percent &\hphantom0\SI{8.00}\percent \\
\midrule
average     &\hphantom0\SI{4.67}\percent &\hphantom0\SI{4.79}\percent &\hphantom0\SI{4.88}\percent &\hphantom0\SI{5.06}\percent &\hphantom0\SI{4.85}\percent \\

        \bottomrule
    \end{tabular}
    \mycaption{%
        Multi-threaded runtime prediction ARE~\Q t{med}{ARE} for blocked
        \lapack algorithms averaged across problem sizes.
        \captiondetails{
            $n = 56, \ldots, 4152$ in steps of~64, $b = 128$,
            \sandybridgeshort: 8~cores, \haswellshort: 12~cores
        }
    }
    \label{tbl:pred:acc:mt}
\end{table}

\Cref{fig:pred:lapack:mt} presents the relative runtime prediction errors~\Q
t{med}{RE} for this scenario, and \cref{tbl:pred:acc:mt} summarizes their
averaged AREs~\Q t{med}{ARE}.  Compared to the single-threaded case, the
prediction errors are across the board around $2.5\times$~larger with a total
average of $\Q t{med}{ARE} = \SI{4.85}\percent$.  The predictions are roughly
equally accurate across the two architectures and the two \blas implementations.

Considering \cref{fig:pred:lapack:mt}, we note fluctuation patterns in the
prediction errors by up to~\SI{10}\percent, most notably for \dsygst[1L] and
\dtrtri[LN] using \mkl on the \haswellshort~(\ref*{plt:hwmkl}).  As observed in
\cref{sec:pred:chol:mt}, these fluctuations are an artefact of the block size~$b
= 128$ interacting with the considered problem sizes in steps of~64:  Between
consecutive problem sizes, the remaining matrix portions in the last step of the
matrix traversal alternate between widths~56 and~120.

As in the single-threaded case, the QR~decomposition's runtime is
underestimated by on average~\SI{8.00}\percent, due to the \dcopy{}s and the
inlined matrix addition.  Since especially the latter cannot make any use of the
multi-threaded parallelism, their impact increases significantly with the number
of available cores.

Furthermore, several individual algorithms and setups are consistently under- or
overestimated:  e.g., \openblas on the \sandybridge~(\ref*{plt:sbopen}) for
\dlauum[L] and \dpotrf[L].  These problems arise from the multi-threaded
implementations of \dgemm, whose irregular performance is not well represented
in our models:  Since \blas implementations distribute computations among
threads along a certain dimension of the operation, for small dimension (such as
the block size), only a subset of the available threads is used.  When the small
dimension is increased, more threads are activated and the performance increases
suddenly.

\subsection{Summary}
\label{sec:pred:acc:sum}

This section has shown that across experiments on two processor architectures,
three \blas implementations, and six blocked \lapack algorithms, our models
yield accurate predictions that are on average within~\SI{1.91}{\percent}
(single-threaded) and \SI{4.85}{\percent} (multi-threaded) of reference
measurements.  Encouraged by these accuracy results, the following sections use
performance predictions to target our main goals of algorithm selection and
block-size optimization.

    \section{Algorithm Selection}
    \label{sec:pred:var}
    This section uses model-based predictions to determine which of several
alternative blocked algorithms for the same operation is the fastest.  To
confirm the correctness of our predictions' selections on a \haswell using
\openblas, we compare them to the optimal algorithms identified by
time-consuming empirical measurements.

\Cref{sec:pred:var:chol} revisits the Cholesky decomposition with only three
alternative blocked algorithms, \cref{sec:pred:var:trinv} considers the
inversion of a triangular matrix with eight alternatives, and
\cref{sec:pred:var:sylv} addresses the solution of the triangular Sylvester
equation with a total of 64~algorithms.

        \subsection{Cholesky Decomposition}
        \label{sec:pred:var:chol}
        The \definition[Cholesky decomposition: 3~algorithms]{three blocked algorithms}
for the lower-triangular Cholesky decomposition
\[
    \dm[lower]L \dm[upper, ']L \coloneqq \dm A
\]
of a symmetric positive definite matrix $\dm A \in \R^{n \times n}$ were
introduced in \cref{ex:intro:chol} on \cpageref{ex:intro:chol}, and
\cref{sec:pred:chol} studied algorithm~3 in detail.

\begin{figure}[p]\figurestyle
    \ref*{leg:chols}

    \pgfplotsset{
        twocolplot, perfplot=52.8,
        xlabel=problem size $n$,
        table/search path={pred/figures/data/varchol},
    }

    \begin{subfigaxis}[
            fig caption=Predictions on 1~core,
            fig label=var:chol:time:meas:1,
            ylabel={performance \Q p{med}{pred}}
        ]
        \foreach \var in {1, 2, 3}
            \addplot table[y=pred] {chol\var.Haswell.1.OpenBLAS/perf.dat};
    \end{subfigaxis}\hfill
    \begin{subfigaxis}[
            fig caption=Measurements on 1~core,
            fig label=var:chol:pred:1,
            ylabel={performance \Q p{med}{meas}}
        ]
        \foreach \var in {1, 2, 3}
            \addplot table[y=meas] {chol\var.Haswell.1.OpenBLAS/perf.dat};
    \end{subfigaxis}

    \pgfplotsset{ymax=480}

    \begin{subfigaxis}[
            fig caption=Predictions on 12~cores,
            fig label=var:chol:time:meas:12,
            ylabel={performance \Q p{med}{pred}}
        ]
        \foreach \var in {1, 2, 3}
            \addplot table[y=pred] {chol\var.Haswell.12.OpenBLAS/perf.dat};
    \end{subfigaxis}\hfill
    \begin{subfigaxis}[
            fig caption=Measurements on 12~cores,
            fig label=var:chol:time:pred:12,
            ylabel={performance \Q p{med}{meas}}
        ]
        \foreach \var in {1, 2, 3}
            \addplot table[y=meas] {chol\var.Haswell.12.OpenBLAS/perf.dat};
    \end{subfigaxis}

    \mycaption{%
        Performance measurements and predictions for the blocked Cholesky
        decomposition algorithms in lower-triangular storage.
        \captiondetails{$b = 128$, \haswell, \openblas}
    }
    \label{fig:pred:var:chol}
\end{figure}

\cref{fig:pred:var:chol} presents the performance predictions and measurements
for the three algorithms with problem sizes~$n = 56, \ldots 4152$ in steps
of~64.  For both the single- and multi-threaded setup, the predictions
accurately indicate that algorithm~3~(\ref*{plt:chol3}) is the fastest among the
three alternatives.  The differences in performance among the three algorithms
is enormous:  On 1 and 12~cores, algorithm~3~(\ref*{plt:chol3}) is faster than
algorithm~1~(\ref*{plt:chol1}) by, respectively, \SIlist{31.17;391.16}\percent.

Although our study reveals that algorithm~3~(\ref*{plt:chol3}) is the fastest
among the three alternatives, \lapack uses the suboptimal
algorithm~2~(\ref*{plt:chol2}) in its \dpotrf[L].

Note that while the reference performance measurements
(\cref{fig:pred:var:chol:time:meas:1,fig:pred:var:chol:time:meas:12}) together
took around 1~minute, our prediction identified the fastest algorithm in just
over \SI{.5}{\second}---over $100\times$~faster.  Since for these predictions we
represented and evaluated our models in \python, we expect that using another
storage format and evaluation system (e.g., in \clang/\cpplang) would further
increased the prediction speed by one or two orders of magnitude.

        \subsection{Triangular Inversion}
        \label{sec:pred:var:trinv}
        \begin{figure}[p]\figurestyle

    Matrix partitioning:
    \begin{tikzpicture}[
        y={(0, -1)}, scale=.64,
        baseline=(n.base)
    ]
        \def\s{5.5} \def\p{2} \def\q{3}
        \filldraw[mat  ] (0,  0 ) rectangle (\s,     \s    ) node (n) {\strut};
        \filldraw[mat00] (0,  0 ) --        (\p-.05, \p-.05) node {$A_{00}$} -| cycle;
        \filldraw[mat10] (0,  \p) rectangle (\p-.05, \q-.05) node {$A_{10}$};
        \filldraw[mat11] (\p, \p) --        (\q-.05, \q-.05) node {$A_{11}$} -| cycle;
        \filldraw[mat20] (0,  \q) rectangle (\p-.05, \s    ) node {$A_{20}$};
        \filldraw[mat21] (\p, \q) rectangle (\q-.05, \s    ) node {$A_{21}$};
        \filldraw[mat22] (\q, \q) --        (\s,     \s    ) node {$A_{22}$} -| cycle;
        \draw[brace ] (\p,     \p) -- (\q-.05, \p    ) node {$b$}; 
        \draw[brace ] (\q-.05, \p) -- (\q-.05, \q-.05) node {$b$}; 
        \draw[bracei] (0,      \s) -- (\s,     \s    ) node {$n$}; 
        \draw[bracei] (0,      0 ) -- (0,      \s    ) node {$n$}; 
    \end{tikzpicture}

    \makeatletter\let\tikz@ensure@dollar@catcode=\relax\makeatother
    \newcommand\Azz{\dm[mat00, lower]{A_{00}}}
    \newcommand\Azzi{\dm[mat00, lower, inv]{A_{00}}}
    \newcommand\Aoz{\dm[mat10, height=.5]{A_{10}}}
    \newcommand\Aoo{\dm[mat11, size=.5, lower]{A_{11}}}
    \newcommand\Aooi{\dm[mat11, size=.5, lower, inv]{A_{11}}}
    \newcommand\Atz{\dm[mat20, height=1.25]{A_{20}}}
    \newcommand\Ato{\dm[mat21, width=.5, height=1.25]{A_{21}}}
    \newcommand\Att{\dm[mat22, size=1.25, lower]{A_{22}}}
    \newcommand\Atti{\dm[mat22, size=1.25, lower, inv]{A_{22}}}

    \newcommand\traversal{
        \begin{tikzpicture}[
                baseline=(x.base),
                y={(0, -1)}, scale=.1
            ]
            \def\s{5.5} \def\p{2} \def\q{3}
            \filldraw[mat  ] (0,  0 ) rectangle (\s, \s);
            \filldraw[mat00] (0,  0 ) --        (\p, \p) -| cycle;
            \filldraw[mat10] (0,  \p) rectangle (\p, \q);
            \filldraw[mat11] (\p, \p) --        (\q, \q) -| cycle;
            \filldraw[mat20] (0,  \q) rectangle (\p, \s);
            \filldraw[mat21] (\p, \q) rectangle (\q, \s);
            \filldraw[mat22] (\q, \q) --        (\s, \s) -| cycle;
            \path (0, 0) -- (\s, \s) node[midway, black] (x) {$A$};
        \end{tikzpicture}
    }

    \begin{subfigure}\subfigwidth
        \begin{alglisting}[]
            traverse $\traversal$ along $\tsearrow$:
              !\dtrmm[RLNN]:! $\Aoz \coloneqq \Aoz \matmatsep \Azz\vphantom\Att$
              !\dtrsm[LLNN]:! $\Aoz \coloneqq -\Aooi \Aoz\vphantom\Att$
              !\dtrti[LN]2:! $\Aoo \coloneqq \Aooi$
        \end{alglisting}
        \caption{Algorithm 1}
        \label{alg:trinv1}
    \end{subfigure}\hfill
    \begin{subfigure}\subfigwidth
        \begin{alglisting}[]
            traverse $\traversal$ along $\tsearrow$:
              !\dtrsm[LLNN]:! $\Ato \coloneqq \Atti \Ato$
              !\dtrsm[RLNN]:! $\Ato \coloneqq -\Ato \Aooi$
              !\dtrti[LN]2:! $\Aoo \coloneqq \Aooi$
        \end{alglisting}
        \caption{Algorithm 2}
        \label{alg:trinv2}
    \end{subfigure}

    \medskip

    \begin{subfigure}\subfigwidth
        \begin{alglisting}[]
            traverse $\traversal$ along $\tsearrow$:
              !\dtrsm[RLNN]:! $\Ato \coloneqq -\Ato \Aooi$
              !\dgemm[NN]:! $\Atz \coloneqq \Atz + \Ato \matmatsep \Aoz$
              !\dtrsm[LLNN]:! $\Aoz \coloneqq \Aooi \Aoz$
              !\dtrti[LN]2:! $\Aoo \coloneqq \Aooi$
        \end{alglisting}
        \caption{Algorithm 3}
        \label{alg:trinv3}
    \end{subfigure}\hfill
    \begin{subfigure}\subfigwidth
        \begin{alglisting}[]
            traverse $\traversal$ along $\tsearrow$:
              !\dtrsm[LLNN]:! $\Ato \coloneqq -\Atti \Ato$
              !\dgemm[NN]:! $\Atz \coloneqq \Atz - \Ato \matmatsep \Aoz$
              !\dtrmm[RLNN]:! $\Aoz \coloneqq \Aoz \matmatsep \Azz$
              !\dtrti[LN]2:! $\Aoo \coloneqq \Aooi$
        \end{alglisting}
        \caption{Algorithm 4 (unstable)}
        \label{alg:trinv4}
    \end{subfigure}

    \medskip

    \begin{subfigure}\subfigwidth
        \begin{alglisting}[]
            traverse $\traversal$ along $\tnwarrow$:
              !\dtrmm[LLNN]:! $\Ato \coloneqq \Att \Ato$
              !\dtrsm[RLNN]:! $\Ato \coloneqq -\Ato \Aooi$
              !\dtrti[LN]2:! $\Aoo \coloneqq \Aooi$
        \end{alglisting}
        \caption{Algorithm 5 (\lapack)}
        \label{alg:trinv5}
    \end{subfigure}\hfill
    \begin{subfigure}\subfigwidth
        \begin{alglisting}[]
            traverse $\traversal$ along $\tnwarrow$:
              !\dtrsm[RLNN]:! $\Aoz \coloneqq \Aoz \Azzi\vphantom\Atti$
              !\dtrsm[LLNN]:! $\Aoz \coloneqq -\Aooi \Aoz\vphantom\Atti$
              !\dtrti[LN]2:! $\Aoo \coloneqq \Aooi$
        \end{alglisting}
        \caption{Algorithm 6}
        \label{alg:trinv6}
    \end{subfigure}

    \medskip

    \begin{subfigure}\subfigwidth
        \begin{alglisting}[]
            traverse $\traversal$ along $\tnwarrow$:
              !\dtrsm[LLNN]:! $\Aoz \coloneqq -\Aooi \Aoz$
              !\dgemm[NN]:! $\Atz \coloneqq \Atz + \Ato \matmatsep \Aoz$
              !\dtrsm[RLNN]:! $\Ato \coloneqq \Ato \Aooi$
              !\dtrti[LN]2:! $\Aoo \coloneqq \Aooi$
        \end{alglisting}
        \caption{Algorithm 7}
        \label{alg:trinv7}
    \end{subfigure}\hfill
    \begin{subfigure}\subfigwidth
        \begin{alglisting}[]
            traverse $\traversal$ along $\tnwarrow$:
              !\dtrsm[RLNN]:! $\Aoz \coloneqq -\Aoz \Azzi$
              !\dgemm[NN]:! $\Atz \coloneqq \Atz - \Ato \matmatsep \Aoz$
              !\dtrmm[LLNN]:! $\Ato \coloneqq \Att \matmatsep \Ato$
              !\dtrti[LN]2:! $\Aoo \coloneqq \Aooi$
        \end{alglisting}
        \caption{Algorithm 8 (unstable)}
        \label{alg:trinv8}
    \end{subfigure}

    \mycaption{%
        Blocked algorithms for the inversion of a lower-triangular matrix.
    }
    \label{algs:trinv}
\end{figure}


\Cref{algs:trinv} presents the \definition[triangular inversion:
8~algorithms]{eight blocked algorithms} for the inversion of a lower-triangular
matrix
\[
    \dm[lower]A \coloneqq \dm[lower, inv]A
\]
with $A \in \R^{n \times n}$ non-singular.  Note that algorithms~5 through~8 are
the mirrors of algorithms~1 through~4 with the opposite traversal
direction---\tnwarrow instead of~\tsearrow.  Furthermore, algorithms~4 and~8 not
only perform around $3\times$~more FLOPs than required, but are also numerically
unstable.\footnote{%
    Further six algorithms can be obtained from algorithms~1 to~3 and~5 to~6, by
    swapping the inversion of the diagonal~$A_{11}$ with the preceding
    \dtrsm[RLNN] and turning the latter into a \dtrmm[RLNN]; however, the
    resulting algorithms are also numerically unstable and thus not further
    discussed.  For further details on the numerical stability of triangular
    inversion see~\cite{trinvstability}.
}  Note that \lapack's \dtrtri[LN] implements algorithm~5 with a default block
size of~$b = 64$.

\begin{figure}[p]\figurestyle
    \ref*{leg:trinvs}

    \pgfplotsset{
        twocolplot, perfplot=52.8,
        xlabel=problem size $n$,
        table/search path={pred/figures/data/vartrinv},
    }

    \begin{subfigaxis}[
            fig caption={Predictions on 1~core},
            fig label=var:trinv:pred:1,
            ylabel={performance \Q p{med}{pred}},
            legend to name=leg:trinvs
        ]
        \addlegendimage{empty legend}
        \addlegendentry{Algorithm:}
        \foreach \var in {1, ..., 8} {
            \addplot table[y=pred] {trinv\var.Haswell.1.OpenBLAS/perf.dat};
            \addlegendentryexpanded{\var}\label{plt:trinv\var}
        }
    \end{subfigaxis}\hfill
    \begin{subfigaxis}[
            fig caption={Measurements on 1~core},
            fig label=var:trinv:meas:1,
            ylabel={performance \Q p{med}{meas}}
        ]
        \foreach \var in {1, ..., 8}
            \addplot table[y=meas] {trinv\var.Haswell.1.OpenBLAS/perf.dat};
    \end{subfigaxis}

    \pgfplotsset{ymax=480}

    \begin{subfigaxis}[
            fig caption={Predictions on 12~cores},
            fig label=var:trinv:pred:12,
            ylabel={performance \Q p{med}{pred}}
        ]
        \foreach \var in {1, ..., 8}
            \addplot table[y=pred] {trinv\var.Haswell.12.OpenBLAS/perf.dat};
    \end{subfigaxis}\hfill
    \begin{subfigaxis}[
            fig caption={Measurements on 12~cores},
            fig label=var:trinv:meas:12,
            ylabel={performance \Q p{med}{meas}}
        ]
        \foreach \var in {1, ..., 8}
            \addplot table[y=meas] {trinv\var.Haswell.12.OpenBLAS/perf.dat};
    \end{subfigaxis}

    \mycaption{%
        Performance measurements and predictions for the eight blocked
        lower-triangular inversion algorithms.
        \captiondetails{$b = 128$, \haswell, \openblas}
    }
    \label{fig:pred:var:trinv}
\end{figure}

\Cref{fig:pred:var:trinv} presents the performance predictions and measurements
for the eight algorithms for problem sizes between~$n = 56$ and~4152 in steps
of~6 on a \haswell using \openblas.

For the single-threaded case, the predictions correctly indicate that for
different problem sizes different algorithms attain the best performance:  While
for small matrices algorithms~1~(\ref*{plt:trinv1}) and~5~(\ref*{plt:trinv5})
are faster than the third-fastest by up to~\SI{12.80}\percent, beyond $n \approx
1500$, algorithms~3~(\ref*{plt:trinv3}) and~7~(\ref*{plt:trinv7}) take the lead
over algorithm~5~(\ref*{plt:trinv5}) in the third place by up
to~\SI{13.16}{\percent} and growing.  However, the predictions cannot
differentiate which of the two algorithms is actually the fastest; e.g., for
larger matrices, algorithm~3~(\ref*{plt:trinv3}) is up to
\SI{1.53}{\percent}~faster than algorithm~7~(\ref*{plt:trinv7}).

Using all of the \haswellshort's 12~cores, the predictions clearly and correctly
identify that algorithms~3~(\ref*{plt:trinv3}) and~7~(\ref*{plt:trinv7}) attain
the same performance, which is up to $2.73\times$~higher than the third-fastest
algorithm an increasing.  Furthermore, the predictions confirm that
algorithms~4~(\ref*{plt:trinv4}) and~8~(\ref*{plt:trinv8}) are indeed
considerably slower than all alternatives --- by up to~$2.96\times$ on 1~core
and~$7.95\times$ on 12~cores.

In summary, although our predictions in some cases cannot differentiate between
algorithms with nearly identical performance, they reliably distinguish and rank
algorithms with different performance.

        \subsection{Sylvester Equation Solver}
        \label{sec:pred:var:sylv}
        {
\newcommand\dmA{\dm[upper]A\xspace}
\newcommand\dmB{\dm[upper, size=.8, bbox height=1]B\xspace}
\newcommand\dmC{\dm[width=.8]C\xspace}
\newcommand\dmX{\dm[width=.8]X\xspace}

The triangular\footnote{%
    The general Sylvester equation with full~$A$ and~$B$ can be reduced to this
    case by means of the Schur decomposition~\cite{sylvred}, which, however,
    results in only quasi-triangular matrices that may contain full
    $2{\times}2$ diagonal blocks, i.e., individual non-zero elements on the
    first sub-diagonal.  Since each $2{\times}2$-blocks is processed as one
    element, it cannot be split across sub-matrices in a blocked
    matrix-traversal.  The resulting technical implications affect neither a
    blocked algorithm's structure at larger nor its performance, and we thus
    avoid such technicalities and assume upper-triangular~$A$ and~$B$.
} Sylvester equation
\[
    \dmA \matmatsep \dmX + \dmX \matmatsep \dmB = \dmC \enspace,
\]
with $\dmA \in \R^{m \times m}$, $\dmB \in \R^{n \times n}$, and $\dmC, \dmX \in
\R^{m \times n}$, to be solved for \dmX, is commonly used in control theory and
to estimate the condition numbers of eigenvalue problems.  Its solution is
typically implemented in-place with the \dmX overwriting \dmC; \lapack's
provides the operation in the form of the purely unblocked
\dtrsyl[NN1].\footnote{%
    \dtrsyl's first two flag arguments indicate transpositions of~$A$ and~$B$,
    and the third allows to turn the operation's left-hand-side sum into a
    difference.
}

\subsubsection{Algorithms}
The solution to the triangular Sylvester equation is computed by traversing \dmC
from the bottom left to the top right.  However, in contrast to the previous
operations, this traversal does not need to follow \dmC's diagonal; in fact \dmC
can be traversed in various different ways:  Two algorithms traverse \dmC
vertically, two horizontally (using $3 \times 1$ and $1 \times 3$ partitions),
and 14~diagonally (exposing $3 \times 3$ sub-matrices), making a total of
18~algorithms.  Furthermore, as detailed in the following, the Sylvester
equation requires two layers of blocked algorithms, resulting in a total of
\definition[Sylvester equation:\\64~``complete'' algorithms]{64~``complete''
algorithms}.

\begin{figure}[p]\figurestyle

    \makeatletter\let\tikz@ensure@dollar@catcode=\relax\makeatother

    \def\ma{.5}\def\mb{.25}\def\mc{.625}\def\mm{1.375}
    \def\na{.375}\def\nb{.1875}\def\nc{.5625}\def\nn{1.125}
    \newcommand\Xxz{\dm[mat10, width=\na, height=\mm]{C_0}}
    \newcommand\Xxo{\dm[mat11, width=\nb, height=\mm]{C_1}}
    \newcommand\Xxt{\dm[mat12, width=\nc, height=\mm]{C_2}}
    \newcommand\Xzx{\dm[mat01, width=\nn, height=\mc]{C_0}}
    \newcommand\Xox{\dm[mat11, width=\nn, height=\mb]{C_1}}
    \newcommand\Xtx{\dm[mat21, width=\nn, height=\ma]{C_2}}
    \newcommand\Xoxs{\dmx[mat11, width=\nn, height=\mb]{C_1^\ast}}
    \newcommand\Xxos{\dmx[mat11, width=\nb, height=\mm]{C_1^\ast}}
    \renewcommand\A{\dm[size=\mm, upper]A}
    \newcommand\Azo{\dm[mat01, width=\mb, height=\mc]{A_{01}}}
    \newcommand\Aoo{\dm[mat11, width=\mb, height=\mb, upper]{A_{11}}}
    \newcommand\Aot{\dm[mat12, width=\ma, height=\mb]{A_{12}}}
    \newcommand\B{\dm[size=\nn, upper]B}
    \newcommand\Bzo{\dm[mat01, width=\nb, height=\na]{B_{01}}}
    \newcommand\Boo{\dm[mat11, width=\nb, height=\nb, upper]{B_{11}}}
    \newcommand\Bot{\dm[mat12, width=\nc, height=\nb]{B_{12}}}

    \newcommand\traversalXm{
        \begin{tikzpicture}[
                baseline=(x.base),
                y={(0, -1)}, scale=.1
            ]
            \def\sa{4.5} \def\pa{1.5} \def\qa{2.25}
            \def\sb{5.5} \def\pb{2.5} \def\qb{3.5}
            \filldraw[mat  ] (0, 0  ) rectangle (\sa, \sb);
            \filldraw[mat01] (0, 0  ) rectangle (\sa, \pb);
            \filldraw[mat11] (0, \pb) rectangle (\sa, \qb);
            \filldraw[mat21] (0, \qb) rectangle (\sa, \sb);
            \path (0, 0) -- (\sa, \sb) node[midway, black] (x) {$C$};
        \end{tikzpicture}
    }
    \newcommand\traversalXn{
        \begin{tikzpicture}[
                baseline=(x.base),
                y={(0, -1)}, scale=.1
            ]
            \def\sa{4.5} \def\pa{1.5} \def\qa{2.25}
            \def\sb{5.5} \def\pb{2.5} \def\qb{3.5}
            \filldraw[mat  ] (0,   0) rectangle (\sa, \sb);
            \filldraw[mat10] (0,   0) rectangle (\pa, \sb);
            \filldraw[mat11] (\pa, 0) rectangle (\qa, \sb);
            \filldraw[mat12] (\qa, 0) rectangle (\sa, \sb);
            \path (0, 0) -- (\sa, \sb) node[midway, black] (x) {$C$};
        \end{tikzpicture}
    }
    \newcommand\traversalA{
        \begin{tikzpicture}[
                baseline=(x.base),
                y={(0, -1)}, scale=.1
            ]
            \def\s{5.5} \def\p{2} \def\q{3}
            \filldraw[mat  ] (0,  0 ) rectangle (\s, \s);
            \filldraw[mat00] (0,  0 ) --        (\p, \p) |- cycle;
            \filldraw[mat01] (\p, 0 ) rectangle (\q, \p);
            \filldraw[mat02] (\q, 0 ) rectangle (\s, \p);
            \filldraw[mat11] (\p, \p) --        (\q, \q) |- cycle;
            \filldraw[mat12] (\q, \p) rectangle (\s, \q);
            \filldraw[mat22] (\q, \q) --        (\s, \s) |- cycle;
            \path (0, 0) -- (\s, \s) node[midway, black] (x) {$A$};
        \end{tikzpicture}
    }
    \newcommand\traversalB{
        \begin{tikzpicture}[
                baseline=(x.base),
                y={(0, -1)}, scale=.1
            ]
            \def\s{4.5} \def\p{1.5} \def\q{2.25}
            \filldraw[mat  ] (0,  0 ) rectangle (\s, \s);
            \filldraw[mat00] (0,  0 ) --        (\p, \p) |- cycle;
            \filldraw[mat01] (\p, 0 ) rectangle (\q, \p);
            \filldraw[mat02] (\q, 0 ) rectangle (\s, \p);
            \filldraw[mat11] (\p, \p) --        (\q, \q) |- cycle;
            \filldraw[mat12] (\q, \p) rectangle (\s, \q);
            \filldraw[mat22] (\q, \q) --        (\s, \s) |- cycle;
            \path (0, 0) -- (\s, \s) node[midway, black] (x) {$B$};
        \end{tikzpicture}
    }

    \tikzset{
        matpic/.style={
            y={(0, -1)}, scale=.75,
            baseline={(0, 0)}
        }
    }

    \begin{subfigure}\textwidth
        \begin{tikzpicture}[matpic]
            \def\s{5.5} \def\p{2.5} \def\q{3.5}
            \filldraw[mat  ] (0,  0 ) rectangle (\s,     \s    );
            \filldraw[mat00] (0,  0 ) --        (\p-.05, \p-.05) node {$A_{00}$} |- cycle;
            \filldraw[mat01] (\p, 0 ) rectangle (\q-.05, \p-.05) node {$A_{01}$};
            \filldraw[mat02] (\q, 0 ) rectangle (\s,     \p-.05) node {$A_{02}$};
            \filldraw[mat11] (\p, \p) --        (\q-.05, \q-.05) node {$A_{11}$} |- cycle;
            \filldraw[mat12] (\q, \p) rectangle (\s,     \q-.05) node {$A_{12}$};
            \filldraw[mat22] (\q, \q) --        (\s,     \s    ) node {$A_{22}$} |- cycle;
            \draw[bracei] (\p, \q-.05) -- (\q-.05, \q-.05) node {$b$}; 
            \draw[bracei] (\p,     \p) -- (\p,     \q-.05) node {$b$}; 
            \draw[bracei] (0,      \s) -- (\s,     \s    ) node {$m$}; 
            \draw[bracei] (0,      0 ) -- (0,      \s    ) node {$m$}; 
        \end{tikzpicture}
        \begin{tikzpicture}[matpic]
            \def\sa{4.5} \def\pa{1.5} \def\qa{2.25}
            \def\sb{5.5} \def\pb{2.5} \def\qb{3.5}
            \filldraw[mat  ] (0,   0  ) rectangle (\sa, \sb    );
            \filldraw[mat01] (0,   0  ) rectangle (\sa, \pb-.05) node {$C_0$};
            \filldraw[mat11] (0,   \pb) rectangle (\sa, \qb-.05) node {$C_1$};
            \filldraw[mat21] (0,   \qb) rectangle (\sa, \sb    ) node {$C_2$};
            \draw[brace ] (\sa, \pb) -- (\sa,     \qb-.05) node {$b$}; 
            \draw[bracei] (0,   \sb) -- (\sa,     \sb    ) node {$n$}; 
            \draw[bracei] (0,   0  ) -- (0,       \sb    ) node {$m$}; 
        \end{tikzpicture}
        \begin{tikzpicture}[matpic]
            \def\s{4.5} \def\p{1.5} \def\q{2.25}
            \filldraw[mat] (0,  0 ) rectangle (\s, \s);
            \filldraw[mat] (0,  0 ) --        (\s, \s) node {$B$} |- cycle;
            \draw[bracei] (0, \s) -- (\s, \s) node {$n$}; 
            \draw[bracei] (0, 0 ) -- (0,  \s) node (n) {$n$}; 
        \end{tikzpicture}
        \caption{Vertical traversal of \dmC: $3 \times 1$ matrix partitioning}
    \end{subfigure}

    \medskip

    \begin{subfigure}\subfigwidth
        \begin{alglisting}[]
            traverse $\traversalA$ along $\tnwarrow$, $\traversalXm$ along $\tnarrow$
              !\dgemm[NN]:! $\Xox \coloneqq \Xox - \Aot \Xtx$
              $\Xox \coloneqq$ sylv($\Aoo \Xoxs + \Xoxs \B = \Xox$)
        \end{alglisting}
        \caption{Algorithm $m1$}
        \label{alg:sylvm1}
    \end{subfigure}\hfill
    \begin{subfigure}\subfigwidth
        \begin{alglisting}[]
            traverse $\traversalA$ along $\tnwarrow$, $\traversalXm$ along $\tnarrow$
              $\Xox \coloneqq$ sylv($\Aoo \Xoxs + \Xoxs \B = \Xox$)
              !\dgemm[NN]:! $\Xzx \coloneqq \Xzx - \Azo \Xox$
        \end{alglisting}
        \caption{Algorithm $m2$}
        \label{alg:sylvm2}
    \end{subfigure}

    \begin{subfigure}\textwidth
        \begin{tikzpicture}[matpic]
            \def\s{5.5} \def\p{2.5} \def\q{3.5}
            \filldraw[mat] (0, 0) rectangle (\s, \s);
            \filldraw[mat] (0, 0) --        (\s, \s) node {$A$} |- cycle;
            \draw[bracei] (0, \s) -- (\s, \s) node {$m$}; 
            \draw[bracei] (0, 0 ) -- (0,  \s) node {$m$}; 
        \end{tikzpicture}
        \begin{tikzpicture}[matpic]
            \def\sa{4.5} \def\pa{1.5} \def\qa{2.25}
            \def\sb{5.5} \def\pb{2.5} \def\qb{3.5}
            \filldraw[mat  ] (0,   0  ) rectangle (\sa,     \sb);
            \filldraw[mat10] (0,   0  ) rectangle (\pa-.05, \sb) node {$C_0$};
            \filldraw[mat11] (\pa, 0  ) rectangle (\qa-.05, \sb) node {$C_1$};
            \filldraw[mat12] (\qa, 0  ) rectangle (\sa,     \sb) node {$C_2$};
            \draw[brace ] (\pa, 0  ) -- (\qa-.05, 0      ) node {$b$}; 
            \draw[bracei] (0,   \sb) -- (\sa,     \sb    ) node {$n$}; 
            \draw[bracei] (0,   0  ) -- (0,       \sb    ) node {$m$}; 
        \end{tikzpicture}
        \begin{tikzpicture}[matpic]
            \def\s{4.5} \def\p{1.5} \def\q{2.25}
            \filldraw[mat  ] (0,  0 ) rectangle (\s,     \s    );
            \filldraw[mat00] (0,  0 ) --        (\p-.05, \p-.05) node {$B_{00}$} |- cycle;
            \filldraw[mat01] (\p, 0 ) rectangle (\q-.05, \p-.05) node {$B_{01}$};
            \filldraw[mat02] (\q, 0 ) rectangle (\s,     \p-.05) node {$B_{02}$};
            \filldraw[mat11] (\p, \p) --        (\q-.05, \q-.05) node {$B_{11}$} |- cycle;
            \filldraw[mat12] (\q, \p) rectangle (\s,     \q-.05) node {$B_{12}$};
            \filldraw[mat22] (\q, \q) --        (\s,     \s    ) node {$B_{22}$} |- cycle;
            \draw[bracei] (\p, \q-.05) -- (\q-.05, \q-.05) node {$b$}; 
            \draw[bracei] (\p,     \p) -- (\p,     \q-.05) node {$b$}; 
            \draw[bracei] (0,      \s) -- (\s,     \s    ) node {$n$}; 
            \draw[bracei] (0,      0 ) -- (0,      \s    ) node {$n$}; 
        \end{tikzpicture}
        \caption{Horizontal traversal of \dmC: $1 \times 3$ matrix partitioning}
    \end{subfigure}

    \medskip

    \begin{subfigure}\subfigwidth
        \begin{alglisting}[]
            traverse $\traversalXn$ along $\tearrow$, $\traversalB$ along $\tsearrow$
              !\dgemm[NN]:! $\Xxo \coloneqq \Xxo - \Xxz \Bzo$
              $\Xxo \coloneqq$ sylv($\A \Xxos + \Xxos \Boo = \Xxo$)
        \end{alglisting}
        \caption{Algorithm $n1$}
        \label{alg:sylvn1}
    \end{subfigure}\hfill
    \begin{subfigure}\subfigwidth
        \begin{alglisting}[]
            traverse $\traversalXn$ along $\tearrow$, $\traversalB$ along $\tsearrow$
              $\Xxo \coloneqq$ sylv($\A \Xxos + \Xxos \Boo = \Xxo$)
              !\dgemm[NN]:! $\Xxt \coloneqq \Xxt - \Xxo \Bot$
        \end{alglisting}
        \caption{Algorithm $n2$}
        \label{alg:sylvn2}
    \end{subfigure}

    \mycaption{%
        Blocked algorithms solving the triangular Sylvester equation with $1
        \times 3$ and $3 \times 1$  matrix partitionings.
        \captiondetails{Output~$X$ overwrites input~$C$.}
    }
    \label{algs:sylv1d}
\end{figure}

\Cref{algs:sylv1d} presents the four algorithms that traverse \dmC vertically or
horizontally, thereby exposing $3 \times 1$ or $1 \times 3$ sub-matrices; each
of these algorithms consists of one call to \dgemm[NN] and the solution of a
sub-problem (another triangular Sylvester equation).  To obtain a ``complete''
algorithm, two of these algorithms with orthogonal traversals are combined---the
first traverses the full~\dmC and invokes the second to solve sub-problem in
each iteration; the second, in turn, solves its small $b \times b$ sub-problem
using \lapack's unblocked \dtrsyl[NN1].  E.g., one can use algorithm~$m1$ to
traverse \dmC vertically and in each step apply algorithm~$n2$ to traverse the
middle panel~\dm[mat11, height=.2, width=.8]{C_1} horizontally.  We call the
resulting ``complete'' algorithm~$m1n2$, and see that eight such combinations
are possible: $m1n1$, $m1n2$, $m2n1$, $m2n2$, $n1m1$, $n1m2$, $n2m1$,
and~$n2m2$.  Note that in principle the block sizes for the two layered blocked
algorithms can be chosen independently; however, we limit our study to a single
block size for both layers.

\begin{figure}[p]\figurestyle

    \makeatletter\let\tikz@ensure@dollar@catcode=\relax\makeatother

    \def\mb{.5}\def\mc{1.25}
    \def\na{.75}\def\nb{.375}\def\nc{1.125}
    \newcommand\Xzo{\dm[mat01, width=\nb, height=\mc]{C_{01}}}
    \newcommand\Xzt{\dm[mat02, width=\nc, height=\mc]{C_{02}}}
    \newcommand\Xoz{\dm[mat10, width=\na, height=\mb]{C_{10}}}
    \newcommand\Xoo{\dm[mat11, width=\nb, height=\mb]{C_{11}}}
    \newcommand\Xot{\dm[mat12, width=\nc, height=\mb]{C_{12}}}
    \newcommand\Xtz{\dm[mat20, width=\na]{C_{20}}}
    \newcommand\Xto{\dm[mat21, width=\nb]{C_{21}}}
    \newcommand\Xtt{\dm[mat22, width=\nc]{C_{22}}}
    \newcommand\Xzos{\dmx[mat01, width=\nb, height=\mc]{C_{01}^\ast}}
    \newcommand\Xozs{\dmx[mat10, width=\na, height=\mb]{C_{10}^\ast}}
    \newcommand\Xoos{\dmx[mat11, width=\nb, height=\mb]{C_{11}^\ast}}
    \newcommand\Xtos{\dmx[mat21, width=\nb]{C_{21}^\ast}}
    \newcommand\Azz{\dm[mat00, width=\mc, height=\mc, upper]{A_{00}}}
    \newcommand\Azo{\dm[mat01, width=\mb, height=\mc]{A_{01}}}
    \newcommand\Azt{\dm[mat02, height=\mc]{A_{02}}}
    \newcommand\Aoo{\dm[mat11, width=\mb, height=\mb, upper]{A_{11}}}
    \newcommand\Aot{\dm[mat12, height=\mb]{A_{12}}}
    \newcommand\Att{\dm[mat22, upper]{A_{22}}}
    \newcommand\Bzz{\dm[mat00, width=\na, height=\na, upper]{B_{00}}}
    \newcommand\Bzo{\dm[mat01, width=\nb, height=\na]{B_{01}}}
    \newcommand\Boo{\dm[mat11, width=\nb, height=\nb, upper]{B_{11}}}
    \newcommand\Bot{\dm[mat12, width=\nc, height=\nb]{B_{12}}}

    \newcommand\traversalX{
        \begin{tikzpicture}[
                baseline=(x.base),
                y={(0, -1)}, scale=.1
            ]
            \def\sa{4.5} \def\pa{1.5} \def\qa{2.25}
            \def\sb{5.5} \def\pb{2.5} \def\qb{3.5}
            \filldraw[mat  ] (0,   0  ) rectangle (\sa, \sb);
            \filldraw[mat00] (0,   0  ) rectangle (\pa, \pb);
            \filldraw[mat01] (\pa, 0  ) rectangle (\qa, \pb);
            \filldraw[mat02] (\qa, 0  ) rectangle (\sa, \pb);
            \filldraw[mat10] (0,   \pb) rectangle (\pa, \qb);
            \filldraw[mat11] (\pa, \pb) rectangle (\qa, \qb);
            \filldraw[mat12] (\qa, \pb) rectangle (\sa, \qb);
            \filldraw[mat20] (0,   \qb) rectangle (\pa, \sb);
            \filldraw[mat21] (\pa, \qb) rectangle (\qa, \sb);
            \filldraw[mat22] (\qa, \qb) rectangle (\sa, \sb);
            \path (0, 0) -- (\sa, \sb) node[midway, black] (x) {$C$};
        \end{tikzpicture}
    }
    \newcommand\traversalA{
        \begin{tikzpicture}[
                baseline=(x.base),
                y={(0, -1)}, scale=.1
            ]
            \def\s{5.5} \def\p{2} \def\q{3}
            \filldraw[mat  ] (0,  0 ) rectangle (\s, \s);
            \filldraw[mat00] (0,  0 ) --        (\p, \p) |- cycle;
            \filldraw[mat01] (\p, 0 ) rectangle (\q, \p);
            \filldraw[mat02] (\q, 0 ) rectangle (\s, \p);
            \filldraw[mat11] (\p, \p) --        (\q, \q) |- cycle;
            \filldraw[mat12] (\q, \p) rectangle (\s, \q);
            \filldraw[mat22] (\q, \q) --        (\s, \s) |- cycle;
            \path (0, 0) -- (\s, \s) node[midway, black] (x) {$A$};
        \end{tikzpicture}
    }
    \newcommand\traversalB{
        \begin{tikzpicture}[
                baseline=(x.base),
                y={(0, -1)}, scale=.1
            ]
            \def\s{4.5} \def\p{1.5} \def\q{2.25}
            \filldraw[mat  ] (0,  0 ) rectangle (\s, \s);
            \filldraw[mat00] (0,  0 ) --        (\p, \p) |- cycle;
            \filldraw[mat01] (\p, 0 ) rectangle (\q, \p);
            \filldraw[mat02] (\q, 0 ) rectangle (\s, \p);
            \filldraw[mat11] (\p, \p) --        (\q, \q) |- cycle;
            \filldraw[mat12] (\q, \p) rectangle (\s, \q);
            \filldraw[mat22] (\q, \q) --        (\s, \s) |- cycle;
            \path (0, 0) -- (\s, \s) node[midway, black] (x) {$B$};
        \end{tikzpicture}
    }

    \begin{subfigure}\textwidth
        \tikzset{
            every picture/.append style={
                y={(0, -1)}, scale=.7,
                baseline={(0, 0)}
            }
        }
        \begin{tikzpicture}
            \def\s{5.5} \def\p{2.5} \def\q{3.5}
            \filldraw[mat  ] (0,  0 ) rectangle (\s,     \s    );
            \filldraw[mat00] (0,  0 ) --        (\p-.05, \p-.05) node {$A_{00}$} |- cycle;
            \filldraw[mat01] (\p, 0 ) rectangle (\q-.05, \p-.05) node {$A_{01}$};
            \filldraw[mat02] (\q, 0 ) rectangle (\s,     \p-.05) node {$A_{02}$};
            \filldraw[mat11] (\p, \p) --        (\q-.05, \q-.05) node {$A_{11}$} |- cycle;
            \filldraw[mat12] (\q, \p) rectangle (\s,     \q-.05) node {$A_{12}$};
            \filldraw[mat22] (\q, \q) --        (\s,     \s    ) node {$A_{22}$} |- cycle;
            \draw[bracei] (\p, \q-.05) -- (\q-.05, \q-.05) node {$b_m$}; 
            \draw[bracei] (\p, \p    ) -- (\p,     \q-.05) node {$b_m$}; 
            \draw[bracei] (0,  \s    ) -- (\s,     \s    ) node {$m$}; 
            \draw[bracei] (0,  0     ) -- (0,      \s    ) node {$m$}; 
        \end{tikzpicture}
        \begin{tikzpicture}
            \def\sa{4.5} \def\pa{1.5} \def\qa{2.25}
            \def\sb{5.5} \def\pb{2.5} \def\qb{3.5}
            \filldraw[mat  ] (0,   0  ) rectangle (\sa,     \sb    );
            \filldraw[mat00] (0,   0  ) rectangle (\pa-.05, \pb-.05) node {$C_{00}$};
            \filldraw[mat01] (\pa, 0  ) rectangle (\qa-.05, \pb-.05) node {$C_{01}$};
            \filldraw[mat02] (\qa, 0  ) rectangle (\sa,     \pb-.05) node {$C_{02}$};
            \filldraw[mat10] (0,   \pb) rectangle (\pa-.05, \qb-.05) node {$C_{10}$};
            \filldraw[mat11] (\pa, \pb) rectangle (\qa-.05, \qb-.05) node {$C_{11}$};
            \filldraw[mat12] (\qa, \pb) rectangle (\sa,     \qb-.05) node {$C_{12}$};
            \filldraw[mat20] (0,   \qb) rectangle (\pa-.05, \sb    ) node {$C_{20}$};
            \filldraw[mat21] (\pa, \qb) rectangle (\qa-.05, \sb    ) node {$C_{21}$};
            \filldraw[mat22] (\qa, \qb) rectangle (\sa,     \sb    ) node {$C_{22}$};
            \draw[brace ] (\pa, 0  ) -- (\qa-.05, 0      ) node {$b_n$}; 
            \draw[brace ] (\sa, \pb) -- (\sa,     \qb-.05) node {$b_m$}; 
            \draw[bracei] (0,   \sb) -- (\sa,     \sb    ) node {$n$}; 
            \draw[bracei] (0,   0  ) -- (0,       \sb    ) node {$m$}; 
        \end{tikzpicture}
        \begin{tikzpicture}
            \def\s{4.5} \def\p{1.5} \def\q{2.25}
            \filldraw[mat  ] (0,  0 ) rectangle (\s,     \s    );
            \filldraw[mat00] (0,  0 ) --        (\p-.05, \p-.05) node {$B_{00}$} |- cycle;
            \filldraw[mat01] (\p, 0 ) rectangle (\q-.05, \p-.05) node {$B_{01}$};
            \filldraw[mat02] (\q, 0 ) rectangle (\s,     \p-.05) node {$B_{02}$};
            \filldraw[mat11] (\p, \p) --        (\q-.05, \q-.05) node {$B_{11}$} |- cycle;
            \filldraw[mat12] (\q, \p) rectangle (\s,     \q-.05) node {$B_{12}$};
            \filldraw[mat22] (\q, \q) --        (\s,     \s    ) node {$B_{22}$} |- cycle;
            \draw[bracei] (\p, \q-.05) -- (\q-.05, \q-.05) node {$b_n$}; 
            \draw[bracei] (\p,     \p) -- (\p,     \q-.05) node {$b_n$}; 
            \draw[bracei] (0,      \s) -- (\s,     \s    ) node {$n$}; 
            \draw[bracei] (0,      0 ) -- (0,      \s    ) node {$n$}; 
        \end{tikzpicture}
        \caption{$3 \times 3$ matrix partitioning}
    \end{subfigure}

    \medskip

    \begin{subfigure}{.75\textwidth}
        \begin{alglisting}[]
            traverse $\traversalA$ along $\tnwarrow$, $\traversalX$ along $\tnearrow$, $\traversalB$ along $\tsearrow$:
              !\dgemm[NN]:! $\Xoz \coloneqq \Xoz - \Aot \Xto$
              $\Xoz \coloneqq $ sylv($\Aoo \Xozs + \Xozs \Bzz = \Xoz$)
              !\dgemm[NN]:! $\Xto \coloneqq \Xto - \Xtz \Bzo$
              $\Xto \coloneqq $ sylv($\Att \Xtos + \Xtos \Boo = \Xto$)
              !\dgemm[NN]:! $\Xoo \coloneqq \Xoo - \Xoz \Bzo$
              !\dgemm[NN]:! $\Xoo \coloneqq \Xoo - \Aot \Xto$
              !\dtrsyl[NN1]:! $\Xoo \coloneqq $ sylv($\Aoo \Xoos + \Xoos \Boo = \Xoo$)
        \end{alglisting}
        \caption{Algorithm 1}
        \label{alg:sylv1}
    \end{subfigure}

    \medskip

    \begin{subfigure}{.75\textwidth}
        \begin{alglisting}[]
            traverse $\traversalA$ along $\tnwarrow$, $\traversalX$ along $\tnearrow$, $\traversalB$ along $\tsearrow$:
              $\Xto \coloneqq $ sylv($\Att \Xtos + \Xtos \Boo = \Xto$)
              !\dgemm[NN]:! $\Xoo \coloneqq \Xoo - \Aot \Xto$
              !\dtrsyl[NN1]:! $\Xoo \coloneqq $ sylv($\Aoo \Xoos + \Xoos \Boo = \Xoo$)
              !\dgemm[NN]:! $\Xzo \coloneqq \Xzo - \Azt \Xto$
              !\dgemm[NN]:! $\Xzo \coloneqq \Xzo - \Azo \Xoo$
              $\Xzo \coloneqq $ sylv($\Azz \Xzos + \Xzos \Boo = \Xzo$)
              !\dgemm[NN]:! $\Xzt \coloneqq \Xzt - \Xzo \Bot$
              !\dgemm[NN]:! $\Xot \coloneqq \Xot - \Xoo \Bot$
              !\dgemm[NN]:! $\Xtt \coloneqq \Xtt - \Xto \Bot$
        \end{alglisting}
        \caption{Algorithm 10}
        \label{alg:sylv10}
    \end{subfigure}

    \mycaption{%
        Sample of blocked algorithms solving the triangular Sylvester
        equation with $3 \times 3$ matrix partitionings.
        \captiondetails{Output~$X$ overwrites input~$C$.}
    }
    \label{algs:sylv2d}
\end{figure}

Beyond the combination of the vertically and horizontally traversing algorithms
above, an additional 14~algorithms traverse the matrix diagonally (with
potentially different block sizes~$b_m$ and~$b_n$ for dimensions~$m$ and~$n$),
and operate on a set of $3 \times 3$ sub-matrices in each iteration;
\cref{algs:sylv2d} presents a sample of two of these algorithms (all
14~algorithms are found in \libflame~\cite{libflameweb}).  Each algorithm
consists of a sequence of \dgemm[NN]{}s and three solutions of sub-problems that
are also triangular Sylvester equations.  While the sub-problem involving
\dm[mat11, size=.5]{B_{11}} of size $b_m \times b_n$ is directly solved by the
unblocked \dtrsyl[NN1], the other two involve potentially large yet thin panels
of~\dmC.  Complete algorithms are constructed by solving each of these sub
problems with an appropriate vertical or horizontal traversal
algorithm.\footnote{%
    Setting one of the block sizes of a diagonally traversing algorithm to the
    corresponding matrix size results in one of the vertical or horizontal
    traversal algorithms.
}  Since each of the 14~algorithms has
two such sub-problems, for each of which we can choose from two algorithms, we
end up with a total of $14 \cdot 2 \cdot 2 = 56$~possible combinations.
Together with the eight combinations of only vertical and horizontal traversal
algorithms, this results in a grand total of 64~different ``complete'' blocked
algorithms.

\subsubsection{Algorithm Selection}

\begin{figure}[p]\figurestyle
    \ref*{leg:sylv}

    \pgfplotsset{
        twocolplot,
        ymax=26.4,
        ylabel={performance \Q p{med}{pred}},
        y unit=\si{\giga\flops\per\second},
        xlabel=problem size $n$,
        table/search path={pred/figures/data/varsylv},
        plotheightsub=.8pt,
    }

    \begin{subfigaxis}[
            fig caption=Predictions on 1~core,
            fig label=var:sylv:pred:1,
            legend columns=5,
            legend to name=leg:sylv
        ]
        \addlegendimage{empty legend}
        \addlegendentry{Algorithms:}
        \addlegendimage{plot1}\addlegendentry{$m1n1$}\label{plt:sylvm1n1}
        \addlegendimage{plot2}\addlegendentry{$m1n2$}\label{plt:sylvm1n2}
        \addlegendimage{plot3}\addlegendentry{$m2n1$}\label{plt:sylvm2n1}
        \addlegendimage{plot4}\addlegendentry{$m2n2$}\label{plt:sylvm2n2}
        \addlegendimage{matrixborder}\addlegendentry{3 x 3 partitioning}
        \addlegendimage{plot5}\addlegendentry{$n1m1$}\label{plt:sylvn1m1}
        \addlegendimage{plot6}\addlegendentry{$n1m2$}\label{plt:sylvn1m2}
        \addlegendimage{plot7}\addlegendentry{$n2m1$}\label{plt:sylvn2m1}
        \addlegendimage{plot8}\addlegendentry{$n2m2$}\label{plt:sylvn2m2}
        \foreach \var in {1, ..., 56}
            \addplot[matrixborder] table
                {sylv_\var.Haswell.1.OpenBLAS/perf_pred.dat};
        \foreach \var [count=\i] in {57, ..., 64} {\edef\temp{
                \noexpand\addplot[plot\i] table
                    {sylv_\var.Haswell.1.OpenBLAS/perf_pred.dat};
        }\temp}
    \end{subfigaxis}\hfill
    \begin{subfigaxis}[
            fig caption=Measurements on 1~core,
            fig label=var:sylv:meas:1,
        ]
        \foreach \a/\b in {
            15/17, 15/18, 16/17, 16/18, 17/15, 17/16, 18/15, 18/16%
        }
            \addplot table
                {sylv\a_\b.Haswell.1.OpenBLAS/perf_meas.dat};
    \end{subfigaxis}

    \begin{subfigaxis}[
            fig caption=Predictions on 12~cores,
            fig label=var:sylv:pred:12,
        ]
        \foreach \var in {1, ..., 56}
            \addplot[matrixborder] table
                {sylv_\var.Haswell.12.OpenBLAS/perf_pred.dat};
        \foreach \var in {57, ..., 64}
            \addplot table
                {sylv_\var.Haswell.12.OpenBLAS/perf_pred.dat};
    \end{subfigaxis}\hfill
    \begin{subfigaxis}[
            fig caption=Measurements on 1~core,
            fig label=var:sylv:meas:12,
        ]
        \foreach \a/\b in {
            15/17, 15/18, 16/17, 16/18, 17/15, 17/16, 18/15, 18/16%
        }
            \addplot table
                {sylv\a_\b.Haswell.12.OpenBLAS/perf_meas.dat};
    \end{subfigaxis}

    \mycaption{%
        Performance predictions and measurements for the blocked triangular
        Sylvester equation solvers.
        \captiondetails{$b = 128$, \haswell, \openblas}
    }
    \label{fig:pred:var:sylv}
\end{figure}

\cref{fig:pred:var:sylv} presents performance predictions and measurements for
the Sylvester equation solver for problem sizes between~$n = 56$ and~4152 in
steps of~64 and block size~$b = 64$ on a \haswell using \openblas.  Since the
executions for this setup take between 40~minutes and 2~hours for each
algorithm, we only measured the eight algorithms based exclusively on orthogonal
matrix traversals.  Our predictions, which are generated up to
$1500\times$~faster at roughly \SI5{\second}~per algorithm, indicate that in
terms of performance these eight algorithms are evenly spread across the entire
64~``complete'' algorithms.

For the single-threaded scenario, the predictions in
\cref{fig:pred:var:sylv:pred:1} suggest that
algorithms~$n2m2$~(\ref*{plt:sylvn2m2}) and $m1n1$~(\ref*{plt:sylvm1n1}) are,
respectively, the fastest and slowest, and differ in performance
by~\SI{9.99}\percent.  The measurements in \cref{fig:pred:var:sylv:meas:1}
confirm that, while algorithm~$n2m2$~(\ref*{plt:sylvn2m2}) is indeed the
fastest, algorithm~~$n1m1$~(\ref*{plt:sylvn1m1}) is the slowest.  While the
performance of algorithms~$m1n1$~(\ref*{plt:sylvm1n1}) and
$n1m1$~(\ref*{plt:sylvn1m2}) is predicted to be almost identical, the
measurements show that $m1n1$~(\ref*{plt:sylvm1n1}) is in fact up to
\SI{3.00}{\percent} faster than $n1m1$~(\ref*{plt:sylvn1m2}).  Furthermore,
while the remaining algorithms are correctly placed between the fastest and the
slowest, they are not accurately ranked.

The predictions and measurements for the multi-threaded scenario in
\cref{fig:pred:var:sylv:pred:12,fig:pred:var:sylv:meas:12} are at first sight
surprising:  Compared to the  single-threaded case the attained performance is
considerably lower.  For matrices of size~$n = 4000$, the algorithms reach
roughly \SI8{\giga\flops\per\second}, which corresponds to
merely~\SI{1.67}{\percent} of the processor's 12-core peak performance of
\SI{480}{\giga\flops\per\second} (without \turboboost).  An analysis revealed
that the source of the drastic increase in runtime is the \blasl1 kernel \dswap,
which the unblocked \dtrsyl\footnote{%
    Technically within \code{dlasy2}, which is called from \dtrsyl.
} uses to swap two vectors of length~4:  Although the workload for this
operation is tiny, with multiple threads \openblas (version~0.2.15) activates
its parallelisation, which for a copy operation on only~\SI{64}{\bytes}
introduces a overhead of over~$200\times$ the kernel's single-threaded runtime.
(The problem was subsequently fixed in \openblas version~0.2.16 (March 2016) and
is not present in \mkl.)

While the multi-threaded predictions for all 64~algorithms indicate virtually
identical performance and thus do not allow a meaningful performance ranking,
they support the crucial revelation that using \openblas~0.2.15 the triangular
Sylvester equation is solved considerably faster on a single core than on
12~cores without exception.
}

        \subsection{Summary}
        \label{sec:pred:var:sum}
        We evaluated performance predictions for blocked algorithms as a means to select
the fastest algorithm from a set of mathematically equivalent alternatives.  We
considered three operations with an increasing number of algorithms and found
our predictions to rank the algorithms with great precision, thereby correctly
identifying the fastest algorithm(s) in all cases.  We also noted that using our
model-based predictions instead of empirical measurements speeds up the
identification process by two to three orders of magnitude.

    \section{Block Size Optimization}
    \label{sec:pred:b}
    We now turn to our second goal for blocked algorithms:  Using model-based
performance predictions to optimize the algorithmic block size~$b$.

To understand how the block size influences an algorithm's performance, recall
that it determines the shape of the sub-matrices exposed in each traversal
step---most notably the width of matrix panels such as~$A_{10}$ and~$A_{21}$ and
the size of the square diagonal block~$A_{11}$ (see \cref{ex:intro:chol} on
\pageref{ex:intro:chol}).  It hence incurs a trade-off between an increase in
performance of \blasl3 kernels for larger operations and a shift of
computational workload to the comparatively inefficient of unblocked \lapack
kernel.

\begin{figure}\figurestyle
    \ref*{leg:cholbreakdown}

    \pgfplotsset{
        twocolplot,
        xlabel=block size $b$,
        table/search path={pred/figures/data/blockedbreakdown/},
        restrict x to domain=0:320,
    }

    \begin{subfigaxis}[
            fig caption=Accumulated kernel runtime,
            fig label=blocksize:breakdown,
            ylabel=runtime, y unit=\ms,
            stack plots=y,
            every axis plot/.append style={fill, fill opacity=.5},
        ]
        \addplot[blue] table[y=tdtrsm] {breakdown.dat} \closedcycle;
        \addplot[green] table[y=tdsyrk] {breakdown.dat} \closedcycle;
        \addplot[red] table[y=tdpotf2] {breakdown.dat} \closedcycle;
    \end{subfigaxis}\hfill
    \begin{subfigaxis}[
            fig caption=Average kernel performance,
            fig label=blocksize:perf,
            perfplot=52.8,
            legend to name=leg:cholbreakdown,
        ]
        \addplot[red] table[y=pdpotf2] {breakdown.dat};
        \addlegendentry{\dpotf[L]2}\label{plt:breakdown:dpotf2}
        \addplot[green] table[y=pdsyrk] {breakdown.dat};
        \addlegendentry{\dsyrk[LN]}\label{plt:breakdown:dsyrk}
        \addplot[blue] table[y=pdtrsm] {breakdown.dat};
        \addlegendentry{\dtrsm[RLTN]}\label{plt:breakdown:dtrsm}
    \end{subfigaxis}

    \mycaption{%
        Breakdown of the blocked Cholesky decomposition algorithm~3 in terms of
        kernel runtime and performance.
        \captiondetails{$n = 1000$, \haswell, 1~threads, \openblas, predictions}
    }
    \label{fig:pred:blocksize}
\end{figure}

\begin{example}{Block size trade-off}{pred:blocksize}
    We study how the kernels within the blocked Cholesky decomposition
    algorithm~3 (\cref{alg:chol3} on \cpageref{alg:chol3}) contribute to its
    runtime for a problem of size~$n = 1000$ and varying block size~$b$ on a
    \haswell using single-threaded \openblas. For this setup
    \cref{fig:pred:blocksize} presents model-based performance estimates of
    \subfigref{fig:pred:blocksize:breakdown}~how much of the algorithm's runtime
    is spent in the kernels \dpotf[LN]2, \dtrsm[RLTN] and \dsyrk[LN], and
    \subfigref{fig:pred:blocksize:perf}~these kernels' average performance.

    For small block sizes~$b$, the arithmetic intensity of \dtrsm[RLTN] and
    \dsyrk[LN] is so low that they are effectively bandwidth-bound, and thus
    fairly inefficient.  As $b$~increases, the operands of all three kernels
    grow in size and so does their performance (\cref{fig:pred:blocksize:perf}):
    \dsyrk~(\ref*{plt:breakdown:dsyrk}) plateaus near
    \SI{43}{\giga\flops\per\second} around~$b = 100$, and while \dtrsm's
    efficiency~(\ref*{plt:breakdown:dtrsm}) steadily rises towards that of
    \dsyrk~(\ref*{plt:breakdown:dsyrk}), \dpotf2~(\ref*{plt:breakdown:dpotf2})
    approaches its peak of only \SI{12}{\giga\flops\per\second} around~$b =
    175$.  On the other hand, with increasing~$b$ more and more computation is
    shifted from the \blasl3 routines to the inefficient
    \dpotf2~(\ref*{plt:breakdown:dpotf2}); beyond~$b = 112$ this kernel's low
    performance causes the overall runtime to increase
    (\cref{fig:pred:blocksize:breakdown}).
\end{example}

In the following analysis of our model-based performance predictions, we once
more consider the lower-triangular Cholesky decomposition and the inversion of a
lower-triangular matrix in, respectively,
\cref{sec:pred:b:chol,sec:pred:b:trinv}, and study three of \lapack's blocked
algorithms in \cref{sec:pred:b:lapack}.

        \subsection{Cholesky Decomposition}
        \label{sec:pred:b:chol}
        \begin{figure}[p]\figurestyle
    \ref*{leg:b}

    \pgfplotsset{
        twocolplot, perfplot=52.8,
        xlabel={block size $b$},
        table/search path={pred/figures/data/bchol/chol3.Haswell.1.OpenBLAS/},
        restrict x to domain=0:320,
    }

    \begin{subfigaxis}[
            fig caption=Predictions on 1~core,
            fig label=b:chol:time:meas:1,
            ylabel={performance \Q p{med}{pred}}
        ]
        \foreach \n/\y/\x in {
            1/33.9246660568/112,
            2/42.1940162643/144,
            3/45.9810887608/160,
            4/48.0816711909/184
        } {\edef\temp{
            \noexpand\draw[plot\n, dashed] (\x, 0) -- (\x, \y);
            \noexpand\filldraw[plot\n] (\x, \y) circle (2pt);
        }\temp}
        \foreach \n in {1000, 2000, 3000, 4000}
            \addplot table[y=pred] {perf\n.dat};
    \end{subfigaxis}\hfill
    \begin{subfigaxis}[
            fig caption=Measurements on 1~core,
            fig label=b:chol:pred:1,
            ylabel={performance \Q p{med}{meas}}
        ]
        \foreach \n/\y/\x in {
            1/34.002049227/96,
            2/42.1764395425/144,
            3/45.9288817014/160,
            4/48.1008517845/184
        } {\edef\temp{
            \noexpand\draw[plot\n, dashed] (\x, 0) -- (\x, \y);
            \noexpand\filldraw[plot\n] (\x, \y) circle (2pt);
        }\temp}
        \foreach \n in {1000, 2000, 3000, 4000}
            \addplot table[y=meas] {perf\n.dat};
    \end{subfigaxis}

    \pgfplotsset{
        ymax=480,
        table/search path={pred/figures/data/bchol/chol3.Haswell.12.OpenBLAS/},
    }

    \begin{subfigaxis}[
            fig caption=Predictions on 12~cores,
            fig label=b:chol:meas:12,
            ylabel={performance \Q p{med}{pred}}
        ]
        \foreach \n/\y/\x in {
            1/157.628411765/72,
            2/244.813547132/88,
            3/304.782429574/72,
            4/330.308651142/96
        } {\edef\temp{
            \noexpand\draw[plot\n, dashed] (\x, 0) -- (\x, \y);
            \noexpand\filldraw[plot\n] (\x, \y) circle (2pt);
        }\temp}
        \foreach \n in {1000, 2000, 3000, 4000}
            \addplot table[y=pred] {perf\n.dat};
    \end{subfigaxis}\hfill
    \begin{subfigaxis}[
            fig caption={Measurements on 12~cores},
            fig label=b:chol:pred:12,
            ylabel={performance \Q p{med}{meas}}
        ]
        \foreach \n/\y/\x in {
            1/165.648808762/56,
            2/254.775870975/96,
            3/317.770809266/120,
            4/349.714619186/112
        } {\edef\temp{
            \noexpand\draw[plot\n, dashed] (\x, 0) -- (\x, \y);
            \noexpand\filldraw[plot\n] (\x, \y) circle (2pt);
        }\temp}
        \foreach \n in {1000, 2000, 3000, 4000}
            \addplot table[y=meas] {perf\n.dat};
    \end{subfigaxis}

    \mycaption{%
        Model-based block size optimization and empirical optima for the
        Cholesky decomposition algorithm~3.
        \captiondetails{\haswell, \openblas}
    }
    \label{fig:pred:b:chol}
\end{figure}

We revisit the Cholesky decomposition with blocked algorithm~3 (\cref{alg:chol3}
on \cpageref{alg:chol3}), which \cref{sec:pred:var:chol} identified as the
fastest.  \Cref{fig:pred:b:chol} presents the algorithm's performance
predictions and measurements for problem sizes~$n = 1000$, 2000, 3000, and~4000
on a \haswell using single- and multi-threaded \openblas; it highlights the
predicted and empirical optimal block sizes~\bpred and~\bopt.

In the single-threaded case, the predicted optimal block sizes~\bpred are
identical to the empirical optima~\bopt for $n = 2000$~(\ref*{plt:n2000}),
3000~(\ref*{plt:n3000}), and 4000~(\ref*{plt:n4000}).  For $n =
1000$~(\ref*{plt:n1000}) the predicted optimum~$\bpred = 112$ is larger than
empirical~$\bopt = 96$, but choosing~$b = 112$ nonetheless yields
\SI{99.92}{\percent}~of the optimal performance.

In the multi-threaded case, the performance predictions do not match the
measurements quite as well, and none of the predicted~\bpred match the
empirical~\bopt.  However, with $b = \bpred$ the algorithm still reaches on
average \SI{98.52}{\percent}~of the optimal performance.

\begin{figure}\figurestyle
    \pgfplotsset{
        twocolplot,
        xlabel={problem size $n$},
        table/search path={pred/figures/data/bchol/},
        every subfigaxis/.append style={fig vertical align=b}
    }

    \begin{subfigaxis}[
            fig caption={Optimal block size\\\mstrut\vspace{1.25pt}},
            fig label=b:chol:optb,
            fig legend=bopt,
            ylabel=block size $b$,
            legend columns=3,
            legend to name=leg:bopt,
        ]
        \addlegendimage{empty legend}
        \addlegendentry{1 core:}
        \addplot[green] table[y=bpred] {chol3.Haswell.1.OpenBLAS/opt.dat};
        \addlegendentry{\bpred}\label{plt:bpred1}
        \addplot[yellowgreen] table[y=bopt] {chol3.Haswell.1.OpenBLAS/opt.dat};
        \addlegendentry{\bopt}\label{plt:bopt1}
        \addlegendimage{empty legend}
        \addlegendentry{12 cores:}
        \addplot[blue] table[y=bpred] {chol3.Haswell.12.OpenBLAS/opt.dat};
        \addlegendentry{\bpred}\label{plt:bpred12}
        \addplot[lightblue] table[y=bopt] {chol3.Haswell.12.OpenBLAS/opt.dat};
        \addlegendentry{\bopt}\label{plt:bopt12}
    \end{subfigaxis}\hfill
    \begin{subfigaxis}[
            fig caption={
                Performance yield of \bpred\\
                (Note: $y$-axis not 0-based)
            },
            fig label=b:chol:yield,
            fig legend=yield,
            ymin=90, ymax=100,
            ylabel=yield, y unit=\percent,
            clip=false,
            legend to name=leg:yield,
            legend columns=1,
        ]
        \addplot[green] table[y=yieldpred] {chol3.Haswell.1.OpenBLAS/opt.dat};
        \addlegendentry{1 core}
        \addplot[blue] table[y=yieldpred] {chol3.Haswell.12.OpenBLAS/opt.dat};
        \addlegendentry{12 cores}
    \end{subfigaxis}

    \mycaption{%
        Predicted and empirical optimal block sizes and prediction yields for
        the Cholesky decomposition algorithm~3.
        \captiondetails{\haswell, \openblas}
    }
    \label{fig:pred:b:chol2}
\end{figure}

We expand our study to a wider range of problem sizes between~$n = 56$ and~4152
in steps of~64 and analyze both how closely our predicted~\bpred match the
empirical~\bopt and how much of the optimal performance~$\Q p{med}{meas}(\bopt)$
the algorithm attains with~\bpred.  We referred to this ratio as \bpred's
\definition{performance yield}:
\begin{equation}
    \text{yield} \defeqq \frac{\Q p{med}{meas}(\bpred)}{\Q p{med}{meas}(\bopt)}
    = \frac{\Q t{med}{meas}(\bopt)}{\Q t{med}{meas}(\bpred)} \enspace.
\end{equation}

\Cref{fig:pred:b:chol:optb} confirms that \bpred matches \bopt slightly better
on one core (\ref*{plt:bpred1}, \ref*{plt:bopt1}) than on 12~cores
(\ref*{plt:bpred12}, \ref*{plt:bopt12}).  This is also reflected in its
performance yield presented in \cref{fig:pred:b:chol:yield}:  On 1~core, the
average yield is~\SI{99.35}{\percent} while on 12~cores it is slightly lower
at~\SI{98.57}\percent.

Note that for this study we measured the runtime of the algorithm 10~times for
each considered problem size~$n$ and block size~$b$, which took almost 2~hours
in the single-threaded case and around 20~minutes with 12~threads---in contrast
the predictions for the same range of sizes were obtained in under a minute in
both cases.

        \subsection{Triangular Inversion}
        \label{sec:pred:b:trinv}
        \begin{figure}\figurestyle
    \pgfplotsset{
        twocolplot,
        xlabel={problem size $n$},
        table/search path={pred/figures/data/btrinv/},
    }

    \begin{subfigaxis}[
            fig caption={Optimal block size},
            fig label=b:trinv:optb,
            fig legend=bopt,
            ylabel=block size $b$, ymax=200
        ]
        \addplot[green]       table[y=bpred] {trinv3.Haswell.1.OpenBLAS/opt.dat};
        \addplot[yellowgreen] table[y=bopt]  {trinv3.Haswell.1.OpenBLAS/opt.dat};
        \addplot[blue]        table[y=bpred] {trinv3.Haswell.12.OpenBLAS/opt.dat};
        \addplot[lightblue]   table[y=bopt]  {trinv3.Haswell.12.OpenBLAS/opt.dat};
    \end{subfigaxis}\hfill
    \begin{subfigaxis}[
            fig caption=Performance yield of $b_{pred}$,
            fig label=b:trinv:yield,
            fig legend=yield,
            ymin=90, ymax=100,
            ylabel=yield, y unit=\percent,
            clip=false
        ]
        \addplot[green] table[y=yieldpred] {trinv3.Haswell.1.OpenBLAS/opt.dat};
        \addplot[blue]  table[y=yieldpred] {trinv3.Haswell.12.OpenBLAS/opt.dat};
    \end{subfigaxis}

    \mycaption{%
        Predicted and empirical optimal block sizes and prediction yields for
        the inversion of a lower-triangular matrix algorithm~3.
        \captiondetails{\haswell, \openblas}
    }
    \label{fig:pred:b:trinv}
\end{figure}

we repeat the above study for the inversion of a lower-triangular matrix with
blocked algorithm~3 (\cref{alg:trinv3}), which was shown to be the fastest for
large problem sizes in \cref{sec:pred:var:chol}.  \Cref{fig:pred:b:trinv}
presents \subfigref{fig:pred:b:trinv:optb}~the predicted and empirical optimal
block sizes~\bpred and~\bopt using single- and multi-threaded \openblas, and
\subfigref{fig:pred:b:trinv:yield}~\bpred's performance yields.

\Cref{fig:pred:b:trinv:optb} shows that, in contrast to the Cholesky
decomposition (\cref{fig:pred:b:chol:optb}), the optimal block sizes for the
single- and multi-threaded inversion of a lower-triangular matrix are fairly
similar, yet slightly lower in the single-threaded case.  However, the
empirical~\bopt exhibits a behavior not well represented by the
predicted~\bpred:  Beyond $n \approx 2000$, the multi-threaded
\bopt~(\ref*{plt:bopt12}) assumes only two values---96 and~192---while our
prediction~\bpred indicates a more gradual transition.  (On 1~core the effect is
similar with $\bopt = 96$ for almost all problem sizes beyond~$n \approx 1700$.)
The cause for this problem is that our models only poorly represent certain
spikes in the performance of the multi-threaded \dsyrk[LN] implementation at the
optimal block sizes.

The sub-optimal choices of block sizes are reflected in the prediction yields:
\Cref{fig:pred:b:trinv:yield} shows that, while on a single core the yield is
almost ideal at~\SI{99.53}\percent, on 12~cores, it drops notably---especially
for larger problems---averaging~\SI{97.13}\percent.

        \subsection{\lapack Algorithms}
        \label{sec:pred:b:lapack}
        To conclude our study on block size optimization we consider three of \lapack's
blocked algorithms on square matrices:
\begin{itemize}
    \item \dsygst[1L]: $\dm A \coloneqq \dm[lower, inv]L \dm A \matmatsep
        \dm[upper, inv']L$ (\cref{alg:dsygst}),
    \item \dgetrf: $\dm[dotted]P \matmatsep \dm[lower]L \dm[upper]U \coloneqq
        \dm A$ (\cref{alg:dgetrf}), and
    \item \dgeqrf: $\dm Q \matmatsep \dm[upper]R \coloneqq \dm A$
        (\cref{alg:dgeqrf}).
\end{itemize}

\begin{figure}[p]\figurestyle
    \pgfplotsset{
        twocolplot, threerowplot,
        xlabel={problem size $n$},
        table/search path={pred/figures/data/blapack/},
        bplot/.style={
            ylabel={block size $b$},
            before end axis/.code={
                \addplot[red]         table[y=bdef]
                    {#1.Haswell.1.OpenBLAS/opt.dat};
                \addplot[green]       table[y=bpred]
                    {#1.Haswell.1.OpenBLAS/opt.dat};
                \addplot[yellowgreen] table[y=bopt]
                    {#1.Haswell.1.OpenBLAS/opt.dat};
                \addplot[blue]        table[y=bpred]
                    {#1.Haswell.12.OpenBLAS/opt.dat};
                \addplot[lightblue]   table[y=bopt]
                    {#1.Haswell.12.OpenBLAS/opt.dat};
            },
            fig caption={Block sizes for \csname#1\endcsname},
            fig label=b:lapack:#1:b,
        },
        yieldplot/.style={
            ymin=0, ymax=100,
            ylabel=yield, y unit=\percent,
            clip=false,
            before end axis/.code={
                \addplot[orange]
                    table[y=edef]  {#1.Haswell.1.OpenBLAS/opt.dat};
                \addplot[red]
                    table[y=edef]  {#1.Haswell.12.OpenBLAS/opt.dat};
                \addplot[green]
                    table[y=epred] {#1.Haswell.1.OpenBLAS/opt.dat};
                \addplot[blue]
                    table[y=epred] {#1.Haswell.12.OpenBLAS/opt.dat};
            },
            fig label=b:lapack:#1:yield,
            fig caption={Performance yield for \csname#1\endcsname},
        },
        plotheightsub=32pt,
    }

    \begin{subfigaxis}[
            bplot=dsygst, fig caption/.append={[1L]},
            fig caption={Block sizes for \dsygst[1L]},
            fig legend=b:lapack:b,
            legend to name=leg:b:lapack:b,
            legend columns=4
        ]
        \addlegendimage{empty legend}\addlegendentry{1~core:\mstrut}
        \addlegendimage{green}\addlegendentry{\bpred}
        \addlegendimage{yellowgreen}\addlegendentry{\bopt}
        \addlegendimage{red}\addlegendentry{\bdef}
        \addlegendimage{empty legend}\addlegendentry{12~cores:\mstrut}
        \addlegendimage{blue}\addlegendentry{\bpred}
        \addlegendimage{lightblue}\addlegendentry{\bopt}
        \addlegendimage{red}\addlegendentry{\bdef}
    \end{subfigaxis}\hfill
    \begin{subfigaxis}[
            yieldplot=dsygst, fig caption/.append={[1L]},
            fig legend=b:lapack:yield,
            legend to name=leg:b:lapack:yield,
            legend columns=3
        ]
        \addlegendimage{empty legend}\addlegendentry{1~core:}
        \addlegendimage{dashdotted 2 color={yellowgreen}{green}}
        \addlegendentry{yield}\label{plt:yieldpred1}
        \addlegendimage{dashdotted 2 color={yellowgreen}{orange}}
        \addlegendentry{yield($\bdef$)}\label{plt:yielddef1}
        \addlegendimage{empty legend}\addlegendentry{12~cores:}
        \addlegendimage{dashdotted 2 color={lightblue}{blue}}
        \addlegendentry{yield}\label{plt:yieldpred12}
        \addlegendimage{dashdotted 2 color={lightblue}{red}}
        \addlegendentry{yield(\bdef)}\label{plt:yielddef12}
    \end{subfigaxis}

    \begin{subfigaxis}[bplot=dgetrf]
    \end{subfigaxis}\hfill
    \begin{subfigaxis}[yieldplot=dgetrf]
    \end{subfigaxis}

    \begin{subfigaxis}[bplot=dgeqrf]
    \end{subfigaxis}\hfill
    \begin{subfigaxis}[yieldplot=dgeqrf]
    \end{subfigaxis}

    \mycaption{%
        Predicted and empirical optimal block sizes and prediction yields for
        \dsygst[1L], \dgetrf, and \dgeqrf.
        \captiondetails{\haswell, \openblas}
    }
    \label{fig:pred:b:lapack}
\end{figure}

\begin{table}\tablestyle
    \begin{tabular}{cl@{ }lccc}
        \toprule
        &&&\dsygst[1L]          &\dgetrf                &\dgeqrf \\
        \midrule
        \arrayrulecolor{lightblue}
                        &yield          &(\ref*{plt:yieldpred1})    &\SI{99.64}\percent  &\SI{99.90}\percent  &\SI{99.05}\percent \\
    1 core      &yield(\bdef)   &(\ref*{plt:yielddef1})     &\SI{93.64}\percent  &\SI{96.92}\percent  &\SI{92.92}\percent \\
                &\multicolumn2l{improvement}                &\SI{6.70}\percent  &\SI{3.41}\percent  &\SI{6.89}\percent \\
    \midrule
                &yield          &(\ref*{plt:yieldpred12})   &\SI{98.93}\percent  &\SI{98.10}\percent  &\SI{90.98}\percent \\
    12 cores    &yield(\bdef)   &(\ref*{plt:yielddef12})    &\SI{70.76}\percent  &\SI{95.23}\percent  &\SI{36.23}\percent \\
                &\multicolumn2l{improvement}                &\SI{45.18}\percent  &\SI{3.08}\percent  &\SI{189.64}\percent \\

        \arrayrulecolor{blue}
        \bottomrule
    \end{tabular}
    \mycaption{
        Average performance yields and improvement over \lapack for \dsygst[1L],
        \dgetrf, and \dgeqrf.
        \captiondetails{
            $n = 56, \ldots, 4152$ in steps of~64, $\bdef = 64$, except \dgeqrf:
            $\bdef = 32$, \haswell, \openblas
        }
    }
    \label{tbl:pred:b:lapack}
\end{table}

For these routines, the left half of \cref{fig:pred:b:lapack} presents the
predicted and empirical optimal block sizes~\bpred and~\bopt, as well as
\lapack's default block size~\bdef (\dsygst[1L], \dgetrf: 64; \dgeqrf: 32); and
the right half shows the performance yields for both \bpred and~\bdef.
Furthermore, \cref{tbl:pred:b:lapack} summarizes the yields for these block
sizes averaged across the chosen problem sizes.

For the single-threaded operations our predicted optimal block
sizes~\bpred~(\ref*{plt:bpred1}) match the empirical
optimum~\bopt~(\ref*{plt:bopt1}) quite well, resulting in an average performance
yield~(\ref*{plt:yieldpred1}) of~\SI{99.53}\percent.  For both \dsygst[1L] and
\dgeqrf, the optimal block size quickly exceed \lapack's default values, leading
to an improved performance of roughly \SI{10}{\percent} (\dsygst[1L]) and
\SI{15}{\percent} (\dgeqrf).  For \dgetrf on the other hand, \lapack's~$\bdef =
64$ is actually ideal between $n = 2500$ and~3700, meaning our predicted~\bpred
only yields improvements for smaller problem sizes.

In the multi-threaded case, the optimal block sizes are across the board larger.
\begin{itemize}
    \item For \dsygst[1L], \bopt~(\ref*{plt:bopt12}) is correctly
        predicted~(\ref*{plt:bpred12}) to jump from~$\approx 100$ to~$\approx
        200$ around~$n = 1500$, and next to~$\approx 290$ at~$n \approx 3000$.
        These predictions yield~\SI{98.93}{\percent}~(\ref*{plt:yieldpred12}) of
        the optimal performance, which is an average
        \SI{45.18}{\percent}~improvement over \lapack's~$\bdef = 64$
        (\ref*{plt:yielddef12})---reaching up to $\approx \SI{90}\percent$
        for~$n \approx 4000$.

    \item For \dgetrf, the optimal block size~\bopt~(\ref*{plt:bopt12})
        fluctuates constantly with a magnitude of~32, which is not represented
        by the prediction~\bpred~(\ref*{plt:bpred12}).  However, the general
        trend is captured fairly well up to~$n \approx 2500$, after which
        \bopt~stagnates, while \bpred increases further.  As a result, the
        performance yield~(\ref*{plt:yieldpred12}) decreases slightly beyond $n
        = 3000$, yet retains a high average of \SI{98.10}\percent.  Since
        \dgetrf is generally less sensitive to the block size, \lapack's~$\bdef
        = 64$ also yields above~\SI{95}\percent~(\ref*{plt:yielddef12}) of the
        optimal performance.

    \item For \dgeqrf, the unblocked \dgeqr2 is faster for small problem sizes
        than the blocked algorithm with any block size, which translates
        to~$\bopt = n$;\footnote{%
            To leverage the performance of optimized \blasl3 through a blocked
            algorithm, \dgeqrf performs \SIvar{O(b n^2)}{\flops} more than
            \dgeqr2.  The fact that \dgeqr2 is faster than \dgeqrf not only for
            small problems indicates that these extra \flops are not easily
            amortized in the multi-threaded scenario.
        } this behavior is for the considered block sizes up to~$b
        = 536$ correctly predicted.  The trend of \bdef's yield in
        \cref{fig:pred:b:lapack:dgeqrf:yield} suggests that \dgeqr2 may continue
        to be faster than \dgeqrf until~$n \approx 1000$.  Beyond this point,
        \bopt~(\ref*{plt:bopt12}) jumps between~$\approx 100$ and~$\approx 200$
        until $n \approx 2000$, after which it remains around $\bopt = 200$.
        Since our predicted~\bpred~(\ref*{plt:bpred12}) indicates a smoother
        increase beyond $n = 2000$, the performance
        yield~(\ref*{plt:yieldpred12}) eventually drops to~$\approx
        \SI{84}\percent$.  Compared to the yield of \lapack's~$\bdef =
        32$~(\ref*{plt:yielddef12}), however, this is still a major improvement
        of up to~$4\times$, averaging~\SI{189.64}\percent.
\end{itemize}

In summary, our model-based predictions of the optimal block size show varying
degrees of accuracy, yet consistently provide performance improvements over
\lapack's default block sizes by up to~\SI{300}\percent.  Note that while the
measurements for the above study took in total almost 4~days, all corresponding
predictions were obtained from our models in under 25~minutes.  While choosing
a coarser set of samples (i.e., fewer problem and block sizes) for the empirical
optimization might reduce its runtime to below 10~hours, our predictions, to
which the same reduction can be applied, would still provide a significant
speedup.  By porting our currently \python-based models to other formats to be
evaluated in a faster language (e.g., in~\clang/\cpplang), we expect that this
prediction time can be reduce mere seconds.

    \section{Summary}
    \label{sec:pred:conclusion}
    This chapter presented performance predictions for blocked algorithms based on
the performance models described in \cref{ch:model}.  These predictions were
found to closely match the measured performance of blocked \lapack algorithms in
a variety of setups.  They allow us to solve two important problems without any
algorithm executions:
\begin{itemize}
    \item We can rank alternative blocked algorithms according to their
        performance, and thereby identify the fastest algorithm for various
        operations.

    \item We can select near-optimal block sizes that lead algorithms to within
        a few percent of their empirically optimal performance; they are often
        an enormous improvement over \lapack's default block sizes.
\end{itemize}
Since our models predict algorithm executions two to three orders of magnitude
faster than corresponding empirical measurements, they make previously
disproportionately time-consuming optimization processes feasible.

}

    \chapter[Cache Modeling and Prediction]
        {Cache Modeling\newline and Prediction}
\chapterlabel{cache}
{
    

\newcommand\refdgeqr{\dgeqr2~(\ref*{plt:dgeqr2})\xspace}
\newcommand\refdlarftFC{\dlarft[FC]~(\ref*{plt:dlarftFC})\xspace}
\newcommand\refdcopy{\dcopy~(\ref*{plt:dcopy})\xspace}
\newcommand\refdtrmmRLNU{\dtrmm[RLNU]~(\ref*{plt:dtrmmRLNU})\xspace}
\newcommand\refdgemmTN{\dgemm[TN]~(\ref*{plt:dgemmTN})\xspace}
\newcommand\refdtrmmRUNN{\dtrmm[RUNN]~(\ref*{plt:dtrmmRUNN})\xspace}
\newcommand\refdgemmNT{\dgemm[NT]~(\ref*{plt:dgemmNT})\xspace}
\newcommand\refdtrmmRLTU{\dtrmm[RLTU]~(\ref*{plt:dtrmmRLTU})\xspace}

\newcommand\refdsyrkUT{\dsyrk[UT]~(\ref*{plt:dsyrkUT})\xspace}
\newcommand\refdpotfU{\dpotf[U]2~(\ref*{plt:dpotf2U})\xspace}

\newcommand\refdtrmmLLNN{\dtrmm[LLNN]~(\ref*{plt:dtrmmLLNN})\xspace}
\newcommand\refdtrtiLN{\dtrti[LN]2~(\ref*{plt:dtrti2LN})\xspace}

\newcommand\kernelplots[1]{
    \foreach \r in {
        dgeqr2_, dlarft_, dcopy_, dtrmm_RLNU,
        dgemm_TN, dtrmm_RUNN, dgemm_NT, dtrmm_RLTU%
    } \addplot table {#1.\r.dat};
}

\newcommand\kernelplotpic[2][]{
    \tikzsetnextfilename{#2}
    \begin{tikzpicture}
        \begin{axis}[dgeqrf, #1]
            \kernelplots{#2}
        \end{axis};
    \end{tikzpicture}%
}

\pgfplotsset{
    kernelplot/.style={
        twocolplot,
        xlabel=kernel invocation,
        ylabel=relative error, y unit=\percent,
        enlarge x limits={value=0, auto},
        every axis plot/.append style={only marks},
    },
    dtrmm/.style={blue},
    dgemm/.style={red},
    dtrsm/.style={lightblue},
    dsryk/.style={orange},
    dcopy/.style={orange, mark=*, mark size=.5pt},
    dgeqr2/.style={green, mark=x},
    dtrti2LN/.style={green, mark=x},
    dpotf2U/.style={green, mark=x},
    dlarftFC/.style={yellowgreen, mark=x},
    dtrmmRLNU/.style={dtrmm, mark=x},
    dgemmTN/.style={dgemm, mark=x},
    dtrmmRUNN/.style={dtrmm, mark=o},
    dgemmNT/.style={dgemm, mark=o},
    dtrmmRLTU/.style={dtrmm, mark=diamond},
    dtrmmLLNN/.style={dtrmm, mark=x},
    dtrsmRLNN/.style={dtrsm, mark=x},
    dsyrkUT/.style={dsryk, mark=x},
    dtrsmLUTN/.style={dtrsm, mark=x},
    dgeqrf/.style={
        fig label=dgeqrf:#1,
        kernelplot,
        cycle list={dgeqr2, dlarftFC, dcopy, dtrmmRLNU, dgemmTN, dtrmmRUNN,
        dgemmNT, dtrmmRLTU},
        table/search path={cache/figures/data/qr/},
        before end axis/.code={
            \foreach \r in {
                dgeqr2_, dlarft_FC, dcopy_, dtrmm_RLNU,
                dgemm_TN, dtrmm_RUNN, dgemm_NT, dtrmm_RLTU%
            } \addplot table {#1.\r.dat};
        }
    },
    dpotrfU/.style={
        kernelplot,
        cycle list={dsyrkUT, dpotf2U, dgemmTN, dtrsmLUTN},
        table/search path={cache/figures/data/chol/},
        before end axis/.code={
            \foreach \r in {dsyrk_UT, dpotf2_U, dgemm_TN, dtrsm_LUTN}
                \addplot table {#1.\r.dat};
        }
    },
    dtrtriLN/.style={
        kernelplot,
        cycle list={dtrmmLLNN, dtrsmRLNN, dtrti2LN},
        before end axis/.code={
            \foreach \r in {dtrmm_LLNN, dtrsm_RLNN, dtrti2_LN}
                \addplot table {#1.\r.dat};
        }
    },
}

\newcommand\sic{\ensuremath{s_\mathrm{ic}}\xspace}
\newcommand\soc{\ensuremath{s_\mathrm{oc}}\xspace}
\newcommand\tic{\ensuremath{t_\mathrm{ic}}\xspace}
\newcommand\toc{\ensuremath{t_\mathrm{oc}}\xspace}
\newcommand\op{\ensuremath{\mathcal Op}\xspace}

    The previous chapter introduced the concept of model-based performance
predictions for dense linear algebra algorithms.  While such predictions are
accurate for many scenarios, we observed a degradation in accuracy for operands
larger than the processor's last-level cache.  This chapter analyzes such
caching effects and explores how they can be accounted for in predictions.

\Cref{sec:cache:qr} presents a case study on \lapack's blocked QR~decomposition
\dgeqrf on a \harpertown using \openblas, and details efforts to accurately
estimate the runtime of the kernel invocations within \dgeqrf by combining
isolated in- and out-of-cache timings.  Next, \cref{sec:cache:algs} applies the
developed approach to two further \lapack algorithms.  Finally,
\cref{sec:cache:new} attempts to employ the same concepts on more recent
hardware, and reveals limitations to how well isolated kernel timings can
predict an algorithm's total runtime.

\paragraph{Publication}
The work presented in this chapter---in particular
\cref{sec:cache:qr,sec:cache:algs}---is in parts based on research previously
published in:
\begin{pubitemize}
    \pubitem{qrcaching}
\end{pubitemize}

    \section[Case Study: QR~Decomposition on a \texorpdfstring\harpertown{Harpertown E5450}]
            {Case Study: QR~Decomposition\newline on a \harpertown}
    \label{sec:cache:qr}
    We focus on a specific, yet exemplary algorithm and setup:  We analyze the
performance of \lapack's QR~decomposition \dgeqrf
\[
    \dm Q \matmatsep \dm[upper] R \coloneqq \dm A
\]
of a square matrix \definition[$n = 1568$\\$b = 32$]{$\dm A \in \R^{1568 \times
1568}$} with \lapack's default block size~$b = 32$ on a \harpertown using
single-threaded \openblas---with a size of about~\SI{18}{\mebi\byte}, \dm A
exceeds the processor's last-level cache~(L2) of~\SI6{\mebi\byte} per 2~cores.

In the following, \cref{sec:cache:qr:alg} presents the blocked algorithm behind
\dgeqrf and instrumentation-based in-algorithm timings that serve as the
reference for our per-kernel runtime predictions.  Next,
\cref{sec:cache:qr:timings} measures the runtime of each kernel invocation in
isolation with cache preconditions, and establishes in- and out-of-cache timings
as, respectively, lower and upper bounds on the in-algorithm timings.
\Cref{sec:cache:qr:cache} combines the in- and out-of-cache timings to estimate
the in-algorithm timing by tracking which parts of the kernel operands reside in
the processor's L2~cache prior to the invocation.  Finally,
\cref{sec:cache:qr:res} expands the introduced methodology beyond the initially
considered instance of the blocked QR~decomposition towards other scenarios on
the \harpertown, including other matrix and block sizes, \blas implementations,
and kernel parallelism.

        \subsection{Timing Kernels in \lapack's \texorpdfstring\dgeqrf{dgeqrf}}
        \label{sec:cache:qr:alg}
        \begin{figure}\figurestyle
    \begin{subfigure}{.53\textwidth}\centering
        \tikzset{
            every picture/.append style={
                y={(0, -1)}, scale=.7,
                baseline={(0, 0)}
            }
        }
        \def\sm{6} \def\sn{5.5} \def\p{2} \def\q{3}
        \begin{tikzpicture}
            \filldraw[mat  ] (0,  0 ) rectangle (\sn,    \sm   );
            \filldraw[mat  ] (0,  0 ) rectangle (\p-.05, \p-.05) node {$A_{00}$};
            \filldraw[mat  ] (\p, 0 ) rectangle (\q-.05, \p-.05) node {$A_{01}$};
            \filldraw[mat  ] (\q, 0 ) rectangle (\sn,    \p-.05) node {$A_{02}$};
            \filldraw[mat  ] (0,  \p) rectangle (\p-.05, \q-.05) node {$A_{10}$};
            \filldraw[mat11] (\p, \p) rectangle (\q-.05, \q-.05) node {$A_{11}$};
            \filldraw[mat12] (\q, \p) rectangle (\sn,    \q-.05) node {$A_{12}$};
            \filldraw[mat  ] (0,  \q) rectangle (\p-.05, \sm   ) node {$A_{20}$};
            \filldraw[mat21] (\p, \q) rectangle (\q-.05, \sm   ) node {$A_{21}$};
            \filldraw[mat22] (\q, \q) rectangle (\sn,    \sm   ) node {$A_{22}$};
            \draw[brace ] (\p,  0  ) -- (\q-.05, 0     ) node {$b$}; 
            \draw[brace ] (\sn, \p ) -- (\sn,    \q-.05) node {$b$}; 
            \draw[bracei] (0,   \sm) -- (\sn,    \sm   ) node {$n$}; 
            \draw[bracei] (0,   0  ) -- (0,      \sm   ) node {$m$}; 
        \end{tikzpicture}
        \begin{tikzpicture}
            \draw[mat   ] (0, 0 ) -- (0, \sn);
            \draw[blue] (0, \p) -- (0, \q ) node[midway] {$\tau_1$};
        \end{tikzpicture}
        \begin{tikzpicture}
            \filldraw[mat  ] (0, 0) rectangle (1, \sn   );
            \filldraw[mat10] (0, 0) --        (1, 1     ) node {$W_1$} |- cycle;
            \filldraw[mat20] (0, 1) rectangle (1, \sn-\p) node {$W_2$};
            \draw[brace ] (1, 0) -- (1, \sn) node {$n$};
            \draw[brace ] (0, 0) -- (1, 0  ) node {$b$};
            \draw[brace2] (1, 0) -- (1, 1  ) node {$b$};
        \end{tikzpicture}

        \caption{Matrix partitioning}
        \label{fig:cache:qralg:traversal}
    \end{subfigure}\hfill
    \begin{subfigure}{.46\textwidth}\centering
        \newcommand\traversalA{
            \begin{tikzpicture}[
                    baseline=(x.base),
                    y={(0, -1)}, scale=.1
                ]
                \def\sm{6} \def\sn{5.5} \def\p{2} \def\q{3}
                \filldraw[mat]   (0,  0 ) rectangle (\sn, \sm);
                \filldraw[mat]   (0,  0 ) rectangle (\p,  \p);
                \filldraw[mat]   (\p, 0 ) rectangle (\q,  \p);
                \filldraw[mat]   (\q, 0 ) rectangle (\sn, \p);
                \filldraw[mat]   (0,  \p) rectangle (\p,  \q);
                \filldraw[mat11] (\p, \p) rectangle (\q,  \q);
                \filldraw[mat12] (\q, \p) rectangle (\sn, \q);
                \filldraw[mat]   (0,  \q) rectangle (\p,  \sm);
                \filldraw[mat21] (\p, \q) rectangle (\q,  \sm);
                \filldraw[mat22] (\q, \q) rectangle (\sn, \sm);
                \path (0, 0) -- (\sn, \sm) node[midway] (x) {$A$};
            \end{tikzpicture}
        }
        \newcommand\traversaltau{
            \begin{tikzpicture}[
                    baseline=(x.base),
                    y={(0, -1)}, scale=.1
                ]
                \def\sm{5.5} \def\sn{5} \def\p{2} \def\q{3}
                \draw[mat   ] (0,  0) rectangle (0, 5);
                \draw[blue] (0,  2) rectangle (0, 3);
                \path (0, 0) -- (0, 5) node[midway] (x) {$\tau$};
            \end{tikzpicture}
        }

        \makeatletter\let\tikz@ensure@dollar@catcode=\relax\makeatother
        \newcommand\Aoo{\dm[mat11, size=.5]{A_{11}}}
        \newcommand\AooL{\dm[mat11, size=.5, lower]{A_{11}}}
        \newcommand\AooLT{\dm[mat11, size=.5, upper, ']{A_{11}}}
        \newcommand\Aot{\dm[mat12, width=1.25, height=.5]{A_{12}}}
        \newcommand\AotT{\dm[mat12, width=.5, height=1.25, ']{A_{12}}}
        \newcommand\Ato{\dm[mat21, width=.5, height=1.5]{A_{21}}}
        \newcommand\Att{\dm[mat22, width=1.25, height=1.5]{A_{22}}}
        \newcommand\AttT{\dm[mat22, width=1.5, height=1.25, ']{A_{22}}}
        \newcommand\Wo{\dm[mat10, size=.5, upper]{W_1}}
        \newcommand\Wt{\dm[mat20, width=.5, height=1.25]{W_2}}
        \newcommand\WtT{\dm[mat20, width=1.25, height=.5, ']{W_2}}
        \newcommand\tauo{\dv[blue]\tau_1}
        \begin{alglisting}[]
            traverse $\traversalA$ along $\tsearrow$, $\traversaltau$ along $\tsarrow$:
              !\llap{\ref*{plt:dgeqr2} }\dgeqr2:! $\dmstack\Aoo\Ato, \tauo \coloneqq$ QR($\dmstack\Aoo\Ato$)
              !\llap{\ref*{plt:dlarftFC} }\dlarft[FC]:! $\Wo \coloneqq$ T($\dmstack\Aoo\Ato, \tauo$)
              !\dlarfb[LTFC]:!
                !\llap{\ref*{plt:dcopy}\,\ }$b\times$\dcopy:! $\Wt \coloneqq \AotT$
                !\llap{\ref*{plt:dtrmmRLNU} }\dtrmm[RLNU]:! $\Wt \coloneqq \Wt \AooL$
                !\llap{\ref*{plt:dgemmTN} }\dgemm[TN]:! $\Wt \coloneqq \Wt + \AttT \Ato$
                !\llap{\ref*{plt:dtrmmRUNN} }\dtrmm[RUNN]:! $\Wt \coloneqq \Wt \Wo$
                !\llap{\ref*{plt:dgemmNT} }\dgemm[NT]:! $\Att \coloneqq \Att - \Ato \WtT$
                !\llap{\ref*{plt:dtrmmRLTU} }\dtrmm[RLTU]:! $\Wt \coloneqq \Wt \AooLT$
                !inline:! $\Aot \coloneqq \Aot - \WtT$
        \end{alglisting}
        \caption{Algorithm}
    \end{subfigure}

    \mycaption{%
        \lapack's blocked algorithm for \dgeqrf.
        \iftableoflist\else\\The routine markers (\ref*{plt:dgemmTN},
        \ref*{plt:dtrmmRLTU}, etc.) are references for following plots.\fi
    }
    \label{alg:dgeqrf2}
\end{figure}

\Cref{alg:dgeqrf} outlines the blocked algorithm employed by \lapack's
QR~decomposition \dgeqrf.  The algorithm overwrites \dm A's upper-triangular
part with \dm[upper]R, and stores \dm Q as the combination of 1) a series of
elementary reflectors in \dm A's strictly lower-triangular portion, and 2) a
separate output vector of scalar factors~\dv\tau.  It furthermore requires an
auxiliary matrix $\dm[width=.3]W \in \R^{n \times b}$ for temporary data.

\dgeqrf itself invokes only three routines: the unblocked QR~decomposition
\dgeqr2, the formation of the triangular block reflector~\dm[upper, size=.5]T
(stored in~\dm[mat10, upper, size=.5]{W_1}) through the unblocked
\dlarft[FC],\footnote{%
    The flags $\code{direct} = \code F$ and $\code{storev} = \code C$ indicate
    that the reflectors are stored in {\em forward} order and as {\em column}
    vectors.
} and the application of the block reflector through \dlarfb[LTFC].  The latter
in turn is implemented largely in terms of the \blasl3 kernels \dtrmm and
\dgemm; it furthermore performs a transposed matrix copy through a series of
$b$~\dcopy{}s, and an inlined transposed matrix subtraction.\footnote{%
    A series of $b$~\daxpy{}s would likely be more efficient.
}

\begin{figure}\figurestyle
    \ref*{leg:dgeqrf}

    \begin{subfigaxis}[
            fig caption=In-algorithm timings,
            dgeqrf=instr,
            ylabel=time, y unit=\ms,
            legend to name=leg:dgeqrf,
            legend columns=4,
        ]
        \addlegendimage{dgeqr2, mark size=3pt, very thick}
        \addlegendentry{\dgeqr2}\label{plt:dgeqr2}
        \addlegendimage{dlarftFC, mark size=3pt, very thick}
        \addlegendentry{\dlarft[FC]}\label{plt:dlarftFC}
        \addlegendimage{dcopy, mark size=.5pt, very thick}
        \addlegendentry{\dcopy}\label{plt:dcopy}
        \addlegendimage{dtrmmRLNU, mark size=3pt, very thick}
        \addlegendentry{\dtrmm[RLNU]}\label{plt:dtrmmRLNU}
        \addlegendimage{dgemmTN, mark size=3pt, very thick}
        \addlegendentry{\dgemm[TN]}\label{plt:dgemmTN}
        \addlegendimage{dtrmmRUNN, mark size=3pt, very thick}
        \addlegendentry{\dtrmm[RUNN]}\label{plt:dtrmmRUNN}
        \addlegendimage{dgemmNT, mark size=3pt, very thick}
        \addlegendentry{\dgemm[NT]}\label{plt:dgemmNT}
        \addlegendimage{dtrmmRLTU, mark size=3pt, very thick}
        \addlegendentry{\dtrmm[RLTU]}\label{plt:dtrmmRLTU}
    \end{subfigaxis}\hfill
    \begin{subfigaxis}[
            fig caption=Error of repeated execution,
            dgeqrf=repeat-i, ymin=-35, ymax=5,
        ]
    \end{subfigaxis}

    \mycaption{%
        In-algorithm timings and error of repeated execution timings with
        respect to these for the 1873 kernel invocations within \dgeqrf.
        \captiondetails{$n = 1568$, $b = 32$, \harpertown, 1~thread, \openblas,
        median of 100~repetitions}
    }
\end{figure}

To measure the runtime of the kernels within the QR~decomposition---henceforth
called \definition{in-algorithm timings}---we manually instrument \dgeqrf and
\dlarft, and collect timestamps (through the x86~instruction \code{rdtsc})
between kernel invocations.  For the studied algorithm execution,
\cref{fig:cache:dgeqrf:instr} presents the in-algorithm timings computed from
these timestamps:  The $x$-axis enumerates the 1873~kernel
invocations,\footnote{%
    $n / b - 1 = 1568 / 32 - 1 = 48$ traversal steps à 39~kernels (\dgeqr2,
    \dlarfb, $b = 32$~\dcopy{}s, 3~\dtrmm{}s, and 2~\dgemm{}s) and 1~final
    \dgeqr2: $48 \times 39 + 1 = 1873$.
} for each of which one data-point presents the kernel runtime.  The total
execution time (\SI{946.68}\ms) is dominated by the two
\dgemm{}s~(\ref*{plt:dgemmTN}, \ref*{plt:dgemmNT}); although the size of their
operands is the same, their runtimes differ significantly.  Our ultimate goal is
to develop a strategy to accurately predict the runtimes for all kernel
invocations without executing \dgeqrf itself.

        \subsection{Cache-Aware Timings}
        \label{sec:cache:qr:timings}
        We begin to predict the in-algorithm timings with an elementary setup:
\definition{repeated execution} of the kernels in isolation.  In these
executions, which are performed one right after the other without any
modifications to the data, we use the same flags and matrix sizes as within
\dgeqrf and well separated memory locations as operands.
\Cref{fig:cache:dgeqrf:repeat-i} shows the relative runtime error for the median
of 100~such independent repetitions with respect to the in-algorithm timings.
While the relative error for \refdcopy is rather large, the total contribution
of its 1536~invocations to the algorithm's runtime is below~\SI1\percent.  Not
considering these \dcopy{}s, the absolute relative error of the repeated
execution runtime estimates relative to the in-algorithm timings averaged across
all kernel invocations---in the following simply referred to as
\definition{error}---is~\SI{4.42}\percent.

For most routines and especially for \refdtrmmRLTU and \refdgeqr, the repeated
execution timings underestimates the in-algorithm timings for the first
1000~kernel invocations.  More surprisingly however, \refdgemmNT is even
overestimated---it is faster within \dgeqrf.

The change around the 1000th~kernel invocation in
\cref{fig:cache:dgeqrf:repeat-i} is directly linked to the cache:  While
traversing the matrix, \dgeqrf only operates on \dm A's bottom-right quadrant,
which becomes smaller in step, and beyond invocation 1000 fits in the L2~cache.
As a result,  the subsequent runtime measurements of repeated executions show
only minimal differences with respect to the in-algorithm timings.  This
confirms caching as the cause of the discrepancies.

To better understand the scope of this influence we manipulate the cache
locality of the kernel operands in our isolated executions.  For this purpose,
we \definition[cache assumption:\\fully associative LRU]{assume} a simplified
cache replacement policy: a {\em fully associative Least Recently Used} (LRU)
algorithm.  We consider the two extreme scenarios in which the operands
immediately required by the kernels are either entirely in the L2~cache or not
cached at all.  These in- and out-of-cache scenarios serve as, respectively,
lower and upper bounds on the in-algorithm timings.

\begin{figure}\figurestyle
    \ref*{leg:dgeqrf}

    \begin{subfigaxis}[
            fig caption=In-cache,
            dgeqrf=ic-i,
            ymin=-35, ymax=5
        ]
    \end{subfigaxis}\hfill
    \begin{subfigaxis}[
            fig caption=Out-of-cache,
            dgeqrf=ooc-i,
            ymin=-5, ymax=100
        ]
    \end{subfigaxis}

    \mycaption{
        Error of in- and out-of-cache timings with respect to in-algorithm
        timings for \dgeqrf.
        \iftableoflist\else%
            The out-of-cache errors for \refdcopy are around~\SI{1000}\percent.
        \fi
        \captiondetails{$n = 1568$, $b = 32$, \harpertown, 1~thread, \openblas,
        median of 100~repetitions}
    }
    \label{fig:cache:i-ic_oc}
\end{figure}

For kernels with operands smaller than \SI6{\mebi\byte}, repeated execution
suffices to guarantee an in-cache setup.  By contrast, when the aggregate size
of the operands exceeds \SI6{\mebi\byte} (as for \refdgemmNT), different kernel
implementations may initially access different memory portions.  An ideal
in-cache setup would place exactly these immediately accessed portions in cache.
However, since we do not assume knowledge of kernel implementation, we restrict
our in-cache setup to fulfill the reasonable assumption that input-only operands
are accessed before input/output and output-only operands.  In order to prepare
the cache accordingly, we load\footnote{%
    Through a simple update to each data element, e.g., $x \coloneqq x +
    \varepsilon$.
} all input-only operands into the cache just before the kernel invocation.
\Cref{fig:cache:dgeqrf:ic-i} compares the such obtained \definition{in-cache
timings} to the in-algorithm timings:  The estimates are in all cases equal to
or underestimating the in-algorithm timings;\footnote{%
    To be precise, the largest overestimation is~\SI{.06}\percent.
} the error is~\SI{4.44}\percent.

To ensure that the operands are not in the cache, it suffices to access a
main memory section larger than the cache size.  \Cref{fig:cache:dgeqrf:ooc-i}
compares this setup's \definition{out-of-cache timings} to the in-algorithm
timings:  Almost all estimates are equal to or overestimating the in-algorithm
timings; the error is~\SI{29.14}\percent.

Not only do the established in- and out-of-cache timings indeed serve as lower
and upper bounds on the in-algorithm timings; for most kernel invocations one of
these two bounds is actually attained (see
\cref{fig:cache:dgeqrf:ic-i,fig:cache:dgeqrf:ooc-i}).  Based on this
observation, the next section introduces a cache model to combine these in- and
out-of-core timings to estimate the in-algorithm timings.

        \subsection{Modeling the Cache}
        \label{sec:cache:qr:cache}
        To predict the state of the cache throughout the execution of \dgeqrf, we
consider which parts of~\dm A and~\dm[width=.3]W are accessed by each kernel
invocation.  We examine the sequence of kernel invocations within \dgeqrf (see
\cref{alg:dgeqrf}), but, due to the lack of information on the implementations
of these kernels, make no assumptions on the patterns in which the kernels
access their operands.

For the assumed fully associative LRU cache replacement policy, identifying if a
kernel operand is in cache boils down to counting how many other data elements
were accessed since its last use.  To determine this count---henceforth referred
to as \definition{access distance}---we scan the sequence of kernel invocations
and keep a history of the memory regions they access.\footnote{%
    The length of this history is restricted to the number of kernel calls
    per iteration of the blocked algorithm.
}  (Note that for our purposes cache lines are the smallest accessible units of
memory:  An access to a single data element means an access to the entire
surrounding cache line.)  For each operand, we go backward through the access
history until (and including) we find its last occurrence; thereby summing the
sizes of the encountered memory regions yields the operand's access distance.
If a previous access is not found, the access distance is set to the total size
of~\dm A and~\dm[width=.3]W.\footnote{%
    This corresponds to the scenario where the entire QR~decomposition is
    repeatedly executed on the same data.
}

\begin{figure}\figurestyle
    \ref*{leg:dgeqrf}

    \begin{subfigaxis}[
            fig caption=Initial estimates,
            dgeqrf=weighted-i,
            ymin=-20, ymax=45
        ]
    \end{subfigaxis}\hfill
    \begin{subfigaxis}[
            fig caption=Splitting estimates,
            dgeqrf=split-i,
            ymin=-20, ymax=45
        ]
    \end{subfigaxis}

    \mycaption{%
        Error of our initial and splitting estimates with respect to
        in-algorithm timings for \dgeqrf.
        \captiondetails{$n = 1568$, $b = 32$, \harpertown, 1~thread, \openblas,
        median of 100~repetitions}
    }
    \label{fig:cache:i-weighted_split}
\end{figure}

By comparing the obtained access distances to the cache size, we determine
whether the corresponding operand is expected to be in the cache or not.  Given
these expectations, we separately sum the sizes of the in- and out-of-cache
operands to, respectively, \sic and~\soc.  These sums are then used to weight
the runtime of the corresponding timings~\tic and~\toc to yield
\definition{initial estimates} of the in-algorithm timings:
\begin{equation}\label{eqn:cache:weight}
    t_\mathrm{est} \coloneqq \frac{\sic \tic + \soc \toc}{\sic + \soc} \enspace.
\end{equation}
Comparing these estimates in \cref{fig:cache:dgeqrf:weighted-i} to
\cref{fig:cache:dgeqrf:ic-i,fig:cache:dgeqrf:ooc-i}, we find that our mechanism
chooses (or weights) the in-cache and out-of-cache timings correctly for most
kernels.  However, the error is~\SI{4.61}\percent, because for \refdtrmmRUNN
out-of-cache is erroneously favored over in-cache.

The reason for this flaw is that (see \cref{alg:dgeqrf}) \refdtrmmRUNN is
preceded by the large \refdgemmTN:  This \dgemm's operands, which are together
larger than the cache, are accumulated into the access distance of
\dtrmm[RUNN]'s operand~\dm[mat20, width=.5, height=1.25]{W_2}.  However, since
\dm[mat20, width=.5, height=1.25]{W_2} happens to be the output of the
matrix-times-vector-shaped \dgemm[TN], it appears to be left in cache.  We use
this insight to extend our cache model with a crucial assumption:  After a
kernel whose (input/)output operand is significantly smaller than its
input-only operands we expect the (input/)output operand to remain in cache.
This assumption is implemented by splitting the memory accesses of such a kernel
into two parts:  The first part contains the large input-only operand(s), while
the second only involves the small (input/)output operand.  Therefore, the
back-traversal of the access history encounters the latter separately, and,
in case it is the sought operand, terminates before processing the
cache-exceeding accesses.  The runtime estimates from this
modifications---called \definition{splitting estimates}---are evaluated in
\cref{fig:cache:dgeqrf:split-i}:  All kernels are chosen correctly from the
in-cache and out-of-cache timings; as a result, the error is reduced
to~\SI{2.24}\percent.

\begin{figure}\figurestyle
    \hfill\ref*{leg:dgeqrf}\hspace{2pt}

    \begin{subfigaxis}[
            fig caption=Smoothing function,
            fig label=smooth,
            twocolplot,
            xmin=-1.5, xmax=1,
            xlabel=$r$\vphantom f,
            ymin={},
            ytick={-1, 0, 1}, yticklabels={$-1$, $0$, $+1$},
            samples=200,
            legend pos=north west,
            legend columns=1
        ]
        \addplot[plot1, domain=-2:1] {-1 + 2 * (x > 0)};
        \addlegendentry{$\operatorname{sgn}(r)$}

        \addlegendimage{plot2};\addlegendentry{$\operatorname{tanh}(4 r)$}
        \addlegendimage{plot3};\addlegendentry{$\operatorname{tanh}(2 r)$}

        \addplot[plot2, domain=-2:1, dashed] {tanh(4 * x)};
        \addplot[plot3, domain=-2:1, dashed] {tanh(2 * x)};
        \addplot[plot2, domain=0:1]  {tanh(4 * x)};
        \addplot[plot3, domain=-2:0] {tanh(2 * x)};
    \end{subfigaxis}\hfill
    \begin{subfigaxis}[
            fig caption=Final estimates,
            dgeqrf=smooth-i,
            ymin=-15, ymax=10
        ]
    \end{subfigaxis}

    \mycaption{%
        Smoothing function and error of final estimates with respect to
        in-algorithm timings for \dgeqrf.
        \captiondetails{$n = 1568$, $b = 32$, \harpertown, 1~thread, \openblas,
        median of 100~repetitions}
    }
    \label{fig:cache:smooth_i-smooth}
\end{figure}

The only remaining deficiency of our estimates is the cluster of spikes around
the transition from out-of-cache to in-cache around the 900th~kernel invocation.
To avoid such spikes, we ``smooth'' the association of operands to in- and
out-of-cache.  To determine whether an operator~\op is in-cache~($+1$) or
out-of-cache~($-1$), we previously used a step function.  In terms of the
relative access distance
\[
    r_\op = \frac{
        (\text{cache size}) - (\text{access distance})_{\mathcal Op}
    }{\text{cache size}} \enspace,
\]
this was the sign function:  Based on the operand sizes~$s_\op$ the weights for
our estimates (\cref{eqn:cache:weight}) were computed as
\[
    \sic \coloneqq \sum_\op \frac{1 + \operatorname{sgn}(r_\op)}2 s_\op
    \quad\text{and}\quad
    \soc \coloneqq \sum_\op \frac{1 - \operatorname{sgn}(r_\op)}2 s_\op
    \enspace.
\]
We now replace the association function with
\[
    f(r) = \begin{cases}
        \operatorname{tanh}(\alpha r) &\text{for } r \geq 0\\
        \operatorname{tanh}(\beta  r) &\text{for } r < 0
    \end{cases} \enspace,
\]
where $\alpha$ and~$\beta$ are smoothing coefficients.  As shown in
\cref{fig:cache:smooth}, $f(r)$ converges toward $\operatorname{sgn}(r)$ for
both large and small values of~$r$, and exhibits a smooth transition from~$-1$
to~$+1$ through the origin.  When applied to our estimates with empirical values
of $\alpha = 4$ and~$\beta = 2$, we obtain the \definition{final estimates}
evaluated in \cref{fig:cache:dgeqrf:smooth-i}.  With all estimates close to the
instrumentation timings, the error further decreases to~\SI{1.80}\percent.

        \subsection{Varying the Setup}
        \label{sec:cache:qr:res}
        \begin{table}\tablestyle
    \begin{tabular}{ccccS[table-space-text-post=\,\percent]S[table-space-text-post=\,\percent]c}
        \toprule
                   &                &           &                   &{repeated}         &{final} \\
        \multirow{-2}*{\#cores}
                    &\multirow{-2}*{\blas}
                                    &\multirow{-2}*{$n$}        &
                                                \multirow{-2}*{$b$}
                                                                    &{execution}        &{estimates}        &\multirow{-2}*{improvement} \\
        \midrule
        1           &\openblas      &1568       &32                 &4.42\,\percent     &1.80\,\percent     &$2.46\times$ \\
        1           &\openblas      &1568       &\bf 64             &3.15\,\percent     &1.64\,\percent     &$1.92\times$ \\
        1           &\openblas      &1568       &\bf 128\hphantom1  &2.68\,\percent     &2.13\,\percent     &$1.26\times$ \\
        1           &\openblas      &\bf 2080   &32                 &5.11\,\percent     &1.84\,\percent     &$2.78\times$ \\
        1           &\openblas      &\bf 2400   &32                 &5.23\,\percent     &1.75\,\percent     &$2.99\times$ \\
        1           &\bf \atlas     &1568       &32                 &3.55\,\percent     &1.98\,\percent     &$1.79\times$ \\
        1           &\bf \atlas     &\bf 2400   &32                 &4.22\,\percent     &2.51\,\percent     &$1.68\times$ \\
        1           &\bf \mkl       &1568       &32                 &8.58\,\percent     &4.40\,\percent     &$1.95\times$ \\
        1           &\bf \mkl       &\bf 2400   &32                 &9.58\,\percent     &6.22\,\percent     &$1.54\times$ \\
        1           &\bf reference  &1568       &32                 &2.31\,\percent     &1.54\,\percent     &$1.50\times$ \\
        \bf 2       &\openblas      &1568       &32                 &9.58\,\percent     &4.63\,\percent     &$2.07\times$ \\
        \bf 4       &\openblas      &1568       &32                 &22.71\,\percent    &19.75\,\percent    &$1.15\times$ \\
        \bottomrule
    \end{tabular}
    \mycaption{%
        Estimation errors and improvements through cache-modeling for \dgeqrf.
        \captiondetails{\harpertown}
    }
    \label{tab:cache:compqr}
\end{table}

In the previous sections we focused on one specific setup for the
QR~decomposition \dgeqrf on a \harpertown:  We factorized a square matrix of
size~$n = 1568$ with block size~$b = 32$ using single-threaded \openblas.  To
demonstrate that our observations and models are more broadly applicable, we now
vary this setup:  For a range of scenarios \cref{tab:cache:compqr} presents the
improvements of our final estimates (e.g., \cref{fig:cache:dgeqrf:smooth-i})
over the repeated execution timings (e.g., \cref{fig:cache:dgeqrf:repeat-i}).

Although the error of our estimates remains above~\SI{1.5}\percent, they are in
many cases an improvement of about~$2\times$ over the repeated execution
timings.  For both increasing block size~$b$ and problem size~$n$ the accuracy
of the repeated executions timings varies, but our estimates reliably yield an
error of around~\SI2\percent.\footnote{%
    Since for larger block sizes the arithmetic intensity of the kernels
    increases, caching plays a smaller role and the repeated execution estimates
    become more accurate on their own.
} Changing the \blas implementation, we can appreciate that with \atlas the
results are much the same as with \openblas.  While with \mkl the error
in both the repeated execution timings and our estimates instead increases
significantly, the estimates are still a good improvement of the repeated
execution timings.  Even for the reference \blas implementation our estimates
improve the already low error further by a factor of~1.5.  When doubling the
number of cores to~2, the errors increase, but our estimates
still provide a $2\times$~improvement over the repeated execution timings.  When
we use all 4~of our processor's cores however, the error increases
drastically---mainly because, while our model is designed for a single
last-level cache, every two cores of the \harpertownshort share a separate
L2~cache.  To account for multiple last-level caches, would require detailed
knowledge of the \blas implementation and thus substantial changes in our
models.

    \section{Application to Other Algorithms}
    \label{sec:cache:algs}
    After studying \lapack's QR~decomposition in great depth, we now consider two
other blocked \lapack algorithms: the upper-triangular Cholesky decomposition
\dpotrf[U] (\cref{sec:cache:dpotrfU}) and the inversion of a lower-triangular
matrix \dtrtri[LN] (\cref{sec:cache:dtrtriLN}).

\subsection{Cholesky Decomposition: \texorpdfstring{\dpotrf[U]}{dpotrf}}
\label{sec:cache:dpotrfU}

\begin{figure}\figurestyle
    \begin{minipage}\subfigwidth
        \begin{subfigure}\textwidth\centering
            \begin{tikzpicture}[y={(0, -1)}, scale=.7]
                \def\s{5.5} \def\p{2} \def\q{3}
                \filldraw[mat  ] (0,  0 ) rectangle (\s,     \s    );
                \filldraw[mat  ] (0,  0 ) --        (\p-.05, \p-.05) node {$A_{00}$} |- cycle;
                \filldraw[mat01] (\p, 0 ) rectangle (\q-.05, \p-.05) node {$A_{01}$};
                \filldraw[mat02] (\q, 0 ) rectangle (\s,     \p-.05) node {$A_{02}$};
                \filldraw[mat11] (\p, \p) --        (\q-.05, \q-.05) node {$A_{11}$} |- cycle;
                \filldraw[mat12] (\q, \p) rectangle (\s,     \q-.05) node {$A_{12}$};
                \filldraw[mat  ] (\q, \q) --        (\s,     \s    ) node {$A_{22}$} |- cycle;
                \draw[bracei] (\p, \q-.05) -- (\q-.05, \q-.05) node {$b$}; 
                \draw[bracei] (\p, \p    ) -- (\p,     \q-.05) node {$b$}; 
                \draw[bracei] (0,  \s    ) -- (\s,     \s    ) node {$n$}; 
                \draw[bracei] (0,  0     ) -- (0,      \s    ) node {$n$}; 
            \end{tikzpicture}
            \caption{Matrix partitioning}
        \end{subfigure}

        \medskip

        \begin{subfigure}\textwidth
            \makeatletter\let\tikz@ensure@dollar@catcode=\relax\makeatother
            \newcommand\Azo{\dm[mat01, width=.5]{A_{01}}}
            \newcommand\AzoT{\dm[mat01, height=.5, ']{A_{01}}}
            \newcommand\Azt{\dm[mat02, width=1.25, height=1]{A_{02}}}
            \newcommand\Aoo{\dm[mat11, size=.5, upper]{A_{11}}}
            \newcommand\AooT{\dm[mat11, size=.5, lower, ']{A_{11}}}
            \newcommand\AooiT{\dm[mat11, size=.5, lower, inv']{A_{11}}}
            \newcommand\Aoofull{\dm[mat11, size=.5]{A_{11}}}
            \newcommand\Aot{\dm[mat12, width=1.25, height=.5]{A_{12}}}
            \newcommand\traversal{%
                \begin{tikzpicture}[
                        baseline=(x.base),
                        y={(0, -1)}, scale=.1
                    ]
                    \def\s{5.5} \def\p{2} \def\q{3}
                    \filldraw[mat  ] (0,  0 ) rectangle (\s, \s);
                    \filldraw[mat  ] (0,  0 ) --        (\p, \p) |- cycle;
                    \filldraw[mat01] (\p, 0 ) rectangle (\q, \p);
                    \filldraw[mat02] (\q, 0 ) rectangle (\s, \p);
                    \filldraw[mat11] (\p, \p) --        (\q, \q) |- cycle;
                    \filldraw[mat12] (\q, \p) rectangle (\s, \q);
                    \filldraw[mat  ] (\q, \q) --        (\s, \s) |- cycle;
                    \path (0, 0) -- (\s, \s) node[midway, black] (x) {$A$};
                \end{tikzpicture}%
            }
            \begin{alglisting}[]
                traverse $\traversal$ along $\tsearrow$:
                  !\llap{\ref*{plt:dsyrkUT} }\dsyrk[UT]:! $\Aoo \coloneqq \Aoo - \AzoT \Azo$
                  !\llap{\ref*{plt:dpotf2U} }\dpotf[U]2:! $\AooT \Aoo \coloneqq \Aoofull$
                  !\llap{\ref*{plt:dgemmTN} }\dgemm[TN]:! $\Aot \coloneqq \Aot - \AzoT \Azt$
                  !\llap{\ref*{plt:dtrsmLUTN} }\dtrsm[LUTN]:! $\Aot \coloneqq \AooiT \Aot$
            \end{alglisting}
            \caption{Algorithm}
        \end{subfigure}
    \end{minipage}\hfill
    \begin{subfigaxis}[
            fig caption=In-algorithm timings,
            fig label=dpotrfU:instr,
            fig legend=dpotrfU:2col,
            dpotrfU=i,
            ymax=17,
            ylabel=time, y unit=\si{\ms},
            legend to name=leg:dpotrfU:2col,
            legend columns=2,
        ]
        \addlegendimage{dsyrkUT, mark size=3pt, very thick}
        \addlegendentry{\dsyrk[UT]}\label{plt:dsyrkUT}
        \addlegendimage{dpotf2U, mark size=3pt, very thick}
        \addlegendentry{\dpotf[U]2}\label{plt:dpotf2U}
        \addlegendimage{dgemmTN, mark size=3pt, very thick}
        \addlegendentry{\dgemm[TN]}
        \addlegendimage{dtrsmLUTN, mark size=3pt, very thick}
        \addlegendentry{\dtrsm[LUTN]}\label{plt:dtrsmLUTN}
    \end{subfigaxis}

    \mycaption{%
        \lapack's blocked algorithm for the upper-triangular Cholesky
        decomposition \dpotrf[U] and in-algorithm timings.
        \captiondetails{$n = 2400$, $b = 32$, \harpertown, 1~thread, \openblas,
        100~repetitions}
    }
    \label{alg:dpotrfU}
\end{figure}

First, we consider \lapack's upper triangular Cholesky decomposition \dpotrf[U]
\[
    \dm[lower, ']U \dm[upper]U \coloneqq \dm A
\]
of a symmetric positive definite $\dm A \in \R^{n \times n}$ in upper triangular
storage.  \Cref{alg:dpotrfU} presents the blocked algorithm employed in this
routine, which is the transpose of \dpotrf's algorithm for lower-triangular
case (\cref{alg:chol2} on \cpageref{algs:chol}).  As the algorithm traverses
\dm A, both the size and shape of~\dm[mat02, width=1.25]{A_{02}} (the largest
operand) change noticeably:  It starts as row panel, then grows to a square
matrix and finally shrinks to a column panel.  \dm[mat02, width=1.25]{A_{02}}'s
size determines the workload performed by the algorithm's large \refdgemmTN,
which is reflected in the in-algorithm timings in
\cref{fig:cache:dpotrfU:instr}.

\begin{figure}\figurestyle
    \ref*{leg:dpotrfU}

    \pgfplotsset{ymin=-30, ymax=30}

    \begin{subfigaxis}[
            fig caption=Repeated execution,
            fig label=dpotrf:repeat,
            legend to name=leg:dpotrfU,
            dpotrfU=repeat-i,
        ]
        \addlegendimage{dsyrkUT, mark size=3pt, very thick}
        \addlegendentry{\dsyrk[UT]}
        \addlegendimage{dpotf2U, mark size=3pt, very thick}
        \addlegendentry{\dpotf[U]2}
        \addlegendimage{dgemmTN, mark size=3pt, very thick}
        \addlegendentry{\dgemm[TN]}
        \addlegendimage{dtrsmLUTN, mark size=3pt, very thick}
        \addlegendentry{\dtrsm[LUTN]}
    \end{subfigaxis}\hfill
    \begin{subfigaxis}[
            fig caption=Smoothed estimates,
            fig label=dpotrf:pred,
            dpotrfU=smooth-i,
        ]
    \end{subfigaxis}

    \mycaption{%
        Error of our final estimates with respect to in-algorithm timings for
        the Cholesky decomposition \dpotrf[U].
        \captiondetails{$n = 2400$, $b = 32$, \harpertown, 1~thread, \openblas,
        median of 100~repetitions}
    }
    \label{fig:cache:dpotrf:res}
\end{figure}

In our experiments, we execute \dpotrf[U] on a \harpertown with single-threaded
\openblas, $\dm A \in \R^{2400 \times 2400}$,\footnote{%
    For $n = 2400$, the upper-triangular portion of~$A$ takes up about
    \SI{12}{\mebi\byte}---twice the size of the L2~cache.
} and block size $b = 32$.  \Cref{fig:cache:dpotrf:res} presents the relative
performance difference with respect to in-algorithm timings for both repeated
execution timings and our final estimates.  Our estimates yield improvements for
the \refdsyrkUT and \refdpotfU involving large matrices in the middle of \dm A's
traversal.  In the beginning of the traversal, the estimates are generally too
pessimistic because some matrices are (partially) brought into cache by
prefetching, which is not accounted for in our estimates.  On average the
relative error is reduced from~\SIrange{11.11}{7.87}{\percent}, i.e., by a
factor of~1.41.

However, note that the improvement is only visible in the averaged per-kernel
relative error:  Since the runtime of large \dgemm[TN]~(\ref*{plt:dgemmTN}) is
overestimated, the accumulated runtime estimate for the entire algorithm
actually becomes less accurate.

\subsection{Inversion of a Triangular Matrix:
\texorpdfstring{\dtrtri[LN]}{dtrtri}}
\label{sec:cache:dtrtriLN}

\begin{figure}\figurestyle
    \begin{minipage}\subfigwidth
        \begin{subfigure}\textwidth\centering
            \begin{tikzpicture}[y={(0, -1)}, scale=.7]
                \def\s{5.5} \def\p{2} \def\q{3}
                \filldraw[mat  ] (0,  0 ) rectangle (\s,     \s    );
                \filldraw[mat  ] (0,  0 ) --        (\p-.05, \p-.05) node {$A_{00}$} -| cycle;
                \filldraw[mat  ] (0,  \p) rectangle (\p-.05, \q-.05) node {$A_{10}$};
                \filldraw[mat11] (\p, \p) --        (\q-.05, \q-.05) node {$A_{11}$} -| cycle;
                \filldraw[mat  ] (0,  \q) rectangle (\p-.05, \s    ) node {$A_{20}$};
                \filldraw[mat21] (\p, \q) rectangle (\q-.05, \s    ) node {$A_{21}$};
                \filldraw[mat22] (\q, \q) --        (\s,     \s    ) node {$A_{22}$} -| cycle;
                \draw[brace ] (\p,     \p) -- (\q-.05, \p    ) node {$b$}; 
                \draw[brace ] (\q-.05, \p) -- (\q-.05, \q-.05) node {$b$}; 
                \draw[bracei] (0,      \s) -- (\s,     \s    ) node {$n$}; 
                \draw[bracei] (0,      0 ) -- (0,      \s    ) node {$n$}; 
            \end{tikzpicture}
            \caption{Matrix partitioning}
        \end{subfigure}

        \medskip

        \begin{subfigure}\textwidth
            \makeatletter\let\tikz@ensure@dollar@catcode=\relax\makeatother
            \newcommand\Aoo{\dm[mat11, size=.5, lower]{A_{11}}}
            \newcommand\Aooi{\dm[mat11, size=.5, lower, inv]{A_{11}}}
            \newcommand\Ato{\dm[mat21, width=.5, height=1.25]{A_{21}}}
            \newcommand\Att{\dm[mat22, width=1.25, height=1.25, lower]{A_{22}}}
            \newcommand\traversal{
                \begin{tikzpicture}[
                        baseline=(x.base),
                        y={(0, -1)}, scale=.1
                    ]
                    \def\s{5.5} \def\p{2} \def\q{3}
                    \filldraw[mat  ] (0,  0 ) rectangle (\s, \s);
                    \filldraw[mat  ] (0,  0 ) --        (\p, \p) -| cycle;
                    \filldraw[mat  ] (0,  \p) rectangle (\p, \q);
                    \filldraw[mat11] (\p, \p) --        (\q, \q) -| cycle;
                    \filldraw[mat  ] (0,  \q) rectangle (\p, \s);
                    \filldraw[mat21] (\p, \q) rectangle (\q, \s);
                    \filldraw[mat22] (\q, \q) --        (\s, \s) -| cycle;
                    \path (0, 0) -- (\s, \s) node[midway, black] (x) {$A$};
                \end{tikzpicture}
            }
            \begin{alglisting}[]
                traverse $\traversal$ along $\tnwarrow$:
                  !\llap{\ref*{plt:dtrmmLLNN} }\dtrmm[LLNN]:! $\Ato \coloneqq \Att \Ato$
                  !\llap{\ref*{plt:dtrsmRLNN} }\dtrsm[RLNN]:! $\Ato \coloneqq -\Ato \Aooi$
                  !\llap{\ref*{plt:dtrti2LN} }\dtrti[LN]2:! $\Aoo \coloneqq \Aooi$
            \end{alglisting}
            \caption{Algorithm}
        \end{subfigure}
    \end{minipage}\hfill
    \begin{subfigaxis}[
            fig caption=In-algorithm timings,
            fig label=dtrtriLN:instr,
            fig legend=dtrtriLN,
            dtrtriLN=i,
            ymax=35,
            ylabel=time, y unit=\ms,
            legend to name=leg:dtrtriLN,
            table/search path={cache/figures/data/trinv/},
        ]
        \addlegendimage{dtrmmLLNN, mark size=3pt, very thick}
        \addlegendentry{\dtrmm[LLNN]}\label{plt:dtrmmLLNN}
        \addlegendimage{dtrsmRLNN, mark size=3pt, very thick}
        \addlegendentry{\dtrsm[RLNN]}\label{plt:dtrsmRLNN}
        \addlegendimage{dtrti2LN, mark size=3pt, very thick}
        \addlegendentry{\dtrti[LN]2}\label{plt:dtrti2LN}
    \end{subfigaxis}

    \mycaption{%
        \lapack's blocked algorithm for the inversion of a lower-triangular
        matrix \dtrtri[LN] and in-algorithms timings.
        \captiondetails{$n = 2400$, $b = 32$, \harpertown, 1~thread, \openblas,
        100~repetitions}
    }
    \label{alg:dtrtriLN2}
\end{figure}

We now take a closer look at \lapack's inversion of a lower-triangular matrix
\dtrtri[LN]
\[
    \dm[lower]A \coloneqq \dm[lower, inv]A
\]
with $\dm A \in \R^{n \times n}$, whose blocked algorithm is presented in
\cref{alg:dtrtriLN2}.  In contrast to the previous operations, this algorithm
traverses \dm A \tnwarrow from the bottom-right to the top-left, thereby
operating on sub-matrices of increasing size.  \Cref{fig:cache:dtrtriLN:instr}
shows the in-algorithm timings for the algorithm, which are dominated by
\refdtrmmLLNN.

\begin{figure}\figurestyle
    \ref*{leg:dtrtriLN}

    \pgfplotsset{
        ymin=-40, ymax=10,
        table/search path={cache/figures/data/trinv/},
    }

    \begin{subfigaxis}[
            fig caption=Repeated execution,
            fig label=dtrtriLN:repeat,
            dtrtriLN=repeat-i,
        ]
    \end{subfigaxis}\hfill
    \begin{subfigaxis}[
            fig caption=Smoothed estimates,
            fig label=dtrtriLN:pred,
            dtrtriLN=smooth-i,
        ]
    \end{subfigaxis}

    \mycaption{%
        Error of our final estimates with respect to in-algorithm timings for
        the inversion of a lower-triangular matrix \dtrtri[LN].
        \captiondetails{$n = 2400$, $b = 32$, \harpertown, 1~thread, \openblas,
        median of 100~repetitions}
    }
    \label{fig:cache:dtrtriLN:res}
\end{figure}

We execute \dtrtri[LN] on a \harpertown with single-threaded \openblas, $\dm A
\in \R^{2400 \times 2400}$, and block size $b = 32$.
\Cref{fig:cache:dtrtriLN:res} compares the performance measurements from
repeated execution and our final estimates to in-algorithm timings:  The
improvements of our estimates are most significant in \refdtrmmLLNN (which
performs the most computation) and \refdtrtiLN; the error is reduced from an
average of~\SIrange{6.70}{3.37}{\percent}---a total improvement of~$1.99\times$.

\subsection{Summary}

We have seen that, on a \harpertown the accuracy of our runtime estimates for
kernels within blocked algorithms is increased by taking the state of the
L2~cache throughout the algorithm execution into consideration.  For different
algorithms, problem sizes, block sizes, \blas implementations, and thread
counts, we have seen improvements between~$1.15\times$ (with all 4~cores)
and~$2.99\times$.

    \section{Feasibility on Modern Hardware}
    \label{sec:cache:new}
    The analysis and cache model in the previous two sections focused on a
\harpertown{}---a fairly old processor released in~2007.  In this section, we
study how well the same approach is applicable to more recent processors, namely
a \sandybridge and a \haswell.

The study reveals that on these systems it is especially challenging to
establishing in- and out-of-cache timings as lower and upper bounds for the
in-algorithm timings (\cref{sec:cache:icoc}).  We present evidence that, while
we can indeed estimate the in-algorithm timings, this is only possible by
replicating the execution context within the algorithms, which is infeasible in
the context of algorithm-independent performance models
(\cref{sec:cache:algaware}).

\subsection{In- and Out-of-Cache Timings}
\label{sec:cache:icoc}

\begin{figure}\figurestyle
    \pgfplotsset{
        every subfigaxis/.style={fig vertical align=b},
    }

    \begin{subfigaxis}[
            fig caption={
                Triangular inverstion \dtrtri[LN]
                \captiondetails{$n = 3200$, $b = 64$, \haswell}
            },
            fig label=ooc:dtrtri:haswell,
            fig legend=dtrtriLN,
            dtrtriLN=setupo-instr,
            table/search path={cache/figures/data/new/trinv.3200.64.Haswell.1.OpenBLAS},
        ]
    \end{subfigaxis}\hfill
    \begin{subfigaxis}[
            fig caption={
                QR~decomposition \dgeqrf
                \captiondetails{$n = 2400$, $b = 32$, \sandybridge}
            },
            fig legend=dgeqrf2col,
            dgeqrf=setupo-instr,
            fig label=ooc:dgeqrf:sandybridge,
            ymax=175,
            table/search path={cache/figures/data/new/qr.2400.32.SandyBridge.1.OpenBLAS},
        ]
    \end{subfigaxis}

    \mycaption{%
        Error of out-of-cache timings with respect to in-algorithm timings for
        \dtrtri[LN] and \dgetrf.
        \captiondetails{1~thread, \openblas, median of 100~repetitions}
    }
    \label{fig:cache:ooc}
\end{figure}

Out-of-core timings are hardware independent, and just as on the
\harpertownshort serve as an upper bound on the \sandybridgeshort and
\haswellshort.  This is illustrated in \cref{fig:cache:ooc} for the inversion of
a lower-triangular matrix $\dm[lower]A \in \R^{3200 \times 3200}$ with
\dtrtri[LN] (\cref{alg:dtrtriLN2}) and block size~$b = 64$ on the \haswellshort,
and the QR~decomposition of $\dm A \in \R^{2400 \times 2400}$ with \dgeqrf
(\cref{alg:dgeqrf}) and $b = 32$  on the \sandybridgeshort{}---the chosen
matrices comprise around \SI{40}{\mebi\byte} and thus exceed the
\sandybridgeshort's and \haswellshort's last-level cache~(L3) of, respectively,
\SIlist{20,30}{\mebi\byte}.  The out-of-cache timings indeed consistently
overestimate the in-algorithm timings---by up to~\SI{347}{\percent} for the last
call to \refdtrmmRUNN in the QR~decomposition \dgeqrf on the \sandybridgeshort
(\cref{fig:cache:ooc:dgeqrf:sandybridge} is clipped at~\SI{175}\percent).  As
such, these measurements serve well as an upper bound on the in-algorithm
timings.

\begin{figure}[p]\figurestyle
    \ref*{leg:dtrtriLN}
    \pgfplotsset{
        dtrtriLN=setupi-instr,
        ymin=-30, ymax=5,
        plotheightsub=18.7pt,
    }

    \begin{subfigaxis}[
            fig caption={\dtrtri[LN] on the \sandybridgeshort},
            fig label=ic:dtrtri:sandybridge,
            table/search path={cache/figures/data/new/trinv.3200.64.SandyBridge.1.OpenBLAS},
        ]
    \end{subfigaxis}\hfill
    \begin{subfigaxis}[
            fig caption={\dtrtri[LN] on the \haswellshort},
            fig label=ic:dtrtri:haswell,
            table/search path={cache/figures/data/new/trinv.3200.64.Haswell.1.OpenBLAS},
        ]
    \end{subfigaxis}

    \medskip

    \ref*{leg:dgeqrf}
    \pgfplotsset{
        ymin=-20, ymax=5,
        every subfigaxis/.style={dgeqrf=setupi-instr}
    }

    \begin{subfigaxis}[
            fig caption=\dgeqrf on the \sandybridgeshort,
            fig label=ic:dgeqrf:sandybridge,
            table/search path={cache/figures/data/new/qr.2400.32.SandyBridge.1.OpenBLAS},
        ]
    \end{subfigaxis}\hfill
    \begin{subfigaxis}[
            fig caption=\dgeqrf on the \haswellshort,
            fig label=ic:dgeqrf:haswell,
            table/search path={cache/figures/data/new/qr.2400.32.Haswell.1.OpenBLAS},
        ]
    \end{subfigaxis}

    \mycaption{%
        Error for attempted in-cache timings with respect to in-algorithm
        timings for \dtrtri[LN] and \dgetrf.
        \captiondetails{
            \dtrtri[LN]: $n = 3200$, $b = 64$; \dgeqrf: $n = 2400$, $b = 32$;
            \sandybridge and \haswell, 1~thread, \openblas, 100~repetitions
        }
    }
    \label{fig:cache:ic}
\end{figure}

Fore the same scenarios \cref{fig:cache:ic} presents the error of our previous
in-cache setup with respect to the in-algorithm timings:  While we expect the
our setup to yield faster kernel executions than the in-algorithm timings, on
the \sandybridge (with \turboboost disabled) the in-cache timings are still up to~\SI{.51}{\percent} slower than
the in-algorithm timings (not accounting for the small unblocked \dgeqr2); on
the \haswell (with \turboboost enabled), the relative errors for \dtrtri[LN] and
\dgeqrf reach, respectively, \SIlist{1.67;3.44}\percent.

\begin{figure}\figurestyle
    \pgfplotsset{
        every subfigaxis/.style={fig vertical align=b}
    }
    \begin{subfigaxis}[
            fig caption={\dtrtri[LN] \captiondetails{$n = 3200$, $b = 64$}},
            fig label=ictb:dtrtri:sandybridge,
            fig legend=dtrtriLN,
            dtrtriLN=setupi-instr,
            ymin=-30, ymax=5,
            table/search path={cache/figures/data/new2/trinv.3200.64.SandyBridge.1.OpenBLAS},
        ]
    \end{subfigaxis}\hfill
    \begin{subfigaxis}[
            fig caption={\dgeqrf \captiondetails{$n = 2400$, $b = 32$}},
            dgeqrf=setupi-instr,
            fig label=ictb:dgeqrf:sandybridge,
            fig legend=dgeqrf2col,
            ymin=-20, ymax=5,
            table/search path={cache/figures/data/new2/qr.2400.32.SandyBridge.1.OpenBLAS},
            legend to name=leg:dgeqrf2col,
            legend columns=2,
        ]
        \addlegendimage{dgeqr2, mark size=3pt, very thick}
        \addlegendentry{\dgeqr2}
        \addlegendimage{dlarftFC, mark size=3pt, very thick}
        \addlegendentry{\dlarft[FC]}
        \addlegendimage{dcopy, mark size=.5pt, very thick}
        \addlegendentry{\dcopy}
        \addlegendimage{dtrmmRLNU, mark size=3pt, very thick}
        \addlegendentry{\dtrmm[RLNU]}
        \addlegendimage{dgemmTN, mark size=3pt, very thick}
        \addlegendentry{\dgemm[TN]}
        \addlegendimage{dtrmmRUNN, mark size=3pt, very thick}
        \addlegendentry{\dtrmm[RUNN]}
        \addlegendimage{dgemmNT, mark size=3pt, very thick}
        \addlegendentry{\dgemm[NT]}
        \addlegendimage{dtrmmRLTU, mark size=3pt, very thick}
        \addlegendentry{\dtrmm[RLTU]}
    \end{subfigaxis}

    \mycaption{%
        Error for attempted in-cache timings with respect to in-algorithm
        timings on a \sandybridge with \turboboost enabled.
        \captiondetails{1~thread, \openblas, median of 100~repetitions}
    }
    \label{fig:cache:ictb}
\end{figure}

Further investigation reveals that the processor's \intel{} \turboboost is a source of
complication for out measurements:  As \cref{fig:cache:ictb} shows, enabling
\turboboost on the \sandybridge leads to overestimations of the \dtrtri[LN]'s
and \dgeqrf's most compute-intensive operations (i.e., the \refdtrmmLLNN and the
two \dgemm{}s (\ref*{plt:dgemmTN}, \ref*{plt:dgemmNT})), by up to, respectively,
\SIlist{3.20;2.79}\percent.

While \turboboost increases the overestimation of individual kernels, this
phenomenon's origin lies in the processor's cache hierarchy:  Within an
algorithm, each kernel is invoked with a distinct cache precondition, i.e., with
only portions of its operands in the processor's caches.  Since our
algorithm-independent measurements do clearly not match such preconditions, we
attempted to construct conditions in which the kernel executes at its absolute
peak performance with different cache setups:
\begin{itemize}
    \item First, we used simple repeated execution of the kernel without any
        modification of the cache in between as before.

    \item Next, we accessed the kernel operands in various
        orders prior to the invocation.  E.g., for a \dgemm $\dm[width=.25]C
        \coloneqq \dm[width=.8]A \matvecsep \dm[width=.25, height=.8]B +
        \dm[width=.25]C$, we attempted all permutations of access orders, such
        as \dm[width=.8]A--\dm[width=.25, height=.8]B--\dm[width=.25]C and
        \dm[width=.25]C--\dm[width=.8]A--\dm[width=.25, height=.8]B.

    \item Finally, we refined the access granularity and attempted to bring
        operands into cache not as a whole but only partially:  For a kernel
        with one operand larger than the cache and the other operand(s) only a
        fraction of that size (e.g., the \dgemm[TN] (\ref*{plt:dgemmTN}) in
        \dgeqrf: $\dm[width=.25]C \coloneqq \dm[width=.8]A \matvecsep \dm
        [width=.25, height=.8]B + \dm[width=.25]C$ where \dm[width=.25]B and
        \dm[width=.25]C are of width~$b$ and close to the problem size~$n$ in
        height), we bring the entire small operand(s) into cache but only
        portions of the large one.

        \begin{figure}\figurestyle
    \tikzset{
        every picture/.append style={
            y={(0, -1)},
            scale=.7,
        },
        every node/.append style=midway
    }

    \begin{subfigure}{.3333\textwidth}\centering
        \begin{tikzpicture}
            \filldraw[mat  ] (0, 0) rectangle (5, 5) node {$A$};
            \filldraw[mat11] (0, 0) rectangle (1, 5);
            \draw[brace] (0, 0) -- (1, 0) node[above] {$s_1$};
            \draw[brace] (0, 5) -- (0, 0) node[left] {$\approx n$};
        \end{tikzpicture}
        \caption{Column panel}
    \end{subfigure}\hfill
    \begin{subfigure}{.3333\textwidth}\centering
        \begin{tikzpicture}
            \filldraw[mat  ] (0, 0) rectangle (5, 5) node[midway] {$A$};
            \filldraw[mat11] (0, 0) rectangle (5, 1);
            \draw[brace] (0, 0) -- (5, 0) node[above] {$\approx n\vphantom{_1}$};
            \draw[brace] (0, 1) -- (0, 0) node[left] {$s_2$};
        \end{tikzpicture}
        \caption{Row panel}
    \end{subfigure}\hfill
    \begin{subfigure}{.3333\textwidth}\centering
        \begin{tikzpicture}
            \filldraw[mat  ] (0, 0) rectangle (5, 5) node {$A$};
            \filldraw[mat11] (0, 0) rectangle (2, 2);
            \draw[brace] (0, 0) -- (2, 0) node[above] {$s_3$};
            \draw[brace] (0, 2) -- (0, 0) node[left] {$s_3$};
        \end{tikzpicture}
        \caption{Square block}
    \end{subfigure}

    \mycaption{%
        Basic operand regions accessed for attempted in-cache setups.
    }
    \label{fig:cache:acc}
\end{figure}

        \Cref{fig:cache:acc} presents which operand portions we chose to load
        into the cache.  These choices are based on the assumption that any
        kernel implementation likely traverses the input matrix somehow form
        from the top-left \tsearrow to the bottom-right.\footnote{%
            Exceptions are, e.g., \dtrsm[RLNN] ($B \coloneqq B A^{-1}$) and
            \dtrsm[LUNN] ($B \coloneqq A^{-1} B$), which must traverse the
            triangular~$A$ from the bottom-right to the top-left---in these
            cases the accessed matrix portions are mirrored accordingly.
        }  Therefore, we bring a column panel of the operand, a row panel, a
        square block, or any combination of these into the processor's caches.
        While doing so, we varied the sizes~$s_1$, $s_2$, and~$s_3$ of the
        accessed operand portions.
\end{itemize}

While in some scenarios changing the in-cache setup for kernel invocations
reduced the runtime overestimation, the effects were not consistent across
different algorithms, kernels, processors, and \blas implementations.
Altogether, it was not possible to determine general, algorithm-independent
in-cache setups that yield a clear lower bound on the in-algorithm timings.

\subsection{Algorithm-Aware Timings}
\label{sec:cache:algaware}

Since our above attempts at algorithm-independent in-cache timings did not yield
the required lower bound on in-algorithm timings, the only alternative is to
tailor the timing setups to individual algorithms.  We might for instance setup
each kernel timing with several preceding kernel invocations from within the
algorithms.  Such obtained \definition{algorithm-aware timings} yield accurate
estimates for the in-algorithm timings, and rid us of the need for combining in-
and out-of-cache estimates.

\begin{figure}\figurestyle
    \pgfplotsset{
        every subfigaxis/.style={fig vertical align=b},
        ymin=-5, ymax=5
    }
    \begin{subfigaxis}[
            fig caption={\dtrtri[LN] \captiondetails{$n = 3200$, $b = 64$}},
            fig label=exact:dtrtri,
            fig legend=dtrtriLN,
            dtrtriLN=exact3-instr,
            table/search path={cache/figures/data/exact/trinv.3200.64.SandyBridge.1.OpenBLAS},
        ]
    \end{subfigaxis}\hfill
    \begin{subfigaxis}[
            fig caption={\dgeqrf \captiondetails{$n = 2400$, $b = 32$}},
            dgeqrf=exact39-instr,
            fig label=exact:dgeqrf,
            fig legend=dgeqrf2col,
            restrict y to domain=-15:100,
            table/search path={cache/figures/data/exact/qr.2400.32.SandyBridge.1.OpenBLAS},
        ]
    \end{subfigaxis}

    \mycaption{%
        Error for algorithm-aware timings with respect to in-algorithm
        timings.
        \captiondetails{\sandybridge, 1~thread, \openblas, median of
        10~repetitions}
    }
    \label{fig:cache:exact}
\end{figure}

\begin{example}{Algorithm-aware timings}{cache:algaware}
    \Cref{fig:cache:exact} presents the accuracy of algorithm-aware timings as
    estimates for in-algorithm timings for the inversion of a lower-triangular
    matrix (\dtrtri[LN]) and the QR~decomposition (\dgeqrf) on a \sandybridge
    (with \turboboost enabled) using single-threaded \openblas.  The
    algorithm-aware timings were created by preceding each measured kernel
    invocation with the calls from the corresponding blocked algorithm that were
    executed since that kernel's last invocation.

    \Cref{fig:cache:exact:dtrtri} shows that for \dtrtri[LN] the algorithm-aware
    timings are with few exceptions within~\SI1{\percent} of the in-algorithm
    timings with an average absolute relative error (ARE) of~\SI{.54}\percent.
    As seen in \Cref{fig:cache:exact:dtrtri}, for the \dgetrf the relative error
    is overall larger yet similarly spread around~\SI0{\percent} with an average
    ARE of~\SI{.84}\percent.
\end{example}

While this approach yields accurate estimates, when the kernel invocations for
each algorithm execution are timed separately and each measurement is preceded
with a setup of one or more kernels, the timing procedure takes effectively
longer than executing and measuring the target algorithm repeatedly.  As a
result, this method is at the same time highly accurate and impractical, which
is why we do not further pursue it.

    \section{Summary}
    \label{sec:cache:conclusion}
    This chapter investigated the possibility of improving the accuracy of
performance predictions for blocked algorithms by accounting for caching
effects.  On a \harpertown, we were able to establish algorithm-independent in-
and out-of-cache kernel timings as, respectively, lower and upper bounds on
in-algorithm timings.  By tracking which (portions of) operands are in-cache
throughout an algorithm's execution, we were able to combine these timings into
more accurate runtime estimates than repeated execution timings.

This approach did not work equally well on more recent processors:  On a
\sandybridge and a \haswell, we concluded that constructing a cache precondition
to yield lower bounds on the in-algorithm timings was only attainable with
algorithm-aware measurements.  Since such measurements are not only incompatible
with our modeling approach but are also less efficient than straightforward
measurements of the target algorithm, we conclude that no efficient strategy to
improve the accuracy for our model-based predictions on modern hardware was
found.

}

    \chapter[Micro-Benchmarks for Tensor Contractions]
        {Micro-Benchmarks for\newline Tensor Contractions}
\chapterlabel{tensor}
{
    \pgfplotsset{
    predplt/.style={
        cycle list={
            {plot1, dashdotted},
            {plot2, dashdotted},
            {plot3, dashdotted},
            {plot4, dashdotted},
            {plot5, dashdotted},
            {plot6, dashdotted},
            {plot1, dashed},
            {plot2, dashed},
            {plot3, dashed},
            {plot4, dashed},
            {plot5, solid},
            {plot6, solid},
        },
        ylabel=performance,
        y unit=\si{\giga\flops\per\second},
        xlabel={tensor size $a = b = c$},
        table/search path={tensor/figures/data/pred/meas/}
    },
    plotdot/.style={plot1},
    plotaxpy/.style={plot2},
    plotgemv/.style={plot3},
    plotger/.style={plot4},
    plotgemm/.style={plot5},
    predplotpred/.style={
        fig caption=Predictions,
        fig label=pred:#1,
        twocolplot, predplt,
        ymax=6,
        before end axis/.code={
            \foreach \var in {25, ..., 36}
                \plot table {tensor/figures/data/pred/#1/var\var.min};
        }
    },
    predplotmeas/.style={
        fig caption=Measurements,
        fig label=pred:meas#1,
        twocolplot, predplt,
        ymax=6,
        before end axis/.code={
            \foreach \var in {25, ..., 36}
                \plot table {tensor/figures/data/pred/meas/var\var.min};
        }
    }
}

\lstdefinelanguage{tensoralgs}{
    keywords=[1]{for},
    keywordstyle=[1]{\bf},
}

\newcommand\tind[2]{\ensuremath{#1\text{\sf[#2]}}\xspace}
\newcommand\op{\ensuremath{\mathcal Op}\xspace}

    This chapter addresses the problem of accurately predicting the performance of
\blas-based algorithms for tensor contractions.  Since in practice, such
contractions are commonly used with skewed dimensions, the previously
developed performance models are unfortunately unsuitable:  For small matrices,
the performance of \blas kernels is quite irregular, and our models are less
accurate.  Furthermore, for small and skewed operations, caching effects can
play an immense role.  Hence, for tensor contractions, we follow a different
approach, and exploit that contraction algorithms are based on repeated
executions of a single kernel operation with fixed operand sizes:  We use
cache-aware micro-benchmarks that perform only a fraction of these executions in
a replica of the algorithm's executions environment, and extrapolate their
runtime to obtain performance predictions.

In the following, \cref{sec:tensor:alggen} discusses the systematic generation
of \blas-based algorithms for tensor contractions, \cref{sec:tensor:pred}
introduces our micro-benchmarks and performance predictions, and
\cref{sec:tensor:results} presents experimental results for a range of
contractions.

\paragraph{Publication}
The work presented in this chapter is based on research previously published in:
\begin{pubitemize}
    \pubitem{tensorpred}
\end{pubitemize}
\noindent In this collaboration, Diego Fabregat-Traver implemented the algorithm
generation presented in \cref{sec:tensor:alggen}, while this author developed
the performance predictions detailed in
\cref{sec:tensor:pred,sec:tensor:results}.

    \section{Algorithm Generation}
    \label{sec:tensor:alggen}
    Following a brief overview of tensor notation and storage, this section explains
the systematical generation of a family of \blas-based algorithms for a tensor
contraction.  For a detailed discussion of the topic, see~\cite{tensorgen}.

We express tensor contractions in Einstein notation:\footnote{%
    For the sake of simplicity and without any loss of generality, we ignore any
    distinction between covariant and contravariant vectors; this means we treat
    any index as a subscript.
} E.g., a matrix-matrix product $\dm C \coloneqq \dm A \matmatsep \dm B$ is
denoted by $C_{ab} \coloneqq A_{ai} B_{ib}$, meaning the entries of \dm C are
computed as $\tind C{a,b} \coloneqq \sum_\code i \tind A{a,i} \tind B{i,b}$. The
\definition[indices:]{indices} that appear in both tensors~$A$ and~$B$---the
summation indices $i, j, \ldots$---are called \definition{contracted}, while
those that only appear in either~$A$ or~$B$ (and thus in~$C$)---$a, b, c,
\ldots$---are called \definition[uncontracted / free]{uncontracted} or {\em
free}.  Without loss of generality, we assume that tensors are stored as
\fortran-style contiguous multidimensional double-precision arrays:  Vectors
(1D~tensors) are stored contiguously, matrices (2D~tensors) are stored as
sequences of column vectors, 3D~tensors (visualized as cubes) are stored as
sequences of matrices (planes of the cube), and so on.

Aware of the extreme level of efficiency inherent to optimized \blas
implementations, our approach for computing a contraction consists in reducing
it to a sequence of calls to one \blas kernel.  Since \blas operates on scalars,
vectors, and matrices (zero-, one- and, two-dimensional objects), tensors must
be expressed in terms of a collection of such objects.  To this end, we
introduce the concept of \definition{slicing}:  With the help of \matlab's
``\code:'' notation,\footnote{%
    In \matlab the index ``\code:'' in a tensor refers to all elements along
    that dimension, e.g., \tind C{:,b} is the \code b-th column of~$C$.
} slicing a $d$-dimensional operand $\mathcal Op \in \R^{n_1 \times n_2 \times
\cdots \times n_d}$ along the $i$-th index (or dimension) means creating the
$n_i$ $(d{-}1)$-dimensional slices 
\smash{\sf$\mathcal Op$[$%
    \underbrace{\text{:,}\ldots\text{,:}}_{i-1}%
    \text{,k,}%
    \underbrace{\text{:,}\ldots\text{,:}}_{d-i}%
$]}, 
where $\code k = 1, \ldots, n_i$.

\begin{example}{Contraction algorithm for \dgemm[NN]}{tensor:gemm}
    Consider the matrix-matrix product $C_{ab} \coloneqq A_{ai} B_{ib}$
    (\dgemm[NN]).  Slicing the matrix~\dm B along dimension~$b$ reduces it to a
    collection of column vectors~\tind B{:,b}; accordingly, the matrix-matrix
    product is reduced to a sequence of matrix-vector operations:\footnotemark

    \medskip

    \noindent
    \begin{alglisting}[width=.6\textwidth]
        for $\code b = 1\code :b$
          !\dgemv[N]:! $\tind C{:,b} \pluseqq \tind A{:,:} \tind B{:,b}$
    \end{alglisting}
    \hfill
    \begin{tikzpicture}[baseline]
        \begin{drawsquare}
            \sliceB
        \end{drawsquare}
        \node at (1, 0) {$\pluseqq$};
        \begin{drawsquare}[(2, 0)]
            \slicenone
        \end{drawsquare}
        \begin{drawsquare}[(3.5, 0)]
            \sliceB
        \end{drawsquare}
        \path (4, 0, -.5);
    \end{tikzpicture}

    \medskip
\end{example}
\footnotetext{%
    The pictogram next to the algorithm visualizes the slicing of the tensors
    that originates the algorithm's sequence of \dgemv[N]{}s.  The {\color{red}
    red} shapes represent the operands of the \blas kernel.
}

Depending on the slicing choices, a tensor contraction is reduced to a number of
\definition[nested loops\\+ one kernel]{nested loops} with one of the following
five {\em kernels} at the innermost loop's body:
\begin{itemize}
    \newcommand\x{\dv x}
    \newcommand\xT{\dm[height=0, ']x}
    \newcommand\y{\dv y}
    \newcommand\yT{\dm[height=0, ']y}
    \renewcommand\A{\dm A}
    \newcommand\B{\dm B}
    \newcommand\C{\dm C}
    \item \blasl1:
        \begin{itemize}
            \item \ddot: vector-vector inner product $\alpha \coloneqq \xT
                \matvecsep \y$,
            \item \daxpy: vector scaling and addition $\y \pluseqq \alpha \x$,
        \end{itemize}
    \item \blasl2:
        \begin{itemize}
            \item \dgemv: matrix-vector product $\y \pluseqq \A \matvecsep \x$,
            \item \dger: vector-vector outer product $\A \pluseqq \x \yT$, and
        \end{itemize}
    \item \blasl3:
        \begin{itemize}
            \item \dgemm: matrix-matrix product $\C \pluseqq \A \matmatsep \B$.
        \end{itemize}
\end{itemize}
Notice that to comply with the \blas interface, the elements in one of the two
dimensions of a matrix must be contiguous.  Therefore, algorithms that rely on
\dgemv, \dger, or \dgemm as their computational kernel may require a
\definition{temporary copy} of slices before and/or after the invocation of the
corresponding \blas routine.

\begin{table}\tablestyle
    \setlength{\tabcolsep}{6pt}
    \begin{tabular}{l>{\quad}cc>{\qquad\ \ }ccl}
        \toprule
        Kernel  &\multicolumn2{>{\quad}c}{Number of indices}
        &\multicolumn3{>{\quad}c}{Examples from $C_{abc} \coloneqq A_{ai} B_{ibc}$}\\\midrule
                &con-   &\multirow2*{free}                      &kernel     &sliced     &resulting\\
                &tracted&                                       &indices    &indices    &algorithm\\
        \midrule
        \ddot   &1      &0                                      &$i$        &$c, a, b$  &\tensoralgname{cab}{dot}\\[2pt]
        \multirow2*{\daxpy} &\multirow2*0
                        &(1 in $A$ $\wedge$ 0 in $B)$ $\vee$    &$a$        &$b, c, i$  &\tensoralgname{bci}{axpy}\\
                &       &(0 in $A$ $\wedge$ 1 in $B)$           &$c$        &$a, i, b$  &\tensoralgname{aib}{axpy}\\[2pt]
        \multirow2*{\dgemv} &\multirow2*1
                        &(1 in $A$ $\wedge$ 0 in $B)$ $\vee$    &$i, a$     &$b, c$     &\tensoralgname{bc}{gemv}\\
                &       &(0 in !$A$ $\wedge$ 1 in $B)$          &$i, b$     &$c, a$     &\tensoralgname{ca}{gemv}\\[2pt]
        \dger   &0      &1 in $A$ $\wedge$ 1 in $B$             &$a, c$     &$i, b$     &\tensoralgname{ib}{ger}\\[2pt]
        \dgemm  &1      &1 in $A$ $\wedge$ 1 in $B$             &$i, a, b$  &$c$        &\tensoralgname{c}{gemm}\\
        \bottomrule
    \end{tabular}

    \caption{%
        Free and contracted indices in \blas kernels, examples of mapping
        them to $C_{abc} \coloneqq A_{ai} B_{ibc}$, and resulting contraction
        algorithms.%
        \iftableoflist\else\\
            $A$ and $B$ refer to, respectively, the first and second kernel
            operand.
        \fi
    }
    \label{tbl:tensor:slicing-examples}
\end{table}

Instead of a blind search for appropriate slicings, we generate algorithms by
following a goal-oriented approach:  We implement a contraction in terms of one
of the five suitable kernels by mapping this kernel's free and contracted
indices (listed in the left part of \cref{tbl:tensor:slicing-examples}) to
corresponding tensor indices, and slicing along all remaining tensor dimensions.
If such a mapping is not possible, the contraction cannot be implemented in
terms of the selected kernel (e.g., the matrix-vector product $C_a \coloneqq
A_{ai} B_i$ cannot be implemented in terms of \dgemm, because $B_i$ has no free
index).

\begin{figure}\figurestyle

    \begin{subfigure}\textwidth\centering
        \begin{alglisting}[width=\subfigwidth]
            for $\code b = 1\code :b$
              !\dgemm[NN]:! $\tind C{:,b,:} \pluseqq \tind A{:,:} \tind B{:,b,:}$
        \end{alglisting}
        \quad
        \begin{tikzpicture}[baseline]
            \begin{drawcube}
                \sliceB
            \end{drawcube}
            \node at (1, 0, 0) {$\pluseqq$};
            \begin{drawsquare}[(2, 0, 0)]
                \slicenone
            \end{drawsquare}
            \begin{drawcube}[(3.5, 0, 0)]
                \sliceB
            \end{drawcube}
        \end{tikzpicture}
        \caption{Algorithm \tensoralgname b{gemm}}
        \label{alg:aiibc:b-gemm}
    \end{subfigure}

    \medskip

    \begin{subfigure}\textwidth\centering
        \begin{alglisting}[width=\subfigwidth]
            for $\code c = 1\code :c$
              !\dgemm[NN]:! $\tind C{:,:,c} \pluseqq \tind A{:,:} \tind B{:,:,c}$
        \end{alglisting}
        \quad
        \begin{tikzpicture}[baseline]
            \begin{drawcube}
                \sliceC
            \end{drawcube}
            \node at (1, 0, 0) {$\pluseqq$};
            \begin{drawsquare}[(2, 0, 0)]
                \slicenone
            \end{drawsquare}
            \begin{drawcube}[(3.5, 0, 0)]
                \sliceC
            \end{drawcube}
        \end{tikzpicture}
        \caption{Algorithm \tensoralgname c{gemm}}
        \label{alg:aiibc:c-gemm}
    \end{subfigure}

    \mycaption{%
        Contraction algorithms for $C_{abc} \coloneqq A_{ai} B_{ibc}$ based on
        \dgemm.
    }
    \label{algs:aiibc:l3}
\end{figure}

\begin{example}{\dgemm-based algorithms for $C_{abc} \coloneqq A_{ai}
    B_{ibc}$}{tensor:aiibc:gemm} Let us consider the contraction $C_{abc}
    \coloneqq A_{ai} B_{ibc}$, which is visualized as 
    \[
        \begin{tikzpicture}[baseline=(c.base)]\begin{drawcube}
            \node[left] at (-1,0,1) {$\scriptstyle a$}; 
            \node[below] at (0,-1,1) {$\scriptstyle b$}; 
            \node[below right] at (1,-1,0) {$\scriptstyle c$}; 
            \node (c) {$C$}; 
        \end{drawcube}\end{tikzpicture}
        \coloneqq 
        \begin{tikzpicture}[baseline=(c.base)]\begin{drawsquare}
            \node[left] at (-1,0,0) {$\scriptstyle a$};
            \node[below] at (0,-1,0) {$\scriptstyle i$}; 
            \node {$A$};
        \end{drawsquare}\end{tikzpicture} 
        \matmatsep
        \begin{tikzpicture}[baseline=(c.base)]\begin{drawcube}
            \node[left] at (-1,0,1) {$\scriptstyle i$};
            \node[below] at (0,-1,1) {$\scriptstyle b$};
            \node[below right] at (1,-1,0) {$\scriptstyle c$}; 
            \node {$B$};
        \end{drawcube}\end{tikzpicture} 
        \enspace, 
    \] 
    and implement it in terms of \dgemm.

    Since \dgemm involves one free index in each of its operands~\dm A and~\dm
    B, and one contracted index (common to both \dm A and~\dm B), in order to
    reduce any contraction to a sequence of \dgemm calls, one must slice all but
    one free index of both~\dm A and~\dm B, and all but one contracted index.
    For the above contraction, this is achieved by slicing either dimension~$b$
    or~$c$, resulting in the two algorithms \tensoralgname b{gemm} and
    \tensoralgname c{gemm}\footnotemark{} shown in \cref{algs:aiibc:l3}.
\end{example}
\footnotetext{
    The name of each algorithm stems from the dimensions its \code{\bf
    for}-loops index and its \blas kernel.  If the algorithm uses
    \code{copy}-kernels, they are indicated by apostrophes~$'$.
}

Since for a given contraction, there is no obvious a-priori choice of kernel and
slicings to maximize performance, we consider all possible combinations.
Moreover, we consider all possible \definition[loop permutations]{permutations
of the loops}, because, due to caching effects, each permutation yields a
different performance.

\begin{figure}\figurestyle

    \begin{subfigure}\textwidth\centering
        \begin{alglisting}[width=\subfigwidth]
            for $\code b = 1\code :b$
              for $\code c = 1\code :c$
                !\dgemv[N]:! $\tind C{:,b,c} \pluseqq \tind A{:,:} \tind B{:,b,c}$
        \end{alglisting}
        \quad
        \begin{tikzpicture}[baseline]
            \begin{drawcube}
                \sliceB[outer]
                \sliceC[outer]
                \sliceBC
            \end{drawcube}
            \node at (1, 0, 0) {$\pluseqq$};
            \begin{drawsquare}[(2, 0, 0)]
                \slicenone
            \end{drawsquare}
            \begin{drawcube}[(3.5, 0, 0)]
                \sliceB[outer]
                \sliceC[outer]
                \sliceBC
            \end{drawcube}
        \end{tikzpicture}
        \caption{Algorithm \tensoralgname{bc}{gemv}}
        \label{alg:aiibc:bc-gemv}
    \end{subfigure}

    \medskip

    \begin{subfigure}\textwidth\centering
        \begin{alglisting}[width=\subfigwidth]
            for $\code c = 1\code :c$
              for $\code a = 1\code :a$
                !\dgemv[N]:! $\tind C{a,:,c} \pluseqq \tind A{a,:} \tind B{:,:,c}$
        \end{alglisting}
        \quad
        \begin{tikzpicture}[baseline]
            \begin{drawcube}
                \sliceA[outer]
                \sliceC[outer]
                \sliceAC
            \end{drawcube}
            \node at (1, 0, 0) {$\pluseqq$};
            \begin{drawsquare}[(2, 0, 0)]
                \sliceA
            \end{drawsquare}
            \begin{drawcube}[(3.5, 0, 0)]
                \sliceC
            \end{drawcube}
        \end{tikzpicture}
        \caption{Algorithm \tensoralgname{ca}{gemv}}
        \label{alg:aiibc:ca-gemv}
    \end{subfigure}

    \medskip

    \begin{subfigure}\textwidth\centering
        \begin{alglisting}[width=\subfigwidth]
            for $\code c = 1\code :c$
              for $\code i = 1\code :i$
                !\dger:! $\tind C{:,:,c} \pluseqq \tind A{:,i} \tind B{i,:,c}$
        \end{alglisting}
        \quad
        \begin{tikzpicture}[baseline]
            \begin{drawcube}
                \sliceC
            \end{drawcube}
            \node at (1, 0, 0) {$\pluseqq$};
            \begin{drawsquare}[(2, 0, 0)]
                \sliceB
            \end{drawsquare}
            \begin{drawcube}[(3.5, 0, 0)]
                \sliceA[outer]
                \sliceC[outer]
                \sliceAC
            \end{drawcube}
        \end{tikzpicture}
        \caption{Algorithm \tensoralgname{ci}{ger}}
        \label{alg:aiibc:ci-ger}
    \end{subfigure}

    \medskip

    \begin{subfigure}\textwidth\centering
        \begin{alglisting}[width=\subfigwidth]
            for $\code b = 1\code :b$
              for $\code i = 1\code :i$
                !\dger:! $\tind C{:,b,:} \pluseqq \tind A{:,i} \tind B{i,b,:}^T$
        \end{alglisting}
        \quad
        \begin{tikzpicture}[baseline]
            \begin{drawcube}
                \sliceB
            \end{drawcube}
            \node at (1, 0, 0) {$\pluseqq$};
            \begin{drawsquare}[(2, 0, 0)]
                \sliceB
            \end{drawsquare}
            \begin{drawcube}[(3.5, 0, 0)]
                \sliceA[outer]
                \sliceB[outer]
                \sliceAB
            \end{drawcube}
        \end{tikzpicture}
        \caption{Algorithm \tensoralgname{bi}{ger}}
        \label{alg:aiibc:bi-ger}
    \end{subfigure}

    \mycaption{%
        Sample of contraction algorithms for $C_{abc} \coloneqq A_{ai} B_{ibc}$
        based on \blasl2.
        \iftableoflist\else All slicings are visualized in {\color{blue}
        blue}; only the kernel operands (the intersections) are in
        {\color{red} red}.\fi
    }
    \label{algs:aiibc:l2}
\end{figure}
 \begin{figure}\figurestyle
    \begin{subfigure}\textwidth\centering
        \begin{alglisting}[width=\subfigwidth]
            for $\code c = 1\code :c$
              for $\code a = 1\code :a$
                for $\code b = 1\code :b$
                  !\ddot:! $\tind C{a,b,c} \pluseqq \tind A{a,:} \tind B{:,b,c}$
        \end{alglisting}
        \quad
        \begin{tikzpicture}[baseline]
            \begin{drawcube}
                \sliceA[outer]
                \sliceB[outer]
                \sliceC[outer]
                \sliceAB[outer]
                \sliceAC[outer]
                \sliceBC[outer]
                \sliceABC
            \end{drawcube}
            \node at (1, 0, 0) {$\pluseqq$};
            \begin{drawsquare}[(2, 0, 0)]
                \sliceA
            \end{drawsquare}
            \begin{drawcube}[(3.5, 0, 0)]
                \sliceB[outer]
                \sliceC[outer]
                \sliceBC
            \end{drawcube}
        \end{tikzpicture}
        \caption{Algorithm \tensoralgname{cab}{dot}}
        \label{alg:aiibc:cab-ddot}
    \end{subfigure}

    \medskip

    \begin{subfigure}\textwidth\centering
        \begin{alglisting}[width=\subfigwidth]
            for $\code b = 1\code :b$
              for $\code c = 1\code :c$
                for $\code i = 1\code :i$
                  !\daxpy:! $\tind C{:,b,c} \pluseqq \tind A{:,i} \tind B{i,b,c}$
        \end{alglisting}
        \quad
        \begin{tikzpicture}[baseline]
            \begin{drawcube}
                \sliceB[outer]
                \sliceC[outer]
                \sliceBC
            \end{drawcube}
            \node at (1, 0, 0) {$\pluseqq$};
            \begin{drawsquare}[(2, 0, 0)]
                \sliceB
            \end{drawsquare}
            \begin{drawcube}[(3.5, 0, 0)]
                \sliceA[outer]
                \sliceB[outer]
                \sliceC[outer]
                \sliceAB[outer]
                \sliceAC[outer]
                \sliceBC[outer]
                \sliceABC
            \end{drawcube}
        \end{tikzpicture}
        \caption{Algorithm \tensoralgname{bci}{axpy}}
        \label{alg:aiibc:bci-axpy}
    \end{subfigure}

    \medskip

    \begin{subfigure}\textwidth\centering
        \begin{alglisting}[width=\subfigwidth]
            for $\code a = 1\code :a$
              for $\code i = 1\code :i$
                for $\code b = 1\code :b$
                  !\daxpy:! $\tind C{a,b,:} \pluseqq \tind A{a,i} \tind B{i,b,:}$
        \end{alglisting}
        \quad
        \begin{tikzpicture}[baseline]
            \begin{drawcube}
                \sliceA[outer]
                \sliceB[outer]
                \sliceAB
            \end{drawcube}
            \node at (1, 0, 0) {$\pluseqq$};
            \begin{drawsquare}[(2, 0, 0)]
                \sliceA[outer]
                \sliceB[outer]
                \sliceAB
            \end{drawsquare}
            \begin{drawcube}[(3.5, 0, 0)]
                \sliceA[outer]
                \sliceB[outer]
                \sliceAB
            \end{drawcube}
        \end{tikzpicture}
        \caption{Algorithm \tensoralgname{aib}{axpy}}
        \label{alg:aiibc:aib-axpy}
    \end{subfigure}

    \mycaption{%
        Sample of contractions algorithms for $C_{abc} \coloneqq A_{ai} B_{ibc}$
        based on \blasl1.
    }
    \label{algs:aiibc:l1}
\end{figure}

\begin{example}{Other algorithms for $C_{abc} \coloneqq A_{ai}
    B_{ibc}$}{intro:aiibc:other} 
    For the contraction $C_{abc} \coloneqq A_{ai} B_{ibc}$ from
    \cref{ex:tensor:aiibc:gemm}, the right part of
    \cref{tbl:tensor:slicing-examples} lists examples of algorithm generations
    for all five suitable \blas kernels:  A selection of the contraction's free
    and contracted indices are mapped to each kernel's indices (column ``kernel
    indices''), the remaining indices can be sliced in any (loop-)order (column
    ``sliced indices'') with each order resulting in a different algorithm.  The
    resulting algorithms are presented in full in
    \cref{algs:aiibc:l1,algs:aiibc:l2,algs:aiibc:l3}.  
\end{example}

We developed a small \definition[algorithm
generator:\\C-implementation\\abstract syntax tree (AST)]{algorithm and code
generator} that produces all algorithms derived in this manner, and constructs
corresponding {\em C-implementation}, as well as {\em abstract syntax trees}
(ASTs) representing their loop-based structure.  These ASTs form the starting
point for the micro-benchmarks introduced in the following section.

    \section{Runtime Prediction}
    \label{sec:tensor:pred}
    This section describes development accurate runtime and performance performance
for the previously introduced type of \blas-based algorithms for tensor
contractions.  Taking advantage of these algorithms loop-based structure, we aim
at estimating each algorithm's runtime through \definition{micro-benchmarks} of
its \blas kernel, i.e., with no direct execution of the algorithm itself.  In
order to obtain reliable estimates, these micro-benchmarks need to be executed
in a setup that mirrors the computing environment (most importantly the cache)
within the contraction algorithm as closely as possible.  In the following, we
incrementally go through the steps required to build meaningful ``replicas'' of
the computing environment.

        \subsection{Example Contraction: \texorpdfstring{$C_{abc} \coloneqq
        A_{ai} B_{ibc}$}{C\_abc := A\_ai B\_ibc}}
        \label{sec:tensor:extc}
        Throughout this section, we track the improvement of various changes to our
predictions by considering the \definition[example contraction]{contraction}
$C_{abc} \coloneqq A_{ai} B_{ibc}$ with $A \in \R^{a \times
i}$ and $B \in \R^{i \times b \times c}$ and sizes $i = 8$ and $a = b = c = 8,
\ldots, 1000$:\[
    \begin{tikzpicture}[baseline=(c.base)]\begin{drawcube}
        \node[anchor=east] at (-1,0,1) {$\scriptstyle a$};
        \node[anchor=north] at (0,-1,1) {$\scriptstyle b$};
        \node[anchor=north west] at (1,-1,0) {$\scriptstyle c$};
        \node (c) {$C$};
    \end{drawcube}\end{tikzpicture}
    \coloneqq
    \begin{tikzpicture}[baseline=(a.base), x={(.08, 0, 0)}]\begin{drawsquare}
        \node[anchor=east] at (-1,0,0) {$\scriptstyle a$};
        \node[anchor=north] at (0,-1,0) {$\scriptstyle i$};
        \node (a) {$A$};
        \end{drawsquare}
    \end{tikzpicture}
    \begin{tikzpicture}[baseline=(b.base), y={(0, .08, 0)}]\begin{drawcube}
        \node[anchor=east] at (-1,0,1) {$\scriptstyle i$};
        \node[anchor=north] at (0,-1,1) {$\scriptstyle b$};
        \node[anchor=north west] at (1,-1,0) {$\scriptstyle c$};
        \node (b) {$B$};
    \end{drawcube}\end{tikzpicture}
    \enspace.
\]
This scenario is deliberately challenging due to the small tensor dimension~$i$,
for which \blas kernels are generally not optimized.

For the selected contraction, our generator produces 36~algorithms, some of
which are shown in \cref{algs:aiibc:l1,algs:aiibc:l2,algs:aiibc:l3}:
\begin{itemize}
    \item 6 \ddot-based,
    \item 18 \daxpy-based,
    \item 6 \dgemv-based:
            \tensoralgname{bc}{gemv}~(\ref*{plt:ai_ibc:bc_gemv}),
            \tensoralgname{cb}{gemv}~(\ref*{plt:ai_ibc:cb_gemv}),
            \tensoralgname{ac}{gemv}~(\ref*{plt:ai_ibc:ac_gemv}),\\
            \tensoralgname{ca}{gemv}~(\ref*{plt:ai_ibc:ca_gemv}),
            \tensoralgname{ab}{gemv}~(\ref*{plt:ai_ibc:ab_gemv}),
            \tensoralgname{ba}{gemv}~(\ref*{plt:ai_ibc:ba_gemv}),
    \item 4 \dger-based:
            \tensoralgname{ci}{ger}~(\ref*{plt:ai_ibc:ci_ger}),
            \tensoralgname{ic}{ger}~(\ref*{plt:ai_ibc:ic_ger}),
            \tensoralgname{bi}{ger}~(\ref*{plt:ai_ibc:bi_ger}),\\
            \tensoralgname{ib}{ger}~(\ref*{plt:ai_ibc:ib_ger}), and
    \item 2 \dgemm-based:
            \tensoralgname c{gemm}~(\ref*{plt:ai_ibc:c_gemm}),
            \tensoralgname b{gemm}~(\ref*{plt:ai_ibc:b_gemm}).
\end{itemize}
However to avoid overloaded performance plots, this section only considers the
algorithms based on \blasl2 and~3, i.e., with the kernels \dgemv, \dger, and
\dgemm.

\begin{figure}\figurestyle
    \ref*{leg:tensor:pred}

    \raggedleft
    \tikzsetnextfilename{ai_ibc_meas}
    \begin{tikzpicture}
        \begin{axis}[
                predplt,
                legend to name=leg:tensor:pred,
                legend columns=4,
                ymax=6,
            ]
            \foreach \var/\loops/\kernel in {
                25/bc/gemv, 26/cb/gemv, 27/ac/gemv, 28/ca/gemv, 29/ab/gemv,
                30/ba/gemv, 31/ci/ger, 32/ic/ger, 33/bi/ger, 34/ib/ger,
                35/c/gemm, 36/b/gemm%
            } {
                \plot table {var\var.min};
                \addlegendentryexpanded{\tensoralgname\loops\kernel}
                \label{plt:ai_ibc:\loops_\kernel}
            }
        \end{axis}
    \end{tikzpicture}

    \mycaption{%
        Performance measurements of algorithms for $C_{abc} \coloneqq A_{ai}
        B_{ibc}$ based on \blasl2 and~3.
        \captiondetails{$i = 8$, \harpertown, 1~thread, \openblas, median of
        10~repetitions}
    }
    \label{fig:tensor:pred:meas}
\end{figure}

\cref{fig:tensor:pred:meas} displays the measured performance of these
algorithms on a \harpertown using single-threaded \openblas.  Our goal in the
following sections is to accurately predict this performance without
executing the algorithms.  Although it is evident that only two of the
algorithms---the \dgemm-based \tensoralgname c{gemm}~(\ref*{plt:ai_ibc:c_gemm})
and \tensoralgname b{gemm}~(\ref*{plt:ai_ibc:b_gemm})---are
competitive\footnote{%
    Due to the extremely small dimension~$i = 8$, they achieve less than half of
    the \harpertownshort's theoretical peak performance of
    \SI{12}{\giga\flops\per\second}.
} we aim to accurately predict all algorithms to develop and demonstrate the
broad applicability of our methodology.

        \subsection{Repeated Execution}
        \label{sec:tensor:repeat}
        The first, most intuitive, attempt to predict the performance of an algorithm
through a micro-benchmark relies on the \definition[repeated measurements in
isolation]{repeated measurement} of its \blas kernel's performance {\em in
isolation}.  We implemented this approach by executing each kernel ten times in
the \sampler, and extracting the median runtime; the corresponding estimate is
then obtained by multiplying this median by the number of kernel invocations
within the algorithm.  In our example, this boils down to multiplying the kernel
runtime with the product of all loop lengths.

\begin{figure}\figurestyle
    \ref*{leg:tensor:pred}
    \pgfplotsset{ymax=6.5}

    \begin{subfigaxis}[predplotpred=step]
    \end{subfigaxis}\hfill
    \begin{subfigaxis}[predplotmeas=2]
    \end{subfigaxis}

    \mycaption{%
        Performance predictions for $C_{abc} \coloneqq A_{ai} B_{ibc}$ based on
        repeated execution.
        \captiondetails{$i = 8$, \harpertown, 1~thread, \openblas, median of
        10~repetitions}
    }
    \label{fig:tensor:pred_repeat_meas}
\end{figure}

The performance predicted by this first, rough approach is shown in
\cref{fig:tensor:pred:step}.  By comparing this figure with the reference
repeated in \cref{fig:tensor:pred:meas2}, it becomes apparent that while the two
fastest algorithms are already correctly identified, the performance of almost
all algorithms is consistently overestimated---the average absolute error with
respect to the measured performance is~\SI{154}\percent.  In other words, when
executed as part of the algorithms, the \blas kernels take longer to complete
than in the isolated micro-benchmarks.  The reason for this discrepancy is that
the micro-benchmarks invoke the kernels repeatedly with the same operands, i.e.,
they operate on cached (``warm'') data.  Within an algorithm, by contrast, at
least one operand varies from one invocation to the next, i.e., the kernel
operates at least partially on ``cold'' data.

        \subsection{Operand Access Distance}
        \label{sec:tensor:accdist}
        In order to improve our predictions' accuracy, we attempt to replicate the state
of the cache within an algorithm prior to the kernel invocation (the ``cache
precondition'') within our micro-benchmarks.  For this purpose, we assume a
fully associative Least Recently Used (LRU) cache replacement policy,\footnote{%
    Due to the regular storage format and memory access strides of dense linear
    algebra operations such as the considered tensor contractions, this
    simplifying assumption does not affect the reliability of the results.
} and, in first instance, consider the case where all loops surrounding the
kernel are somewhere in the middle of their traversal (i.e., not in their first
iteration); this second assumption will be lifted later.

To determine if an operand is cached and to place it in the correct cache level
for the micro-benchmark, we determine how much data was used in any operations
since its last access, referred to as its \definition{access distance}.  Once
this access distance is known for all kernel operands, we create an artificial
sequence of memory accesses to reconstruct the cache precondition.  Using this
cache setup, our micro-benchmark's measurement of the kernel closely resembles
the actual execution of the algorithm.  As before, the median runtime of ten
micro-benchmark repetitions multiplied with the number of kernel invocations
yields the algorithm's runtime prediction.

We now describe how to obtain the access distance for each operand.  While the
presented method allows for any combinations of loops and multiple kernels
(e.g., a \blas kernel and a \code{copy} kernel), for the sake of clarity, we
limit the discussion to abstract syntax trees (ASTs) that consist of only a one
or more nested loops with a single \blas kernel at their core.

To determine the access distance for an operand~\op, we examine an algorithm's
AST (see \cref{sec:tensor:alggen}) starting at the kernel, and traverse it
backwards until the previous access to~\op (or the AST's root) is found.  While
doing so, we collect the operands of all encountered kernels in an initially
empty set~$M$, whose total data volume---the sum of the collected operands'
sizes---ultimately determines the access distance.  Going up the AST, three
different cases can be encountered.
\begin{enumerate}
    \item {\usekomafont{disposition}\op does not vary across the surrounding
        loop.}\\
        In this case \op referred to the same operand in the previous iteration
        of the surrounding loop.  The back-traversal therefore terminates, and
        the operands collected in~$M$ so far determine the access distance.

        \begin{example}{Loop-independent operand}{tensor:accdist:indep}
            In algorithm \tensoralgname{ca}{gemv}~(\ref*{plt:ai_ibc:ca_gemv})
            the operand~\tind B{:,:,c} does not depend on the surrounding loop's
            iterator~\code a:

            \smallskip

            \noindent
            \begin{alglisting}[width=.59\textwidth]
                for $\code c = 1\code :c$
                  for $\code a = 1\code :a$
                    !\dgemv[N]:! $\tind C{a,:,c} \pluseqq \tind A{a,:} \tind B{:,:,c}$
            \end{alglisting}
            \hfill
            \begin{tikzpicture}[baseline]
                \begin{drawcube}
                    \sliceA[outer]
                    \sliceC[outer]
                    \sliceAC
                \end{drawcube}
                \node at (1,0,0) {$\pluseqq$};
                \begin{drawsquare}[(2,0,0)]
                    \sliceA
                \end{drawsquare}
                \begin{drawcube}[(3.5,0,0)]
                    \sliceC
                \end{drawcube}
            \end{tikzpicture}

            \smallskip

            \noindent
             Hence, $M = \emptyset$ and \tind B{:,:,c}'s access distance is~0.
        \end{example}

    \item {\usekomafont{disposition}\op varies across the surrounding loop.}\\
        In this case \op referred to a different operand in the previous
        iteration of the loop.  As a result, it is safe to assume that at least
        all kernel operands throughout this loop's iterations were accessed
        since the last access to~\op.  Hence, all operands are added to~$M$ and
        they are symbolically joined along the dimensions the loop iterates
        over.

        Since a previous access to~\op was not yet detected, the traversal
        proceeds by going up one level in the AST and applying the method
        recursively:  The surrounding loop now takes the role of the starting
        node and we look for a previous access to~\op joined across this loop.

        \begin{example}{Loop-dependent operand}{tensor:accdist:depend1}
            In algorithm \tensoralgname{ca}{gemv}~(\ref*{plt:ai_ibc:ca_gemv})
            the operand~\tind A{a,:} depends on the surrounding loop's
            iterator~\code a.  The algorithm's kernel operates on~\tind A{a,:},
            \tind B{:,:,c}, and~\tind C{a,:,c}, which joint across the
            index~\code a yields the collection
            \[
                M = \{\tind A{:,:}, \tind B{:,:,c}, \allowbreak \tind C{:,:,c}\}
                \enspace.
            \]

            The backward-traversal of the AST continues and now looks for a
            previous access to~\tind A{:,:}---\tind A{a,:} joint across~\code
            a---in the second-innermost loop.  Since this operand is independent
            of this loop's iterator~\code c, case~1 above applies and \tind
            A{a,:}'s access distance is computed from the set~$M$ above.
        \end{example}

    \item {\usekomafont{disposition}The parent node is the AST's root.}\\
        In this case, \op is accessed only once (and for the first time).  Since
        we do not know how the contraction is used (within a surrounding
        application), we can generally not make any assertions on the access
        distance.  For the purpose of this study, in which we execute the
        contraction repeatedly to measure its performance, however, we assume
        that no other data was accessed since the last invocation of the
        contraction; hence, we compute the access distance from the
        collection~$M$.

        \begin{example}{No loops remaining}{tensor:accdist:noloop}
            In algorithm \tensoralgname{ca}{gemv}~(\ref*{plt:ai_ibc:ca_gemv}),
            the operand~\tind C{a,:,c} depends on both of the surrounding loops'
            iterators~\code a and~\code c.  Therefore, the back-traversal
            encounters case~2 above in both its first and second step, and
            joining the kernel's operands~\tind A{a,:}, \tind B{:,:,c},
            and~\tind C{a,:,c} across first~\code a and then~\code c, yields
            \[
                M = \{\tind A{:,:}, \tind B{:,:,:}, \tind C{:,:,:}\} \enspace.
            \]
            In the third step of the back-traversal, the outermost loop is
            already the starting point---the AST's root is reached.  Assuming
            repeated executions of the entire contraction, \tind C{a,:,c}'s
            access distance is computed from the set~$M$ above.
        \end{example}
\end{enumerate}

Based on access distance for each operand of an algorithm's kernel, we construct
a micro-benchmark that \definition[emulate accesses]{emulates the accesses}
within the algorithm prior to the kernel's execution.  This micro-benchmark
consists of accesses to the kernel's operands interleaved with accesses to
remote memory regions that \definition[cache flushing]{flush} portions of the
{\em cache} corresponding to the access distances:  First, we access the operand
with the largest access distance, and then a remote region that accounts for the
difference to the next smaller access distance; this is repeated until the
operand with the smallest access distance is loaded followed by a remote access
of this size.  If the access distances to the first operand in this list s
larger than $\frac54$~times the cache size, the list is truncated to this limit
at the front.

\begin{table}
    \tablestyle
    \begin{tabular}{lS[table-format=4.0]lS[table-format=8.0]}
        \toprule
        operand         &{size} &collection of operands $M$                      &{access distance}\\
                        &{[\doubles{}]} & &{[\doubles{}]}\\
        \midrule
        \tind B{:,:,c}  &3200   &$\emptyset$                                        &0\\
        \tind A{a,:}    &8      &$\{\tind A{:,:}, \tind B{:,:,c}, \tind C{:,:,c}\}$ &166400\\
        \tind C{a,:,c}  &400    &$\{\tind A{:,:}, \tind B{:,:,:}, \tind C{:,:,:}\}$ &65283200\\
        \bottomrule
    \end{tabular}
    \mycaption{%
        Operand sizes and access distances in \tensoralgname{ca}{gemv} for
        $C_{abc} \coloneqq A_{ai} B_{ibc}$.
        \captiondetails{$a = b = c = 400$, $i = 8$, sizes in \doubles}
    }
    \label{tbl:tensor:accdist}
\end{table}

\begin{example}{Cache access emulation}{tensor:cacheacc}
    For algorithm \tensoralgname{ca}{gemv}~(\ref*{plt:ai_ibc:ca_gemv}),
    \cref{tbl:tensor:accdist} summarizes the operands, their sizes, the
    corresponding collections~$M$, and the implicated access distances for
    tensor sizes $a = b = c = 400$ and $i = 8$.  From these distances, we
    get the following list of memory accesses as a setup for the
    \dgemv[N]-kernel, where the {\sf[$s$]} correspond to remote memory accesses
    of $\SIvar s{\doubles}$ ($= \SIvar{8 s}{\bytes}$):
    \[\text{%
        \tind C{a,:,c},
        {\sf[\num{65116792}]},
        \tind A{a,:},
        {\sf[\num{163200}]},
        \tind B{:,:,c}
    }\enspace.\]
    Note that the remote accesses do not directly correspond to the access
    distances; instead, this distance is reached for each operand as the sum of
    the sizes of all accesses to its right in this list. (e.g., the access
    distances of~\tind A{a,:} is reached as $\SI{163200}{\doubles} +
    \mathrm{sizeof}(\tind B{:,:,c}) = \SI{166400}{\doubles}$).

    The largest access distance of \SI{65283200}{\doubles} is considerably
    larger than \SI{983040}{\doubles} ($= \frac54 \times \SI6{\mebi\byte} =
    \frac54 \times \text{L2~cache size}$).  Hence, the list is cut at this size,
    yielding the final setup for this algorithm's micro-benchmark:
    \[\text{
        {\sf[\num{816632}]},
        \tind A{a,:},
        {\sf[\num{163200}]},
        \tind B{:,:,c}
    }\enspace.\]
\end{example}

The thus obtained benchmark, consisting of the setup followed by the kernel
invocation, is as before executed ten~times, and the resulting median runtime is
used to compute our second runtime and performance predictions.

\begin{figure}\figurestyle
    \ref*{leg:tensor:pred}

    \begin{subfigaxis}[predplotpred=step+s]
    \end{subfigaxis}\hfill
    \begin{subfigaxis}[predplotmeas=3]
    \end{subfigaxis}

    \mycaption{%
        Performance predictions for $C_{abc} \coloneqq A_{ai} B_{ibc}$ with
        cache emulation based on access distances.
        \captiondetails{$i = 8$, \harpertown, 1~thread, \openblas, median of
        10~repetitions}
    }
    \label{fig:tensor:pred_accdist_meas}
\end{figure}

\Cref{fig:tensor:pred:step+s} presents our new performance predictions:
Compared to our initial estimates (\cref{fig:tensor:pred:step}), these
predictions are already much closer to the measured performance
(\cref{fig:tensor:pred:meas3}); the average error is reduced
to~\SI{26.3}\percent.  For several algorithms (such as
\tensoralgname{ic}{ger}~(\ref*{plt:ai_ibc:ic_ger})), the error is already within
a few percent; for many others instead, the predictions are still off.  In
particular, the performance of some algorithms---for instance,
\tensoralgname{bi}{ger}~(\ref*{plt:ai_ibc:bi_ger})---is now underestimated; this
is due to the fact that based on the access distance, certain operands are
placed out of cache, while in practice they are (partially) brought into cache
through either prefetching or because they share cache-lines across the
innermost loop's iterations.  We address this disparity by further refining our
micro-benchmarks.

        \subsection{Cache Prefetching}
        \label{sec:tensor:prefetch}
        In the considered type of tensor contraction algorithms, prefetching of operands
and sharing of cache-lines across loop iterations occur frequently.

\begin{example}{Prefetching and shaed cache-lines}{tensor:prefetch}
    In algorithm \tensoralgname{bi}{ger}~(\ref*{plt:ai_ibc:bi_ger}), the vector
    operand~\tind A{:,i} points to a different memory location in each
    iteration~\code i of the innermost loop:

    \smallskip

    \noindent
    \begin{alglisting}[width=.59\textwidth]
        for $\code b = 1\code :b$
          for $\code i = 1\code :i$
            !\dger:! $\tind C{:,b,:} \pluseqq \tind A{:,i} \tind B{i,b,:}^T$
    \end{alglisting}
    \hfill
    \begin{tikzpicture}[baseline]
            \begin{drawcube}
                \sliceA[outer]
                \sliceC[outer]
                \sliceAC
            \end{drawcube}
            \node at (1,0,0) {$\pluseqq$};
            \begin{drawsquare}[(2,0,0)]
                \sliceA
            \end{drawsquare}
            \begin{drawcube}[(3.5,0,0)]
                \sliceC
            \end{drawcube}
    \end{tikzpicture}

    \smallskip

    \noindent
    However, since these vectors are consecutive
    in memory, when the end of~\tind A{:,i} is reached, the prefetcher likely
    already loads the next elements, which constitute~\tind A{:,i} in the
    next iteration.  At the same time, the innermost loop over~\code i indexes
    \tind B{i,b,:}'s first dimension, and hence 8~consecutive operands~\tind
    B{i,b,:} occupy the same cache-line\footnotemark{} (e.g., $\tind B{0,b,:},
    \ldots, \tind B{7,b,:}$).
\end{example}
\footnotetext{%
    Each cache-line fits $\SI{64}{\bytes} = \SI8{\doubles}$.
}

Such prefetching situations occur when the following conditions are met:
\begin{enumerate}
    \item the operand varies across the directly surrounding loop, and
    \item this loop's iterator indexes either
        \begin{itemize}
            \item the first dimension of the operand,
            \item or its second dimension, while the first is accessed entirely
                or fits in a single cache-line.
        \end{itemize}
\end{enumerate}
We test these conditions as part of our AST-based algorithm analysis, and when
both are fulfilled, we use a slight modification of the previously introduced
back-traversal of the AST to compute the \definition{prefetch distance}, i.e.,
how long ago the prefetching occurred.  These prefetch distances are then
integrated into the micro-benchmark's setup just like the access distances, only
that the prefetch accesses are limited to one cache-line along an operand's
first dimension.

\begin{example}{Cache emulation with prefetch distances}{tensor:prefetchacc}
    In algorithm \tensoralgname{ca}{gemv}~(\ref*{plt:ai_ibc:ca_gemv}), for which
    \cref{ex:tensor:cacheacc} constructed a cache-aware setup, operands~\tind
    A{a,:} and~\tind C{a,:,b} meet both prefetching conditions: 1) they vary
    with the surrounding loop's iterator~\code a, and 2) \code a~indexes their
    first dimensions (sharing of cache-lines).  As a result, their prefetch
    distances are~\SI0\bytes, and since their extent along the first,
    contiguously stored dimension is~1, the prefetching access loads them
    entirely.  Since the remaining operand~\tind B{:,:,c} has an access distance
    of~\SI0{\bytes}, all operands are now accessed immediately before the kernel
    invocation; the setup is reduced to the accesses
    \[\text{%
        \tind C{a,:,c},
        \tind A{a,:},
        \tind B{:,:,c}%
    }\enspace.\]
    Since this setup consists only of accesses to the operands, it becomes
    redundant in our micro-benchmarks, because each of the ten repetitions
    already touches all operands for the next repetition; hence, in such a case,
    we omit the setup altogether.
\end{example}

\begin{figure}\figurestyle
    \ref*{leg:tensor:pred}

    \begin{subfigaxis}[predplotpred=step+sp]
    \end{subfigaxis}\hfill
    \begin{subfigaxis}[predplotmeas=4]
    \end{subfigaxis}

    \mycaption{%
        Performance predictions for $C_{abc} \coloneqq A_{ai} B_{ibc}$
        with cache emulation including prefetch distances.
        \captiondetails{$i = 8$, \harpertown, 1~thread, \openblas, median of
        10~repetitions}
    }
    \label{fig:tensor:pred_prefdist_meas}
\end{figure}

Accounting for prefetching, we obtain the performance predictions presented in
\cref{fig:tensor:pred:step+sp}.  Here, several algorithms, such as
\tensoralgname b{gemm}~(\ref*{plt:ai_ibc:b_gemm}) and
\tensoralgname{ba}{gemv}~(\ref*{plt:ai_ibc:ba_gemv}), are estimated closer to
their measured performance, leading to a reduced average error
of~\SI{19.1}\percent.  Note that this improvement also has a major influence on
the fastest algorithm \tensoralgname b{gemm}~(\ref*{plt:ai_ibc:b_gemm}):  Since
its matrix operands~\tind B{:b:} of size~$8 \times n$ are prefetched entirely by
each preceding loop iteration, both of the \dgemm[NN]'s input operands are
in-cache.

However, the new micro-benchmarks now overestimates the performance of several
other algorithms, including
\tensoralgname{ca}{gemv}~(\ref*{plt:ai_ibc:ca_gemv}); i.e., the runtime is
underestimated.  There are two separate causes for this discrepancy:
\begin{itemize}
    \item In several algorithms, such as
        \tensoralgname{ca}{gemv}~(\ref*{plt:ai_ibc:ca_gemv}), where prefetching
        is implicit due to operands sharing cache-lines, the prefetcher fails
        once a new cache-line is reached.

    \item In other algorithms, such as
        \tensoralgname{bi}{ger}~(\ref*{plt:ai_ibc:bi_ger}), the innermost loop
        is so short (here: 8~iterations) that each first iteration of the loop
        significantly impacts performance.
\end{itemize}
These two causes are treated separately in the following sections.

        \subsection{Prefetching Failures}
        \label{sec:tensor:prefetchfail}
        When operands are identified as prefetched because they share cache-lines across
iterations (i.e., the surrounding loop indexes their first dimension), the
processor should prefetch the next cache-line every 8~iterations
($1\,\text{cache-line} = \SI8\doubles$).  However, as a detailed analysis of
instrumented algorithms has shown, it \definition[prefetching failure]{fails to
do so}.  As a result, in every 8th~iteration of the innermost loop, the operand
is not available and the kernel may take significantly longer.

We account for this prefetching-artefact by performing
\definition[2~micro-benchmarks]{two separate micro-benchmarks}: one simulating
the 7~iterations in which the operand is available in cache as before, and one
for the 8th~iteration.  In this second micro-benchmark we account for the
``prefetching failures'', and do not emulate a corresponding prefetching access.
The prediction for the total runtime is now obtained by \definition{weighting}
these two benchmark timings according to their number of occurrences in the
algorithm and summing their contributions.

\begin{example}{Benchmarks for prefetch failures}{tensor:preffail}
    In algorithm \tensoralgname{ca}{gemv}~(\ref*{plt:ai_ibc:ca_gemv}), the
    memory regions of both~\tind A{a,:} and~\tind C{a,:,c} each share
    cache-lines across iterations of the innermost loops over~\code a.  Hence,
    in every 8th~iteration the kernel accesses a new cache line and its runtime
    increases drastically by a about~$4.5\times$.  To account for these
    ``prefetching failures'', we introduce a second set of micro-benchmarks
    without the emulated prefetching accesses.  For $a = b = c = 400$ and $i =
    8$ this results in the same setup as without prefetching:
    \[\text{%
        {\sf[\num{816632}]},
        \tind A{a,:},
        {\sf[\num{163200}]},
        \tind B{:,:,c}%
    } \enspace.\]
\end{example}

\begin{figure}\figurestyle
    \ref*{leg:tensor:pred}

    \begin{subfigaxis}[predplotpred=step+spx]
    \end{subfigaxis}\hfill
    \begin{subfigaxis}[predplotmeas=5]
    \end{subfigaxis}

    \mycaption{%
        Performance predictions for $C_{abc} \coloneqq A_{ai} B_{ibc}$
        accounting for pre\-fetching failures.
        \captiondetails{$i = 8$, \harpertown, 1~thread, \openblas, median of
        10~repetitions}
    }
    \label{fig:tensor:pred_preffail_meas}
\end{figure}

\cref{fig:tensor:pred:step+spx} shows the predictions obtained after this
improvement:  The error is reduced to~\SI{14.7}\percent.  Most apparent in
\tensoralgname{ca}{gemv}~(\ref*{plt:ai_ibc:ca_gemv}), the overestimation of
algorithms whose iterations share cache-lines are now corrected.

        \subsection{First Loop Iterations}
        \label{sec:tensor:firstiter}
        The predictions for several algorithms, such as
\tensoralgname{ci}{ger}~(\ref*{plt:ai_ibc:ci_ger}), are still severely off,
because the innermost loop of these algorithms is extremely short (in our
example 8~iterations long).  In such a case, the predictions are only accurate
for all but the \definition{first iteration}.  Due to vastly different cache
preconditions for this first iteration, however, its performance can differ
significantly; e.g., in \tensoralgname{ci}{ger}~(\ref*{plt:ai_ibc:ci_ger}) it is
up to $10\times$~lower, which combined with the low total iteration count
results in predictions that are off by up to~$2\times$.

To treat such situations, we introduce separate micro-benchmarks to predict the
performance of the first iterations of the innermost loop (and further loops if
their first iterations account for more than \SI1{\percent}~of the total kernel
invocations).  For this purpose, the access distance evaluation is slightly
modified:  Instead of the kernel itself, the starting point is now the loop
whose first iteration is considered, and the set~$M$ already contains all of the
kernel's memory regions joined across this loop.

\begin{example}{First loop iterations}{tensor:firstloop}
    In algorithm \tensoralgname{ci}{ger}~(\ref*{plt:ai_ibc:ci_ger}), the
    innermost loop over~$i$ is in our example only 8~iterations long.  All
    but the first iteration use the same operand~\tind C{:,:,c}, and \tind
    A{:,i} and~\tind B{i,:,c} are prefetched, leading to optimal conditions for
    performance.  In the first iteration (i.e., the next \code c~iteration)
    however, \tind C{:,:,c} refers to a different memory location and
    prefetching fails for both \tind A{:,i} and~\tind B{i,:,c}, leading to
    severely lower performance.
\end{example}

Based on these improved access distances, the cache setup and micro-benchmark
are performed just as before.  As before, the prediction for the total runtime
is obtained from weighting all relevant benchmark timings with the corresponding
number of occurrences within the algorithm.

\begin{figure}\figurestyle
    \ref*{leg:tensor:pred}

    \begin{subfigaxis}[predplotpred=pred]
    \end{subfigaxis}\hfill
    \begin{subfigaxis}[predplotmeas=6]
    \end{subfigaxis}

    \mycaption{%
        Final performance predictions for $C_{abc} \coloneqq A_{ai} B_{ibc}$.
        \captiondetails{$i = 8$, \harpertown, 1~thread, \openblas, median of
        10~repetitions}
    }
    \label{fig:tensor:pred_pred_meas}
\end{figure}

In \cref{fig:tensor:pred:pred}, we present the improved performance
predictions obtained from this modification.  The performance of all algorithms
is now predicted with satisfying accuracy---the average absolute error
is~\SI{9.47}\percent.

    \section{Results}
    \label{sec:tensor:results}
    In order to showcase the applicability and effectiveness of our predictions,
this section applies them to other contractions: \Cref{sec:ai_ibc2} revisits
$C_{abs} \coloneqq A_{ai} B_{ibc}$ with entirely different problem sizes and a
changed hardware and software setup, \cref{sec:noblas3} considers a contraction
that only allows the use of \blasl1 and~2 kernels, and \cref{sec:ijb_jcid}
studies a more complex contraction with numerous alternative algorithms and
multi-threading.

\subsection{Changing the Setup for \texorpdfstring{$C_{abc} \coloneqq A_{ai}
B_{ibc}$}{C\_abc := A\_ai B\_ibc}}
\label{sec:ai_ibc2}

\begin{figure}\figurestyle
    \ref*{leg:tensor:pred}

    \medskip

    \ref*{leg:tensor:ai_ibc_other}

    \pgfplotsset{twocolplot, perfplot=28.8, predplt, xlabel=$i$}

    \begin{subfigaxis}[
            fig caption=Predictions,
            fig label=ai_ibc:ivy:pred,
            table/search path={tensor/figures/data/ai_ibc2/pred/}
        ]
        \foreach \var in {25, ..., 36}
            \addplot table {var\var.min};
        \foreach \var in {1, ..., 6}
            \addplot[plot7, dotted] table {var\var.min};
        \foreach \var in {7, ..., 24}
            \addplot[plot8, dotted] table {var\var.min};
    \end{subfigaxis}\hfill
    \begin{subfigaxis}[
            fig caption=Measurements,
            fig label=ai_ibc:ivy:meas,
            table/search path={tensor/figures/data/ai_ibc2/meas/},
            legend to name=leg:tensor:ai_ibc_other
        ]
        \addlegendimage{plot7, dotted}
        \addlegendentry{\ddot-based (6)}
        \addlegendimage{plot8, dotted}
        \addlegendentry{\daxpy-based (18)}
        \foreach \var in {25, ..., 36}
            \addplot table {var\var.min};
        \foreach \var in {1, ..., 6}
            \addplot[plot7, dotted] table {var\var.min};
        \foreach \var in {7, ..., 24}
            \addplot[plot8, dotted] table {var\var.min};
    \end{subfigaxis}

    \mycaption{%
        Performance predictions and measurements for $C_{abc} \coloneqq A_{ai}
        B_{ibc}$ with $a = b = c = 128$ fixed.
        \captiondetails{\ivybridge, 1~thread, \mkl, median of 10~repetitions}
    }
    \label{fig:tensor:ai_ibc2}
\end{figure}

We consider the previously studied contraction with an entirely different setup:
We use $a = b = c = 128$ and $i = 8, \ldots, 1000$ in steps of~8 on an
\ivybridge with single-threaded \mkl.  For this scenario,
\cref{fig:tensor:ai_ibc2} presents the performance predictions and measurements
for all 36~algorithms (see \cref{sec:tensor:extc}).  Although everything,
ranging from the problem sizes to the machine and \blas library was changed in
this setup, the predictions are of equivalent quality and our tool correctly
determines that the \dgemm-based algorithms (\ref*{plt:ai_ibc:c_gemm}),
\ref*{plt:ai_ibc:b_gemm}) not only perform best and equally well but also reach
over~\SI{75}{\percent} of the \ivybridgeshort's theoretical peak performance of
\SI{28.8}{\giga\flops\per\second}.

\subsection{Vector Contraction: \texorpdfstring{$C_a \coloneqq A_{iaj}
B_{ji}$}{C\_a := A\_iaj B\_ji}}
\label{sec:noblas3}

\begin{figure}\figurestyle

    \begin{subfigure}\textwidth\centering
        \begin{alglisting}[width=\subfigwidth]
            for $\code j = 1\code :j$
              !\dgemv[T]:! $\tind C: \pluseqq \tind A{:,:,j}^T \tind B{j,:}^T$
        \end{alglisting}
        \quad
        \begin{tikzpicture}[baseline]
            \draw[thick, red, line cap=rounded, scale=.5] (0, -1, 0) -- (0, 1, 0);
            \node at (.5, 0, 0) {$\pluseqq$};
            \begin{drawcube}[(1.5, 0, 0)]
                \sliceC
            \end{drawcube}
            \begin{drawsquare}[(3, 0, 0)]
                \sliceA
            \end{drawsquare}
        \end{tikzpicture}
        \caption{%
            Algorithm \tensoralgname j{gemv} (\ref*{plt:iaj_ji:j_gemv})
        }
        \label{alg:iaj_ji:j-gemv}
    \end{subfigure}

    \medskip

    \begin{subfigure}\textwidth\centering
        \begin{alglisting}[width=\subfigwidth]
            for $\code i = 1\code :i$
              $\tind{\widetilde A}{:,:} \coloneqq \tind A{i,:,:}$
              !\dgemv[N]:! $\tind C: \pluseqq \widetilde A \tind B{:,i}$
        \end{alglisting}
        \quad
        \begin{tikzpicture}[baseline]
            \draw[thick, red, line cap=rounded, scale=.5] (0, -1, 0) -- (0, 1, 0);
            \node at (.5, 0, 0) {$\pluseqq$};
            \begin{drawcube}[(1.5, 0, 0)]
                \sliceA
            \end{drawcube}
            \begin{drawsquare}[(3, 0, 0)]
                \sliceB
            \end{drawsquare}
        \end{tikzpicture}
        \caption{%
            Algorithm \tensoralgname{i'}{gemv} (\ref*{plt:iaj_ji:i'_gemv})
        }
        \label{alg:iaj_ji:i'-gemv}
    \end{subfigure}

    \mycaption{%
        \dgemv-based algorithms for $C_a \coloneqq A_{iaj} B_{ji}$.
    }
    \label{algs:iaj_ji}
\end{figure}

\begin{figure}\figurestyle
    \ref*{leg:tensor:iaj_ji}

    \pgfplotsset{
        twocolplot,
        ymax=4.5,
        ylabel=performance, y unit=\si{\giga\flops\per\second},
        xlabel={tensor size $a = i = j$},
    }

    \begin{subfigaxis}[
            fig caption=Predictions,
            fig label=iaj_ji:pred,
        ]
        \foreach \var in {1, ..., 8}
            \addplot file {tensor/figures/data/iaj_ji/pred/var\var.min};
    \end{subfigaxis}\hfill
    \begin{subfigaxis}[
            fig caption=Measurements,
            fig label=iaj_ji:meas,
            legend to name=leg:tensor:iaj_ji,
            legend columns=4,
        ]
        \foreach \var/\loops/\kernel in {%
            1/aj/dot, 2/ja/dot, 3/ai/dot, 4/ia/dot, 5/ij/axpy,
            6/ji/axpy, 7/j/gemv, 8/i'/gemv%
        }{
            \addplot file {tensor/figures/data/iaj_ji/meas/var\var.min};
            \addlegendentryexpanded{\tensoralgname{\loops}{\kernel}}
            \label{plt:iaj_ji:\loops_\kernel}
        }
    \end{subfigaxis}

    \mycaption{%
        Performance predictions and measurements for $C_{a} \coloneqq A_{iaj}
        B_{ji}$.
        \captiondetails{\harpertown, 1~thread, \openblas, median of
        10~repetitions}
    }
    \label{fig:tensor:iaj_ji}
\end{figure}

For certain contractions (e.g., those involving vectors), \dgemm cannot be
used as a compute kernel, and algorithms can only be based on \blasl1 or~2
kernels.  One such scenario is encountered in the contraction $C_a \coloneqq
A_{iaj} B_{ji}$, for which our generator yields 8~algorithms:
\begin{itemize}
    \item 4 \ddot-based:
        \tensoralgname{aj}{dot}~(\ref*{plt:iaj_ji:aj_dot}),
        \tensoralgname{ja}{dot}~(\ref*{plt:iaj_ji:ja_dot}),\\
        \tensoralgname{ai}{dot}~(\ref*{plt:iaj_ji:ai_dot}),
        \tensoralgname{ia}{dot}~(\ref*{plt:iaj_ji:ia_dot});
    \item 2 \daxpy-based:
        \tensoralgname{ij}{axpy}~(\ref*{plt:iaj_ji:ij_axpy}),
        \tensoralgname{ji}{axpy}~(\ref*{plt:iaj_ji:ji_axpy}), and
    \item 2 \dgemv-based (see \cref{algs:iaj_ji}):
        \tensoralgname j{gemv}~(\ref*{plt:iaj_ji:j_gemv}),
        \tensoralgname{i'}{gemv}~(\ref*{plt:iaj_ji:i'_gemv}).
\end{itemize}
Note that since last algorithm operates on slices \tind A{i,:,:}, which do not
have contiguously-stored dimension, a \code{copy} kernel (indicated by the
apostrophe in the algorithm name) is required before each \dgemv[N]
(\cref{alg:iaj_ji:i'-gemv}).

\Cref{fig:tensor:iaj_ji} presents the predicted and measured performance for
these algorithms.  Our predictions clearly identify the fastest algorithm
\tensoralgname j{gemv}~(\ref*{plt:iaj_ji:j_gemv}) across the board.
Furthermore, the next group of four algorithms is also correctly recognized, and
the low performance of the second \dgemv[N]-based algorithm
\tensoralgname{i'}{gemv}~(\ref*{plt:iaj_ji:i'_gemv}) (due to the overhead of the
involved copy operation) is correctly predicted as well.

\subsection{Challenging Contraction: \texorpdfstring{$C_{abc} \coloneqq A_{ija}
B_{jbic}$}{C\_abc := A\_ija B\_jbic}}
\label{sec:ijb_jcid}

\begin{figure}\figurestyle
    \begin{subfigure}\subfigwidth
        \begin{alglisting}[width=\textwidth]
            for $\code c = 1\code :c$
              for $\code j = 1\code :j$
                $\tind{\widetilde B}{:,:} \coloneqq \tind B{j,:,:,c}$
                !\dgemm[TT]:! $\tind C{:,:,c} \pluseqq \tind A{:,j,:}^T \widetilde B^T$
        \end{alglisting}
        \caption{%
            Algorithm \tensoralgname{cj'}{gemm}
            (\ref*{plt:ijb_jcid:cj'_gemm})
        }
    \end{subfigure}\hfill
    \begin{subfigure}\subfigwidth
        \begin{alglisting}[width=\textwidth]
            for $\code j = 1\code :j$
              for $\code c = 1\code :c$
                $\tind{\widetilde B}{:,:} \coloneqq \tind B{j,:,:,c}$
                !\dgemm[TT]:! $\tind C{:,:,c} \pluseqq \tind A{:,j,:}^T \widetilde B^T$
        \end{alglisting}
        \caption{%
            Algorithm \tensoralgname{jc'}{gemm}
            (\ref*{plt:ijb_jcid:cj'_gemm})
        }
    \end{subfigure}

    \medskip

    \begin{subfigure}\subfigwidth
        \begin{alglisting}[width=\textwidth]
            for $\code c = 1\code :c$
              for $\code i = 1\code :i$
                $\tind{\widetilde A}{:,:} \coloneqq \tind A{i,:,:}$
                !\dgemm[TN]:! $\tind C{:,:,c} \pluseqq \widetilde A^T \tind B{:,:,i,c}$
        \end{alglisting}
        \caption{%
            Algorithm \tensoralgname{ci'}{gemm}
            (\ref*{plt:ijb_jcid:ci'_gemm})
        }
    \end{subfigure}\hfill
    \begin{subfigure}\subfigwidth
        \begin{alglisting}[width=\textwidth]
            for $\code i = 1\code :i$
              $\tind{\widetilde A}{:,:} \coloneqq \tind A{i,:,:}$
              for $\code c = 1\code :c$
                !\dgemm[TN]:! $\tind C{:,:,c} \pluseqq \widetilde A^T \tind B{:,:,i,c}$
        \end{alglisting}
        \caption{%
            Algorithm \tensoralgname{i'c}{gemm}
            (\ref*{plt:ijb_jcid:i'c_gemm})
        }
    \end{subfigure}

    \medskip

    \begin{subfigure}\subfigwidth
        \begin{alglisting}[width=\textwidth]
            for $\code b = 1\code :b$
              for $\code j = 1\code :j$
                $\tind{\widetilde B}{:,:} \coloneqq \tind B{j,b,:,:}$
                !\dgemm[TN]:! $\tind C{:,b,:} \pluseqq \tind A{:,j,:}^T \widetilde B$
        \end{alglisting}
        \caption{%
            Algorithm \tensoralgname{bj'}{gemm}
            (\ref*{plt:ijb_jcid:bj'_gemm})
        }
    \end{subfigure}\hfill
    \begin{subfigure}\subfigwidth
        \begin{alglisting}[width=\textwidth]
            for $\code j = 1\code :j$
              for $\code b = 1\code :b$
                $\tind{\widetilde B}{:,:} \coloneqq \tind B{j,b,:,:}$
                !\dgemm[TN]:! $\tind C{:,b,:} \pluseqq \tind A{:,j,:}^T \widetilde B$
        \end{alglisting}
        \caption{%
            Algorithm \tensoralgname{jb'}{gemm}
            (\ref*{plt:ijb_jcid:jb'_gemm})
        }
    \end{subfigure}

    \medskip

    \begin{subfigure}\subfigwidth
        \begin{alglisting}[width=\textwidth]
            for $\code b = 1\code :b$
              for $\code i = 1\code :i$
                $\tind{\widetilde A}{:,:} \coloneqq \tind A{i,:,:}$
                !\dgemm[TN]:! $\tind C{:,b,:} \pluseqq \widetilde A^T \tind B{:,b,i,:}$
        \end{alglisting}
        \caption{%
            Algorithm \tensoralgname{bi'}{gemm}
            (\ref*{plt:ijb_jcid:bi'_gemm})
        }
    \end{subfigure}\hfill
    \begin{subfigure}\subfigwidth
        \begin{alglisting}[width=\textwidth]
            for $\code i = 1\code :i$
              $\tind{\widetilde A}{:,:} \coloneqq \tind A{i,:,:}$
              for $\code b = 1\code :b$
                !\dgemm[TN]:! $\tind C{:,b,:} \pluseqq \widetilde A^T \tind B{:,b,i,:}$
        \end{alglisting}
        \caption{%
            Algorithm \tensoralgname{i'b}{gemm}
            (\ref*{plt:ijb_jcid:i'b_gemm})
        }
    \end{subfigure}

    \mycaption{%
        \dgemm-based algorithms for $C_{abc} \coloneqq A_{ija} B_{jbic}$.
    }
    \label{algs:ijb_jcid}
\end{figure}

We now turn to a more complex example inspired by space-time continuum
computations in the field general relativity~\cite{generalrelativity}: $C_{abc}
\coloneqq A_{ija} B_{jbic}$.  For this contraction, we generated a total of
176~different algorithms:
\begin{itemize}
    \item 48 \ddot-based~(\ref*{plt:ijb_jcid:dot}),
    \item 72 \daxpy-based~(\ref*{plt:ijb_jcid:axpy}),
    \item 36 \dgemv-based~(\ref*{plt:ijb_jcid:gemv}),
    \item 12 \dger-based~(\ref*{plt:ijb_jcid:ger}), and
    \item 8 \dgemm-based:\\
        \tensoralgname{cj'}{gemm}~(\ref*{plt:ijb_jcid:cj'_gemm}),
        \tensoralgname{jc'}{gemm}~(\ref*{plt:ijb_jcid:jc'_gemm}),
        \tensoralgname{ci'}{gemm}~(\ref*{plt:ijb_jcid:ci'_gemm}),
        \tensoralgname{i'c}{gemm}~(\ref*{plt:ijb_jcid:i'c_gemm}),\\
        \tensoralgname{bj'}{gemm}~(\ref*{plt:ijb_jcid:bj'_gemm}),
        \tensoralgname{jb'}{gemm}~(\ref*{plt:ijb_jcid:jb'_gemm}),
        \tensoralgname{bi'}{gemm}~(\ref*{plt:ijb_jcid:bi'_gemm}),
        \tensoralgname{i'b}{gemm}~(\ref*{plt:ijb_jcid:i'b_gemm}).
\end{itemize}
All \dgemm-based (see \cref{algs:ijb_jcid}) and several of the \dgemv-based
algorithms involve copy operations to ensure that each matrix has a
contiguously-stored dimension as required by the \blas interface.  Once again,
we consider a challenging scenario where both contracted indices are of size $i
= j = 8$ and the free indices $a = b = c$ vary between~8 and~1000.

\begin{figure}\figurestyle
    \ref*{leg:tensor:ijb_jcid}

    \pgfplotsset{
        twocolplot,
        ymax=20,
        ylabel=performance, y unit=\si{\giga\flops\per\second},
        xlabel={tensor size $a = b = c$},
    }

    \begin{subfigaxis}[
            fig caption=Predictions,
            fig label=ijb_jcid:pred,
            legend to name=leg:tensor:ijb_jcid,
            legend columns=4,
            table/search path={tensor/figures/data/ijb_jcid/pred/}
        ]
        \foreach \var/\loops in {%
            169/cj', 170/jc', 171/ci', 172/i'c, 173/bj', 174/jb',
            175/bi', 176/i'b%
        }{
            \addplot table {var\var.min};
            \addlegendentryexpanded{\tensoralgname\loops{gemm}}
            \label{plt:ijb_jcid:\loops_gemm}
        }
        \addlegendimage{plotdot, dotted}\addlegendentry{\ddot-based}\label{plt:ijb_jcid:dot}
        \addlegendimage{plotaxpy, dotted}\addlegendentry{\daxpy-based}\label{plt:ijb_jcid:axpy}
        \addlegendimage{plotgemv, dashed}\addlegendentry{\dgemv-based}\label{plt:ijb_jcid:gemv}
        \addlegendimage{plotger, dashed}\addlegendentry{\dger-based}\label{plt:ijb_jcid:ger}
        \foreach \var in {1, ..., 48}
            \addplot[plotdot, dotted] table {var\var.min};
        \foreach \var in {49, ..., 120}
            \addplot[plotaxpy, dotted] table {var\var.min};
        \foreach \var in {121, ..., 156}
            \addplot[plotgemv, dashed] table {var\var.min};
        \foreach \var in {157, ..., 168}
            \addplot[plotger, dashed] table {var\var.min};
    \end{subfigaxis}\hfill
    \begin{subfigaxis}[
            fig caption=Measurements,
            fig label=ijb_jcid:meas,
            table/search path={tensor/figures/data/ijb_jcid/meas/}
        ]
        \foreach \var in {169, ..., 176}
            \addplot table {var\var.min};
        \foreach \var in {1, ..., 48}
            \addplot[plotdot, dotted] table {var\var.min};
        \foreach \var in {49, ..., 120}
            \addplot[plotaxpy, dotted] table {var\var.min};
        \foreach \var in {121, ..., 156}
            \addplot[plotgemv, dashed] table {var\var.min};
        \foreach \var in {157, ..., 168}
            \addplot[plotger, dashed] table {var\var.min};
    \end{subfigaxis}

    \mycaption{%
        Performance predictions and measurements for $C_{abc} \coloneqq A_{ija}
        B_{jbic}$.
        \captiondetails{$i = j = 8$, \ivybridge, 1~thread, \openblas, median of
        10~repetitions}
    }
    \label{fig:tensor:ijb_jcid}
\end{figure}

\Cref{fig:tensor:ijb_jcid:pred} presents the predicted performance of the
176~algorithms, where algorithms based on \blasl1 and~2 are grouped by kernel.
Even with the copy operations, the \dgemm-based algorithms are the fastest.
However, within these 8~algorithms, the performance differs by more
than~\SI{20}\percent.  \Cref{fig:tensor:ijb_jcid:meas} compares our predictions
with corresponding performance measurements\footnote{%
    Slow tensor contraction algorithms were stopped before reaching the largest
    problem size by limiting the total measurement time per algorithm
    to~\SI{15}\min.
}:  Among the \dgemm-based algorithms, our predictions clearly separate the bulk
of fast algorithms from the slightly less efficient ones.

\begin{figure}\figurestyle
    \ref*{leg:tensor:ijb_jcid}

    \pgfplotsset{
        twocolplot,
        ymax=180,
        ylabel=performance, y unit=\si{\giga\flops\per\second},
        xlabel={tensor size $a = b = c$},
    }

    \begin{subfigaxis}[
            fig caption=Predictions,
            fig label=ijb_jcid10_pred,
            table/search path={tensor/figures/data/ijb_jcid10/pred/}
        ]
        \foreach \var in {169, ..., 176}
            \addplot table {var\var.min};
        \foreach \var in {1, ..., 48}
            \addplot[plotdot, dotted] table {var\var.min};
        \foreach \var in {49, ..., 120}
            \addplot[plotaxpy, dotted] table {var\var.min};
        \foreach \var in {121, ..., 156}
            \addplot[plotgemv, dotted] table {var\var.min};
        \foreach \var in {157, ..., 168}
            \addplot[plotger, dotted] table {var\var.min};
    \end{subfigaxis}\hfill
    \begin{subfigaxis}[
            fig caption=Measurements,
            fig label=ijb_jcid10_meas,
            table/search path={tensor/figures/data/ijb_jcid10/meas/}
        ]
        \foreach \var in {169, ..., 176}
            \addplot table {var\var.min};
        \foreach \var in {1, ..., 48}
            \addplot[plotdot, dotted] table {var\var.min};
        \foreach \var in {49, ..., 120}
            \addplot[plotaxpy, dotted] table {var\var.min};
        \foreach \var in {121, ..., 156}
            \addplot[plotgemv, dotted] table {var\var.min};
        \foreach \var in {157, ..., 168}
            \addplot[plotger, dotted] table {var\var.min};
    \end{subfigaxis}

    \mycaption{%
        Performance predictions and measurements for $C_{abc} \coloneqq A_{ija}
        B_{jbic}$ on 10~cores.
        \captiondetails{$i = j = 32$, \ivybridge, \openblas, median of
        10~repetitions}
    }
    \label{fig:tensor:ijb_jcid10}
\end{figure}

\paragraph{Multi-Threading}
Our contraction algorithms can profit from shared memory parallelism through
multi-threaded \blas kernels.  To focus on the impact of parallelism, we
increase the contracted tensor dimension sizes to~$i = j = 32$ and use all
10~cores of the \ivybridge with multi-threaded \openblas.
\Cref{fig:tensor:ijb_jcid10} presents performance predictions and measurements
for this setup:  Our predictions accurately distinguish the three groups of
\dgemm-based implementations, and algorithms
\tensoralgname{i'c}{gemm}~(\ref*{plt:ijb_jcid:i'c_gemm}) and
\tensoralgname{i'b}{gemm}~(\ref*{plt:ijb_jcid:i'b_gemm}) (see
\cref{algs:ijb_jcid}), which reach \SI{170}{\giga\flops\per\second}, are
correctly identified as the fastest.
\tensoralgname{jb'}{gemm}~(\ref*{plt:ijb_jcid:jb'_gemm}) on the other hand
merely reaches \SI{60}{\giga\flops\per\second}.  This $3\times$~difference in
performance among \dgemm-based algorithms emphasizes the importance of selecting
the right algorithm.

\subsection{Efficiency Study}

\begin{figure}\figurestyle
    \ref*{leg:tensor:eff}

    \raggedleft
    \tikzsetnextfilename{eff}
    \begin{tikzpicture}
        \begin{semilogyaxis}[
            ymin=1,
            xlabel={tensor size $a = b = c$},
            ylabel={speedup},
            ytickten={0, ..., 9},
            yticklabels={
                \num1, \num{10}, \num{100}, \num{e3}, \num{e4},
                \num{e5}, \num{e6}, \num{e7}, \num{e8}, \num{e9}
            },
            legend to name=leg:tensor:eff,
        ]
            \addlegendimage{empty legend}\addlegendentry{kernel:}
            \addlegendimage{plot1}\addlegendentry{\ddot}\label{plt:eff:dot}
            \addlegendimage{plot2}\addlegendentry{\daxpy}\label{plt:eff:axpy}
            \addlegendimage{plot3}\addlegendentry{\dgemv}\label{plt:eff:gemv}
            \addlegendimage{plot4}\addlegendentry{\dger}\label{plt:eff:ger}
            \addlegendimage{plot5}\addlegendentry{\dgemm}\label{plt:eff:gemm}
            \foreach \var in {1, ..., 6}
                \plot[plot1] file {tensor/figures/data/eff/var\var.dat};
            \foreach \var in {7, ..., 24}
                \plot[plot2] file {tensor/figures/data/eff/var\var.dat};
            \foreach \var in {25, ..., 30}
                \plot[plot3] file {tensor/figures/data/eff/var\var.dat};
            \foreach \var in {31, ..., 34}
                \plot[plot4] file {tensor/figures/data/eff/var\var.dat};
            \foreach \var in {35, ..., 36}
                \plot[plot5] file {tensor/figures/data/eff/var\var.dat};
        \end{semilogyaxis}
    \end{tikzpicture}

    \mycaption{%
        Speedup of predictions over algorithm executions for $C_{abc} \coloneqq
        A_{ai} B_{ibc}$.
        \captiondetails{$i = 8$, \ivybridge, 1~thread, \openblas, median of
        10~repetitions}
    }
    \label{fig:tensor:eff}
\end{figure}

The above study provided evidence that our automated approach successfully
identifies the most efficient algorithm(s).  In the following we show how much
faster this approach is compared to empirical measurements.  For this purpose, we
once more consider the contraction $C_{abc} \coloneqq A_{ai} B_{ibc}$ with $i =
8$ and varying $a = b = c$ on a \harpertown with \openblas.
\Cref{fig:tensor:eff} presents the speedup of our micro-benchmark over
corresponding algorithm measurements:  Generally our predictions are several
orders of magnitude faster than such algorithm executions.  For $a = b = c =
1000$, this relative improvement is smallest for the \dgemm-based
algorithms~(\ref*{plt:eff:gemm}) at $1000\times$, because each \dgemm performs a
significant portion of the computation; for the \dger-based
algorithms~(\ref*{plt:eff:ger}), it lies between 6000 and \num{10000} and for
the \dgemv-based algorithms~(\ref*{plt:eff:gemv}) the gain is $\num{5e5}\times$
to $\num{e6}\times$; finally, for the \blasl1-based
algorithms~(\ref*{plt:eff:axpy}, \ref*{plt:eff:dot}), where each kernel
invocation only performs a tiny fraction of the contraction, our predictions are
\num{1e6} to \num{1e9}~times faster than the algorithm executions.

    \section{Summary}
    \label{sec:tensor:conclusion}
    This chapter focused on the performance prediction of automatically-generated
\blas-based algorithms for tensors contractions.  We tackled the problem of
selecting the fastest algorithm without ever executing it.  Instead, our
approach is based on timing the \blas kernels in a small set of micro-benchmarks
that emulate the execution context of the algorithms.  Thanks to careful
treatment of cache-locality and a model of the cache prefetcher's behavior, our
performance predictions are capable of identifying the best-performing algorithm
in a tiny fraction of the time required to actually run any of the alternatives.

The quality of the predictions was showcased for a number of challenging
scenarios, including contractions among tensors with small dimensions,
contractions that can only be cast in terms of \blasl1 and~2 kernels, and
multi-threaded computations.

}

    \chapter{Conclusion}
\label{ch:conclusion}

This dissertation set out to predict the performance of dense linear algebra
algorithms.  It targeted two types of algorithms that require different
prediction approaches: blocked algorithms and tensor contractions.

For blocked algorithms, we accomplished accurate performance predictions through
automatically generated performance models for compute kernels.  Our
predictions both reliably identify the fastest blocked algorithm from
potentially large numbers of available alternatives, and select a block size
for near-optimal algorithm performance.  Our approach's main advantage is its
separation of the model generation and the performance prediction:  While the
generation may take several hours, thousands of algorithm executions are
afterwards predicted within seconds.  A discussed downside to the approach,
however, is that it does not account for algorithm-dependent caching effects.

For tensor contractions, we established performance predictions that identify
the fastest among potentially hundreds of alternative \blas-based contraction
algorithms.  By using cache-aware micro-benchmarks instead of our performance
models, our solution is highly accurate even for contractions with severely
skewed dimensions.  Furthermore, since these micro-benchmarks only execute a
tiny fraction of each tensor contraction, they provide performance predictions
orders of magnitude faster than empirical measurements.

Together, our model generation framework and micro-benchmarks form a solid
foundation for accurate and fast performance prediction for dense linear algebra
algorithms.

\section{Outlook}
The techniques presented in this dissertations offer numerous opportunities for
applications and extensions:
\begin{itemize}
    \item Our methods can be applied to predict the performance various types of
        algorithms and operations, such as recursive algorithms and
        algorithms-by-blocks.

    \item For dense eigenvalue solvers, our models can predict the two most
        computationally intensive stages:  The reduction to tridiagonal form and
        the back-transformation.  By additionally estimating the data-dependent
        performance of tridiagonal eigensolvers, one can predict the solution of
        complete eigenproblems.

    \item Beyond individual operations, our predictions can be applied to
        composite operations and algorithms, such as matrix chain
        multiplications or least squares solvers.

    \item Our models were designed to provide estimates for configurable yet
        limited ranges of problem sizes.  For extrapolations to larger problems
        they should be revised to ensure that local performance phenomena do not
        distort faraway estimates.

    \item Computations on distributed memory systems, accelerators, and graphics
        cards can be predicted by combining our techniques with models for data
        movement and communication.
\end{itemize}

    \appendix

    \chapter[Terminology: Performance and Efficiency]
        {Terminology:\newline Performance and Efficiency}
\label{app:term}
{
    \tikzsetexternalprefix{externalized/term-}

    \newcommand\pperf{\text{peak performance}}
\newcommand\pbw{\text{peak bandwidth}}

    In a nutshell, performance is the rate at which a software---such as a code
segment, a routine, or an entire application---performs useful work, and
efficiency is the ratio of the attained performance to the used processor's
theoretical peak performance.

This appendix introduces these concepts in detail and thereby provides the
terminology used throughout this work.  It is intended for readers new to the
high-performance computing and as a small reference.  It covers the following
material:
\begin{itemize}
    \item \cref{sec:term:workload} describes an operation's
        implementation-independent {\em workload}  in terms of {\em
        floating-point operations}, {\em data volume} and {\em movement}, and
        {\em arithmetic intensity}.

    \item \cref{sec:term:time} details {\em cycle accurate timing}, which allows
        to measure the {\em runtime} of a computation with high precision.

    \item \cref{sec:term:perf} defines a computation's {\em attained
        performance} and {\em bandwidth} based on its workload and runtime.

    \item \cref{sec:term:hw} briefly introduces the {\em hardware capabilities}
        relevant to dense linear algebra computations, such as {\em peak
        performance} and {\em peak bandwidth}.

    \item \cref{sec:term:eff} differentiates between {\em bandwidth-} or {\em
        compute-bound} computations by relating the attained performance to the
        hardware capabilities, and evaluates a computation's {\em
        efficiency}---the most meaningful metric to quantify how well a piece of
        software performs its work.

    \item \cref{sec:term:other} gives an overview of other performance-related
        measures, such as {\em hardware counters} and {\em energy metrics}.
\end{itemize}

    \section{Workload}
    \label{sec:term:workload}
    In scientific computing, the ultimately most desirable measure of a
computation's work is the ``amount of new science performed'', which, however,
is impractical to quantify---not least because it may well be opinion-based.
Instead, we resort to simpler, computation-oriented metrics, namely the {\em
number of arithmetic operations} required to perform a operation
(\cref{sec:term:flops}), and the involved {\em data volume} and {\em movement}
(\cref{sec:term:datamovement}).  Furthermore, useful characterization of an
operation's workload is its ratio of arithmetic operations to memory
accesses---called {\em arithmetic intensity}---is a useful characterization of
its workload (\cref{sec:term:ai}).

\subsection{Floating-Point Operations}
\label{sec:term:flops}

Most scientific computations, as complex as they may be, perform their work
through a small set of elementary arithmetic operations on floating-point
representations of real numbers, such as scalar additions or
multiplications\footnote{%
    Exceptions that work on integer data or other structures include graph
    algorithms and discrete optimization.
}---These the so-called \definition[\flops: floating-point
operations\\single- and double-precision]{floating-point operations}
({\em\flops}).\footnote{%
    Not to be confused with floating-point operations {\em per second}
    (\si{\flops\per\second}).
}

Contemporary hardware offers two floating-point precisions standardized in
IEEE~754~\cite{ieee754}: {\em single-precision}, and {\em double-precision}.
They differ in the range of representable numbers, their representation
accuracy, and their implementation in hardware.  While we distinguish between
single-precision \flops and double-precision \flops, throughout this work we are
mostly concerned with double-precision computations. Hence we use ``\flops''
without a specification refers to double-precision floating-point operations,
and \R is used to denote double-precision numbers.

As commonly practiced in dense linear algebra, we assume that the multiplication
of two $n \times n$ matrices requires \SIvar{2 n^3}\flops{}---it has an
asymptotic \definition[matrix-matrix multiplication: $O(n^3)$]{complexity} of
$O(n^3)$.  While algorithms with lower asymptotic complexities (such as the {\em
Strassen algorithm} with a complexity of $O(n^{2.807})$~\cite{strassen} or the
{\em Coppersmith-Winograd algorithm} with a complexity of
$O(n^{2.376})$~\cite{coppersmith}) were already known in the 1970s, due to
considerably higher constant factors they found little to no application in
high-performance computing until recently~\cite{blisstrassen}.

The \flop-count of most dense liner algebra operations such as the matrix-matrix
multiplication is \definition[data-independence]{data-independent}, i.e., the
operand entries do not affect what arithmetic operations are
performed.\footnote{%
    Exceptions may be caused by corrupted input, such as \code{NaN}s, or
    floating-point exceptions, such as division by~0 or under-/overflows.
}  In particular, this means that all multiplications with 0's are explicitly
performed no matter how sparse an operand is (i.e., how few non-zero entries
it has).  A notable exception to the data-independence are numerical
eigensolvers, whose FLOP-counts depend on the eigenspectrum of the input matrix;
however, we do not study eigensolvers in further detail in this work.

Assuming the cubic complexity of the matrix-matrix multiplication, the
data-independence allows us to compute the \definition[cost = minimal
FLOP-count]{minimal FLOP-count}---also referred to as {\em cost}---for most
operations solely based on their operands' sizes.

\begin{example}{Minimal \flop-counts}{term:flops}
    The vector inner product $\alpha \coloneqq \dm[height=0, ']x \matvecsep \dv
    y$ (\ddot) with $\dv x, \dv y \in \R^n$ costs \SIvar{2 n}\flops: one
    multiplication and one addition per vector entry.

    The solution of a triangular linear system with multiple right-hand-sides
    $\dm[width=.4]B \coloneqq \dm[lower, inv]A \dm[width=.4]B$ (\dtrsm) with
    $\dm[lower]A\lowerpostsep \in \R^{n \times n}$ and $\dm[width=.4]B \in \R^{n
    \times m}$ requires \SIvar{n^2 m}\flops.

    The Cholesky decomposition of a symmetric positive definite (SPD) matrix
    $\dm[lower]L \dm[upper, ']L \coloneqq \dm A$ (\dpotrf) with $\dm A \in \R^{n
    \times n}$ costs
    \[
        \SIvar{\frac16 n (n + 1) (2 n + 1)}\flops
        \approx \SIvar{\frac13 n^3}\flops \enspace.
    \]
\end{example}

Note that an operation's minimal \flop-count only provides a lower bound for
routines implementing it; reasons for exceeding this bound range from technical
limitations to cache-aware data movement patterns and algorithmic schemes that
perform extra \flops to use faster compute kernels.

\subsection{Data Volume and Movement}
\label{sec:term:datamovement}

The largest portion of a scientific computation's memory footprint is typically
occupied by its numerical data consisting of floating-point numbers.  A real
number in single- and double-precision requires, respectively, 4 and~\SI8\bytes,
whereas complex numbers are represented as two consecutive real numbers
and thus require twice the space.  Since throughout this work we mostly use
double-precision numbers---conventionally called ``\definition[$\SI1\double =
\SI8\bytes$]{doubles}''---we can proceed with the assumption that each number
takes up \SI8\bytes.

In dense linear algebra, the \definition[data volume in \bytes]{data volume}
(in~\bytes) involved in a computation is determined almost exclusively by the
involved matrix operands.  For instance, a square matrix of size $1000 \times
1000$ consists of $\SI{e6}\doubles = \SI{8e6}\bytes \approx
\SI{7.63}{\mebi\byte}$;\footnote{%
    We use the 1024-based binary prefixes for data volumes: $\SI{1024}\bytes =
    \SI1{\kibi\byte}$ (``kibibyte''), $\SI{1024}{\kibi\byte} = \SI1{\mebi\byte}$
    (``mebibyte''), and $\SI{1024}{\mebi\byte} = \SI1{\gibi\byte}$
    (``gibibyte'').
} vector and scalar operands in comparison take up little space:  A vector of
size~1000  requires $\SI{8000}\bytes = \SI{7.81}{\kibi\byte}$, and a scalar fits
in just \SI8\bytes.

While a computation's data volume describes how much data is involved in an
operation, it says nothing about how often it is accessed.  For this purpose we
introduce the concept of \definition{data movement} that quantifies how much
data is read from or written to memory.  A computation's data movement is
commonly higher than its data volume, because (parts of) the data are accessed
multiple times.

While the actual data movement of any dense linear algebra operation is highly
implementation dependent, we can easily derive the \definition{minimal data
movement} from the operation's mathematical formulation by summing the size of
all input and output operands, counting the operands that are both input and
output twice.

\begin{example}{Data volume and movement}{term:datamov}
    The vector inner product $\alpha \coloneqq \dm[height=0, ']x \matvecsep \dv
    y$ (\ddot) with $\dv x, \dv y \in \R^n$ involves a data volume of $\SIvar{2
    n}\doubles = \SIvar{16 n}\bytes$ (ignoring the scalar $\alpha$); since both
    \dv x and \dv y need only be read once the data movement is also \SIvar{16
    n}\bytes.

    The matrix-matrix product $\dm C \coloneqq \dm A \matmatsep \dm B + \dm C$
    (\dgemm[NN]) with $\dm A,\allowbreak \dm B,\allowbreak \dm C \in \R^{n
    \times n}$ involves a data volume of $\SIvar{3 n^2}\doubles = \SIvar{24
    n^2}\bytes$, however, since $\dm C$ is updated, the minimal data movement is
    $\SIvar{4 n^2}\doubles = \SIvar{32 n^2}\bytes$.

    The Cholesky decomposition $\dm[lower]L \dm[upper, ']L \coloneqq \dm A$
    (\dpotrf) with $\dm A \in \R^{n \times n}$ uses only the lower-triangular
    part of the symmetric matrix \dm A,\footnotemark{} and \dm A is decomposed
    in place, i.e., it is overwritten by \dm[lower]L\lowerpostsep upon
    completion.  Hence the data volume is $\SIvar{\frac12 n (n + 1)}\doubles
    \approx \SIvar{4 n^2}\bytes$, while the minimal data movement is at least
    $\SIvar{2 \cdot \frac12 n (n + 1)}\doubles \approx \SIvar{8 n^2}\bytes$.
\end{example}
\footnotetext{%
    Space for the whole matrix is allocated, but the strictly upper-triangular
    part is not accessed.
}

Note that the minimal data movement is a strict lower bound when none of the
involved data is in any of the processor's caches.  Furthermore, depending on
the operation and the cache sizes, it may not be attainable in implementations.

\subsection{Arithmetic Intensity}
\label{sec:term:ai}

Dividing an operation's minimal flop count by its minimal data movement yields
its \definition{arithmetic intensity}:
\begin{equation}
    \label{eq:term:ai}
    \text{arithmetic intensity}
    \defeqq \frac{\text{minimal \flop-count}}{\text{minimal data movement}}
    \enspace.
\end{equation}
A low arithmetic intensity means that few operations are performed per memory
access, thus making the data movement a likely bottleneck; a high arithmetic
intensity on the other hand indicates that a lot of work is performed per data
element, thus making the floating-point computations the potential bottleneck.
Arithmetic intensity divides dense linear algebra operations into two groups:
While for \blasl1 (vector-vector) and~2 (matrix-vector) operations the intensity
is quite small and independent of the problem size, it is considerably larger
for \blasl3 (matrix-matrix) and dense \lapack-level operations, for which
increases linearly with the problem size.

\begin{example}{Arithmetic intensity}{term:ai}
    The vector inner product $\alpha \coloneqq \dm[height=0, ']x \dv y$ (\ddot)
    with $\dv x, \dv y \in \R^n$ is a \blasl1 operation that performs \SIvar{2
    n}{\flops} over \SIvar{2 n}{\doubles} of data movement.  Hence its
    arithmetic intensity is
    \[
        \frac{\text{minimal \flop-count}}{\text{minimal data movement}}
        = \frac{\SIvar{2 n}\flops}{\SIvar{2 n}\doubles}
        = \SIvar{\frac18}{\flops\per\Byte} \enspace.
    \]

    The matrix-vector multiplication $\dv y \coloneqq \dm A \matvecsep \dv x +
    \dm y$ (\dgemv[N]) with $\dm A \in \R^{n \times n}$ and $\dv x, \dv y \in
    \R^n$ is a \blasl2 operation that performs \SIvar{2 n^2}{\flops} over
    \SIvar{n^2 + 3 n}{\doubles} of data movement ($\dv y$ is both read and
    written).  Therefore, its arithmetic intensity is
    \[
        \frac{\text{minimal \flop-count}}{\text{minimal data movement}}
        = \frac{\SIvar{2 n^2}\flops}{\SIvar{n^2 + 3 n}\doubles}
        \approx \SIvar{\frac14}{\flops\per\Byte} \enspace.
    \]

    The matrix-matrix multiplication $\dm C \coloneqq \dm A \matmatsep \dm B +
    \dm C$ (\dgemm[NN]) with $\dm A,\allowbreak \dm B,\allowbreak \dm C \in
    \R^{n \times n}$ is a \blasl3 that performs \SIvar{2 n^3}{\flops} over
    \SIvar{4 n^2}{\doubles} of data movement ($\dm C$ is both read and written).
    Hence, its arithmetic intensity
    \[
        \frac{\text{minimal \flop-count}}{\text{minimal data movement}}
        = \frac{\SIvar{2 n^3}\flops}{\SIvar{4 n^2}\doubles}
        = \SIvar{\frac n{16}}{\flops\per\Byte}
    \]
    grows linearly with the problem size~$n$ and already exceeds the intensity
    of \dgemv for matrices as small as $5 \times 5$.
\end{example}

We revisit the arithmetic intensity in \cref{sec:term:eff}, where it determines
whether a computation's performance is limited by the processor's memory
subsystem or its floating-point units.

    \section{Runtime}
    \label{sec:term:time}
    Since performance describes the amount of work performed per time unit, it is a
critical requirement to accurately measure a calculation's \definition{runtime},
i.e., the duration of its execution.  This can be achieved in any number of
ways, such as the \software{UNIX} command \code{time}, the \software{UNIX}
function \code{gettimeofday()}, or the \openmp routine \code{omp\_get\_wtime()}.
While all of these measure time in seconds or fractions thereof, we are
interested in a \definition{cycle-accurate timer} that counts exactly how many
processor cycles a computation took.

On {\namestyle x86} and {\namestyle x86\_64} machines, the assembly instruction
\definition{\code{rdtsc}} (\code r{}ea\code d \code t{}ime \code s{}tamp \code
c{}ounter) returns the value of the Time Stamp Counter, a 64-bit register that
is incremented once per cycle at the processor's \definition{base
frequency}.\footnote{%
    Technically, it is only guaranteed to be incremented at a constant rate,
    which we observed to be  to the processor's base frequency on all systems
    used in this work.  However, one could easily adapt to any other frequency
    through multiplication with a constant factor.
}  It provides a cycle-accurate timer with minimal overhead, and its cycle count
can be converted to seconds through multiplication with the base frequency.

While we use \code{rdtsc} as a cycle accurate timer throughout most of this
work, we need to be aware that it does not necessarily count actual core cycles.
Althrough the increment rate of the Time Stamp Counter is fixed at the
processor's base frequency, the individual cores may run at \definition{varying
frequencies}---both lower and higher to adapt to their current workload: While
during idle times, the frequency is reduced to save energy, during peak loads,
exceeding the base frequency provides a performance boost.  On \intel
processors, the \definition{\namestyle SpeedStep} technology (or {\namestyle
Enhanced Intel SpeedStep}---EIST) to dynamically scale the frequency was
introduced in~2005 (AMD's counterpart is called {\namestyle AMD PowerTune}), and
in~2008 \definition{\intel \turboboost} added the ability to scale beyond the
base frequency---often called ``dynamic overclocking''---up to a model-dependent
\definition{maximum turbo frequency}.  However, this peak frequency can
typically not be maintained indefinitely, since it increases the processor's
power consumption and temperature, which cannot exceed certain model-specific
limits (see \cref{ex:meas:turbo}).

If we are specifically interested in counting core cycles at the dynamic
frequency, the {\namestyle Performance Application Programming Interface}
(\papi)~\cite{papi, papiweb} offers a solution in the form of the hardware
performance counter \definition{\code{PAPI\_TOT\_CYC}}.  However, \papi, which
is integrated into our performance measurement tool and framework presented in
\cref{sec:meas:elaps}, not only introduces a significantly larger overhead than
\code{rdtsc}, but is also not available on all systems (e.g., \software{macOS}).

    \section{Performance and Attained Bandwidth}
    \label{sec:term:perf}
    In scientific computing the central metric that describes at what rate a
computation performs its work is \definition[floating-point performance
in~\si{\giga\flops\per\second}]{floating-point performance}---or simply {\em
performance}---measured in \si{\giga\flops\per\second} (giga-\flops per second,
sometimes abbreviated as GFLOPS).  For a dense linear algebra computation, it is
the result of dividing the operation's minimal \flop-count (cost) by the
measured runtime:
\begin{equation}
    \label{eq:term:perf}
    \text{performance} \defeqq
    \frac{\text{minimal \flop-count}}{\text{runtime}} \enspace.
\end{equation}

\begin{example}{Performance}{term:perf}
    The matrix-matrix multiplication $\dm C \coloneqq \dm A \matmatsep \dm B +
    \dm C$ (\dgemm[NN]) with $\dm A,\allowbreak \dm B,\allowbreak \dm C \in
    \R^{1000 \times 1000}$ requires $\SIvar{2 \times 1000^3}\flops =
    \SI{2e9}\flops$.  If it is computed in \SI{102}\ms, it attained a
    floating-point performance of
    \[
        \frac{\text{minimal \flop-count}}{\text{runtime}}
        = \frac{\SI{2e9}\flops}{\SI{102}\ms}
        \approx \SI{19.61}{\giga\flops\per\second} \enspace.
    \]
\end{example}

Similarly, dividing an operation's minimal data movement by its measured runtime
yields its \definition[attained bandwidth
in~\si{\gibi\byte\per\second}]{attained bandwidth} measured in
\si{\gibi\byte\per\second}:
\begin{equation}
    \label{eq:term:attainedbw}
    \text{attained bandwidth}
    \defeqq \frac{\text{minimal data movement}}{\text{runtime}} \enspace.
\end{equation}
Note that we define the attained bandwidth independent of whether the hardware's
available bandwidth is a computation's limiting factor.  Performance and
attained bandwidth are related to the hardware capabilities in
\cref{sec:term:eff}.

\begin{example}{Attained bandwidth}{term:bw}
    The vector inner product $\alpha \coloneqq \dm[height=0, ']x \dv y$ (\ddot)
    with $\dv x, \dv y \in \R^{\num{100000}}$ has a minimal data movement of $2
    \times \SI{100000}\doubles \approx \SI{1.53}{\mebi\byte}$.  If it is
    performed in \SI{.13}{\ms} while loading both \dv x and \dv y from main
    memory (i.e., they were not in any of the processor's caches; see also
    \cref{sec:meas:effects:caching}), it attained a bandwidth of
    \[
        \frac{\text{minimal data movement}}{\text{runtime}}
        = \frac{\SI{1.53}{\mebi\byte}}{\SI{.13}\ms}
        \approx \SI{11.49}{\gibi\byte\per\second} \enspace.
    \]
\end{example}

Both performance and the attained bandwidth were so far not put into the context
of the used hardware and its capabilities.  As such, they provide an idea how
fast a computation was, yet not how well it used the available resources.  To
evaluate how efficiently the hardware was used, we first need to understand the
hardware capabilities and limitations.

    \section{Hardware Constraints}
    \label{sec:term:hw}
    Dense linear algebra operations on shared-memory systems are generally
constrained by the processor's capabilities in terms of floating-point
performance and bandwidth, which are covered in this section.

A quick overview of what hardware resources perform floating-point operations
allows us to easily determine the physical limitations to floating-point
performance: Within a processor's core, floating-point operations are performed
in the form of \definition[vectorized floating-point
instructions]{floating-point instructions}.  In particular, contemporary
processors offer so-called {\em vectorized} instructions that operate on vectors
of 2~to 16~floating-point numbers simultaneously.   Both the length of these
vectors and how many vectorized instructions can be issued each cycle are
determined by a processor's floating-point hardware and instruction set.
Multiplying the total number of scaler operations per cycle with the frequency
and number cores yields the processor's \definition[peak floating-point
performance in~\si{\giga\flops\per\second}]{peak floating-point performance} in
\si{\giga\flops\per\second}:
\begin{equation}
    \label{eq:term:peakperf}
    \pperf \defeqq \frac\flops{\text{cycle and core}} \times
    \text{frequency} \times \text{\#cores} \enspace .
\end{equation}

\begin{example}{Peak floating-point performance}{maes:peakperf}
    A \sandybridge can operate on vectors of 4~doubles with its {\namestyle
    Advanced Vector Extensions} (AVX).  It is capable of one vectorized addition
    and one vectorized multiplication instruction per cycle and core, i.e., a
    total of \SI8{\flops\per\cycle\per\core}.  At the processor's base frequency
    of \SI{2.6}{\GHz} each of its cores has a peak double-precision
    floating-point performance of
    \[
        \SI8{\flops\per\cycle\per\core} \times \SI{2.6e9}{\cycles\per\second}
        = \SI{20.8}{\giga\flops\per\second\per\core} \enspace.
    \]
    Hence, the total peak performance of the processor is
    \[
        \SI{20.8}{\giga\flops\per\second\per\core} \times \SI8\cores
        = \SI{166.4}{\giga\flops\per\second} \enspace.
    \]
    At the processor's maximum turbo frequency of \SI{3.5}\GHz, the peak
    performance is about \SI{35}{\percent} higher:
    \SI{28}{\giga\flops\per\second\per\core} and
    \SI{224}{\giga\flops\per\second} in total.

    The AVX registers and instructions also allow to operate on vectors of
    8~single-precision numbers while still offering one vector addition and one
    vector multiplication each cycle.  Hence the peak single-precision
    floating-point performance is twice the peak double-precision performance,
    i.e., \SI{448}{\giga\flops\per\second} using all 8~cores and \turboboost.
\end{example}

A computation's data movement is limited by a processor's \definition[peak
bandwidth]{peak main-memory bandwidth}, i.e., how much data it can load from and
store to main memory per second.  This theoretical peak can be computed from the
I/O~bus frequency, the bus width, and the number of memory channels, but is
usually easily found in the manufacturer's specifications.  Note that this
nominal peak bandwidth always assumes that the processor is equipped with the
fastest compatible main-memory.

A system's peak bandwidth can only be attained using multiple cores; using a
single core, the bandwidth is determined by the memory latency and the maximum
number of pending (``in-flight'') cache-misses plus the rate at which the
prefetcher loads cache-lines~\cite{stbandwidth}.  Unfortunately, since
especially the prefetcher is typically not well documented, it is difficult to
determine a theoretical single-core peak bandwidth.

In practice, the peak bandwidth is commonly measured with benchmarks such as
STREAM~\cite{stream,streamweb}, {\namestyle LIKWID}~\cite{likwid,likwidweb}, or
a highly tuned \blasl1 kernel (e.g., \daxpy).  While such benchmarks do not
report the theoretical peak bandwidth, they give an excellent estimate of the
practically attainable bandwidth.

\begin{example}{Peak bandwidth}{term:peakbw}
    A \sandybridge has a documented peak bandwidth of
    \SI{51.2}{\giga\byte\per\second}.  This bandwidth is the result of a
    4~memory channels each loading 8~bytes simultaneously from a DDR3-1600
    main-memory module over a bus running at \SI{800}\MHz:
    \[
        \SI4{channels} \times \SI8{\bytes\per{channel}} \times \SI{800}\MHz
        = \SI{51.2}{\giga\byte\per\second}
        = \SI{47.68}{\gibi\byte\per\second} \enspace.
    \]
    However, the \code{load}-benchmark from the {\namestyle likwid} suite
    only reports a practical peak bandwidth for the entire processor of
    \SI{37.65}{\mebi\byte\per\second}.
    To determine the single-threaded peak bandwidth, we used the highly tuned
    \openblas kernel \daxpy ($\dv y \coloneqq \alpha \dv x + \dv y$) with
    vectors of size \num{10000000} (\SI{76.29}{\mebi\byte} per vector).  Since
    \daxpy's minimal memory movement is 3~vectors (load \dv x, update \dv y) and
    it took \SI{14.77}{\ms} in our measurements it attained a bandwidth of
    \[
        \frac{3 \times \SI{76.29}{\mebi\byte}}{\SI{14.77}\ms}
        = \SI{16.25}{\gibi\byte\per\second} \enspace.
    \]
\end{example}

    \section{Efficiency}
    \label{sec:term:eff}
    To determine a computation's theoretical attainable peak performance on a
specific processor, we compare the computation's arithmetic intensity to the
hardware's ratio of peak floating-point performance to peak bandwidth (both in
\si{\flops\per\Byte}).  If the arithmetic intensity is higher than this ratio,
the computation is limited by the peak floating-point performance and said to be
\definition{compute-bound}; if it is lower, it is limited by the bandwidth and
said to be \definition{bandwidth-bound}.  While in the compute-bound case, a
computation's efficiency is the ratio of its attained performance to the
processor's peak floating-point performance (\cref{sec:term:eff:compbound}), in
the bandwidth-bound case it is the ratio of the attained bandwidth to the
processor's peak bandwidth (\cref{sec:term:eff:bwbound}).  Finally, the Roofline
Model (\cref{sec:term:roofline}) provides a visualization combining the
arithmetic intensity and both types of efficiency.

\subsection{Compute-Bound Efficiency}
\label{sec:term:eff:compbound}

A computation is compute-bound on a hardware platform if the memory operations
to load and store the involved data can be amortized by floating-point
operations, i.e., the available memory bandwidth is sufficient for all transfers
and the speed at which the processor performs \flops  is the bottleneck.  An
operation is theoretically bandwidth bound when
\[
    \text{arithmetic intensity} \geq \frac\pperf\pbw \enspace.
\]
Furthermore, a computation's
\definition[(compute-bound)\\efficiency]{compute-bound efficiency} (or simply
{\em efficiency}) is given by
\begin{equation}\label{eq:term:eff}
    \text{compute-bound efficiency}
    \defeqq \frac{\text{attained performance}}\pperf \enspace.
\end{equation}
This unit-less metric between 0 and~1 indicates how well the available hardware
resources are utilized:  While a value close to~1 corresponds to near-optimal
utilization, lower values indicate untapped resource potential.

\begin{example}{Compute-bound efficiency}{term:eff}
    The matrix-matrix multiplication $\dm C \coloneqq \dm A \matmatsep \dm B +
    \dm C$ (\dgemm[NN]) with $\dm A,\allowbreak \dm B,\allowbreak \dm C \in
    \R^{1000 \times 1000}$ has an arithmetic intensity of (see
    \cref{ex:term:ai})
    \[
        \SIvar{1000 \times \frac1{16}}{\flops\per\Byte} 
        = \SI{62.5}{\flops\per\Byte} \enspace.
    \]
    On a single core of a \sandybridge with a peak floating-point performance of
    \SI{20.8}{\giga\flops\per\second} (\turboboost disabled) and peak bandwidth
    of \SI{51.2}{\gibi\byte\per\second} this operation is clearly compute bound:
    \[
        \frac
            {\SI{20.8}{\giga\flops\per\second}}
            {\SI{16.25}{\gibi\byte\per\second}}
        \approx \SI{1.28}{\flops\per\Byte}
        < \SI{62.5}{\flops\per\Byte} \enspace.
    \]
    If the \dgemm[NN] runs at \SI{19.61}{\giga\flops\per\second}
    (\cref{ex:term:perf}), it reached an efficiency of
    \[
        \frac{\text{attained performance}}\pperf
        = \frac
            {\SI{19.61}{\giga\flops\per\second}}
            {\SI{20.8}{\giga\flops\per\second}}
        \approx \SI{94.27}\percent \enspace.
    \]
\end{example}

There are many different ways to look at efficiency other than the ratio of
attained performance to peak performance.  Rewriting the definition of
efficiency as
\begin{align*}
    \text{efficiency}
    &= \frac{\text{attained performance}}\pperf \\
    &= \frac
        {\text{cost} / \text{runtime}}
        {\text{cost} / \text{optimal runtime}} \\
    &= \frac{\text{optimal runtime}}{\text{runtime}} \enspace,
\end{align*}
it is expressed as the ratio of the minimum time required to perform the
operation's minimal \flops on the given hardware to the computation's runtime.
If we reorganize it as
\begin{align*}
    \text{efficiency}
    &= \frac{\text{attained performance}}\pperf \\
    &= \frac{\text{cost} / \text{runtime}}\pperf \\
    &= \frac{\text{cost}}{\text{runtime} \times \pperf} \\
    &= \frac{\text{cost}}{\text{available \flops}} \enspace,
\end{align*}
it can be seen as the ratio of the operation's minimal \flop-count to how many
\flops the processor could theoretically perform during the computation's
runtime.

\begin{example}{Expressing compute-bound efficiency}{term:eff2}
    In \cref{ex:term:eff} the \dgemm[NN] took \SI{102}\ms, while the
    \sandybridge with a peak performance of \SI{20.8}{\giga\flops\per\second}
    (\turboboost disabled) could have performed the required $\SIvar{2 \times
    1000^3}\flops = \SI{2e9}\flops$ in
    \[
        \frac{\SI{2e9}\flops}{\SI{20.8}{\giga\flops\per\second}}
        \approx \SI{96.15}\ms \enspace .
    \]
    Hence, the computation's efficiency can be computed as
    \[
        \frac{\text{optimal runtime}}{\text{runtime}}
        = \frac{\SI{96.15}\ms}{\SI{102}\ms} \approx \SI{94.26}\percent \enspace.
    \]

    We can also consider that in the \SI{102}{\ms} that the \dgemm[NN] took, the
    \sandybridgeshort core could have performed
    \[
        \SI{102}\ms \times \SI{20.8}{\giga\flops\per\second}
        \approx \SI{2.12e9}\flops \enspace.
    \]
    Once again we obtain the same efficiency, as a \flop-count ratio:
    \[
        \frac{\text{cost}}{\text{available \flops}}
        = \frac{\SI{2e9}\flops}{\SI{2.12e9}\flops}
        \approx \SI{94.26}\percent
        \enspace.
    \]
\end{example}

\subsection{Bandwidth-Bound Efficiency}
\label{sec:term:eff:bwbound}

A computation is bandwidth-bound on a hardware platform if the memory operations
cannot load and store the involved data as fast as the processor's
floating-point units can process it, i.e., the memory bandwidth is the
bottleneck and the compute units are partially idle.  An operation is
theoretically bandwidth-bound when
\[
    \text{arithmetic intensity} \leq \frac\pperf\pbw \enspace.
\]
Furthermore, a computation's \definition{bandwidth-bound efficiency} is defined
as
\begin{equation}
    \label{eq:term:eff:bwbound}
    \text{bandwidth-bound efficiency} \defeqq
    \frac{\text{attained bandwidth}}\pbw \enspace.
\end{equation}
A bandwidth-bound efficiency close to~1 indicates a good utilization of the
processor's main-memory bandwidth, while smaller values signal underutilization.

\begin{example}{Bandwidth-bound efficiency}{term:bwbeff}
    The vector inner product $\alpha \coloneqq \dm[height=0, ']x \matvecsep \dv
    y$ (\ddot) with $\dv x, \dv y \in \R^{\num{100000}}$ has an arithmetic
    intensity of \SIvar{\frac18}{\flops\per\Byte} (\cref{ex:term:ai}) and is
    thus clearly bandwidth-bound.  If on one core of a \sandybridge, it attains
    a bandwidth of \SI{11.49}{\gibi\byte\per\second} (\cref{ex:term:bw}),
    relative to the processor's empirical peak bandwidth of
    \SI{16.25}{\gibi\byte\per\second} (\cref{ex:term:peakbw}), it performed at a
    bandwidth-bound efficiency of
    \[
        \frac{\text{attained bandwidth}}\pbw
        = \frac
            {\SI{11.49}{\gibi\byte\per\second}}
            {\SI{16.25}{\gibi\byte\per\second}}
        \approx \SI{70.71}\percent \enspace.
    \]
\end{example}

\subsection{The Roofline Model}
\label{sec:term:roofline}

The \definition{Roofline model}~\cite{roofline1} plots the performance of
computations (in \si{\giga\flops\per\second}) against their arithmetic intensity
(in \si{\flops\per\Byte}).  In addition to data-points from measurements, two
lines are added to such a plot to indicate the theoretically attainable
performance depending on the arithmetic intensity: The product of peak bandwidth
and arithmetic intensity (in units: $\si{\gibi\byte\per\second} \times
\si{\flops\per\Byte} = \si{\gibi\flops\per\second} \approx
\SI{.93}{\giga\flops\per\second}$) constitutes a straight line through the
origin with the bandwidth as a gradient (visually: \tikz\draw[thick, darkred]
(0, 0) -- (1.5ex, 1.5ex);) that represents the bandwidth-bound performance limit;
and the peak floating-point performance is a constant line (\tikz\draw[thick,
darkred] (0,0) (0, 1.5ex) -- (3ex, 1.5ex);).  Together these two lines form the
roofline-shaped performance limit (\tikz\draw[thick, darkred] (0, 0) -- (1.5ex,
1.5ex) -- (4.5ex, 1.5ex);) that gives the visualization its name:
\begin{equation}\label{eq:term:roofline}
    \text{performance limit} =
    \min\left(\begin{array}c
            \pbw \times \text{intensity},\\
            \pperf
    \end{array}\right) \enspace.
\end{equation}
Comparing the attained performance of a computation to this limit yields the
computation's efficiency---bandwidth-bound below the left part of the ``roof''
and compute-bound below the right part.

\begin{figure}[t]
    \figurestyle

    \ref*{leg:term:roofline}

    \tikzsetnextfilename{roofline}
    \begin{tikzpicture}
        \begin{axis}[
                xlabel={arithmetic intensity}, x unit=\si{\flops\per\Byte},
                ylabel={performance}, y unit=\si{\giga\flops\per\second},
                ymax=25,
                legend columns=2,
                legend to name=leg:term:roofline,
                table/search path={appterm/figures/data/roofline},
            ]
            \addplot[orange, mark=*] table[x=ci, y=perf] {ddot.dat};
            \label{plt:term:roofline:ddot}
            \addlegendentry{\ddot}
            \addlegendimage{empty legend}
            \addlegendentry{$n = 1000, \num{10000}, \num{100000}$}
            \addplot[green, mark=*] table[x=ci, y=perf] {dgemv.dat};
            \label{plt:term:roofline:dgemv}
            \addlegendentry{\dgemv}
            \addlegendimage{empty legend}
            \addlegendentry{$n = 100, 500, 2000$}
            \addplot[blue, mark=*] table[x=ci, y=perf] {dgemm.dat};
            \label{plt:term:roofline:dgemm}
            \addlegendentry{\dgemm}
            \addlegendimage{empty legend}
            \addlegendentry{$n = 5, 10, \ldots, 100$}

            \fill[red, fill opacity=.1]
                (0, 0) -- (1.28, 20.8) coordinate (cross)
                -- (10, 20.8) -- (10, 35) -| (0, 0);
            \addplot[red, thick, update limits=false] coordinates
                {(0, 0) (1.28, 20.8) (10, 20.8)};
            \label{plt:term:roofline:peak}
            \path (0, 0) -- (cross)
                node[black, midway, sloped, above] {bandwidth bound}
                -- (10, 20.8 -| current axis.east)
                node[black, midway, above] {compute bound};
        \end{axis}
    \end{tikzpicture}

    \mycaption{%
        \ddot, \dgemv, and \dgemm in the Roofline Model.
        \captiondetails{\sandybridge, 1~thread, \openblas}
    }
    \label{fig:term:roofline}
\end{figure}

\begin{example}{The roofline model}{term:roofline}
    \Cref{fig:term:roofline} presents the Roofline model for one core of a
    \sandybridge.  This processor has a  single-core peak performance of
    \SI{20.8}{\giga\flops\per\cycle} (\turboboost disabled), and we use the
    measured single-core peak bandwidth of \SI{16.25}{\gibi\byte\per\second}
    (\cref{ex:term:peakbw}).  Together these two factors impose the performance
    limit~(\ref*{plt:term:roofline:peak})
    \[
        \min(\SI{16.25}{\gibi\byte\per\second} \times \text{arithmetic
        intensity}, \SI{20.8}{\giga\flops\per\second})
    \]

    \cref{fig:term:roofline} also contains the measured performance of
    representative \blasl1, 2, and~3 operations, whose arithmetic intensity was
    determined in \cref{ex:term:ai}.
    \begin{itemize}
        \item The vector inner product $\alpha \coloneqq \dm[height=0, ']x
            \matvecsep \dv y$ (\ddot) with $\dv x, \dv y \in \R^n$
            (\ref*{plt:term:roofline:ddot}) has a arithmetic intensity of
            \SIvar{\frac18}{\flops\per\Byte}, making it clearly bandwidth-bound
            below the left part of the ``roofline''.  The attained
            (bandwidth-bound) efficiency, which is given by the ratio of the
            measured performance~(\ref*{plt:term:roofline:ddot}) to the
            attainable peak performance~(\ref*{plt:term:roofline:peak}), is
            quite high at~\SI{87.93}\percent.

        \item The matrix-vector multiplication $\dv y \coloneqq \dm A \matvecsep
            \dv x + \dm y$ (\dgemv) with $\dm A \in \R^{n \times n}$ and $\dv x,
            \dv y \in \R^n$ (\ref*{plt:term:roofline:dgemv}) has a computation
            intensity of $\approx \SIvar{\frac14}{\flops\per\Byte}$, making it
            also bandwidth-bound.  The (bandwidth-bound) efficiency
            (\ref*{plt:term:roofline:dgemv} divided by
            \ref*{plt:term:roofline:peak}) is between~\SI{45.32}{\percent} (for
            $n = 100$) and \SI{76.66}{\percent} (for~$n = 2000$).

        \item The matrix-matrix multiplication $\dm C \coloneqq \dm A \matmatsep
            \dm B + \dm C$ (\dgemm[NN]) with $\dm A, \dm B, \dm C \in \R^{n
            \times n}$ (\ref*{plt:term:roofline:dgemm}) has a higher arithmetic
            intensity of \SIvar{\frac n{16}}{\flops\per\Byte}, which makes it
            theoretically compute-bound on our system for~$n \geq 21$.  In the
            memory-bound domain it reaches its peak (memory-bound) efficiency
            (\ref*{plt:term:roofline:dgemm} divided
            by~\ref*{plt:term:roofline:peak}) of \SI{50.15}{\percent} at~$n =
            20$.  Within the compute-bound domain, its (compute-bound)
            efficiency grows towards \SI{74.32}{\percent} for our largest
            problem size~$n = 100$.  Beyond this size the efficiency keeps
            growing and converge to its peak of \SI{93.70}{\percent} for
            matrices of size~$n = 2000$.
    \end{itemize}
\end{example}

    \section{Other Metrics}
    \label{sec:term:other}
    In addition to the fundamental metrics of workload, time, performance, and
efficiency, a range of other metrics provides further insights into the
execution and behavior of computations.  For instance, many hardware
events---such as various types of cache misses, interrupts, and branch
prediction failures---can be counted via a processor's performance counters,
which are easily accessed through tools such as \papi~\cite{papi, papiweb} and
\intel {\swstyle VTune}~\cite{vtuneweb}.  While our performance measurement
framework introduced in \cref{sec:meas:elaps} also provides access to these
counters, they only play a minor role throughout this work; the central metric
for our modeling and prediction efforts of \blas-based algorithms are runtime
and its derivatives.

Another noteworthy class of performance metrics that play an increasingly
important role quantify a computation's energy consumption and efficiency.  The
commonly used metric of \si{\flops\per\second\per\watt} for instance is used to
rank the {\namestyle TOP500}~\cite{top500} supercomputers according to their
energy efficiency in the {\namestyle Green500} list~\cite{green500}.

}

    \chapter[Dense Linear Algebra Routines and Libraries]
        {Dense Linear Algebra\newline Routines and Libraries}
\label{app:libs}
{
    \tikzsetexternalprefix{externalized/applibs-}
    \def\gobbletwo#1#2{}
\newlength\argwidth
\settowidth\argwidth{{\small\codestyle ldWork}}
\pgfkeys{
    /routine/.style={
        /routine/.cd,
        name/.initial,
        description/.initial,
        note/.initial,
        flops/.initial,
        datavol/.initial,
        datamov/.initial,
        opstex/.initial=,
        arglist/.initial=\gobbletwo,
        argstex/.initial=,
    },
    /routine,
    .unknown/.code={
        \let\name\pgfkeyscurrentname
        \pgfkeysalso{name/.expand once=\name}
    },
    argsparse/.style args={#1=#2}{
        arglist/.append={,\ #1},
        argstex/.append={%
            \settowidth\hangindent{\hspace\argwidth: }%
            \makebox[\argwidth]{\hfill\codestyle#1}: #2\par
        },
    },
    arguments/.style={argsparse/.list={#1}},
    opsparse/.style={opstex/.append={#1\par}},
    operations/.style={opsparse/.list={#1}},
}

\newlength\sechangindent
\newcommand\routinedoc[1]{{
    \pgfkeys{
        /routine,
        #1,
        name/.get=\routine,
        arglist/.get=\arglist,
        description/.get=\description,
        note/.get=\note,
        argstex/.get=\arguments,
        opstex/.get=\operations,
        flops/.get=\flops,
        datavol/.get=\datavol,
        datamov/.get=\datamov,
    }
    \renewcommand\columnseprulecolor{\color{blue}}
    \setlength\columnseprule{.5pt}
    \setlength\multicolsep{0pt}
    \setlength\columnsep{1em}
    \setlength\parindent{0pt}
    \setlength\premulticols{5\baselineskip}
    \settowidth\sechangindent{\large\codestyle\bf\routine(}
    \RedeclareSectionCommand[indent=\sechangindent]{subsection}
    \RedeclareSectionCommand[%
        beforeskip=-6.0pt plus -2.0pt minus -2.0pt,%
        afterskip=1sp%
    ]{paragraph}
    \begin{multicols}2[
            \subsection*{\codestyle\bf\llap{\routine(}\arglist)}
            \label{routine:\routine}
            {\it\description}
        ]
        \def\empty{}\small\singlespacing
        \expandafter\ifx\note\pgfkeysnovalue\else
            \paragraph{Note\strut}
            \note
        \fi

        \ifx\operations\empty\else
            \paragraph{Operations\strut}
            \operations
        \fi

        {
            \raggedright
            \hbadness=10000
            \hangafter=1
            \renewcommand\newline{
                \par
                \settowidth\hangindent{\hspace\argwidth: }
                \makebox[\hangindent]{}%
            }
            \paragraph{Arguments\strut}
            \arguments
        }

        \expandafter\ifx\flops\pgfkeysnovalue\else
            \paragraph{Minimal FLOP-count\strut}
            \flops
        \fi

        \expandafter\ifx\datavol\pgfkeysnovalue\else
            \paragraph{Data volume\strut}
            \datavol
        \fi

        \expandafter\ifx\datamov\pgfkeysnovalue\else
            \paragraph{Minimal data movement\strut}
            \datamov
        \fi
    \end{multicols}

    \filbreak
}}

\newcommand\routinedocforward[2]{
    \pgfkeys{
        /routine,
        #2,
        name/.get=\routine,
        arglist/.get=\arglist,
        description/.get=\description,
    }
    \subsection*{\codestyle\bf\routine(\arglist)}
    \label{routine:\routine}
    {\it\description.}
    See #1.

    \filbreak
}

    This appendix gives an overview of the core dense linear algebra
libraries used throughout this work: \blas and \lapack.
\begin{itemize}
    \item The {\namestyle Basic Linear Algebra Subprograms}
        (\blas)~\cite{blasl1, blasl2, blasl3} provide kernels for various vector
        and matrix multiplications, as well as triangular linear system solvers
        (back substitution).

    \item On top of \blas, the {\namestyle Linear Algebra PACKage}
        (\lapack)~\cite{lapack, lapackweb} offers more advanced operations, such
        as matrix decompositions, inversions, linear-system and least-squares
        solvers, and eigensolvers.
\end{itemize}
While \blas and \lapack are sometimes referred to as ``libraries'', they should
be seen as standardized interface specifications with fully functional, yet
unoptimized reference implementations.

In the following, \cref{app:libs:store} introduces the operand storage format
expected by both \blas and \lapack, \cref{app:libs:blas,app:libs:lapack} give an
overview of these interfaces and their routines used in this work, and
\cref{app:libs:libs} discusses significant \blas and \lapack implementations.

    \section{Storage Format}
    \label{app:libs:store}
    This section describes how operands are stored and passed as arguments to \blas
and \lapack routines.  Note that due to the interfaces' roots in \fortran, all
arguments are passed by reference.

\subsection{Scalars}
Each scalar operand (e.g., $\alpha \in \R$) is passed as a single argument,
 (e.g., \code{double *alpha}).  Complex scalars are stored as two consecutive
 elements of the basis data-type (\code{float} or \code{double}) that represent
 the real and imaginary parts.

\subsection{Vectors}
Each vector operand (e.g., $\dv x \in \R^n$) is specified by three arguments:
\begin{itemize}
    \item A size argument (e.g., \code{int *n}) determines the length of the
        vector.  One size argument can describe multiple vectors (and/or
        matrices) with the same size.

    \item A data argument (e.g., \code{double *x}) points to the vector's first
        element in memory.

    \item An increment argument (e.g., \code{int *incx}) identifies the stride
        between consecutive elements of the vector.  For instance, a
        contiguously stored vector has an increment of~1.

        Note that most routines allow negative increments.  In this case, the
        vector is stored in reverse, and the data argument points to the
        vector's last element---the first memory location.
\end{itemize}
To summarize, vector element~$x_i$ is stored at \code{x[i * incx]} if
\code{incx} is positive and \code{x[(i - n + 1) * incx]} otherwise.

\subsection{Matrices}
Each matrix (e.g., $\dm[width=.7]A \in \R^{m \times n}$) is specified by four
arguments:
\begin{itemize}
    \item Two size arguments (e.g., \code{int *m} and \code{int *n}) determine
        the matrix height~($m$) and width~($n$).  One size argument can describe
        the dimensions of multiple matrices (and/or vectors), or both dimensions
        of a square matrix.

    \item A data argument (e.g., \code{double *A}) points to the first matrix
        element in memory (e.g., $a_{00}$).  The following elements of the
        first column (e.g., $a_{i0}$) are stored consecutively in memory as
        vector with increment~1.

    \item A leading dimension argument (e.g., \code{int *ldA}) describes the
        distance in memory between matrix columns.  It can hence be understood
        and used as the increment argument for the matrix rows as vectors.  The
        term ``leading dimension'' comes from the concept that a referenced
        matrix is part of a larger, contiguously stored ``leading'' matrix. It
        allows to operate on sub-matrices or tensor panels as shown throughout
        this work.

        Leading dimensions must be at least equal to the height of the matrix
        (e.g., $m$).
\end{itemize}
To summarize, matrix element~$a_{ij}$ is stored at \code{A[i + j * ldA]}.

    \section{\namestyle Basic Linear Algebra Subprograms}
    \label{app:libs:blas}
    The {\namestyle Basic Linear Algebra Subprograms} (\blas) cover fundamental
dense vector and matrix operations, such as various types of multiplications and
triangular linear system solvers.  \blas is structured in three levels:
\begin{itemize}
    \item \blasl1~\cite{blasl1} provides vector operations, such as copying,
        scaling, additions, norms, and inner products.

    \item \blasl2~\cite{blasl2} provides matrix-vector operations, such as outer
        products, matrix-vector multiplications and solvers for triangular
        linear systems.

    \item \blasl3~\cite{blasl3} provides matrix-matrix operations, such as
        various multiplications, and solvers for triangular linear system with
        multiple right-hand-sides.
\end{itemize}
The following details the \blas routines used throughout this work; a complete
reference can be found online~\cite{blasweb}.

Note that for \blasl2 and~3 kernels the minimal \flop-counts assume that all
scalars are $\alpha = \beta = 1$.

\subsection{\blasl1}

\routinedoc{dcopy,
    arguments={
        n=dimension $n$,
        x=vector $\dv x \in \R^n$,
        incx=increment for \dv x,
        y=vector $\dv y \in \R^n$,
        incy=increment for \dv y
    },
    description={double-precision vector copy},
    operations={$\dv y \coloneqq \alpha \dv x$},
    flops=0,
    datavol=$2 n$,
    datamov=$2 n$,
}

\routinedoc{dswap,
    arguments={
        n=dimension $n$,
        x=vector $\dv x \in \R^n$,
        incx=increment for \dv x,
        y=vector $\dv y \in \R^n$,
        incy=increment for \dv y
    },
    description={double-precision vector swap},
    operations={${\dv x, \dv y \coloneqq \dv y, \dv x}$},
    flops=0,
    datavol=$2 n$,
    datamov=$4 n$,
}

\routinedoc{daxpy,
    arguments={
        n=dimension $n$,
        alpha=scalar $\alpha$,
        x=vector $\dv x \in \R^n$,
        incx=increment for \dv x,
        y=vector $\dv y \in \R^n$,
        incy=increment for \dv y
    },
    description={double-precision scaled vector addition},
    operations={$\dv y \coloneqq \alpha \dv x + \dv y$},
    flops=$2 n$,
    datavol=$2 n$,
    datamov=$3 n$,
}

\routinedoc{ddot,
    arguments={
        n=dimension $n$,
        x=vector $\dv x \in \R^n$,
        incx=increment for \dv x,
        y=vector $\dv y \in \R^n$,
        incy=increment for \dv y
    },
    description={double-precision inner vector product},
    operations={${\alpha \coloneqq \dm[height=0, ']x \dv x}$},
    flops=$2 n$,
    datavol=$2 n$,
    datamov=$2 n$,
}

\subsection{\blasl2}

\routinedoc{dgemv,
    arguments={
        trans=\dm A is transposed,
        m=dimension $m$,
        n=dimension $n$,
        alpha=scalar $\alpha$,
        A=matrix $\dm A \in \R^{m \times n}$,
        ldA=leading dimension for \dm A,
        x={vector $\dv x \in \begin{cases}
            \R^n &\text{if } \code{trans} = \code N\\
            \R^m &\text{else}
        \end{cases}$},
        incx=increment for \dv x,
        beta=scalar $\beta$,
        y={vector $\dv y \in \begin{cases}
            \R^m &\text{if } \code{trans} = \code N\\
            \R^n &\text{else}
        \end{cases}$},
        incy=increment for \dv y
    },
    description={double-precision matrix-vector product},
    operations={
        {$\dv y \coloneqq \alpha \dm A \matmatsep \dv x + \beta\dv y$},
        {$\dv y \coloneqq \alpha \dm[']A \dv x + \beta\dv y$}
    },
    flops=$2 m n$,
    datavol={$\begin{array}{ll}
        m n + m &\text{if } \code{trans} = \code N\\
        m n + n &\text{else}
    \end{array}$},
    datamov={$\begin{array}{ll}
        m n + 2 m &\text{if } \code{trans} = \code N\\
        m n + 2 n &\text{else}
    \end{array}$},
}

\routinedoc{dger,
    arguments={
        m=dimension $m$,
        n=dimension $n$,
        alpha=scalar $\alpha$,
        x=vector $\dv x \in \R^m$,
        incx=increment for \dv x,
        y=vector $\dv y \in \R^n$,
        incy=increment for \dv y,
        A=matrix $\dm A \in \R^{m \times n}$,
        ldA=leading dimension for \dm A
    },
    description={double-precision vector outer product},
    operations={${\dm A \coloneqq \alpha \dv x \dm[height=0, ']y + \dm A}$},
    flops=$2 m n$,
    datavol=$m n + m + n$,
    datamov=$2 m n + m + n$,
}

\routinedoc{dtrsv,
    arguments={
        uplo=\dm[lower]A is lower- or upper-triangular,
        trans=\dm[lower]A is transposed,
        diag=\dm[lower]A is unit triangular,
        n=dimension $n$,
        A=matrix $\dm[lower]A \in \R^{n \times n}$,
        ldA=leading dimension for \dm[lower]A,
        x=vector $\dv x \in \R^n$,
        incX=increment for \dv x
    },
    description={double-precision triangular linear system solve},
    operations={
        {$\dv x \coloneqq \dm[lower, inv]A \dv x$},
        {$\dv x \coloneqq \dm[lower, inv']A \dv x$}
    },
    flops=$n^2$,
    datavol={$\frac12 n (n + 1) + n$},
    datamov={$\frac12 n (n + 1) + 2 n$}
}

\subsection{\blasl3}

\routinedoc{dgemm,
    arguments={
        transA=\dm A is transposed,
        transB=\dm B is transposed,
        m=dimension $m$,
        n=dimension $n$,
        k=dimension $k$,
        alpha=scalar $\alpha$,
        A={matrix $\dm A \in \begin{cases}
            \R^{m \times k} &\text{if } \code{transA} = \code N\\
            \R^{k \times m} &\text{else}
        \end{cases}$},
        ldA=leading dimension for \dm A,
        B={matrix $\dm B \in \begin{cases}
            \R^{k \times n} &\text{if } \code{transB} = \code N\\
            \R^{n \times k} &\text{else}
        \end{cases}$},
        ldB=leading dimension for \dm B,
        beta=scalar $\beta$,
        C={matrix $\dm C \in \R^{m \times n}$},
        ldC=leading dimension for \dm C
    },
    description={double-precision matrix-matrix product},
    operations={
        {$\dm C \coloneqq \alpha \dm A \matmatsep \dm B + \beta \dm C$},
        {$\dm C \coloneqq \alpha \dm A \matmatsep \dm[']B + \beta \dm C$},
        {$\dm C \coloneqq \alpha \dm[']A \matmatsep \dm B + \beta \dm C$},
        {$\dm C \coloneqq \alpha \dm[']A \matmatsep \dm[']B + \beta \dm C$}
    },
    flops=$2 m n k$,
    datavol=$m k + k n + m n$,
    datamov=$m k + k n + 2 m n$,
}

\routinedoc{dsymm,
    arguments={
        side=\dm A is on the left or right of \dm B,
        uplo=\dm A is in lower- or upper-triangular storage,
        m=dimension $m$,
        n=dimension $n$,
        alpha=scalar $\alpha$,
        A={matrix $\dm A \in \begin{cases}
            \R^{m \times m} &\text{if } \code{side} = \code L\\
            \R^{n \times n} &\text{else}
        \end{cases}$},
        ldA=leading dimension for \dm A,
        B={matrix $\dm B \in \R^{m \times n}$},
        ldB=leading dimension for \dm B,
        beta=scalar $\beta$,
        C={matrix $\dm C \in \R^{m \times n}$},
        ldC=leading dimension for \dm C
    },
    description={double-precision symmetric matrix-matrix product},
    operations={
        {$\dm C \coloneqq \alpha \dm A \matmatsep \dm B + \beta \dm C$},
        {$\dm C \coloneqq \alpha \dm B \matmatsep \dm A + \beta \dm C$}
    },
    flops={$\begin{array}{ll}
        2 m^2 n &\text{if } \code{side} = \code L\\
        2 m n^2 &\text{else}
    \end{array}$},
    datavol={$\begin{array}{ll}
        \frac12 m (m + 1) + 2 m n &\text{if } \code{side} = \code L\\
        \frac12 n (n + 1) + 2 m n &\text{else}
    \end{array}$},
    datamov={$\begin{array}{ll}
        \frac12 m (m + 1) + 3 m n &\text{if } \code{side} = \code L\\
        \frac12 n (n + 1) + 3 m n &\text{else}
    \end{array}$},
}

\routinedoc{dtrmm,
    arguments={
        side=\dm[lower]A is on the left or right of \dm B,
        uplo=\dm[lower]A is lower- or upper-triangular,
        transA=\dm[lower]A is transposed,
        diag=\dm[lower]A is unit triangular,
        m=dimension $m$,
        n=dimension $n$,
        alpha=scalar $\alpha$,
        A={matrix $\dm[lower]A \in \begin{cases}
            \R^{m \times m} &\text{if } \code{side} = \code L\\
            \R^{n \times n} &\text{else}
        \end{cases}$},
        ldA=leading dimension for \dm[lower]A,
        B={matrix $\dm B \in \R^{m \times n}$},
        ldB=leading dimension for \dm B
    },
    description={double-precision triangular matrix-matrix product},
    operations={
        {$\dm B \coloneqq \alpha \dm[lower]A \matmatsep \dm B$},
        {$\dm B \coloneqq \alpha \dm[lower, ']A \dm B$},
        {$\dm B \coloneqq \alpha \dm[upper]A \matmatsep \dm B$},
        {$\dm B \coloneqq \alpha \dm[upper, ']A \dm B$},
        {$\dm B \coloneqq \alpha \dm B \matmatsep \dm[lower]A$},
        {$\dm B \coloneqq \alpha \dm B \matmatsep \dm[lower, ']A$},
        {$\dm B \coloneqq \alpha \dm B \matmatsep \dm[upper]A$},
        {$\dm B \coloneqq \alpha \dm B \matmatsep \dm[upper, ']A$}
    },
    flops={$\begin{array}{ll}
        m^2 n &\text{if } \code{side} = \code L\\
        m n^2 &\text{else}
    \end{array}$},
    datavol={$\begin{array}{ll}
        \frac12 m (m + 1) + m n &\text{if } \code{side} = \code L\\
        \frac12 n (n + 1) + m n &\text{else}
    \end{array}$},
    datamov={$\begin{array}{ll}
        \frac12 m (m + 1) + 2 m n &\text{if } \code{side} = \code L\\
        \frac12 n (n + 1) + 2 m n &\text{else}
    \end{array}$},
}

\routinedocforward\dsyrk{ssyrk,
    arguments={uplo=, trans=, n=, k=, alpha=, A=, ldA=, beta=, C=, ldB=},
    description={single-precision symmetric rank-k update},
}

\routinedoc{dsyrk,
    arguments={
        uplo=\dm C has lower- or upper-triangular storage,
        trans=\dm A is transposed,
        n=dimension $n$,
        k=dimension $k$,
        alpha=scalar $\alpha$,
        A={matrix $\dm A \in \begin{cases}
            \R^{n \times k} &\text{if } \code{trans} = \code N\\
            \R^{k \times n} &\text{else}
        \end{cases}$},
        ldA=leading dimension for \dm A,
        beta=scalar $\beta$,
        C={symmetric matrix $\dm C \in \R^{n \times n}$},
        ldB=leading dimension for \dm C
    },
    description={double-precision symmetric rank-k update},
    operations={
        {$\dm C \coloneqq \alpha \dm A \matmatsep \dm[']A + \dm C$},
        {$\dm C \coloneqq \alpha \dm[']A \dm A + \dm C$},
    },
    flops={$n (n + 1) k$},
    datavol={$\frac12 n (n + 1) + n k$},
    datamov={$n (n + 1) + n k$},
}

\routinedocforward\dsyrk{cherk,
    arguments={uplo=, trans=, n=, k=, alpha=, A=, ldA=, beta=, C=, ldB=},
    description={single-precision complex Hermitian rank-k update},
}

\routinedocforward\dsyrk{zherk,
    arguments={uplo=, trans=, n=, k=, alpha=, A=, ldA=, beta=, C=, ldB=},
    description={double-precision complex Hermitian rank-k update},
}

\routinedoc{dsyr2k,
    arguments={
        uplo=\dm C has lower- or upper-triangular storage,
        trans=\dm A is transposed,
        n=dimension $n$,
        k=dimension $k$,
        alpha=scalar $\alpha$,
        A={matrix $\dm A \in \begin{cases}
            \R^{n \times k} &\text{if } \code{trans} = \code N\\
            \R^{k \times n} &\text{else}
        \end{cases}$},
        ldA=leading dimension for \dm A,
        B={matrix $\dm B \in \begin{cases}
            \R^{n \times k} &\text{if } \code{trans} = \code N\\
            \R^{k \times n} &\text{else}
        \end{cases}$},
        ldB=leading dimension for \dm B,
        beta=scalar $\beta$,
        C={symmetric matrix $\dm C \in \R^{n \times n}$},
        ldC=leading dimension for \dm C
    },
    description={double-precision symmetric rank-2k update},
    operations={
        {$\dm C \coloneqq \alpha \dm A \matmatsep \dm[']B + \alpha \dm B \matmatsep \dm[']A + \dm C$},
        {$\dm C \coloneqq \alpha \dm[']A \dm B + \alpha \dm[']B \dm A + \dm C$}
    },
    flops={$2 n (n + 1) k$},
    datavol={$\frac12 n (n + 1) + 2 n k$},
    datamov={$n (n + 1) + 2 n k$},
}

\routinedocforward\dtrsm{strsm,
    arguments={side=, uplo=, transA=, diag=, m=, n=, alpha=, A=, ldA=, B=, ldB=},
    description={single-precision triangular linear system solve with multiple
    right hand sides},
}

\routinedoc{dtrsm,
    arguments={
        side=\dm[lower]A is on the left or right of \dm B,
        uplo=\dm[lower]A is lower- or upper-triangular,
        transA=\dm[lower]A is transposed,
        diag=\dm[lower]A is unit triangular,
        m=dimension $m$,
        n=dimension $n$,
        alpha=scalar $\alpha$,
        A={matrix $\dm[lower]A \in \begin{cases}
            \R^{m \times m} &\text{if } \code{side} = \code L\\
            \R^{n \times n} &\text{else}
        \end{cases}$},
        ldA=leading dimension for \dm[lower]A,
        B={matrix $\dm B \in \R^{m \times n}$},
        ldB=leading dimension for \dm B
    },
    description={double-precision triangular linear system solve with multiple
    right hand sides},
    operations={
        {$\dm B \coloneqq \alpha \dm[lower, inv]A \dm B$},
        {$\dm B \coloneqq \alpha \dm[lower, inv']A \dm B$},
        {$\dm B \coloneqq \alpha \dm[upper, inv]A \dm B$},
        {$\dm B \coloneqq \alpha \dm[upper, inv']A \dm B$},
        {$\dm B \coloneqq \alpha \dm B \matmatsep \dm[lower, inv]A$},
        {$\dm B \coloneqq \alpha \dm B \matmatsep \dm[lower, inv']A$},
        {$\dm B \coloneqq \alpha \dm B \matmatsep \dm[upper, inv]A$},
        {$\dm B \coloneqq \alpha \dm B \matmatsep \dm[upper, inv']A$}
    },
    flops={$\begin{array}{ll}
        m^2 n &\text{if } \code{side} = \code L\\
        m n^2 &\text{else}
    \end{array}$},
    datavol={$\begin{array}{ll}
        \frac12 m (m + 1) + m n &\text{if } \code{side} = \code L\\
        \frac12 n (n + 1) + m n &\text{else}
    \end{array}$},
    datamov={$\begin{array}{ll}
        \frac12 m (m + 1) + 2 m n &\text{if } \code{side} = \code L\\
        \frac12 n (n + 1) + 2 m n &\text{else}
    \end{array}$},
}

\routinedocforward\dtrsm{ctrsm,
    arguments={side=, uplo=, transA=, diag=, m=, n=, alpha=, A=, ldA=, B=, ldB=},
    description={single-precision complex triangular linear system solve with
    multiple right hand sides},
}

\routinedocforward\dtrsm{ztrsm,
    arguments={side=, uplo=, transA=, diag=, m=, n=, alpha=, A=, ldA=, B=, ldB=},
    description={double-precision complex triangular linear system solve with
    multiple right hand sides},
}

    \section{\namestyle Linear Algebra PACKage}
    \label{app:libs:lapack}
    The {\namestyle Linear Algebra PACKage} (\lapack) is a collection of advanced
dense matrix operations, such as various factorizations and equation solvers.  A
large portion of \lapack casts the majority of its computations in terms of
\blas routines; it thereby extends the high performance of \blas implementations
to its operations.

The following gives an overview of the \lapack routines employed and studied
this work; a complete reference can be found online~\cite{lapackweb}.

\routinedoc{ilaenv,
    arguments={
        ispec={queried parameter\newline(e.g., \code1 for block size)},
        name=name of calling routine,
        opts=calling routine's concatenated flag arguments,
        n1=problem size~1,
        n2=problem size~2,
        n3=problem size~3,
        n4=problem size~4\newline(according to calling routine)
    },
    description={query algorithmic parameters (\lapack internal)},
    note={Provides various algorithm parameters (e.g., block sizes).  Should be
    modified for each architecture and \blas implementation to optimize
    performance.},
}

\routinedoc{dlauum,
    arguments={
        uplo=\dm[lower]L is lower- or upper-triangular,
        n=dimension $n$,
        A=matrix $\dm A \in \R^{n \times n}$\newline input: \dm[lower]L,
        ldA=leading dimension of \dm A,
        info=return error and info codes
    },
    description={double-precision triangular matrix multiplication with its
    transpose},
    operations={
        {$\dm A \coloneqq \dm[upper, ']L \matmatsep \dm[lower]L$},
        {$\dm A \coloneqq \dm[upper]U \matmatsep \dm[lower, ']U$}
    },
    flops=$\frac16 n (n + 1) (2 n + 1) \approx \frac{n^3}3$,
    datavol=$\frac12 n (n + 1) \approx \frac{n^2}2$,
    datamov=$n (n + 1) \approx n^2$,
}

\routinedocforward\dlauum{dlauu2,
    arguments={uplo=, n=, A=, ldA=, info=},
    description={unblocked double-precision triangular matrix multiplication
    with its transpose},
}

\routinedoc{dsygst,
    arguments={
        itype=wether to invert \dm[lower]B,
        uplo=\dm[lower]B is lower- or upper-triangular,
        n=dimension $n$,
        A=matrix $\dm A \in \R^{n \times n}$,
        ldA=leading dimension of \dm A,
        B=matrix $\dm[lower]BA \in \R^{n \times n}$,
        ldB=leading dimension of \dm[lower]B,
        info=return error and info codes
    },
    description={double-precision symmetric linear system solve},
    operations={
        {$\dm A \coloneqq \dm[lower, inv]B \dm A \matmatsep \dm[upper, inv'] B$},
        {$\dm A \coloneqq \dm[lower, inv']B \dm A \matmatsep \dm[upper, inv] B$},
        {$\dm A \coloneqq \dm[upper, ']B \dm A \matmatsep \dm[lower]B$},
        {$\dm A \coloneqq \dm[upper]B \matmatsep \dm A \matmatsep \dm[lower, ']B$}
    },
    flops=$n * (n + 1)^2 \approx n^3$,
    datavol=$n (n + 1) \approx n^2$,
    datamov=$\frac32 n (n + 1) \approx \frac32 n^2$,
}

\routinedocforward\dsygst{dsygs2,
    arguments={itype=, uplo=, n=, A=, ldA=, B=, ldB=, info=},
    description={unblocked double-precision symmetric linear system solve},
}

\routinedoc{dtrtri,
    arguments={
        uplo=\dm[lower]A is lower- or upper-triangular,
        diag=\dm[lower]A is unit diagonal,
        n=dimension $n$,
        A=matrix $\dm[lower]A \in \R^{n \times n}$\newline output:
        {$\dm[lower, inv]A$},
        ldA=leading dimension of \dm[lower]A,
        info=return error and info codes
    },
    description={double-precision triangular matrix inversion},
    operations={
        {$\dm[lower]A \coloneqq \dm[lower, inv]A$},
        {$\dm[upper]A \coloneqq \dm[upper, inv]A$}
    },
    flops=$\frac16 n (n + 1) (2 n + 1) \approx \frac{n^3}3$,
    datavol=$\frac12 n (n + 1) \approx \frac{n^2}2$,
    datamov=$n (n + 1) \approx n^2$,
}

\routinedocforward{\dtrtri}{dtrti2,
    arguments={uplo=, diag=, n=, A=, ldA=, info=},
    description={unblocked double-precision triangular matrix inversion},
}

\routinedoc{dpotrf,
    arguments={
        uplo=\dm A is in lower- or upper-triangular storage,
        n=dimension $n$,
        A=SPD matrix $\dm A \in \R^{n \times n}$\newline output: \dm[lower]L or
        \dm[upper]U,
        ldA=leading dimension of \dm A,
        info=return error and info codes
    },
    description={double-precision Cholesky decomposition},
    operations={
        {$\dm[lower]L \dm[upper, ']L \coloneqq \dm A$},
        {$\dm[lower, ']U \dm[upper]U \coloneqq \dm A$}
    },
    flops=$\frac16 n (n + 1) (2 n + 1) \approx \frac{n^3}3$,
    datavol=$\frac12 n (n + 1) \approx \frac{n^2}2$,
    datamov=$n (n + 1) \approx n^2$,
}

\routinedocforward{\dpotrf}{spotf2,
    arguments={uplo=, n=, A=, ldA=, info=},
    description={unblocked single-precision Cholesky decomposition},
}

\routinedocforward{\dpotrf}{dpotf2,
    arguments={uplo=, n=, A=, ldA=, info=},
    description={unblocked double-precision Cholesky decomposition},
}

\routinedocforward{\dpotrf}{cpotf2,
    arguments={uplo=, n=, A=, ldA=, info=},
    description={unblcoked single-precision complex Cholesky decomposition},
}

\routinedocforward{\dpotrf}{zpotf2,
    arguments={uplo=, n=, A=, ldA=, info=},
    description={unblocked double-precision complex Cholesky decomposition},
}

{
    \newcommand\dmA{\dm[width=.6]A\xspace}
    \newcommand\dmL{\dm[lower, width=.6]L\xspace}
    \newcommand\dmU{\dm[upper, size=.6, bbox height=1]U\xspace}
    \newcommand\dmP{\dm[dashed]P\xspace}

    \routinedoc{dgetrf,
        note={The matrix \dmP is represented as a list of single-row swaps; see
        \dlaswp.},
        arguments={
            m=dimension $m$,
            n=dimension $n$,
            A={matrix $\dmA \in \R^{m \times n}$\newline output: \dmL and
            \dmU},
            ldA=leading dimension of \dmA,
            ipiv=permuation matrix \dmP,
            info=return error and info codes
        },
        description={double-precision LU decomposition with partial pivoting},
        operations={$\dmP \matmatsep \dmL \matmatsep \dmU \coloneqq \dmA$},
        flops={$\frac23 m n \min(m, n)$},
        datavol=$m n$,
        datamov=$2 m n$,
    }

    \routinedocforward\dgetrf{dgetf2,
        arguments={m=, n=, A=, ldA=, ipiv=, info=},
        description={unblocked double-precision LU decomposition with partial
        pivoting},
    }

    \routinedoc{dlaswp,
        note={The matrix \dmP is represented as a list of indices \code{ipiv}:
        For each $\code i = k_1, \ldots, k_2$, row~\code i is swapped with row
        \code{ipiv[i]}.},
        arguments={
            n=dimension $n$,
            A=matrix $\dmA \in \R^{m \times n}$,
            ldA=leading dimension of \dmA,
            k1=index $k_1$,
            k2=index $k_2$,
            ipiv=permutation matrix \dmP,
            incx=vector increment for \dmP
        },
        description={double-precision multiplication with permutation matrix
        from the left},
        operations={{$\dmA \coloneqq \dmP \matmatsep \dmA$}}
    }
}

{
    \newcommand\dmA{\dm[width=.6666]A\xspace}
    \newcommand\dmQ{\dm[width=.6666]Q\xspace}
    \newcommand\dmR{\dm[upper, size=.6666, bbox height=1]R\xspace}
    \newcommand\dmRc{\dm[upper, size=.6666]R\xspace}
    \newcommand\dvtau{\dv[height=.6666]\tau\xspace}

    \routinedoc{dgeqrf,
        note={The matrix \dmQ is represented as a series of elementary
        reflectors in \dmA and scalar factors in \dvtau.},
        arguments={
            m=dimension $m$,
            n=dimension $n$,
            A={matrix $\dmA \in \R^{m \times n}$\newline output: part of
            \dmQ and \dmRc},
            ldA=leading dimension of \dmA,
            tau=vector $\dvtau \in \R^{\min(m, n)}$,
            Work=auxiliary buffer $W \in \R^l$,
            lWork=buffer size $l \geq n$,
            info=return error and info codes
        },
        description={double-precision QR~decomposition},
        operations={$\dmQ \matmatsep \dmR \coloneqq \dmA$},
        flops={$\approx \begin{cases}
            2 m^2 (n - \frac13 m) &\text{if } $m < n$\\
            2 n^2 (m - \frac13 n) &\text{else}\\
        \end{cases}$},
        datavol={$m n + \min(m, n)$},
        datamov={$2 m n + \min(m, n)$},
    }
}
\routinedocforward{\dgeqrf}{dgeqr2,
    arguments={m=, n=, A=, ldA=, tau=, Work=, info=},
    description={unblocked double-precision QR~decomposition},
}

\routinedoc{dlarft,
    arguments={
        direct=order of elementary reflectors,
        storev=elementary reflectors are stored as rows or columns,
        n=dimension $n$,
        k=dimension $k$,
        alpha=scalar $\alpha$,
        V={matrix $\dm V \in \begin{cases}
            \R^{n \times k} &\text{if } \code{storev} = \code C\\
            \R^{k \times n} &\text{else}
        \end{cases}$},
        ldV=leading dimension for \dm V,
        tau=vector $\dv \tau \in \R^k$,
        T=triangular factor $\dm[upper]T \in \R^{k \times k}$,
        ldT=leading dimension for \dm[upper]T
    },
    description={double-precision construction of the triangular factor for a
    block reflector},
}

\routinedoc{dlarfb,
    arguments={
        side=\dm H is applied form the left or right,
        trans=\dm H is transposed,
        direct=order of elementary reflectors,
        storev=elementary reflectors stored as rows or columns,
        n=dimension $m$,
        n=dimension $n$,
        k=dimension $k$,
        alpha=scalar $\alpha$,
        V={matrix $\dm V \in \begin{cases}
            \R^{m \times k} &\text{if }  \code{storev} = \code C\\[-5pt]
                                   &\text{and } \code{side}   = \code L\\
            \R^{n \times k} &\text{if }  \code{storev} = \code C\\[-5pt]
                                   &\text{and } \code{side}   = \code R\\
            \R^{k \times m} &\text{if }  \code{storev} = \code R\\[-5pt]
                                   &\text{and } \code{side}   = \code L\\
            \R^{k \times n} &\text{if }  \code{storev} = \code R\\[-5pt]
                                   &\text{and } \code{side}   = \code R
        \end{cases}$},
        ldV=leading dimension for \dm V,
        T=triangular factor $\dm[upper]T \in \R^{k \times k}$,
        ldT=leading dimension for \dm[upper]T,
        C=triangular factor $\dm T \in \R^{m \times n}$,
        ldC=leading dimension for \dm C,
        Work={auxiliary buffer $\dm W \in \begin{cases}
            \R^{n \times k} &\text{if }  \code{side} = \code L\\
            \R^{m \times k} &\text{else}
        \end{cases}$},
        ldWork=leading dimension for $\dm W$
    },
    description={double-precision block reflector application to a matrix},
}

{
    \newcommand\dmA{\dm[upper]A\xspace}
    \newcommand\dmAT{\mathrlap{\dm[upper, ']A}\hphantom\dmA\xspace}
    \newcommand\dmBc{\dm[upper, size=.6666]B\xspace}
    \newcommand\dmB{\dm[upper, size=.6666, bbox height=1]B\xspace}
    \newcommand\dmBT{\dm[upper, size=.6666, bbox height=1, ']B\xspace}
    \newcommand\dmC{\dm[width=.6666]C\xspace}
    \newcommand\dmX{\dm[width=.6666]X\xspace}

    \routinedoc{dtrsyl,
        arguments={
            tranA=\dmA is transposed,
            tranB=\dmBc is transposed,
            isgn=sign in the equation,
            m=dimension $m$,
            n=dimension $n$,
            A=matrix $\dmA \in \R^{m \times m}$,
            ldA=leading dimension of \dmA,
            B=matrix $\dmBc \in \R^{n \times n}$,
            ldB=leading dimension of \dmBc,
            C=matrix $\dmC \in \R^{m \times n}$\newline output: \dmX,
            ldC=leading dimension of \dmC,
            scale=output scalar $\gamma$,
            info=return error and info codes
        },
        description={double-precision triangular Sylvester equation solver},
        operations={
            solve for \dmX:,
            {$\dmA  \matmatsep \dmX + \dmX \matmatsep \dmB  = \gamma \dmC$},
            {$\dmA  \matmatsep \dmX + \dmX \matmatsep \dmBT = \gamma \dmC$},
            {$\dmAT \matmatsep \dmX + \dmX \matmatsep \dmB  = \gamma \dmC$},
            {$\dmAT \matmatsep \dmX + \dmX \matmatsep \dmBT = \gamma \dmC$},
            {$\dmA  \matmatsep \dmX - \dmX \matmatsep \dmB  = \gamma \dmC$},
            {$\dmA  \matmatsep \dmX - \dmX \matmatsep \dmBT = \gamma \dmC$},
            {$\dmAT \matmatsep \dmX - \dmX \matmatsep \dmB  = \gamma \dmC$},
            {$\dmAT \matmatsep \dmX - \dmX \matmatsep \dmBT = \gamma \dmC$}
        },
        flops=$m n (m + n + 4)$,
        datavol=$\frac12 m (m + 1) + \frac12 n (n + 1) + m n$,
        datamov=$\frac12 m (m + 1) + \frac12 n (n + 1) + 2 m n$
    }
}

    \section{Implementations}
    \label{app:libs:libs}
    All dense linear algebra algorithms and libraries---including \lapack{}---only
reach high-performance through optimized \blas implementations.  While some
highly tuned \blas implementations are open-source (e.g., \openblas and \blis),
others are provided by hardware vendors (e.g., \mkl and \accelerate).  This
section gives an overview of the implementations used throughout this work.

\subsection*{Reference Implementations}

The \blas and \lapack reference implementations~\cite{blasweb, lapackweb} are
fully functional and well-documented and thus of great value as references for
routine interfaces and semantics.  However, on their own they only attain poor
performance, and should therefore not be used in production codes.

All routines in the \blas reference implementation are single-threaded
and unoptimized.  The central kernel \dgemm, for instance, is realized as a
simple triple loop that reaches around \SI6{\percent} of modern processors'
single-threaded theoretical peak performance---optimized implementations are
commonly $15\times$~faster on a single core and provide excellent multi-threaded
scalability.

Since \lapack primarily relies on a tuned \blas implementation for speed,
the reference implementation can in principle reach good performance.  However,
as its documentation states, this requires careful tuning of its block sizes,
whose default values are generally too low on contemporary processors.
Optimized implementations may further improve \lapack's performance through
faster algorithms, tuned unblocked kernels (e.g, \dtrti2, \dpotf2), and
algorithm-level parallelism (e.g., task-based algorithms-by-blocks).

Throughout this work, we use reference \blas and \lapack version~3.5.0.

\subsection*{\namestyle OpenBLAS}

{\namestyle OpenBLAS}~\cite{openblasweb} is a high-performance open-source \blas
and \lapack implementation that is currently developed and maintained at the
{\namestyle Massachusetts Institute of Technology}.  It provides optimized and
multi-threaded \blas kernels for a wide range of architectures, and offers tuned
version of core \lapack routines, such as the \dlauum, \dtrtri, \dpotrf, and
\dgetrf.  {\namestyle OpenBLAS} is based on the discontinued {\namestyle
GotoBLAS2}, adopting its approach and much of its source-code; it includes
assembly kernels for more recent architectures, such as \sandybridgeshort and
\haswellshort, as well {\namestyle AMD} processors.

Throughout this work, we use {\namestyle OpenBLAS} version~0.2.15.

\subsection*{\namestyle BLIS}

The {\namestyle BLAS-like Library Instantiation Software} ({\namestyle
BLIS})~\cite{blis1, blis2, blis3, blisweb} is a fairly recent framework for
dense linear algebra libraries that is actively developed at the {\namestyle
University of Texas at Austin}.  While it comes with its own API, which is a
superset, generalization, and extension of the \blas, it contains a
compatibility layer offering the original de-factor standard \blas interface.
{\namestyle BLIS} builds upon the {\namestyle GotoBLAS} approach, yet
restructures and solidifies it to make all but a tiny ``micro-kernel''
architecture-independent.  While its performance is so far generally lower than
that of \openblas (see examples in \cref{sec:model:args}), its ambitious goal is
to significantly speed up both the development of new application-specific
kernels, and the adaptation to other architectures.

Although multi-threading was introduced into {\namestyle BLIS}~\cite{blis3} soon
after its inception, its flexible threading model lacked a simple end-user
interface (such as following the environment variable \code{OMP\_NUM\_THREADS})
until November~2016 (commit
\href{https://github.com/flame/blis/commit/6b5a4032d2e3ed29a272c7f738b7e3ed6657e556}{\sf
6b5a403}).  As a result, we only presents single-threaded results for
{\namestyle BLIS}.

Throughout this work we use {\namestyle BLIS} version~0.2.0.

\subsection*{\namestyle MKL}

\intel's {\namestyle Math Kernel Library} ({\namestyle MKL})~\cite{mklweb} is a
high-performance library for \intel processors that covers \blas and
\lapack, as well as other high-performance computations, such as for Fast
Fourier Transforms (FFT) and Deep Neural Networks (DNN).  While {\namestyle MKL}
is a closed-source library, it recently began offering free developer licenses.
In terms of performance, it is in most scenarios superior to open-source
libraries such as \openblas and \blis (see examples in \cref{sec:model:args}).

Throughout this work we use {\namestyle MKL} version~11.3.

\subsection*{\namestyle Accelerate}

\apple's framework {\namestyle Accelerate}~\cite{accelerateweb} is a
high-performance library that ships with {\namestyle macOS} and, among others,
provides full \blas and \lapack functionality.  Its performance is for many
cases comparable to \openblas or slightly better.

\subsection*{Other Implementations}

The following notable \blas and \lapack implementations are not used throughout
this work:
\begin{itemize}
    \item The {\namestyle Automatically Tuned Linear Algebra Software}
        ({\namestyle ATLAS})~\cite{atlas1, atlas2, atlas3, atlasweb} is a
        high-performance \blas implementation that relies on auto-tuning.  While
        {\namestyle ATLAS} kernels typically don not reach the performance of
        hand-tuned implementations such as \openblas, \blis, and \mkl, it
        provides good performance for new and exotic architectures with little
        effort.

    \item {\namestyle GotoBLAS2}~\cite{gotoblas1, gotoblas2, gotoblasweb} is a
        high-performance \blas implementation that was developed at the
        {\namestyle Texas Advanced Computing Center}.  Since its
        discontinuation, much of its code-base was picked up by its successor
        \openblas in~2011, and its approach was refined and generalized in
        \blis.

    \item {\namestyle IBM}'s {\namestyle Engineering and Scientific Subroutine
        Library} ({\namestyle ESSL}) \cite{esslweb} provides a high-performance
        \blas implementation and parts of \lapack for {\namestyle POWER}-based
        systems, such as {\namestyle Blue Gene} supercomputers.
%
\end{itemize}

}

    \newenvironment{hwtable}{%
    \vspace{-\bigskipamount}
    \bgroup\noindent\small\singlespacing%
    \begin{longtable}{p{.3\textwidth}p{.6\textwidth}}
        \toprule
}{%
        \bottomrule
    \end{longtable}
    \egroup%
}

\chapter{Hardware}
\label{app:hardware}

This appendix gives an overview of the processors used throughout this work and
their relevant properties.

Note that, while the single-threaded peak performance is, where appropriate,
based on the processors' maximum turbo frequency, the multi-threaded peak
performance is instead computed from the base frequency.  Furthermore, we only
list the vector instructions that allow to reach a processor's theoretical peak
performance.

\section{\hwstyle Harpertown E5450}
\label{hardware:E5450}

\href{http://ark.intel.com/products/33083/Intel-Xeon-Processor-E5450-12M-Cache-3_00-GHz-1333-MHz-FSB}{\nolinkurl{http://ark.intel.com/products/33083/Intel-Xeon-Processor-E5450-}\\\nolinkurl{12M-Cache-3_00-GHz-1333-MHz-FSB}}

Our {\namestyle Harpertown E5450}s were part of our compute cluster. Because
they were disposed of in mid~2016, they are only used in a part of this work's
performance analyses.

\begin{hwtable}
    Name                &\namestyle Intel\textsuperscript\textregistered{} Xeon\textsuperscript\textregistered{} Processor E5450 \\
    Codename            &\namestyle Harpertown \\
    Lithography         &\SI{45}{\nano\meter} \\
    Release             &Q4 2007 \\
    Cores / Threads     &4 / 4 \\
    Base Frequency      &\SI{3.00}{\giga\hertz} \\
    Peak Performance    &\SI{12}{\giga\flops\per\second} (single-threaded) \\\nopagebreak
                        &\SI{48}{\giga\flops\per\second} (all cores) \\
    Peak Bandwidth      &\SI{10.6}{\giga\byte\per\second} \\
    L2~cache            &\SI6{\mebi\byte} {\em per 2~cores}, 24-way set associative\\
    L1d~cache           &\SI{32}{\kibi\byte} per core, 8-way set-associative \\
    Vector Instructions &1 SSE FMUL + 1 SSE FADD per cycle \\\nopagebreak
                        &$= \SI4{\flops\per\cycle}$ \\
\end{hwtable}

\section{\hwstyle Sandy Bridge-EP E5-2670}
\label{hardware:E5-2670}

\href{http://ark.intel.com/products/64595/Intel-Xeon-Processor-E5-2670-20M-Cache-2_60-GHz-8_00-GTs-Intel-QPI}{\nolinkurl{http://ark.intel.com/products/64595/Intel-Xeon-Processor-E5-2670-}\\\nolinkurl{20M-Cache-2_60-GHz-8_00-GTs-Intel-QPI}}

Our {\namestyle Sandy Bridge E5-2680 v2}s are part of our compute cluster.
\intel{} \turboboost is disabled on these machines unless otherwise stated.

\begin{hwtable}
    Name                &\namestyle Intel\textsuperscript\textregistered{} Xeon\textsuperscript\textregistered{} Processor E5-2670 \\
    Codename            &\namestyle Sandy Bridge-EP \\
    Lithography         &\SI{32}{\nano\meter} \\
    Release             &Q1 2012 \\
    Cores / Threads     &8 / 16 \\
    Base Frequency      &\SI{2.60}{\giga\hertz} \\
    Max Turbo Frequency &\SI{3.30}{\giga\hertz} ({\em disabled unless otherwise stated})\\
    Peak Performance    &\SI{20.8}{\giga\flops\per\second} (single-threaded) \\\nopagebreak
                        &\SI{166.4}{\giga\flops\per\second} (all cores) \\
    Peak Bandwidth      &\SI{51.2}{\giga\byte\per\second} \\
    L3~cache            &\SI{20}{\mebi\byte} shared, 20-way set associative \\
    L2~cache            &\SI{256}{\kibi\byte} per core, 8-way set associative \\
    L1d~cache           &\SI{32}{\kibi\byte} per core, 8-way set-associative \\
    Vector Instructions &1 AVX FMUL + 1 AVX FADD per cycle \\\nopagebreak
                        &$= \SI8{\flops\per\cycle}$ \\
\end{hwtable}

\section{\hwstyle Ivy Bridge-EP E5-2680 v2}
\label{hardware:E5-2680 v2}

\href{http://ark.intel.com/products/75277/Intel-Xeon-Processor-E5-2680-v2-25M-Cache-2_80-GHz}{\nolinkurl{http://ark.intel.com/products/75277/Intel-Xeon-Processor-E5-2680-}\\\nolinkurl{v2-25M-Cache-2_80-GHz}}

Our {\namestyle Ivy Bridge E5-2680 v3}s are part of our compute cluster.

\begin{hwtable}
    Name                &\namestyle Intel\textsuperscript\textregistered{} Xeon\textsuperscript\textregistered{} Processor E5-2680 v2\\
    Codename            &\namestyle Ivy Bridge-EP \\
    Lithography         &\SI{22}{\nano\meter} \\
    Release             &Q3 2013 \\
    Cores / Threads     &10 / 20 \\
    Base Frequency      &\SI{2.80}{\giga\hertz} \\
    Max Turbo Frequency &\SI{3.60}{\giga\hertz} \\
    Peak Performance    &\SI{28.8}{\giga\flops\per\second} (single-threaded) \\\nopagebreak
                        &\SI{224}{\giga\flops\per\second} (all cores) \\
    Peak Bandwidth      &\SI{59.7}{\giga\byte\per\second} \\
    L3~cache            &\SI{25}{\mebi\byte} shared, 20-way set associative \\
    L2~cache            &\SI{256}{\kibi\byte} per core, 8-way set associative \\
    L1d~cache           &\SI{32}{\kibi\byte} per core, 8-way set-associative \\
    Vector Instructions &1 AVX FMUL + 1 AVX FADD per cycle \\\nopagebreak
                        &$= \SI8{\flops\per\cycle}$ \\
\end{hwtable}

\section{\hwstyle Haswell-EP E5-2680 v3}
\label{hardware:E5-2680 v3}

\href{http://ark.intel.com/products/81908/Intel-Xeon-Processor-E5-2680-v3-30M-Cache-2_50-GHz}{\nolinkurl{http://ark.intel.com/products/81908/Intel-Xeon-Processor-E5-2680-}\\\nolinkurl{v3-30M-Cache-2_50-GHz}}

Our {\namestyle Haswell-EP E5-2680 v3}s are part of our compute cluster.

\begin{hwtable}
    Name                &\namestyle Intel\textsuperscript\textregistered{} Xeon\textsuperscript\textregistered{} Processor E5-2680 v3\\
    Codename            &\namestyle Haswell-EP \\
    Lithography         &\SI{22}{\nano\meter} \\
    Release             &Q3 2014 \\
    Cores / Threads     &12 / 24 \\
    Base Frequency      &\SI{2.50}{\giga\hertz} \\
    Max Turbo Frequency &\SI{3.30}{\giga\hertz} \\
    Peak Performance    &\SI{52.8}{\giga\flops\per\second} (single-threaded) \\\nopagebreak
                        &\SI{480}{\giga\flops\per\second} (all cores) \\
    Peak Bandwidth      &\SI{68}{\giga\byte\per\second} \\
    L3~cache            &\SI{30}{\mebi\byte} shared, 20-way set associative \\
    L2~cache            &\SI{256}{\kibi\byte} per core, 8-way set associative \\
    L1d~cache           &\SI{32}{\kibi\byte} per core, 8-way set-associative \\
    Vector Instructions &2 AVX FMA per cycle \\\nopagebreak
                        &$= \SI{16}{\flops\per\cycle}$ \\
\end{hwtable}

\section{\hwstyle Broadwell i7-5557U}
\label{hardware:i7-5557U}

\href{https://ark.intel.com/products/84993/Intel-Core-i7-5557U-Processor-4M-Cache-up-to-3_40-GHz}{\nolinkurl{https://ark.intel.com/products/84993/Intel-Core-i7-5557U-}\\\nolinkurl{Processor-4M-Cache-up-to-3_40-GHz}}

Our {\namestyle Broadwell i7-5557U} is part of a {\namestyle MacBook Pro}.

\begin{hwtable}
    Name                &\namestyle Intel\textsuperscript\textregistered{} Core\texttrademark{} i7-5557U Processor \\
    Codename            &\namestyle Broadwell-U \\
    Lithography         &\SI{14}{\nano\meter} \\
    Release             &Q1 2015 \\
    Cores / Threads     &2 / 4 \\
    Base Frequency      &\SI{3.10}{\giga\hertz} \\
    Max Turbo Frequency &\SI{3.40}{\giga\hertz} \\
    Peak Performance    &\SI{54.4}{\giga\flops\per\second} (single-threaded) \\\nopagebreak
                        &\SI{99.2}{\giga\flops\per\second} (all cores) \\
    Peak Bandwidth      &\SI{25.6}{\giga\byte\per\second} \\
    L3~cache            &\SI4{\mebi\byte} shared, 16-way set associative \\
    L2~cache            &\SI{256}{\kibi\byte} per core, 8-way set associative \\
    L1d~cache           &\SI{32}{\kibi\byte} per core, 8-way set-associative \\
    Vector Instructions &2 AVX FMA per cycle \\\nopagebreak
                        &$= \SI{16}{\flops\per\cycle}$ \\
\end{hwtable}

    \backmatter

    \tcblistof[\chapter]{loe}{List of Examples}

\listoffigures

\listoftables

\printbibheading[title={\label{bib}Bibliography}]

The bibliography is split into three parts:  Papers of which I am a (co-)author
are listed under \nameref{bib:pub} below, other scientific publications are
collected in \nameref{bib:ref} on \cpageref{bib:ref}, and websites and
repositories are found under \nameref{bib:web} on \cpageref{bib:web}.

\newrefcontext[sorting=ymdnt]
\printbibliography[
    keyword=own,
    heading=subbibliography,
    title={\label{bib:pub}Publications}
]

\newrefcontext[sorting=nyt]
\printbibliography[
    notkeyword=own,
    nottype=online,
    heading=subbibliography,
    title={\label{bib:ref}References}
]

\defbibnote{online}{%
    All websites and repositories were accessible as of May~1, 2017.
}
\printbibliography[
    prenote=online,
    notkeyword=own,
    type=online,
    heading=subbibliography,
    title={\label{bib:web}Online Resources}
]

\vfill

\section*{About This Document}

\def\gettexliveversion#1, #2 (#3)#4\relax{#2}
\newcommand\pdftexver{\expandafter\gettexliveversion\pdftexbanner\relax\xspace}

This document was written in \href{https://www.latex-project.org/}{\LaTeXe} and
typeset with \href{http://www.tug.org/applications/pdftex/}{pdfTeX} \pdftexver
on \today.

It relies on the following packages:
\href{http://ctan.org/pkg/microtype}{\code{microtype}} for micro-typography;
\href{http://ctan.org/pkg/listings}{\code{listings}} and
\href{http://ctan.org/pkg/tcolorbox}{\code{tcolorbox}} for algorithms, listings,
and examples;  \href{http://ctan.org/pkg/pgf}{\code{tikz}} and
\href{http://ctan.org/pkg/pgfplots}{\code{pgfplots}} for graphics and plots;
\href{http://ctan.org/pkg/drawmatrix}{\code{drawmatrix}} for matrix
visualizations; \href{http://ctan.org/pkg/cleveref}{\code{cleveref}} and
\href{http://ctan.org/pkg/hyperref}{\code{hyperref}} for references and
hyperlinks; and \href{http://ctan.org/pkg/biblatex}{\code{biblatex}} for the
bibliography.

\end{document}